\documentclass{elsarticle}

\usepackage{amssymb}
\usepackage{amsmath,amsfonts,amssymb,bm,graphicx}
\usepackage{dsfont}
\usepackage[FIGTOPCAP]{subfigure}
\usepackage{url}
\usepackage{multirow}
\usepackage{rotating}
\usepackage{arydshln} %\hdashline

\usepackage{color}
\usepackage[colorlinks=true, linkcolor=blue, filecolor=blue, citecolor=blue]{hyperref}

\definecolor{gray97}{gray}{.97}
\definecolor{gray75}{gray}{.75}
\definecolor{gray45}{gray}{.45}
\definecolor{gray35}{gray}{.35}

\journal{Applied Mathematics and Computation}

\begin{document}

\begin{frontmatter}

\title{SABR/LIBOR market models: pricing and calibration for some interest rate derivatives\tnoteref{t1}}
\tnotetext[t1]{Partially financed by MICINN (MTM2010-21135-C02-01) and by Xunta de Galicia (Grant CN2011/004 cofunded with FEDER funds). Third author has also been funded by a FPU Spanish grant. The authors are very grateful to Nicol\'as G\'omez Sell\'es and Mar\'ia Rodr\'iguez Nogueiras for their collaboration in the development of this work.}

\author[auth]{A.\,M. Ferreiro}\ead{aferreiro@udc.es}

\author[auth]{J.\,A. Garc\'{\i}a}\ead{jagrodriguez@udc.es}

\author[auth]{J.\,G. L\'opez-Salas}\ead{jose.lsalas@udc.es}

\author[auth]{C. V\'azquez\corref{cor1}}\ead{carlosv@udc.es}

\cortext[cor1]{Corresponding author. Tel.: +34 981167000; fax: +34 981167160.}

\address[auth]{Department of Mathematics, Faculty of Informatics, Campus Elvi\~na s/n, 15071-A Coru\~na (Spain)}

\begin{abstract}
\noindent In order to overcome the drawbacks of assuming deterministic volatility coefficients in the standard LIBOR market models to capture volatility smiles and skews in real markets, several extensions of LIBOR models to incorporate stochastic volatilities have been proposed. The efficient calibration to market data of these more complex models becomes a relevant target in practice. The main objective of the present work is to efficiently calibrate some recent SABR/LIBOR market models to real market prices of caplets and swaptions. For the calibration we propose a parallelized version of the  simulated annealing algorithm for multi-GPUs. The  numerical results clearly illustrate the advantages of using the proposed multi-GPUs tools when applied to real market data and popular SABR/LIBOR models.
\end{abstract}


\end{frontmatter}

\section{Introduction}

Since the seminal papers by Brace, Gatarek and Musiela \cite{braceLMM}, Jamshidian \cite{jamshidianLMM} and Miltersen, Sandmann and Sondermann \cite{miltersenLMM} to introduce the \textit{Libor Market Model} (LMM), several authors extended it to reproduce volatility smiles appearing in real markets.

The basic LMM has some desirable features: it is flexible, supports multiple factors and rich volatility structures and it justifies the use of Black's formula for caplet prices, which is the standard formula employed in the cap market (see \cite{brigoMercurio}). The last one constitutes its major advantage since it allows an implicit calibration of at-the-money caps volatilities. In addition, it is possible to calibrate at-the-money swaption volatilities via closed formula approximations with high accuracy (e.g. the Rebonato swaption approximation, see \cite{rebonatoLMM}). These reasons explain the success of the model and why it has been widely accepted by the financial industry.

Nevertheless, the standard LMM presents the same drawbacks as the classical Black-Scholes theory. The major disadvantage comes from the assumption of deterministic volatility coefficients that prevents matching cap and swaption volatility smiles and skews observed in the markets. This implies that after calibrating the model to at-the-money options, the model underprices the off-the-money options.

In order to overcome this drawback, there has been great research in extending the standard LMM to correctly capture market volatility smiles and skews. Different extended LMMs were suggested and can mainly fall into three categories: local volatility models, stochastic volatility models and jumps-diffusion models.

In the local volatility models the volatility is a function of the underlying asset price (forward rate) and the time. These models were introduced by Dupire \cite{dupire} and Derman and Kani \cite{dermanKani}, who proposed this extension for equity  and foreign-exchange options. Andersen and Andreasen \cite{andersenAndreasen} introduced a special case of local volatility models, the Constant Elasticity of Variance (CEV), to develop an extension of the LMM for capturing the skew. They showed how to obtain swaption smile asymptotically. Their method is still based in the Rebonato ``freezing'' argument, see \cite{rebonatoLMM}, which is not completely mathematically justified. CEV model can generate a monotone skew of implied volatilities but fails to reproduce a smile, which is often the case in reality.

Jump-diffusion models for assets were introduced by Merton \cite{merton} and Eberlein \cite{eberleinKeller}. Jamshidian \cite{jamshidianJumps}, Glasserman and Kou \cite{glassermanKou}, Glasserman and Merener \cite{glassermanMerener} and Belomestny and Schoenmakers \cite{belomestny} proposed alternative extensions of the LMM by adding jumps in the forward rate dynamics.  L\'evy LIBOR models have been studied by Eberlein and \"Ozkan \cite{eberlein}. With these models one can manipulate the slope and the curvature of a skewed smile by changing jump intensity and jump sizes. While this jump approach can generate stationary nonmonotonic volatility smiles, it involves several technical difficulties to develop numerical schemes for the resulting model. Moreover, these models result unsuitable to generate asymmetric smiles and skews, since the jump component of the forward rate dynamics typically needs to be of substantial magnitude. While such dynamics are probably reasonable for equity prices (see \cite{andersenAndreasenJumps}) they might be less natural for the term structure of interest rate forwards.

In order to correctly capture the stochastic behaviour of the volatility and to reproduce market smiles, different stochastic volatility models have been proposed. The main examples are Hull and White \cite{hullWhite} and Heston \cite{heston} models. In the Hull and White model, lognormal variance process is modelled. When the correlation between spot and variance is zero, by using a mixing approach, the authors obtained asymptotic expansions for options prices. However, the main drawback of this model comes from its inability to capture nonsymmetric smiles. Heston proposed a model where the volatility is a mean-reverting square-root process. By using Fourier transforms he derived a closed form formula for option prices. The main advantages of this model are its nice empirical properties and its analytical tractability. An application of the Heston model to the LMM appears in Wu and Zhang \cite{wuZhang}. They adopt a multiplicative stochastic factor to the volatility functions of all relevant forward rates. The stochastic factor follows a square-root diffusion process, and it can be correlated to the forward rates. They also develop a closed-form formula for swaptions in terms of Fourier transforms. Other extensions of the LIBOR market model allowing stochastic volatility are those we mention hereafter. Andersen and Brotherton-Ratcliffe \cite{andersenBrotherton} proposed a general framework for extending the LIBOR market model. Their model allows for nonparametric volatility structures and includes a multiplicative perturbation of the forward volatility surface by a one-dimensional mean reverting volatility process. This volatility process is driven by a Brownian motion independent of the Brownian motions driving the forward rates, so that, under different numeraires, the dynamics of the volatilities remains the same. Using asymptotic expansion techniques, they provided closed-form pricing formulas for caplets and swaptions prices. In \cite{piterbarg}, Piterbarg has extended this approach with a model where forward rates follow shifted-lognormal diffusion processes with stochastic volatility. The volatility is a mean reverting square-root process uncorrelated with the Brownian motions governing the dynamics of the forward rates. Using Markovian projection and parameter averaging, Piterbarg derives fast and accurate European option pricing techniques under general time-dependent parameters. In \cite{joshiRebonato}, Joshi and Rebonato proposed a shifted-lognormal LIBOR model with a volatility parameterization based on a functional form with stochastic coefficients. This model has very similar properties to the Andersen and Andreasen \cite{andersenAndreasen} one, among them its major problem being the monotonicity of implied volatility curves. All the stochastic volatility models presented so far have one single volatility factor.

In \cite{hagan}, Hagan, Kumar, Lesniewski and Woodward proposed a stochastic volatility model known as the SABR model (acronym for stochastic, alpha, beta and rho, three of the four model parameters), arguing that local volatility models could not reproduce market volatility smiles and that their predicted volatility dynamics contradicts market smiles and skews. The forward price of an asset follows, under the asset’s canonical measure, a CEV type process with stochastic volatility driven by a driftless process. The Brownian motion driving the volatility can be correlated with the one associated to the forward price. The main advantages of the model are the following. Firstly, it is able to correctly capture market volatility smiles. Secondly, its parameters, which play specific roles in the generation of smiles and skews, have an intuitive meaning. Thirdly, the authors obtained an analytical approximation for the implied volatility (known as Hagan formula) through singular perturbation techniques, thus allowing an easy calibration of the model. Finally, it has become the market standard for interpolating and extrapolating prices of plain vanilla caplets and swaptions (see \cite{rebonatoBOOK}). In \cite{obloj} Obl\'oj improved Hagan formula.

Several authors have recently tried to unify SABR and LIBOR market models. In the more standard LIBOR market model \cite{braceLMM}, the dynamics of each LIBOR forward rate under the corresponding terminal measure are assumed to be martingales with constant volatility. When adding the SABR model, the forward rates and volatility processes satisfy the following coupled dynamics
\begin{align*}
 dF_i(t) &= V_i(t) F_i(t)^{\beta_i} dW_i(t), \\
 dV_i(t) &= \sigma_i V_i(t) dZ_i(t).
\end{align*}
We note that if the interest rate derivative only depends on one particular forward rate, then it is convenient to use the corresponding terminal measure. However, when derivatives depend on several forward rates, a common measure needs to be used. Thus, in the case of pricing complex derivatives a change of measure produces the appearance of drift terms in forward rates and volatilities dynamics.

In \cite{labordereSSRN, labordereRisk}, Labord\`ere presents a unification of LIBOR and SABR models using hyperbolic geometry and heat kernel expansion to fit Taylor expansions for swaption implied volatilities. In \cite{haganSABRLIBOR}, Hagan {\it et al.}  studied the natural extension of both the LMM and the SABR model. They used the technique of low noise expansions in order to produce accurate and workable approximations to swaption volatilities. Mercurio and Morini, arguing that a number of volatility factors lower than the number of state variables is often chosen, proposed in \cite{mercurioMorini} a SABR/LIBOR market model with one single volatility factor. They designed a LIBOR market model starting from the reference SABR dynamics, with the purpose of preserving the SABR closed formula. In \cite{rebonatoBOOK,rebonatoRisk,rebonatoWhite}, Rebonato {\it et al.} designed a time-homogeneous SABR-consistent extension of the LMM. More precisely, they specified financially motivated dynamics for the LMM forward rates and volatilities that match the SABR prices very close. They also suggested a simple financially justifiable and computationally affordable way to calibrate the model. In this work we focus in these last three different SABR/LIBOR market models.

By using heuristic, empirical or very qualitative arguments, in all the here presented extensions of the LMM, the authors obtain accurate analytical approximations for caps/swaptions to calibrate the model. In general, swaptions cannot be priced in closed form in the LMM and the main challenge of these works comes from the analytical approximations to price these swaptions.

All the previous papers argue that the ``brute-force'' approach, which consists in calibrating the models using Monte Carlo simulation to price swaptions, is not a practical choice, because each Monte Carlo evaluation results computationally very expensive. However, in this article we propose the use of relatively old Simulated Annealing type algorithms \cite{Kirkpatrick-1983}, which reveal as highly efficient when implemented using High Performance Computing techniques. This combination makes possible the calibration in a reasonable computational time. Such algorithms have already been successfully applied in other related contexts, see \cite{carlosVazquezSA,carlosVazquezSA_SABR} for more details.

In this work we propose an efficient calibration strategy to some market prices for the parameters appearing in the three selected SABR/LIBOR market models. More precisely, we consider the market prices of caplets and swaptions and we pose the corresponding global optimization problems to calibrate the model parameters. Moreover, we use a simulated annealing algorithm to solve the problem. In order to speed up the algorithm we propose a parallel implementation in GPUs.

The paper is organized as follows. In Section \ref{sec:SABR_LIBOR_market_models}, the SABR/LIBOR market models proposed by Hagan, Mercurio \& Morini and Rebonato are introduced. In Section \ref{sec:model_calibrationChapter3}, the calibration procedures are explained. In Section \ref{sec:numerical_resultsChapter3}, the obtained numerical results are shown. Finally, in Section \ref{sec:conclusions} some conclusions are discussed.

\section{SABR/LIBOR market models} \label{sec:SABR_LIBOR_market_models}

\subsection{Hagan model}
This model arises as the natural coupling between SABR and LMM models \cite{haganSABRLIBOR}. Thus, for each $i=1, \ldots, M$ let $F_i$ and $V_i$ be the $i$-th forward rate that matures at time $T_i$ and its corresponding stochastic volatility, respectively. Then, under a common measure their dynamics are given by
\begin{align}
 dF_i(t) &= \mu^{F_i}(t)dt + V_i(t) F_i(t)^{\beta_i} dW_i(t), \label{eq:dinamicasHagan1} \\
 dV_i(t) &= \mu^{V_i}(t)dt + \sigma_i V_i(t) dZ_i(t), \label{eq:dinamicasHagan2}
\end{align}
with the associated correlations denoted by
$$\mathds{E}[dW_i(t) \cdot dW_j(t)] = \rho_{i,j} dt, \quad \mathds{E}[dW_i(t) \cdot dZ_j(t)] = \phi_{i,j} dt, \quad \mathds{E}[dZ_i(t) \cdot dZ_j(t)] = \theta_{i,j} dt, $$
and the initial given values $\alpha_i = V_i(0)$ and $F_i(0)$. Thus, the correlation structure is given by the block-matrix
\begin{displaymath}
\boldsymbol{P} = \left[
\begin{array}{ll}
\boldsymbol{\rho} & \boldsymbol{\phi}  \\
\boldsymbol{\phi^\top} & \boldsymbol{\theta}
\end{array}
\right],
\end{displaymath}
where the submatrix $\boldsymbol{\rho} = (\rho_{i,j})$ contains all the correlations between the forward rates $F_i$ and $F_j$, the submatrix $\boldsymbol{\phi} = (\phi_{i,j})$ includes the correlations between the forward rates $F_i$ and the instantaneous volatilities $V_j$, and the submatrix $\boldsymbol{\theta} = (\theta_{i,j})$ contains the correlations between the instantaneous volatilities $V_i$ and $V_j$.

More precisely, if we introduce the bank-account numeraire $\beta(t)$, defined by
$$\beta(t) = \displaystyle\prod_{j=0}^{i-1} \big(1 + \Delta t F_j(T_j)\big)\quad \mbox{if } t \in [T_i,T_{i+1}],$$ then, under the associated spot probability measure, the drift terms of the processes defined in \eqref{eq:dinamicasHagan1} and \eqref{eq:dinamicasHagan2} are
$$
 \mu^{F_i}(t) =  V_i(t) F_i(t)^{\beta_i} \displaystyle\sum_{j=h(t)}^i \dfrac{\tau_j \rho_{i,j} V_j(t) F_j(t)^{\beta_j}}{1+\tau_j F_j(t)},  \quad
 \mu^{V_i}(t) = \sigma_i V_i(t) \displaystyle\sum_{j=h(t)}^i \dfrac{\tau_j \phi_{i,j} V_j(t) F_j(t)^{\beta_j}}{1+\tau_j F_j(t)},
$$
where $h(t)$ denotes the index of the first unfixed $F_i$, i.e., \begin{equation} \label{eq:h_t}h(t) = j \mbox{, if } t \in [T_{j-1},T_j).\end{equation}

In terms of the moneyness\footnote{Moneyness measures the ratio between the strike price, $K$, and the current value of the underlying, $F_i(0)$. Thus, if $K=F_i(0)$ then the call or put options are said to be {\em at the money} (moneyness is equal zero). If $K < F_i(0)$ then a call option is said to be {\it in the money} (moneyness is negative) and if $K>F_i(0)$ then the call option is said to be {\em out of the money} (moneyness is positive). For put options,  {\it out of the  money} and {\it in the money} correspond to negative and positive moneyness, respectively.}, defined as $\text{ln}\Bigl(\dfrac{K}{F_i(0)}\Bigr)$, the implied volatility\footnote{The implied volatility is the one that reproduces the market price when inserted in Black-Scholes formula.} for this model is given by the Hagan second order approximation formula (also including the correction of Obl\'oj in \cite{obloj}):
\begin{align} \label{eq:sigmaHagan}
 \sigma\big(K,F_i(0)\big) & \approx  \dfrac{\alpha_i}{F_i(0)^{(1-\beta_i)}} \times \Bigg\{ 1 - \dfrac{1}{2}(1-\beta_i-\phi_{i,i}\sigma_i\omega_i) \cdot  \ln \Bigl(\dfrac{K}{F_i(0)}\Bigr)  \nonumber \\
 &  \hspace{-0.5cm} + \dfrac{1}{12} \Big( (1-\beta_i)^2 + (2-3\phi_{i,i}^2)\sigma_i^2\omega_i^2 + 3\big( (1-\beta_i) - \phi_{i,i}\sigma_i\omega_i \big) \Big) \cdot \left[ \ln \Bigl( \dfrac{K}{F_i(0)} \Bigr) \right]^2 \Bigg\},
\end{align}
where $\omega_i = \alpha_i^{-1} F_i(0)^{(1-\beta_i)}.$

For the correlations, we consider the following functional parameterizations:
\begin{align}
 \rho_{i,j} &= \eta_1 + (1-\eta_1) \exp[-\lambda_1|T_i-T_j|], \label{eq:rhoij}\\
 \theta_{i,j} &= \eta_2 + (1-\eta_2) \exp[-\lambda_2|T_i-T_j|], \label{eq:thetaij} \\
 \phi_{i,j} &= \text{sign} (\phi_{i,i}) \sqrt{|\phi_{i,i} \phi_{j,j}|} \exp \left[ -\lambda_3(T_i-T_j)^+ - \lambda_3(T_j-T_i)^+\right ], \label{eq:phiij}
\end{align}
where the terms $\phi_{i,i}$ have been previously calibrated using \eqref{eq:sigmaHagan} for the whole volatili\-ties surfaces. Moreover, parameters $\eta_i$, $\lambda_i$ and $\phi_{ij}$ are calibrated to fit the smiles of swap rates.

\subsection{Mercurio $\&$ Morini model}

For this model \cite{mercurioMorini}, the existence of a lognormal common volatility process to all forward rates is assumed, while each $F_i$ satisfies a particular SDE. More precisely, we have
\begin{align}
 dF_i(t) &= \mu^{F_i}(t) dt + \alpha_i V(t) F_i(t)^{\beta} dW_i(t), \label{eq:dinamicasMercurio1} \\
 dV(t) &= \sigma V(t) dZ(t), \label{eq:dinamicasMercurio2}
\end{align}
with
$$ \mathds{E}[dW_i(t) \cdot dW_j(t)] = \rho_{i,j} dt, \quad \mathds{E}[dW_i(t) \cdot dZ(t)] = \phi_i dt,$$
and the initial given values $V(0) = 1$ and $F_i(0)$. In this case, the correlation block-matrix is
\begin{displaymath}
\boldsymbol{P} = \left[
\begin{array}{ll}
\boldsymbol{\rho} & \boldsymbol{\phi}  \\
\boldsymbol{\phi}^\top & 1
\end{array}
\right],
\end{displaymath}
where $\boldsymbol{\phi} = (\phi_1,\ldots,\phi_M)^\top$. Under the spot probability measure, the drift terms in equation \eqref{eq:dinamicasMercurio1} are
$$
 \mu^{F_i}(t) = \alpha_i V(t) F_i(t)^{\beta} \displaystyle\sum_{j=h(t)}^i \dfrac{\tau_j \rho_{i,j} \alpha_j V(t) F_j(t)^{\beta}}{1+\tau_j F_j(t)},
$$ where $h(t)$ is given by the expression \eqref{eq:h_t}.

The calibration is similar to the previous case. By using \textit{SABR} superindexes, the parameters of the Hagan implied volatility formula \eqref{eq:sigmaHagan} are
\begin{align}
 \beta_i^{SABR} &= \beta, \quad \phi_{i,i}^{SABR} = \phi_i, \quad  \sigma_i^{SABR} = \sigma, \nonumber \\
 \alpha_i^{SABR} &= \alpha_i \left[ \text{e}^{\int_0^{T_i} M_i(s) ds} \right], \mbox{ where } M_i(t) = -\sigma \displaystyle\sum_{j=h(t)}^i \dfrac{\tau_j \phi_j \alpha_j F_j(0)^{\beta}}{1+\tau_j F_j(0)}. \label{eq:haganForMercurio}
\end{align}

Note that in this case we only need to consider \eqref{eq:rhoij} for the forward rates correlations.

\subsection{Rebonato model}

This model is analogous to Hagan one, except for the dynamics of the volatilities. More precisely, this model assumes the following dynamics \cite{rebonatoRisk}:
\begin{align} \label{eq:dinamicasRebonato}
 dF_i(t) &= \mu^{F_i}(t) dt + V_i(t) F_i(t)^{\beta_i} dW_i(t),\\
 V_i(t) &= \kappa_i(t) g_i(t),\\
 d \kappa_i(t) &= \mu^{\kappa_i}(t) dt + \kappa_i(t) h_i(t) dZ_i(t),
\end{align}
where
$$
 g_i(t) = \big(a + b(T_i-t)\big) \exp \big(-c(T_i-t)\big) + d, \quad
 h_i(t) = \big(\alpha + \beta(T_i-t)\big) \exp \big(-\gamma(T_i-t)\big) + \delta,
$$
and the correlation structure is given by the parameterizations \eqref{eq:rhoij}-\eqref{eq:phiij}.

Again, using the spot probability measure, the drift terms of the previous processes are
$$
 \mu^{F_i}(t) = V_i(t) F_i(t)^{\beta_i} \displaystyle\sum_{j=h(t)}^i \dfrac{\tau_j \rho_{i,j} V_j(t) F_j(t)^{\beta_j}}{1+\tau_j F_j(t)}, \quad
 \mu^{\kappa_i}(t) = \kappa_i(t) h_i(t) \displaystyle\sum_{j=h(t)}^i \dfrac{\tau_j \phi_{i,j} V_j(t) F_j(t)^{\beta_j}}{1+\tau_j F_j(t)}.
$$

Furthermore, in this model the parameters of the Hagan implied volatility formula \eqref{eq:sigmaHagan} are
\begin{align}
 \beta_i^{SABR} &= \beta_i, \quad \phi_{i,i}^{SABR} = \phi_{i,i}, \quad \alpha_i^{SABR} = \kappa_i(0) \left( \dfrac{1}{T_i} \int_0^{T_i} g_i(t)^2 dt \right)^{\frac{1}{2}}, \nonumber \\
 \sigma_i^{SABR} &= \dfrac{\kappa_i(0)}{\alpha_i^{SABR} T_i} \left( 2 \int_0^{T_i} g_i(t)^2 \hat{h}_i(t)^2 t dt\right)^{\frac{1}{2}}, \mbox{ where } \hat{h}_i(t) = \sqrt{\frac{1}{t} \int_0^t \left(h_i(s)\right)^2 ds}. \label{eq:haganForRebonato}
\end{align}

\section{Model calibration} \label{sec:model_calibrationChapter3}

Model parameters are calibrated in two stages, firstly to caplets\footnote{A caplet is a basic interest rate derivative which mainly consists in a call option that pays the positive difference between a floating rate and a fixed one (strike). A cap contract is a set of caplets associated with related maturity dates (tenor structure). See \cite{brigoMercurio}, for example.} and secondly to swaptions\footnote{A swap contract is an interest rate derivative that exchanges two different interest rates. A swaption is an option giving the right to enter in a swap contract at a given future time. See \cite{brigoMercurio}, for example.}. We note that model parameters can be classified into two categories (volatility and correlation parameters):
\begin{itemize}
 \item The volatility parameters for each model are given by:
   \begin{itemize}
   \item Hagan: $\pmb x = (\phi_{ii}, \sigma_i, \alpha_i)$.
   \item Mercurio \& Morini: $\pmb x = (\phi_i, \sigma, \alpha_i)$.
   \item Rebonato: $\pmb x = (\phi_{ii}, \kappa_i, \mbox{parameters of the volatility functions } g \mbox{ and } h)$.
  \end{itemize}
 \item The correlation parameters for each model are given by:
  \begin{itemize}
   \item Hagan: $\pmb y = (\eta_1, \lambda_1, \eta_2, \lambda_2, \lambda_3)$.
   \item Mercurio \& Morini: $\pmb y = (\eta_1, \lambda_1)$.
   \item Rebonato: $\pmb y = (\eta_1, \lambda_1, \eta_2, \lambda_2, \lambda_3)$.
  \end{itemize}
\end{itemize}

According to the previous classification, the cost functions to be minimized in the calibration process are the following:
\begin{itemize}
 \item Function to calibrate the market prices of caplets:
\vspace{-0.25cm}
$$f_c(\pmb x) = \displaystyle\sum_{i=1}^{M} \displaystyle\sum_{j=1}^{numK} \Big(\sigma\big(K_j, F_i(0)\big) - \sigma_{market}\big(K_j, F_i(0)\big)\Big)^2(\pmb x),$$
where $\sigma$ is given by Hagan formula (\eqref{eq:sigmaHagan}, \eqref{eq:haganForMercurio} or \eqref{eq:haganForRebonato}, depending on the model), $\sigma_{market}$ are the market volatilities and $\pmb x$ is the vector containing the volatility parameters of the model. Moreover, $M$ and $numK$ denote the number of maturities and strikes of the caplets, respectively.

 \item Function to calibrate the market prices of swaptions:
\vspace{-0.25cm}
$$f_s(\pmb y) = \displaystyle\sum_{i=1}^{numSws} \left( S_{\mbox{\footnotesize{Black}}}(swaption_i) - S_{MC}(swaption_i) \right)^2 (\pmb y),$$
where $swaption_i$ denotes the $i$-th swaption, $S_{\mbox{\footnotesize{Black}}}$ represents the Black formula for swaptions and $S_{MC}(swaption_i)$ denotes the value of the \textit{i-th} swaption computed with Monte Carlo method. Moreover, the vector $\pmb y$ contains the correlation parameters and $numSws$ is the number of swaptions.
\end{itemize}

In this work, the calibration of the parameters has been performed with a Simulated Annealing (SA) global optimization algorithm \cite{Kirkpatrick-1983}. The idea of the algorithm mimics the annealing process used in metal formation. The metal is heated so that at high temperature, and therefore with a high energy, atoms can move freely. Then the metal is cooled down and as temperature decreases the movement of the atoms is constrained, until they reach an equilibrium state when metal gets cold.

Thus, the algorithm consists in an external decreasing temperature loop. At each fixed temperature a Metropolis process, that can be seen as a Markov chain, is performed to compute the equilibrium state at this temperature level. This Markov chain consists of randomly choosing points in the search domain: if the value of the cost function at a new point decreases, the point is accepted; otherwise the point is randomly accepted following the Metropolis criterion, where the probability of accepting points with higher cost function value decreases with temperature. This process is repeated at each temperature level until temperature is low enough. The pseudocode of the algorithm can be sketched as follows, where $f$ denotes the cost function and $\pmb x_0$ the starting point; the parameters of the algorithm are the following, initial temperature $T_0$, minimum temperature $T_{min}$, decreasing factor of the temperature $0<\rho<1$ and length of the Markov chains $N$:

\framebox[0.95\linewidth]{
% {\footnotesize
\begin{minipage}{\linewidth}
\begin{tabbing}
$\pmb x= \pmb x_0$; $T=T_0$;\\ %$\leftarrow$ GetInitialSolution()
do \= \\
\> for \=$j = 1$ to $N$ do\\
\>\>$ \pmb x'=$ ComputeNeighbour($\pmb x$);\\
\>\>$\Delta E = f( \pmb x') - f( \pmb x)$;  \color{gray35} // \textit{Energy increment} \\
\>\>if \=$\big(\Delta E < 0$ or AcceptWithProbability $\exp(-\Delta E/T)\big)$\\
\>\>\> $ \pmb x =  \pmb x'$;  \color{gray35} // \textit{The trial is accepted} \\
\> end for\\
\>$T = \rho T$; \color{gray35} // with $0 < \rho < 1$\\
while ($T>T_{min}$);
\end{tabbing}
\end{minipage}
% }
}

In real applications the hybrid approaches (in which SA provides a starting point for a local minimization algorithm) are widely used. In this work we have considered the Nelder-Mead algorithm as the local minimizer.

\begin{figure}[!htb]
\begin{center}
\includegraphics[height=5.3cm]{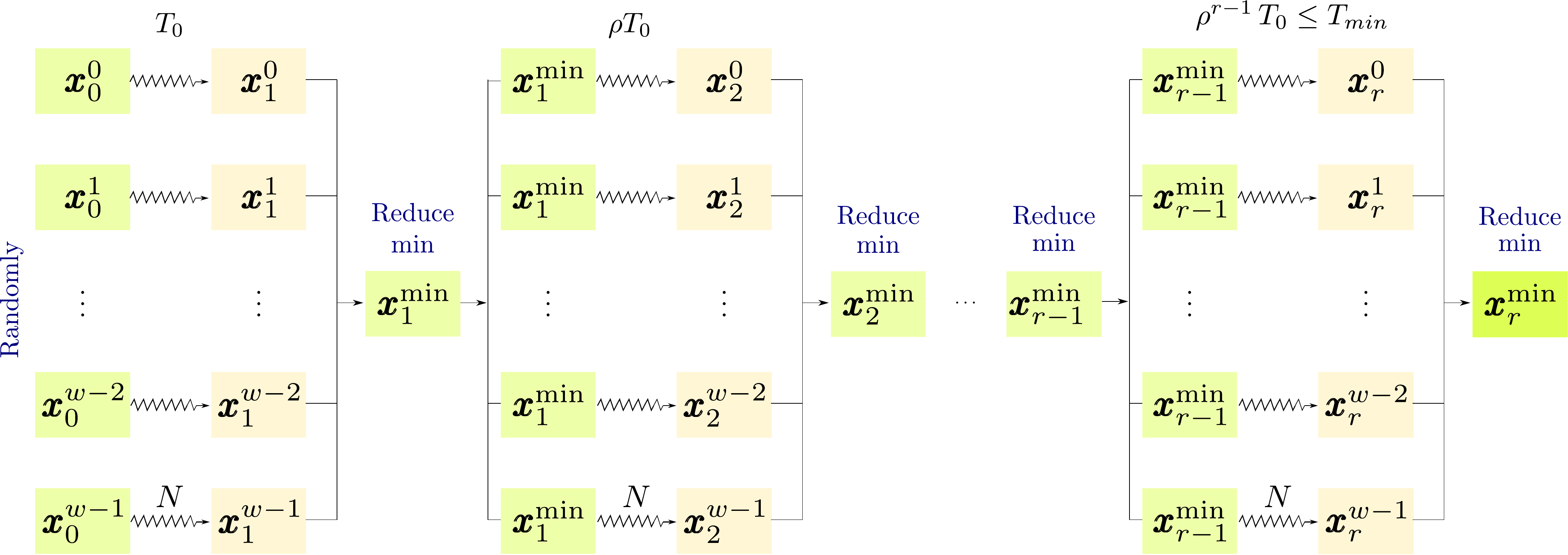}
\caption{Sketch of the parallel SA algorithm using one GPU.}
\label{fig:sincrono}
\end{center}
\end{figure}

As it is well known in the literature, SA involves a great computational cost. In \cite{carlosVazquezSA}, a parallelization of the SA algorithm has been performed for GPUs. The idea is that at each temperature level the Markov chains are distributed among the GPU threads. Among all the final reached points of the threads, the one with the minimum cost function value is selected, thus performing a reduction operation. The selected point is the starting one for all the threads in the next temperature level. The process is repeated until reaching a certain value of temperature, see Figure \ref{fig:sincrono} for more details. In \cite{carlosVazquezSA}, the algorithm was also tested against a benchmark of classical problems in optimization literature and details of the implementation can be found. The code has been integrated in the CUSIMANN library \cite{page:cusimann}.

\begin{figure}[!htb]
\begin{center}
\includegraphics[height=7.5cm]{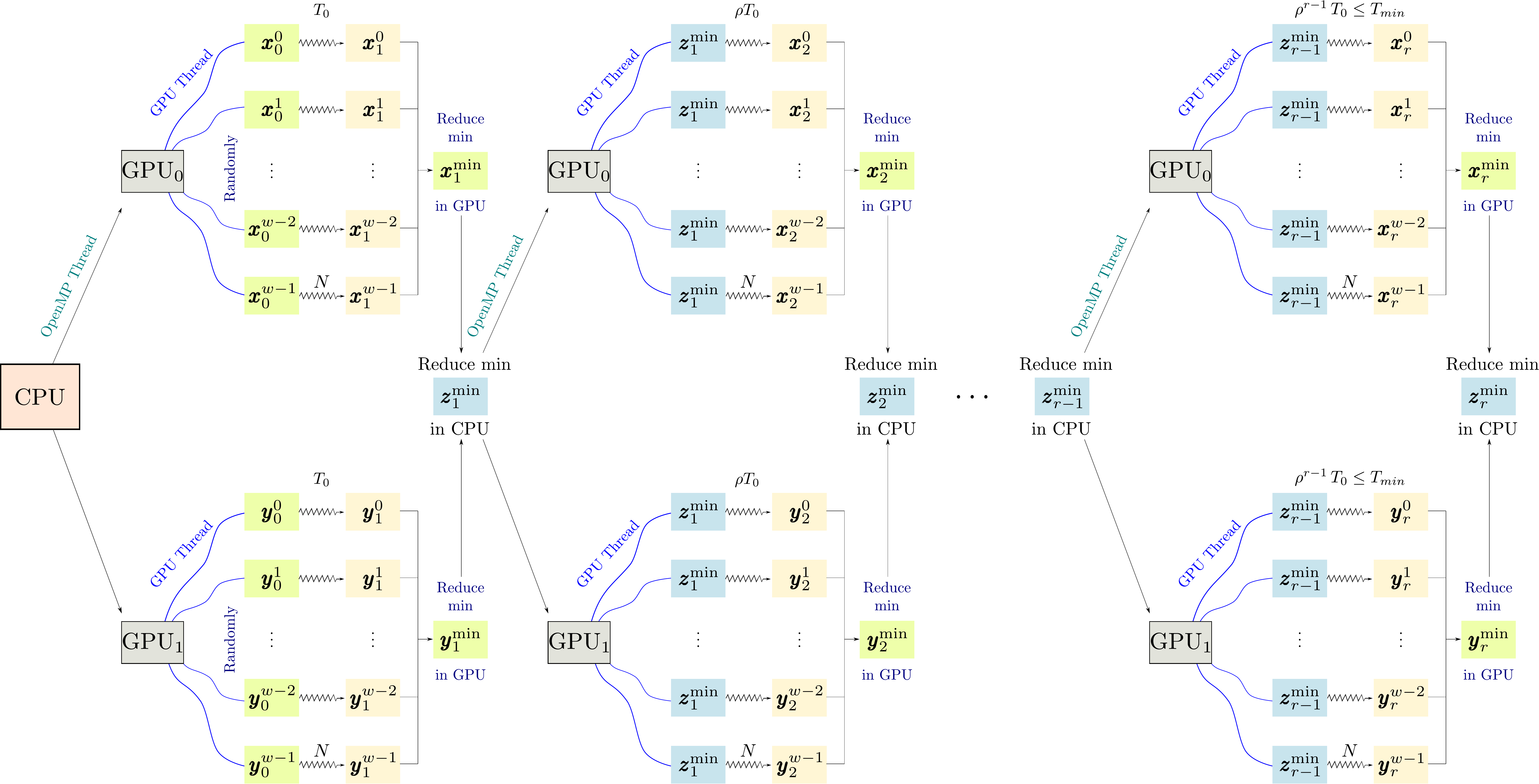}
\caption{Sketch of the parallel SA algorithm using two GPUs and OpenMP.}
\label{fig:sincronoMultiGPU}
\end{center}
\end{figure}

The previous implementation can also be improved using multi-GPUs. In this case, the Markov chains are firstly distributed among GPUs (for example, if we have two GPUs, half of the chains are computed by each GPU) and inside each GPU the chains are distributed among the threads. Before advancing to the next temperature level the best point must be computed in each GPU and then the best point of all GPUs is computed and used as starting point for all the upcoming threads of the new temperature level (see Figure \ref{fig:sincronoMultiGPU}). This multi-GPU algorithm was presented in \cite{carlosVazquezSA_SABR}, where it was used to calibrate some SABR models to a volatility surface.

In order to calibrate the models with fewer parameters (Hagan and Mercurio \& Morini), the mono-GPU version results to be enough. However, in order to calibrate models with more parameters (Rebonato), the multi-GPU version becomes more suitable, since the minimization process is much more costly.

Section \ref{sec:numerical_resultsChapter3} contains the achieved speedups when implied volatility formulas are available.

\begin{figure}[!htb]
\begin{center}
\includegraphics[height=6.7cm]{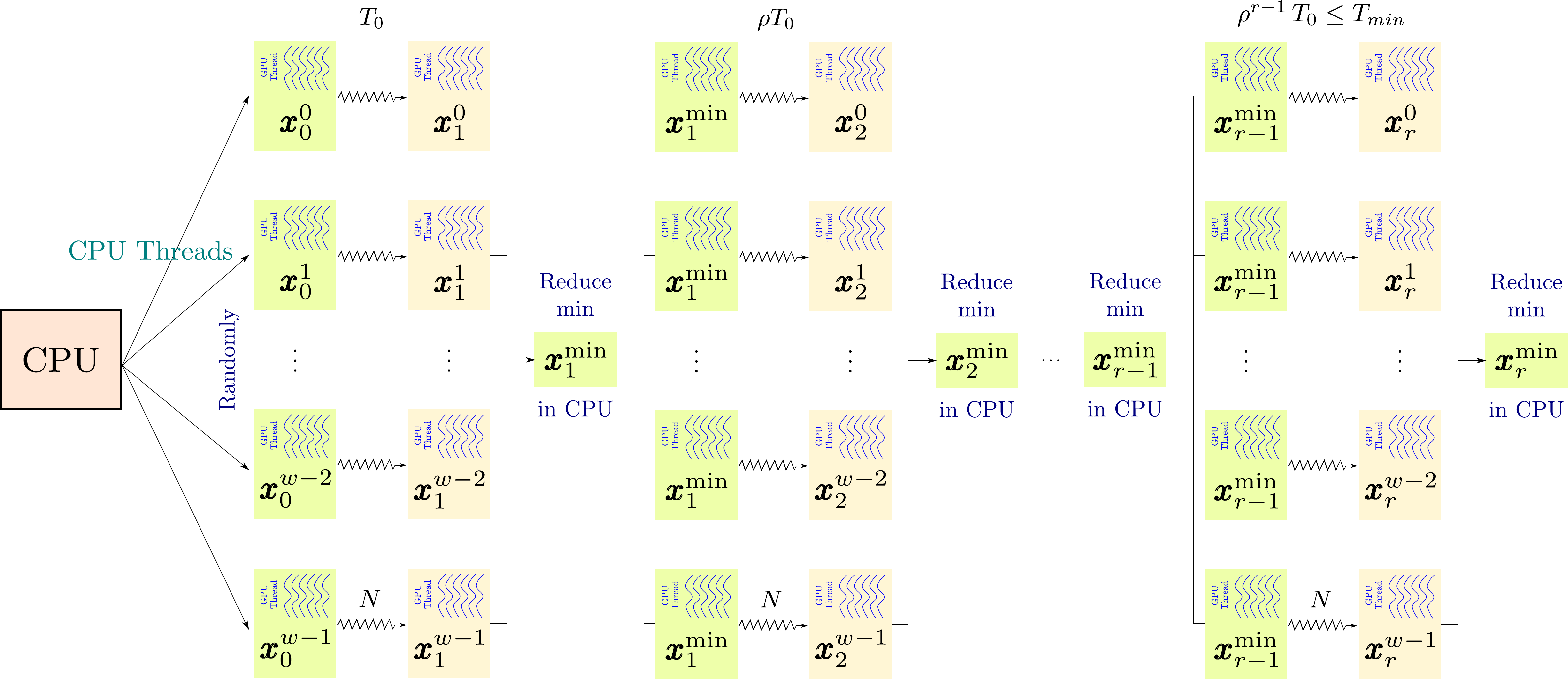}
\caption{Sketch of the parallel SA using OpenMP and considering a Monte Carlo method in the cost function.}
\label{fig:sincrono_hibrido}
\end{center}
\end{figure}

In the SABR/LIBOR market models, for the general calibration to swaption market prices an explicit formula to price swaptions is not available. Therefore, we use Monte Carlo simulation technique to price swaptions, thus leading to two nested Monte Carlo loops: one for the SA and the other one for the swaption pricer. So, as the Monte Carlo swaption pricer is carried out inside the GPU, the SA minimization algorithm is run on CPU. At this point we illustrate in Table \ref{tab:haganMCpricer} the obtained speedups in the LIBOR/SABR pricing with Monte Carlo simulation for different number of paths and values of $\Delta t$. Notice that speedups around $200$ are obtained for $10^6$ paths. In order to use all available GPUs in the system, we propose a CPU SA parallelization using OpenMP \cite{openMP}. So, each OpenMP SA thread uses a GPU to evaluate the Monte Carlo objective function (see Figure \ref{fig:sincrono_hibrido}). This approach could be easily extrapolated to a cluster of GPUs using MPI \cite{MPI}. Notice that in this case the sequential Monte Carlo pricing with CPU leads to prohibited  times for the whole calibration procedure.

\begin{table}[!htp]
\scriptsize{
	\begin{center}
		\begin{tabular}{|c|c| r r r|}
		\hline
		Number of paths & $\Delta t$ & CPU ($s$) & GPU ($s$) & \textit{Speedup} \\
		\hline
		         & $10^{-1}$ & $0.558$ & $0.094$ & $\times 5.936$ \\
		$10^{3}$ & $10^{-2}$ & $5.580$ & $0.222$ & $\times 25.135$ \\
		         & $10^{-3}$ & $55.956$ & $1.406$ & $\times 39.798$ \\
		\hline
		         & $10^{-1}$ & $5.572$ & $0.119$ & $\times 46.823$ \\
		$10^{4}$ & $10^{-2}$ & $55.740$ & $0.390$ & $\times 142.923$ \\
		         & $10^{-3}$ & $557.698$ & $3.081$ & $\times 181.012$ \\
		\hline
		         & $10^{-1}$ & $55.692 $ & $0.323$ & $\times 172.421$ \\
		$10^{5}$ & $10^{-2}$ & $558.331$ & $2.886$ & $\times 193.462$ \\
		         & $10^{-3}$ & $5601.292 $ & $28.550$ & $\times 196.192$ \\
		\hline
		         & $10^{-1}$ & $557.696$ & $2.375$ & $\times 234.819$ \\
		$10^{6}$ & $10^{-2}$ & $5588.070$ & $27.950$ & $\times 199.931$ \\
		         & $10^{-3}$ & $55904.184$ & $283.970$ & $\times 196.866$ \\
		\hline
		\end{tabular}
	\end{center}
\caption{Execution times (in seconds) and speedups in the pricing of caplets with Monte Carlo and using single precision (Hagan model).}
\label{tab:haganMCpricer}
}
\end{table}

\section{Numerical results} \label{sec:numerical_resultsChapter3}

In this section we present a test where we calibrate Hagan, Mercurio \& Morini and Rebonato models to real market data. Market data correspond to the 6 months EURIBOR rate. We show in Table \ref{tab:curvaFD} the discount factor curve, in Table \ref{tab:smilesFws} the smiles of the forward rates and in Table \ref{tab:smilesTS} the smiles of the swap rates.

Numerical experiments have been performed with the following hardware and software configurations: two GPUs Nvidia Geforce GTX470, two quad-core CPUs Xeon E5620 clocked at 2.4 Ghz with 16 GB of RAM, CentOS Linux, Nvidia CUDA SDK 4.0 and GNU C/C++compilers 4.1.2.

\begin{table}[!htb]
\scriptsize{
\begin{center}
\begin{tabular}{|c|c||c|c||c|c|}
\hline
Date & $P(0,t)$ & Date & $P(0,t)$ & Date & $P(0,t)$ \\
\hline
\hline
21/11/2011 & 1 & 19/09/2013 & 0.97713559399 & 23/11/2021 & 0.77845715189 \\
\hline
22/11/2011 & 0.99998041705 & 25/11/2013 & 0.97412238564 & 23/11/2023 & 0.72565274014 \\
\hline
23/02/2012 & 0.99622554093 & 24/11/2014 & 0.95730277130 & 23/11/2026 & 0.65317498182 \\
\hline
21/03/2012 & 0.99575871128 & 23/11/2015 & 0.93611709432 & 24/11/2031 & 0.56564376817 \\
\hline
21/06/2012 & 0.99263851754 & 23/11/2016 & 0.91144251116 & 24/11/2036 & 0.50321672724 \\
\hline
20/09/2012 & 0.98966227733 & 23/11/2017 & 0.88505982818 & 25/11/2041 & 0.45392855927 \\
\hline
19/12/2012 & 0.98673874563 & 23/11/2018 & 0.85798260233 & 23/11/2051 & 0.34982415774 \\
\hline
19/03/2013 & 0.98372608449 & 25/11/2019 & 0.83116001862 & 23/11/2061 & 0.26125094146 \\
\hline
20/06/2013 & 0.98048414547 & 23/11/2020 & 0.80486541573 & 23/11/2071 & 0.19657659346 \\
\hline
\end{tabular}
\caption{Discount factor curve.} \label{tab:curvaFD}
\end{center}
}
\end{table}

\begin{table}[!htb]
\scriptsize{
\begin{center}
\begin{tabular}{|c|ccccccccc|}
\hline
   & -80\% & -60\% & -40\% & -20\% & 0\% & 20\% & 40\% & 60\% & 80\%  \\
\hline
21-05-12 & 142.61\% & 117.05\% & 97.26\% & 82.58\% & 72.29\% & 70.89\% & 69.49\% & 68.08\% & 66.67\% \\
21-11-12 & 112.74\% & 99.23\% & 88.27\% & 79.62\% & 73.03\% & 71.95\% & 70.87\% & 69.77\% & 68.69\% \\
21-05-13 & 91.55\% & 83.75\% & 77.09\% & 71.50\% & 67.93\% & 67.10\% & 66.41\% & 65.88\% & 65.49\% \\
21-11-13 & 64.82\% & 60.95\% & 57.08\% & 53.21\% & 52.49\% & 51.34\% & 50.61\% & 50.30\% & 50.46\% \\
21-05-14 & 66.96\% & 61.84\% & 56.69\% & 52.43\% & 50.32\% & 48.72\% & 47.70\% & 47.14\% & 46.97\% \\
21-11-14 & 69.20\% & 62.75\% & 56.30\% & 51.65\% & 48.19\% & 46.19\% & 44.91\% & 44.12\% & 43.66\% \\
21-05-15 & 71.49\% & 63.67\% & 55.92\% & 50.89\% & 46.19\% & 43.83\% & 42.32\% & 41.35\% & 40.64\% \\
21-11-15 & 73.89\% & 64.61\% & 55.54\% & 50.13\% & 44.25\% & 41.56\% & 39.84\% & 38.71\% & 37.78\% \\
21-05-16 & 76.34\% & 65.56\% & 55.16\% & 49.39\% & 42.40\% & 39.43\% & 37.54\% & 36.26\% & 35.15\% \\
21-11-16 & 78.90\% & 66.53\% & 54.78\% & 48.65\% & 40.61\% & 37.38\% & 35.34\% & 33.94\% & 32.68\% \\
21-05-17 & 81.50\% & 67.50\% & 54.41\% & 47.94\% & 38.93\% & 35.47\% & 33.30\% & 31.81\% & 30.42\% \\
21-11-17 & 84.24\% & 68.50\% & 54.03\% & 47.22\% & 37.29\% & 33.63\% & 31.36\% & 29.78\% & 28.28\% \\
21-05-18 & 87.02\% & 69.50\% & 53.67\% & 46.53\% & 35.74\% & 31.92\% & 29.55\% & 27.90\% & 26.32\% \\
\hline
\end{tabular}
\caption{Smiles of forward rates. Fixing dates (first column) and moneyness (first row).} \label{tab:smilesFws}
\end{center}
}
\end{table}

\begin{table}[!htb]
\scriptsize{
\begin{center}
\begin{tabular}{|c|c|ccccccccc|}
\hline
 &  & -80\% & -60\% & -40\% & -20\% & 0\% & 20\% & 40\% & 60\% & 80\%  \\
% \hline
% & & \multicolumn{9}{|c|}{1 year underlying} \\
\hline
\multirow{4}{*}{\begin{sideways} 1 year \end{sideways}} & 21/05/2012 & 122.30\% & 102.40\% & 87.12\% & 76.45\% & 70.40\% & 66.47\% & 64.20\% & 63.03\% & 62.56\% \\
 & 21/11/2012 & 102.86\% & 89.97\% & 79.85\% & 72.49\% & 67.90\% & 64.58\% & 62.16\% & 60.39\% & 59.19\% \\
& 21/05/2013 & 95.64\% & 83.17\% & 73.42\% & 66.40\% & 62.10\% & 59.03\% & 56.84\% & 55.26\% & 54.18\% \\
& 21/11/2013 & 88.11\% & 76.06\% & 66.69\% & 60.00\% & 56.00\% & 53.18\% & 51.22\% & 49.84\% & 48.87\% \\
\hline
% & \multicolumn{9}{|c|}{2 years underlying} \\
% \hline
\multirow{4}{*}{\begin{sideways}2 years\end{sideways}} & 21/05/2012 & 111.50\% & 91.60\% & 76.32\% & 65.65\% & 59.60\% & 55.67\% & 53.40\% & 52.23\% & 51.76\% \\
& 21/11/2012 & 89.66\% & 76.77\% & 66.65\% & 59.29\% & 54.70\% & 51.38\% & 48.96\% & 47.19\% & 45.99\% \\
& 21/05/2013 & 82.94\% & 70.47\% & 60.72\% & 53.70\% & 49.40\% & 46.33\% & 44.14\% & 42.56\% & 41.48\% \\
& 21/11/2013 & 77.81\% & 65.76\% & 56.39\% & 49.70\% & 45.70\% & 42.88\% & 40.92\% & 39.54\% & 38.57\% \\
\hline
% & \multicolumn{9}{|c|}{3 years underlying} \\
% \hline
\multirow{4}{*}{\begin{sideways}3 years\end{sideways}} & 21/05/2012 & 106.40\% & 86.50\% & 71.22\% & 60.55\% & 54.50\% & 50.57\% & 48.30\% & 47.13\% & 46.66\% \\
& 21/11/2012 & 83.66\% & 70.77\% & 60.65\% & 53.29\% & 48.70\% & 45.38\% & 42.96\% & 41.19\% & 39.99\% \\
& 21/05/2013 & 78.34\% & 65.87\% & 56.12\% & 49.10\% & 44.80\% & 41.73\% & 39.54\% & 37.96\% & 36.88\% \\
& 21/11/2013 & 73.61\% & 61.56\% & 52.19\% & 45.50\% & 41.50\% & 38.68\% & 36.72\% & 35.34\% & 34.37\% \\
\hline
% & \multicolumn{9}{|c|}{4 years underlying} \\
% \hline
\multirow{4}{*}{\begin{sideways}4 years\end{sideways}} & 21/05/2012 & 101.90\% & 82.00\% & 66.72\% & 56.05\% & 50.00\% & 46.07\% & 43.80\% & 42.63\% & 42.16\% \\
& 21/11/2012 & 80.26\% & 67.37\% & 57.25\% & 49.89\% & 45.30\% & 41.98\% & 39.56\% & 37.79\% & 36.59\% \\
& 21/05/2013 & 75.24\% & 62.77\% & 53.02\% & 46.00\% & 41.70\% & 38.63\% & 36.44\% & 34.86\% & 33.78\% \\
& 21/11/2013 & 70.91\% & 58.86\% & 49.49\% & 42.80\% & 38.80\% & 35.98\% & 34.02\% & 32.64\% & 31.67\% \\
\hline
% & \multicolumn{9}{|c|}{5 years underlying} \\
% \hline
\multirow{4}{*}{\begin{sideways}5 years\end{sideways}} & 21/05/2012 & 96.15\% & 74.25\% & 58.83\% & 49.88\% & 47.40\% & 45.74\% & 44.61\% & 43.76\% & 43.05\% \\
& 21/11/2012 & 89.58\% & 68.82\% & 54.14\% & 45.54\% & 43.00\% & 39.36\% & 37.33\% & 36.15\% & 35.37\% \\
& 21/05/2013 & 83.91\% & 64.51\% & 50.71\% & 42.51\% & 39.90\% & 36.48\% & 34.59\% & 33.50\% & 32.76\% \\
& 21/11/2013 & 79.13\% & 61.09\% & 48.17\% & 40.37\% & 37.70\% & 34.50\% & 32.74\% & 31.75\% & 31.05\% \\
\hline
\end{tabular}
\caption{Smiles of swap rates. Length of the underlying swaps (first column), swaptions maturities (second column) and moneyness (first row).} \label{tab:smilesTS}
\end{center}
}
\end{table}
% \clearpage

\subsection{Hagan model}

\subsubsection{Calibration to caplets}

In Table \ref{tab:paramsHaganVolas} the calibrated parameters with SABR formula \eqref{eq:sigmaHagan} are shown. The execution time was $8.739$ seconds, $8.565$ seconds employed by the mono-GPU SA (launched with a relaxed configuration, specifically, $T_0=10$, $T_{min}=0.01$, $\rho=0.99$, $N=10$, $w=256\times64$, the cost function was evaluated $112738304$ times) and $0.174$ seconds to the Nelder-Mead algorithm. The sequential time of the minimization with SA is $971.960$ seconds. Thus, the speedup of the proposed SA parallelization is $113.480$ times.

\begin{table}[!htb]
\scriptsize{
\begin{center}
\begin{tabular}{|c|r|c|c||c|r|c|c|}
\hline
   & $\phi_{ii}$ & $\sigma_i$ & $\alpha_i$ & & $\phi_{ii}$ & $\sigma_i$ & $\alpha_i$ \\
\hline
\hline
$F_1$ & $-0.4712$ & $1.0000$ & $0.0847$ & $F_8$ & $-0.4552$ & $0.4658$ & $0.0723$ \\
\hline
$F_2$ & $-0.1879$ & $0.7354$ & $0.0830$ & $F_9$ & $-0.5215$ & $0.5369$ & $0.0703$ \\
\hline
$F_3$ & $0.0719$ & $0.5260$ & $0.0822$ & $F_{10}$ & $-0.5663$ & $0.6116$ & $0.0706$ \\
\hline
$F_4$ & $0.2636$ & $0.3329$ & $0.0686$ & $F_{11}$ & $-0.5973$ & $0.6858$ & $0.0684$ \\
\hline
$F_5$ & $0.0273$ & $0.3242$ & $0.0662$ & $F_{12}$ & $-0.6204$ & $0.7609$ & $0.0674$ \\
\hline
$F_6$ & $-0.1942$ & $0.3505$ & $0.0714$ & $F_{13}$ & $-0.6378$ & $0.8337$ & $0.0652$ \\
\hline
$F_7$ & $-0.3514$ & $0.4008$ & $0.0696$ &  &  &  &  \\
\hline
\end{tabular}
\caption{Hagan model, calibration to caplets with SABR formula \eqref{eq:sigmaHagan}: calibrated parameters.} \label{tab:paramsHaganVolas}
\end{center}
}
\end{table}

In Table \ref{tab:haganVolas} market vs. model volatilities (both in $\%$) for the first twelve smiles and the moneyness varying from $-40 \%$ to $40 \%$ are shown. In addition, the mean relative error ($MRE$) considering the whole set of smiles is presented.

\begin{table}[!htb]
\scriptsize{
\centering
\begin{tabular}{|r|| c|r|c ||c|c|c |}
\hline
Moneyness & \multicolumn{3}{|c||}{Smile of $F_{1}$}& \multicolumn{3}{|c|}{Smile of $F_{2}$} \\
\hline
& $\sigma_{market}$ & $\sigma_{model}$ & $\frac{|\sigma_{market}-\sigma_{model}|}{\sigma_{market}}$ & $\sigma_{market}$ & $\sigma_{model}$ &$\frac{|\sigma_{market}-\sigma_{model}|}{\sigma_{market}}$ \\ \hline
$-40\%$ & $97.26$ & $100.61$ & $3.44 \times 10^{-2}$ & $88.27$ & $89.06$ & $8.97 \times 10^{-3}$ \\
\hline
$-20\%$ & $82.58$ & $87.53$ & $6.00 \times 10^{-2}$ & $79.62$ & $80.85$ & $1.55 \times 10^{-2}$ \\
\hline
$0\%$ & $72.29$ & $77.45$ & $7.13 \times 10^{-2}$ & $73.03$ & $74.70$ & $2.28 \times 10^{-2}$ \\
\hline
$20\%$ & $70.89$ & $70.36$ & $7.48 \times 10^{-3}$ & $71.95$ & $70.61$ & $1.85 \times 10^{-2}$ \\
\hline
$40\%$ & $69.49$ & $66.26$ & $4.64 \times 10^{-2}$ & $70.87$ & $68.59$ & $3.21 \times 10^{-2}$ \\
\hline
\hline
Moneyness & \multicolumn{3}{|c||}{Smile of $F_{3}$}& \multicolumn{3}{|c|}{Smile of $F_{4}$} \\
\hline
& $\sigma_{market}$ & $\sigma_{model}$ & $\frac{|\sigma_{market}-\sigma_{model}|}{\sigma_{market}}$ & $\sigma_{market}$ & $\sigma_{model}$ &$\frac{|\sigma_{market}-\sigma_{model}|}{\sigma_{market}}$ \\ \hline
$-40\%$ & $77.09$ & $77.41$ & $4.08 \times 10^{-3}$ & $57.08$ & $57.14$ & $1.05 \times 10^{-3}$ \\
\hline
$-20\%$ & $71.50$ & $72.46$ & $1.34 \times 10^{-2}$ & $53.21$ & $54.37$ & $2.18 \times 10^{-2}$ \\
\hline
$0\%$ & $67.93$ & $68.77$ & $1.24 \times 10^{-2}$ & $52.49$ & $52.29$ & $3.82 \times 10^{-3}$ \\
\hline
$20\%$ & $67.10$ & $66.34$ & $1.13 \times 10^{-2}$ & $51.34$ & $50.90$ & $8.56 \times 10^{-3}$ \\
\hline
$40\%$ & $66.41$ & $65.17$ & $1.88 \times 10^{-2}$ & $50.61$ & $50.19$ & $8.23 \times 10^{-3}$ \\
\hline
\hline
Moneyness & \multicolumn{3}{|c||}{Smile of $F_{5}$}& \multicolumn{3}{|c|}{Smile of $F_{6}$} \\
\hline
& $\sigma_{market}$ & $\sigma_{model}$ & $\frac{|\sigma_{market}-\sigma_{model}|}{\sigma_{market}}$ & $\sigma_{market}$ & $\sigma_{model}$ &$\frac{|\sigma_{market}-\sigma_{model}|}{\sigma_{market}}$ \\ \hline
$-40\%$ & $56.69$ & $56.92$ & $4.05 \times 10^{-3}$ & $56.30$ & $56.74$ & $7.69 \times 10^{-3}$ \\
\hline
$-20\%$ & $52.43$ & $53.22$ & $1.51 \times 10^{-2}$ & $51.65$ & $52.10$ & $8.76 \times 10^{-3}$ \\
\hline
$0\%$ & $50.31$ & $50.37$ & $1.07 \times 10^{-3}$ & $48.19$ & $48.48$ & $6.03 \times 10^{-3}$ \\
\hline
$20\%$ & $48.72$ & $48.36$ & $7.29 \times 10^{-3}$ & $46.19$ & $45.89$ & $6.52 \times 10^{-3}$ \\
\hline
$40\%$ & $47.70$ & $47.21$ & $1.03 \times 10^{-2}$ & $44.91$ & $44.32$ & $1.31 \times 10^{-2}$ \\
\hline
\hline
Moneyness & \multicolumn{3}{|c||}{Smile of $F_{7}$}& \multicolumn{3}{|c|}{Smile of $F_{8}$} \\
\hline
& $\sigma_{market}$ & $\sigma_{model}$ & $\frac{|\sigma_{market}-\sigma_{model}|}{\sigma_{market}}$ & $\sigma_{market}$ & $\sigma_{model}$ &$\frac{|\sigma_{market}-\sigma_{model}|}{\sigma_{market}}$ \\ \hline
$-40\%$ & $55.92$ & $56.59$ & $1.19 \times 10^{-2}$ & $55.54$ & $56.47$ & $1.68 \times 10^{-2}$ \\
\hline
$-20\%$ & $50.89$ & $51.04$ & $3.00 \times 10^{-3}$ & $50.13$ & $50.01$ & $2.35 \times 10^{-3}$ \\
\hline
$0\%$ & $46.19$ & $46.70$ & $1.09 \times 10^{-2}$ & $44.25$ & $44.95$ & $1.59 \times 10^{-2}$ \\
\hline
$20\%$ & $43.83$ & $43.56$ & $6.33 \times 10^{-3}$ & $41.56$ & $41.28$ & $6.80 \times 10^{-3}$ \\
\hline
$40\%$ & $42.32$ & $41.61$ & $1.67 \times 10^{-2}$ & $39.84$ & $39.00$ & $2.12 \times 10^{-2}$ \\
\hline
\hline
Moneyness & \multicolumn{3}{|c||}{Smile of $F_{9}$}& \multicolumn{3}{|c|}{Smile of $F_{10}$} \\
\hline
& $\sigma_{market}$ & $\sigma_{model}$ & $\frac{|\sigma_{market}-\sigma_{model}|}{\sigma_{market}}$ & $\sigma_{market}$ & $\sigma_{model}$ &$\frac{|\sigma_{market}-\sigma_{model}|}{\sigma_{market}}$ \\ \hline
$-40\%$ & $55.16$ & $56.39$ & $2.22 \times 10^{-2}$ & $54.78$ & $56.34$ & $2.85 \times 10^{-2}$ \\
\hline
$-20\%$ & $49.39$ & $49.04$ & $7.16 \times 10^{-3}$ & $48.65$ & $48.09$ & $1.15 \times 10^{-2}$ \\
\hline
$0\%$ & $42.40$ & $43.28$ & $2.07 \times 10^{-2}$ & $40.61$ & $41.65$ & $2.54 \times 10^{-2}$ \\
\hline
$20\%$ & $39.43$ & $39.11$ & $8.06 \times 10^{-3}$ & $37.38$ & $37.00$ & $1.02 \times 10^{-2}$ \\
\hline
$40\%$ & $37.54$ & $36.53$ & $2.68 \times 10^{-2}$ & $35.34$ & $34.15$ & $3.36 \times 10^{-2}$ \\
\hline
\hline
Moneyness & \multicolumn{3}{|c||}{Smile of $F_{11}$}& \multicolumn{3}{|c|}{Smile of $F_{12}$} \\
\hline
& $\sigma_{market}$ & $\sigma_{model}$ & $\frac{|\sigma_{market}-\sigma_{model}|}{\sigma_{market}}$ & $\sigma_{market}$ & $\sigma_{model}$ &$\frac{|\sigma_{market}-\sigma_{model}|}{\sigma_{market}}$ \\ \hline
$-40\%$ & $54.41$ & $56.33$ & $3.52 \times 10^{-2}$ & $54.03$ & $56.35$ & $4.28 \times 10^{-2}$ \\
\hline
$-20\%$ & $47.94$ & $47.20$ & $1.54 \times 10^{-2}$ & $47.22$ & $46.34$ & $1.87 \times 10^{-2}$ \\
\hline
$0\%$ & $38.93$ & $40.09$ & $2.99 \times 10^{-2}$ & $37.29$ & $38.57$ & $3.44 \times 10^{-2}$ \\
\hline
$20\%$ & $35.47$ & $35.00$ & $1.33 \times 10^{-2}$ & $33.63$ & $33.04$ & $1.76 \times 10^{-2}$ \\
\hline
$40\%$ & $33.30$ & $31.92$ & $4.16 \times 10^{-2}$ & $31.36$ & $29.75$ & $5.12 \times 10^{-2}$ \\
\hline
% \hline
% Moneyness & \multicolumn{3}{|c||}{Smile of $F_{13}$}& \multicolumn{3}{|c|}{ } \\
% \hline
% & $\sigma_{market}$ & $\sigma_{model}$ & $\frac{|\sigma_{market}-\sigma_{model}|}{\sigma_{market}}$ &  &  & \\ \hline
% $-40\%$ & $53.67$ & $56.40$ & $5.09 \times 10^{-2}$ &  &  & \\
% \hline
% $-20\%$ & $46.53$ & $45.52$ & $2.16 \times 10^{-2}$ &  &  & \\
% \hline
% $0\%$ & $35.74$ & $37.12$ & $3.85 \times 10^{-2}$ &  &  & \\
% \hline
% $20\%$ & $31.91$ & $31.18$ & $2.31 \times 10^{-2}$ &  &  & \\
% \hline
% $40\%$ & $29.55$ & $27.71$ & $6.23 \times 10^{-2}$ &  &  & \\
% \hline
\hline
\multicolumn{7}{|c|}{$MRE=1.80 \times 10^{-2}$} \\
\hline
\end{tabular}
\caption{Hagan model, calibration to caplets, $\sigma_{market}$ vs. $\sigma_{model}$.}
\label{tab:haganVolas}
}
\end{table}

In order to validate the algorithm we also performed the equivalent calibration with Monte Carlo simulation thus obtaining the same parameters as in Table \ref{tab:paramsHaganVolas}, except $\phi_{11}=0.0287$. Moreover the computational time is approximately $2$ hours. We note that with formula \eqref{eq:sigmaHagan} the mean absolute error ($MAE$, in $\%$) in prices is $4.14 \times 10^{-2}$, while using Monte Carlo simulation the obtained $MAE$ is $4.08 \times 10^{-2}$.

\begin{figure}[!htb]
\centering
  \subfigure {\includegraphics[height=3.7cm]{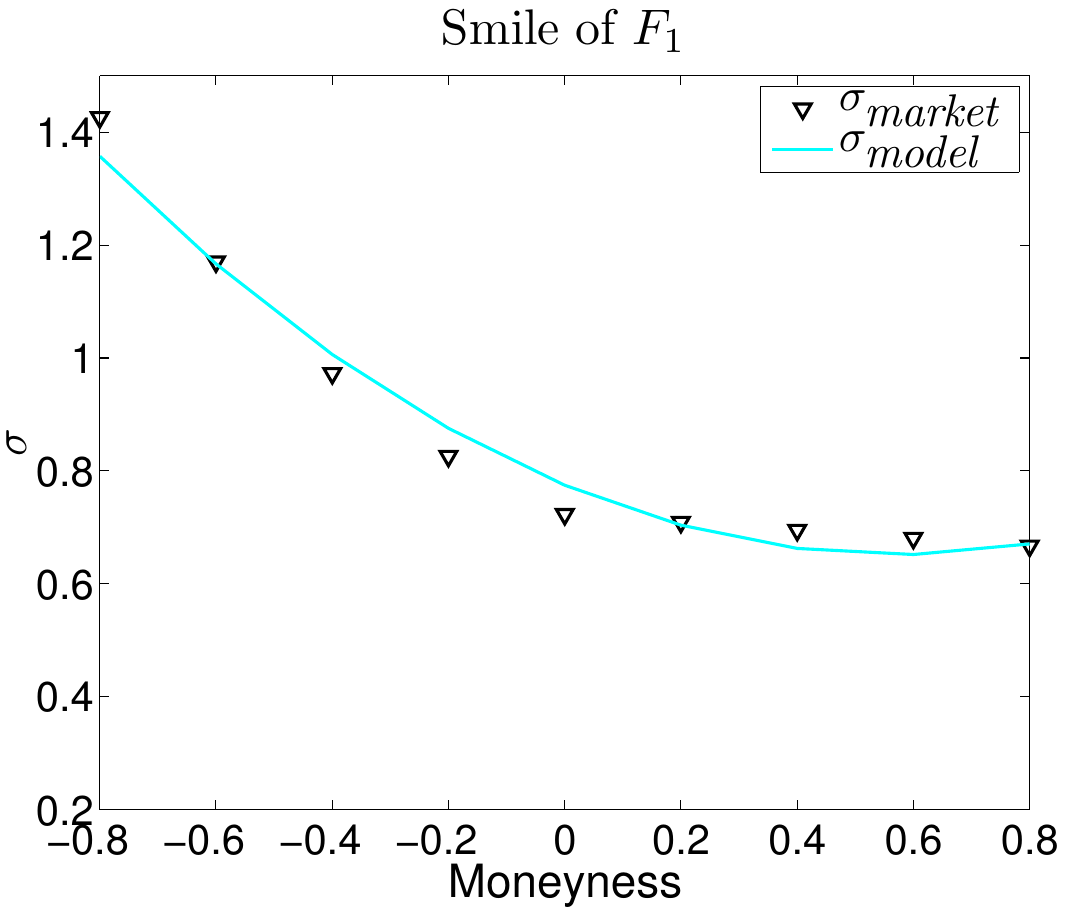}}
  \subfigure {\includegraphics[height=3.7cm]{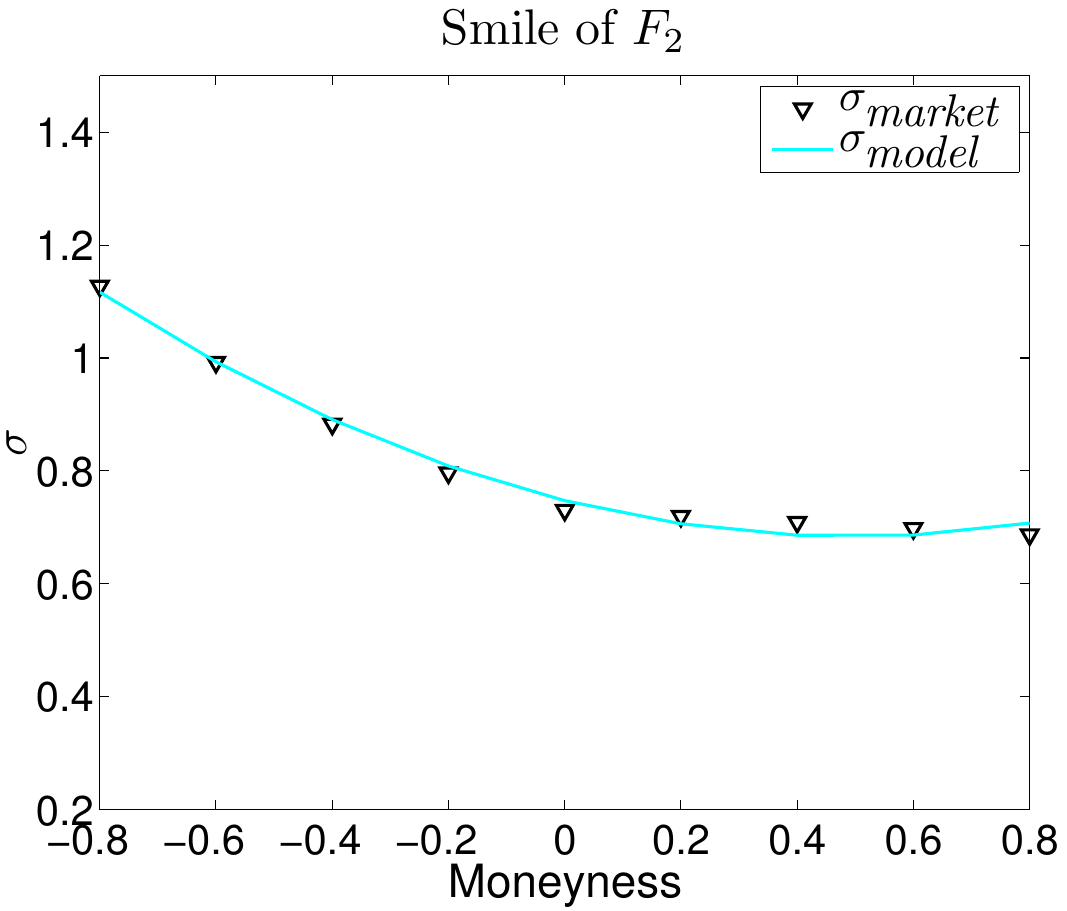}}
  \subfigure {\includegraphics[height=3.7cm]{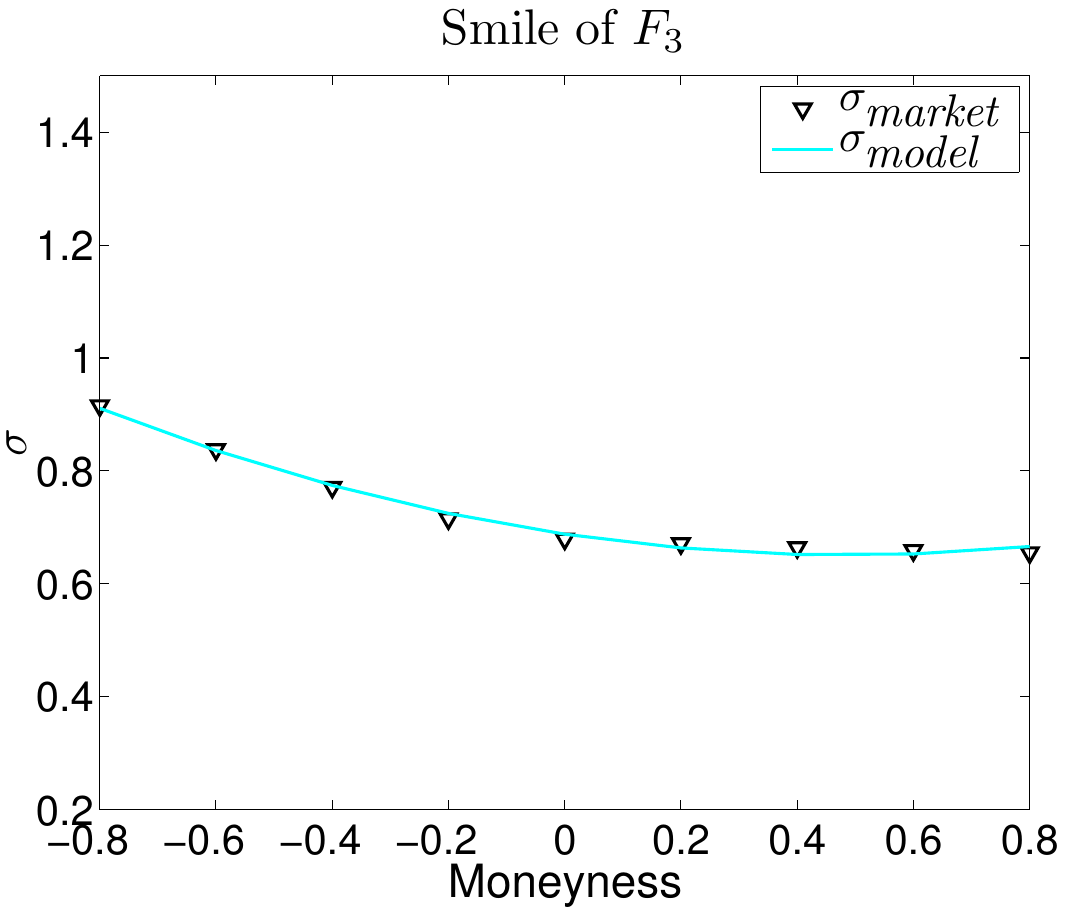}}
  \subfigure {\includegraphics[height=3.7cm]{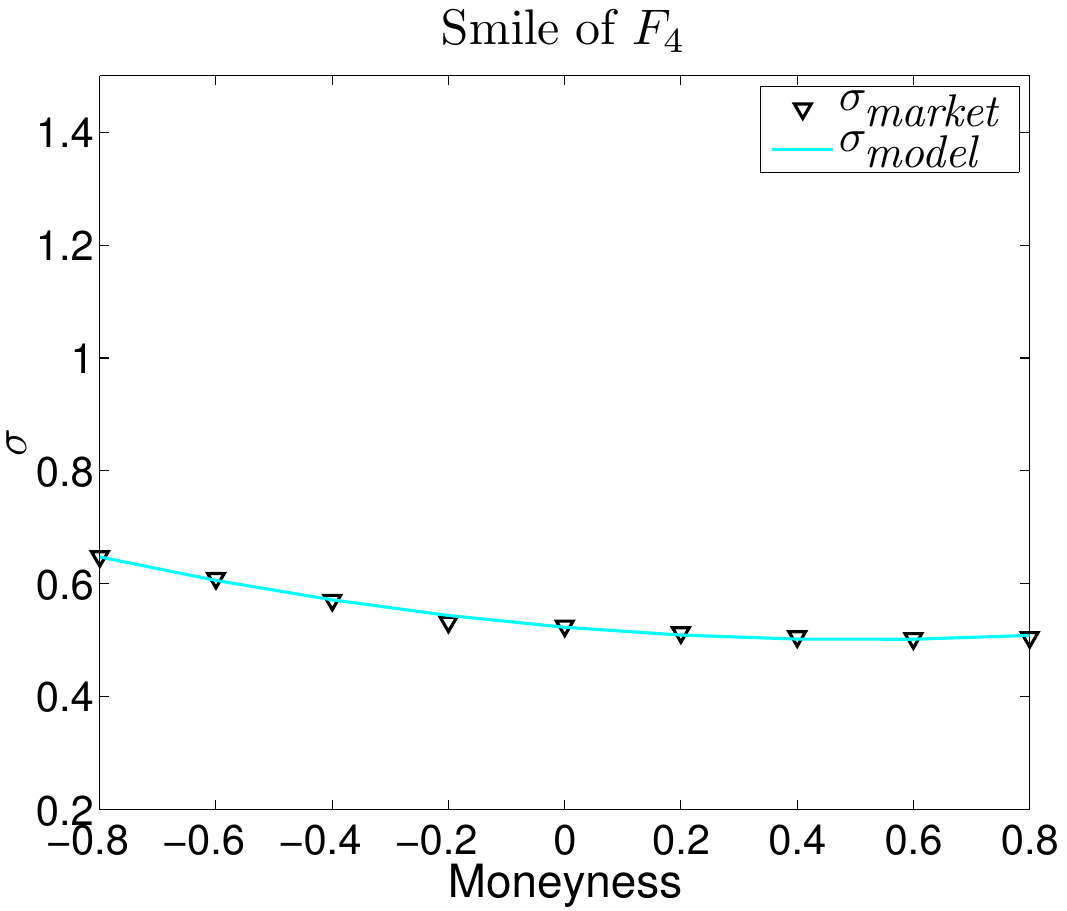}}
  \subfigure {\includegraphics[height=3.7cm]{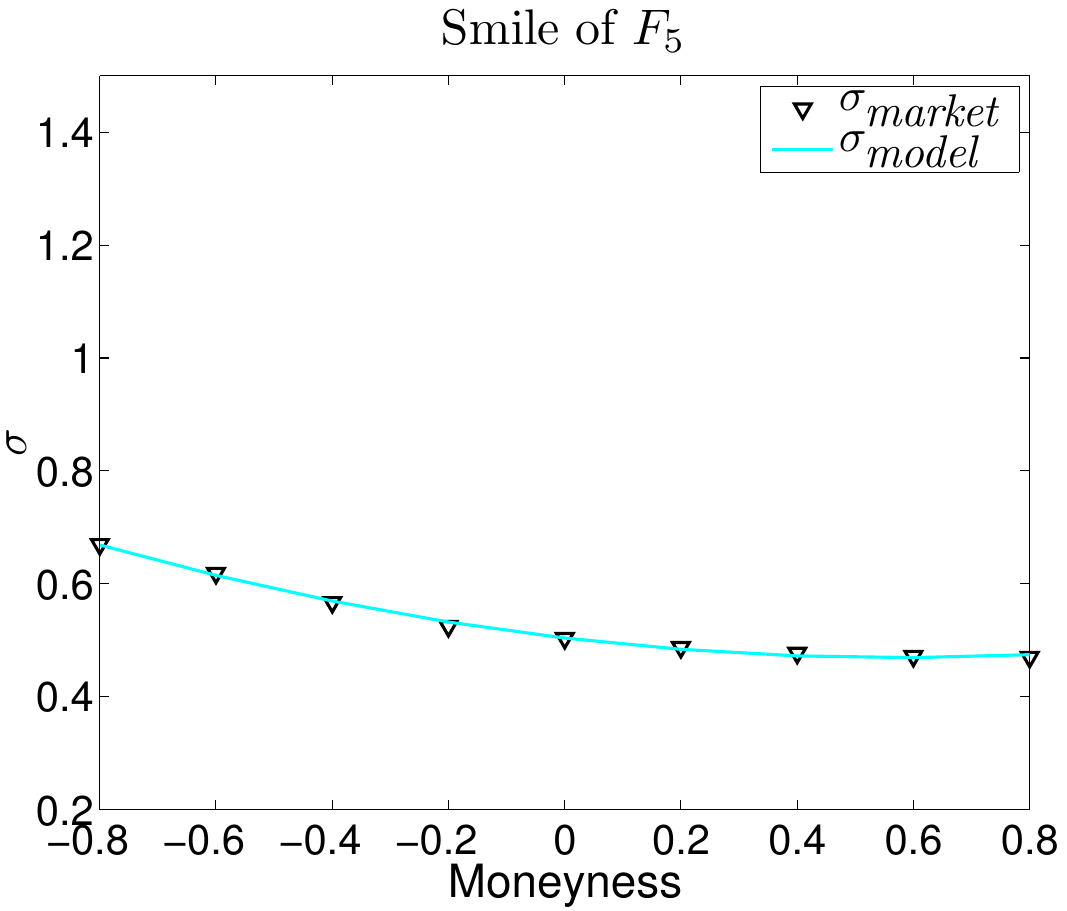}}
  \subfigure {\includegraphics[height=3.7cm]{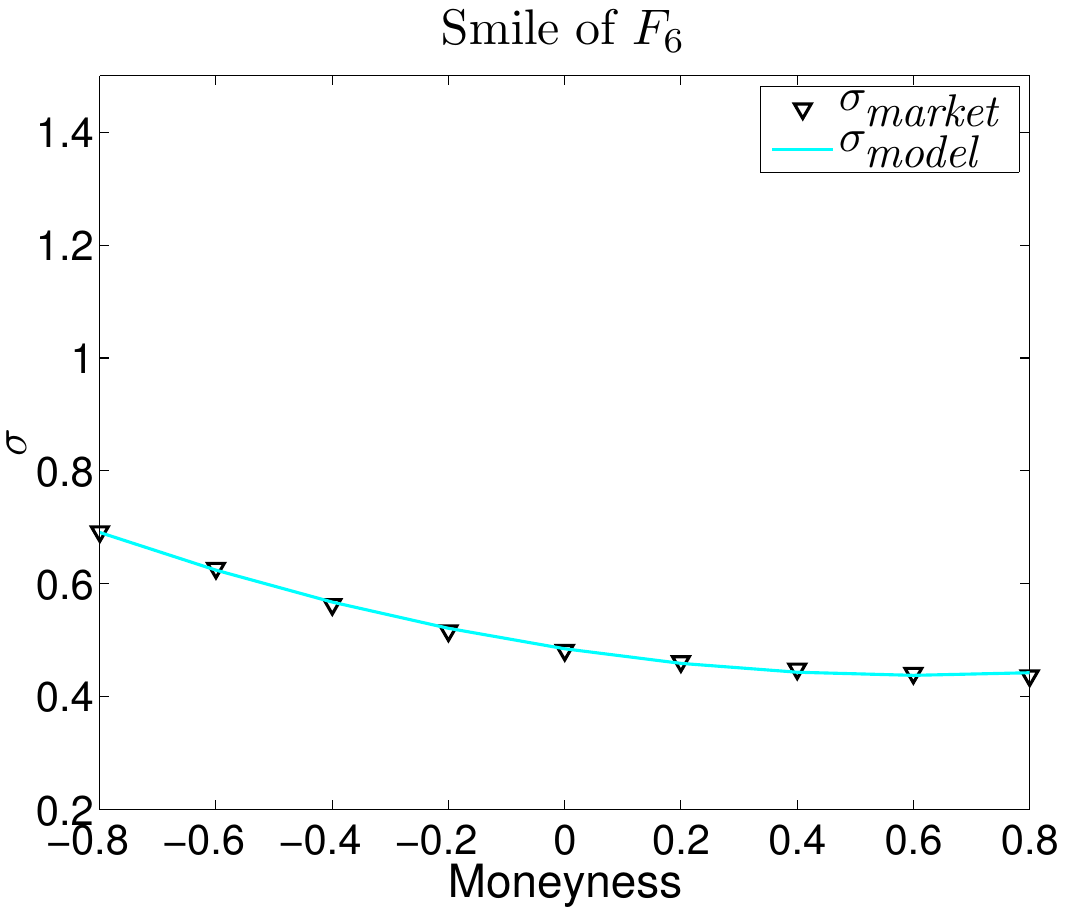}}
  \subfigure {\includegraphics[height=3.7cm]{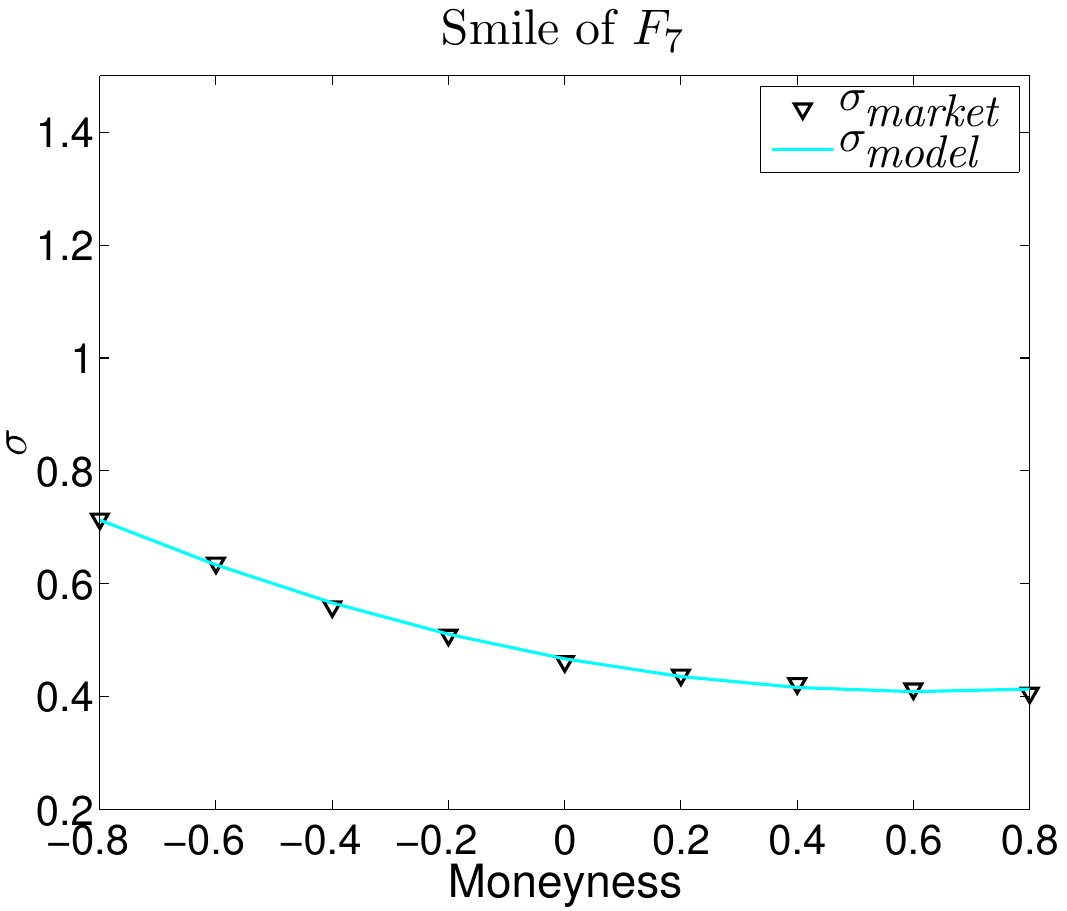}}
  \subfigure {\includegraphics[height=3.7cm]{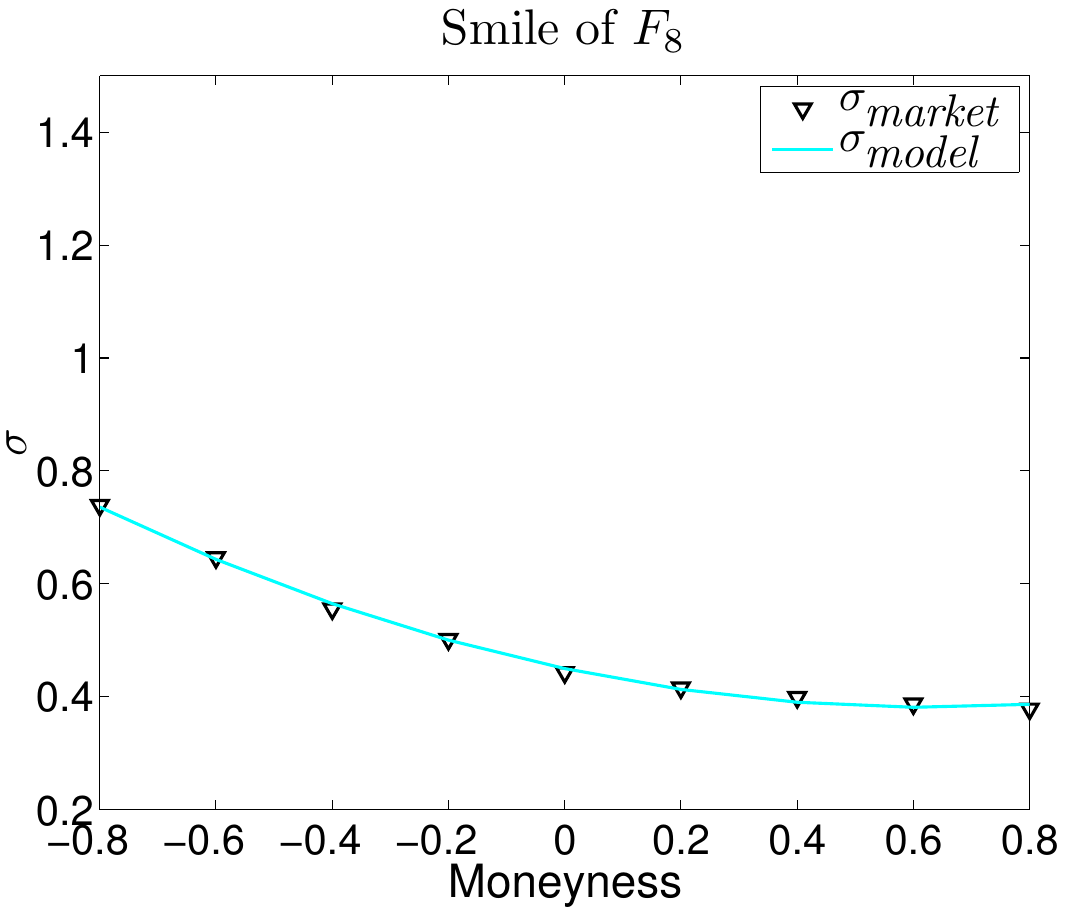}}
  \subfigure {\includegraphics[height=3.7cm]{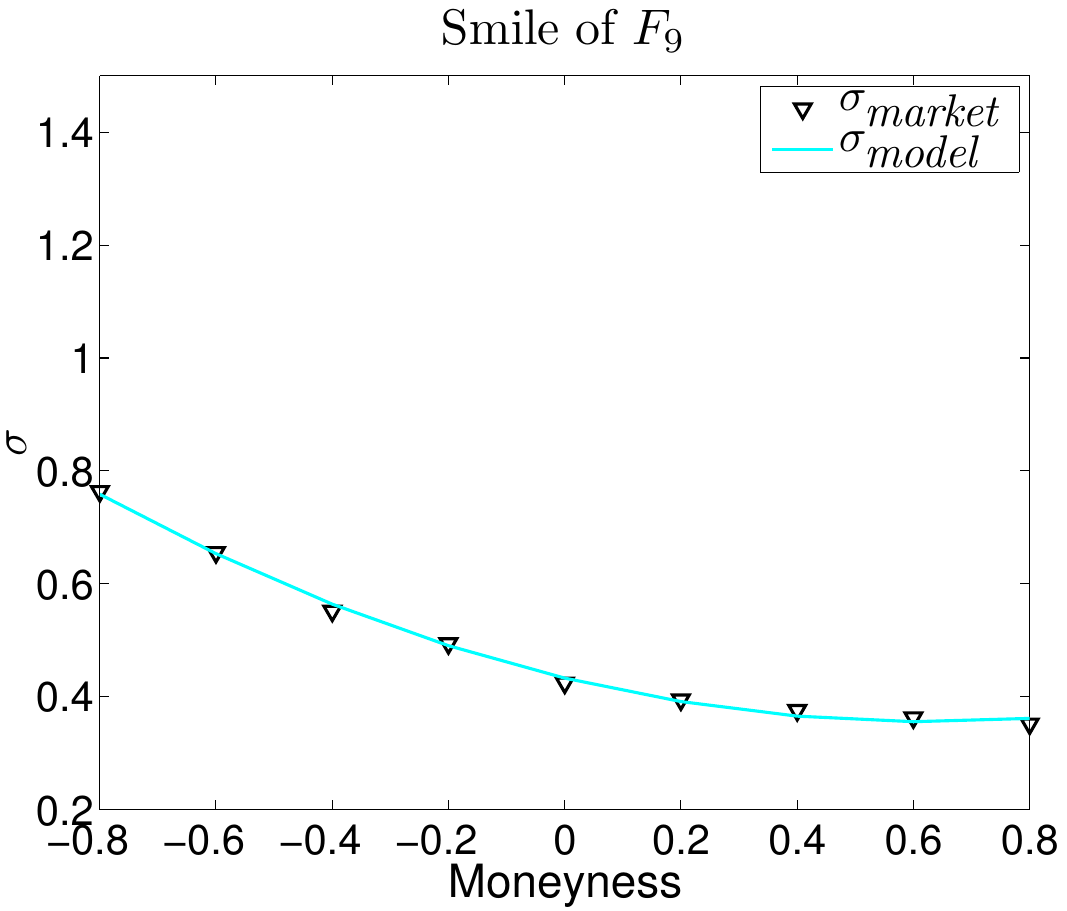}}
  \subfigure {\includegraphics[height=3.7cm]{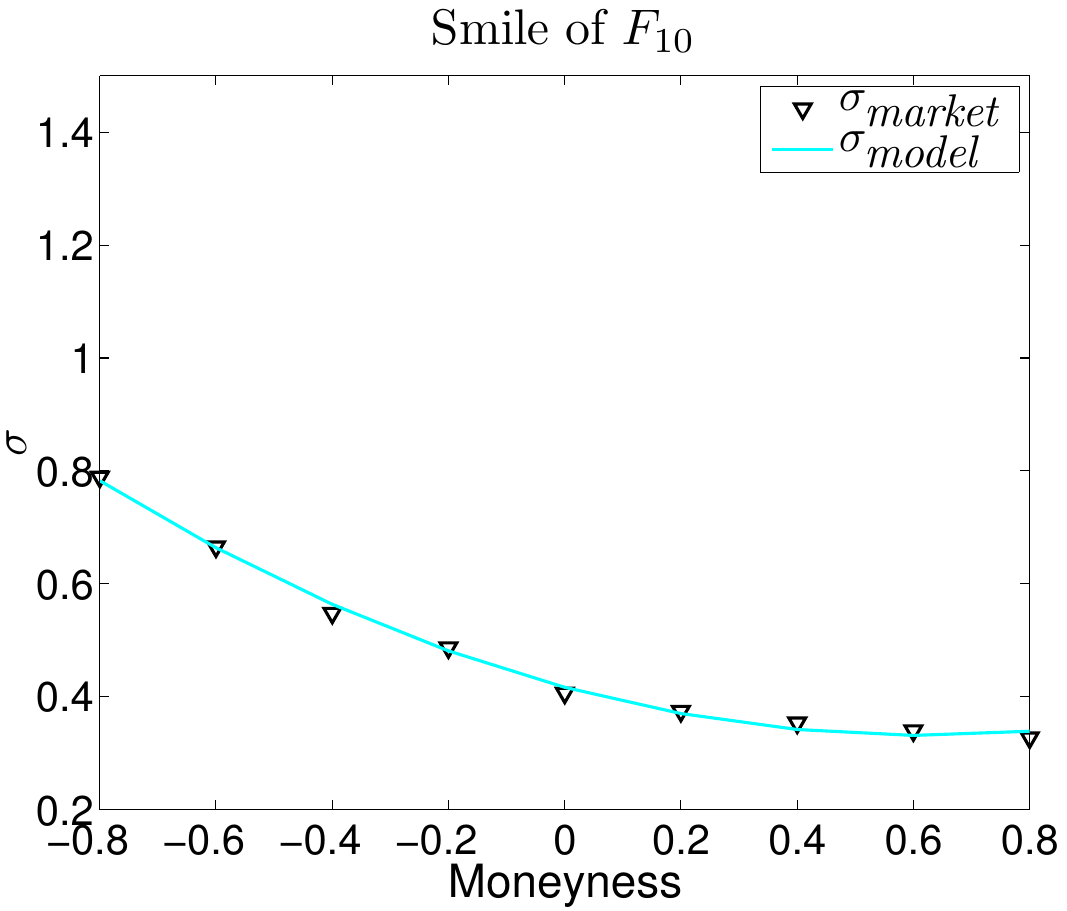}}
  \subfigure {\includegraphics[height=3.7cm]{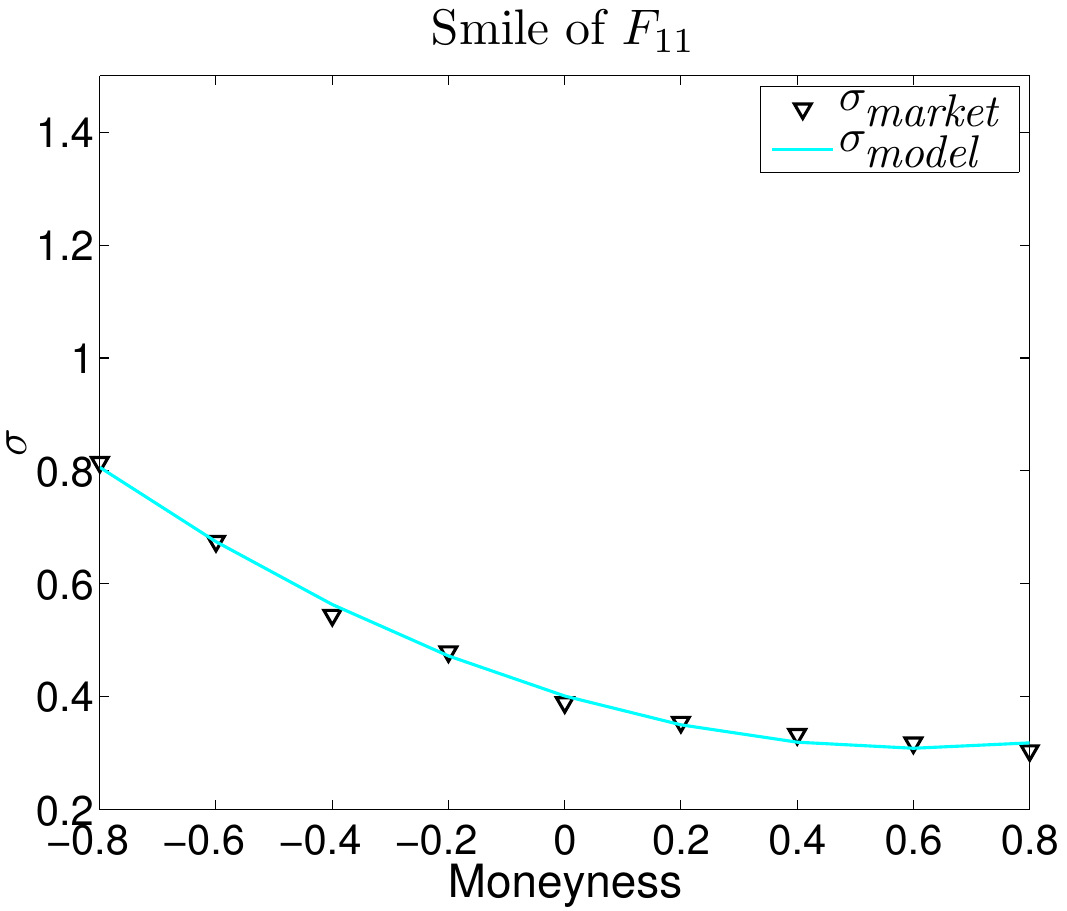}}
  \subfigure {\includegraphics[height=3.7cm]{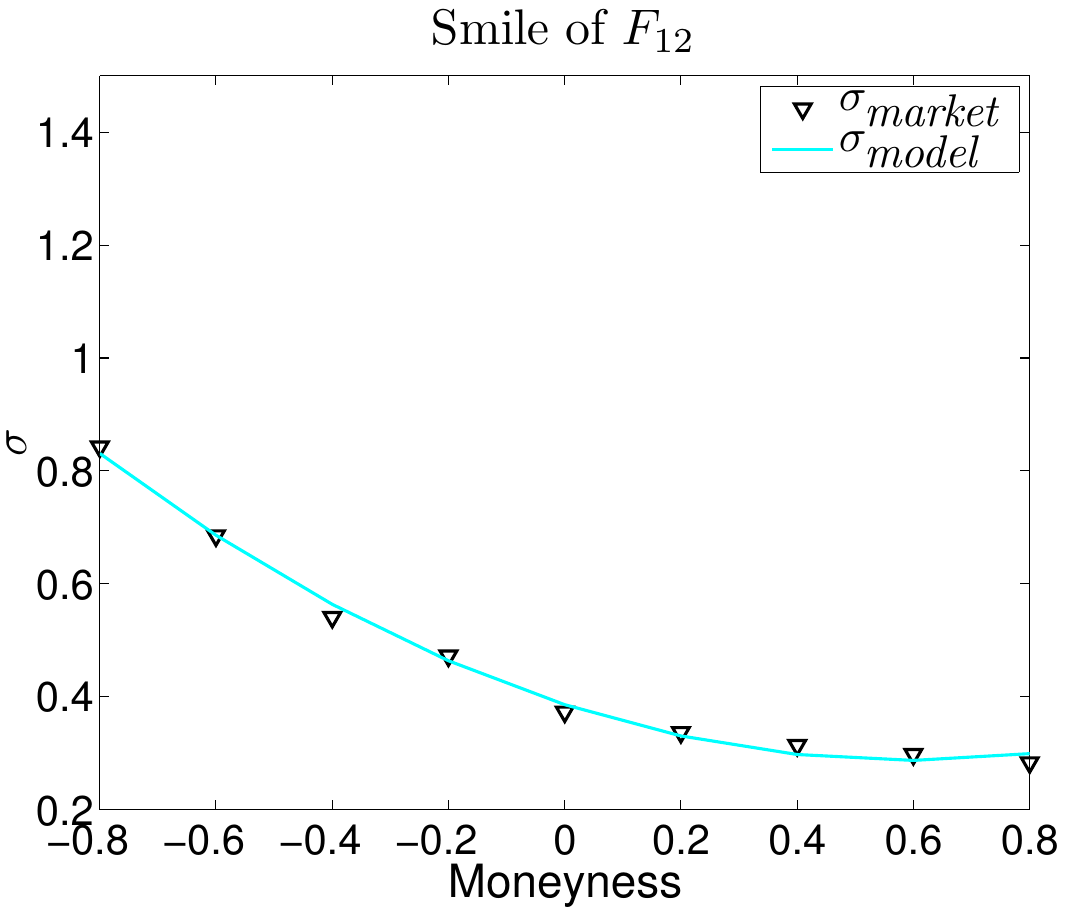}}
  \subfigure {\includegraphics[height=3.7cm]{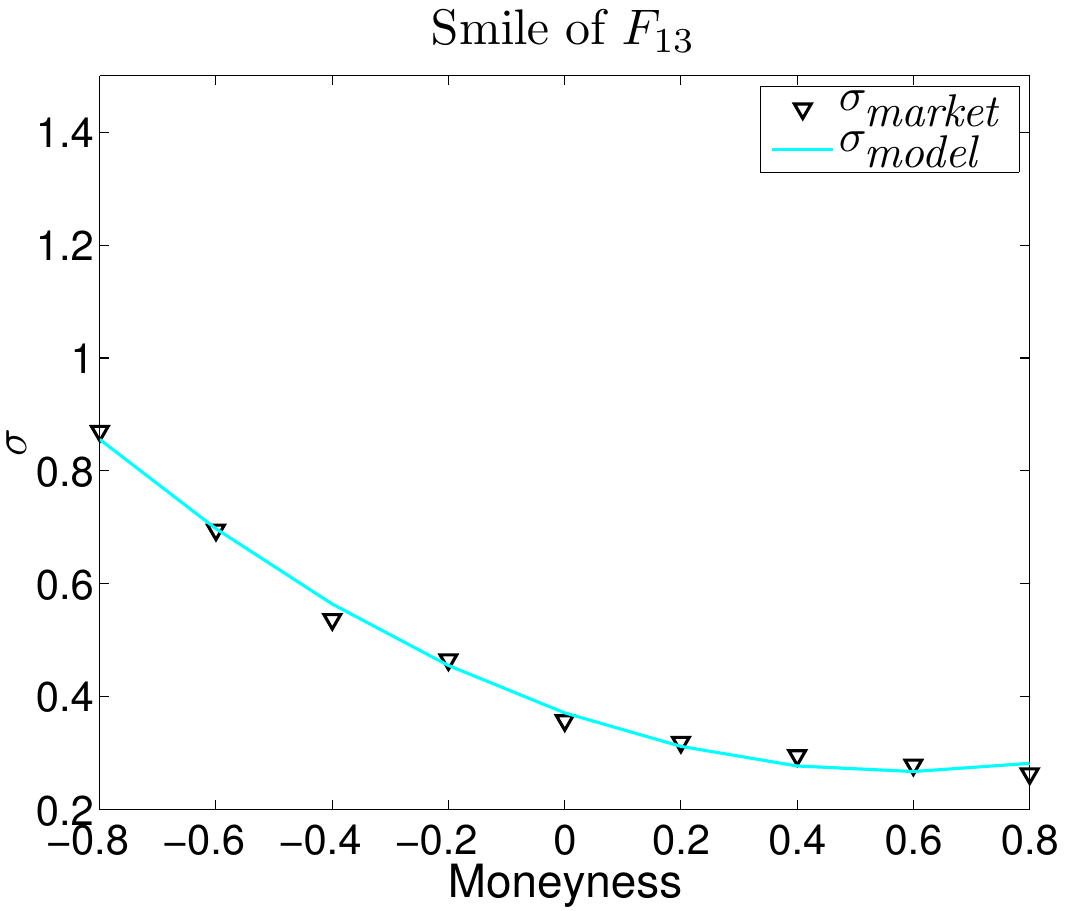}}
  \caption{Hagan model, $\sigma_{market}$ vs. $\sigma_{model}$, smiles of $F_1,\ldots,F_{13}$.}
  \label{fig:haganVolas}
\end{figure}

In Figure \ref{fig:haganVolas} the model fitting for the smiles of all forward rates is shown. Market volatilities are plotted with triangles, while model volatilities are shown in continuous line.

\subsubsection{Calibration to swaptions}

The calibrated parameters are $\eta_1 = 0.814904$, $\lambda_1 = 3.378797$, $\eta_2 = 0.975928$, $\lambda_2 = 3.777324$ and $\lambda_3 = 0.013940$. In Table \ref{tab:haganSwaptions} market vs. model swaptions prices (in $\%$) for the first fourteen swaptions and the the moneyness varying from $-40\%$ to $40\%$ are shown, each pair with its corresponding absolute error. In addition, for the whole set of swaptions the mean absolute error ($MAE$) is presented.

\begin{table}[!htb]
\scriptsize{
\centering
\begin{tabular}{|r|| c|c|c ||c|c|c |}
\hline
Moneyness & \multicolumn{3}{|c||}{$0.5 \times 1$ swaptions} & \multicolumn{3}{|c|}{$1 \times 1$ swaptions} \\
\hline
& $S_{Black}$ & $S_{MC}$ & $|S_{Black}-S_{MC}|$ & $S_{Black}$ & $S_{MC}$ & $|S_{Black}-S_{MC}|$ \\ \hline
$-40\%$ & $0.4866$ & $0.4842$ & $2.40 \times 10^{-3}$ & $0.5917$ & $0.5758$ & $1.59 \times 10^{-2}$ \\
\hline
$-20\%$ & $0.3562$ & $0.3628$ & $6.60 \times 10^{-3}$ & $0.4661$ & $0.4602$ & $5.90 \times 10^{-3}$ \\
\hline
$0\%$ & $0.2356$ & $0.2427$ & $7.10 \times 10^{-3}$ & $0.3467$ & $0.3450$ & $1.70 \times 10^{-3}$ \\
\hline
$20\%$ & $0.1363$ & $0.1390$ & $2.70 \times 10^{-3}$ & $0.2394$ & $0.2399$ & $5.00 \times 10^{-4}$ \\
\hline
$40\%$ & $0.0680$ & $0.0659$ & $2.10 \times 10^{-3}$ & $0.1517$ & $0.1539$ & $2.20 \times 10^{-3}$ \\
\hline
\hline
Moneyness & \multicolumn{3}{|c||}{$1.5 \times 1$ swaptions} & \multicolumn{3}{|c|}{$2 \times 1$ swaptions} \\
\hline
& $S_{Black}$ & $S_{MC}$ & $|S_{Black}-S_{MC}|$ & $S_{Black}$ & $S_{MC}$ & $|S_{Black}-S_{MC}|$ \\ \hline
$-40\%$ & $0.7357$ & $0.6840$ & $5.17 \times 10^{-2}$ & $0.8184$ & $0.7490$ & $6.94 \times 10^{-2}$ \\
\hline
$-20\%$ & $0.5908$ & $0.5548$ & $3.60 \times 10^{-2}$ & $0.6603$ & $0.6068$ & $5.35 \times 10^{-2}$ \\
\hline
$0\%$ & $0.4536$ & $0.4270$ & $2.66 \times 10^{-2}$ & $0.5118$ & $0.4651$ & $4.67 \times 10^{-2}$ \\
\hline
$20\%$ & $0.3277$ & $0.3095$ & $1.82 \times 10^{-2}$ & $0.3754$ & $0.3340$ & $4.14 \times 10^{-2}$ \\
\hline
$40\%$ & $0.2213$ & $0.2101$ & $1.12 \times 10^{-2}$ & $0.2587$ & $0.2229$ & $3.58 \times 10^{-2}$ \\
\hline
\hline
Moneyness & \multicolumn{3}{|c||}{$0.5 \times 2$ swaptions} & \multicolumn{3}{|c|}{$1 \times 2$ swaptions} \\
\hline
& $S_{Black}$ & $S_{MC}$ & $|S_{Black}-S_{MC}|$ & $S_{Black}$ & $S_{MC}$ & $|S_{Black}-S_{MC}|$ \\ \hline
$-40\%$ & $1.0570$ & $1.0144$ & $4.26 \times 10^{-2}$ & $1.2427$ & $1.1963$ & $4.64 \times 10^{-2}$ \\
\hline
$-20\%$ & $0.7440$ & $0.7275$ & $1.65 \times 10^{-2}$ & $0.9322$ & $0.9163$ & $1.59 \times 10^{-2}$ \\
\hline
$0\%$ & $0.4555$ & $0.4573$ & $1.80 \times 10^{-3}$ & $0.6394$ & $0.6460$ & $6.60 \times 10^{-3}$ \\
\hline
$20\%$ & $0.2299$ & $0.2418$ & $1.19 \times 10^{-2}$ & $0.3886$ & $0.4116$ & $2.30 \times 10^{-2}$ \\
\hline
$40\%$ & $0.0925$ & $0.1046$ & $1.21 \times 10^{-2}$ & $0.2037$ & $0.2343$ & $3.06 \times 10^{-2}$ \\
\hline
\hline
Moneyness & \multicolumn{3}{|c||}{$1.5 \times 2$ swaptions} & \multicolumn{3}{|c|}{$2 \times 2$ swaptions} \\
\hline
& $S_{Black}$ & $S_{MC}$ & $|S_{Black}-S_{MC}|$ & $S_{Black}$ & $S_{MC}$ & $|S_{Black}-S_{MC}|$ \\ \hline
$-40\%$ & $1.4884$ & $1.4260$ & $6.24 \times 10^{-2}$ & $1.6938$ & $1.6160$ & $7.78 \times 10^{-2}$ \\
\hline
$-20\%$ & $1.1367$ & $1.1168$ & $1.99 \times 10^{-2}$ & $1.3077$ & $1.2732$ & $3.45 \times 10^{-2}$ \\
\hline
$0\%$ & $0.8059$ & $0.8141$ & $8.20 \times 10^{-3}$ & $0.9466$ & $0.9320$ & $1.46 \times 10^{-2}$ \\
\hline
$20\%$ & $0.5154$ & $0.5446$ & $2.92 \times 10^{-2}$ & $0.6269$ & $0.6229$ & $4.00 \times 10^{-3}$ \\
\hline
$40\%$ & $0.2919$ & $0.3304$ & $3.85 \times 10^{-2}$ & $0.3736$ & $0.3748$ & $1.20 \times 10^{-3}$ \\
\hline
\hline
Moneyness & \multicolumn{3}{|c||}{$0.5 \times 3$ swaptions} & \multicolumn{3}{|c|}{$1 \times 3$ swaptions} \\
\hline
& $S_{Black}$ & $S_{MC}$ & $|S_{Black}-S_{MC}|$ & $S_{Black}$ & $S_{MC}$ & $|S_{Black}-S_{MC}|$ \\ \hline
$-40\%$ & $1.7380$ & $1.6538$ & $8.42 \times 10^{-2}$ & $2.0341$ & $1.9648$ & $6.93 \times 10^{-2}$ \\
\hline
$-20\%$ & $1.1980$ & $1.1506$ & $4.74 \times 10^{-2}$ & $1.4851$ & $1.4628$ & $2.23 \times 10^{-2}$ \\
\hline
$0\%$ & $0.7011$ & $0.6838$ & $1.73 \times 10^{-2}$ & $0.9696$ & $0.9812$ & $1.16 \times 10^{-2}$ \\
\hline
$20\%$ & $0.3242$ & $0.3277$ & $3.50 \times 10^{-3}$ & $0.5413$ & $0.5748$ & $3.35 \times 10^{-2}$ \\
\hline
$40\%$ & $0.1128$ & $0.1214$ & $8.60 \times 10^{-3}$ & $0.2479$ & $0.2882$ & $4.03 \times 10^{-2}$ \\
\hline
\hline
Moneyness & \multicolumn{3}{|c||}{$1.5 \times 3$ swaptions} & \multicolumn{3}{|c|}{$2 \times 3$ swaptions} \\
\hline
& $S_{Black}$ & $S_{MC}$ & $|S_{Black}-S_{MC}|$ & $S_{Black}$ & $S_{MC}$ & $|S_{Black}-S_{MC}|$ \\ \hline
$-40\%$ & $2.3898$ & $2.3012$ & $8.86 \times 10^{-2}$ & $2.6885$ & $2.6037$ & $8.48 \times 10^{-2}$ \\
\hline
$-20\%$ & $1.7850$ & $1.7586$ & $2.64 \times 10^{-2}$ & $2.0311$ & $2.0155$ & $1.56 \times 10^{-2}$ \\
\hline
$0\%$ & $1.2175$ & $1.2288$ & $1.13 \times 10^{-2}$ & $1.4178$ & $1.4296$ & $1.18 \times 10^{-2}$ \\
\hline
$20\%$ & $0.7304$ & $0.7676$ & $3.72 \times 10^{-2}$ & $0.8856$ & $0.9064$ & $2.08 \times 10^{-2}$ \\
\hline
$40\%$ & $0.3749$ & $0.4203$ & $4.54 \times 10^{-2}$ & $0.4832$ & $0.5044$ & $2.12 \times 10^{-2}$ \\
\hline
\hline
Moneyness & \multicolumn{3}{|c||}{$0.5 \times 4$ swaptions} & \multicolumn{3}{|c|}{$1 \times 4$ swaptions} \\
\hline
& $S_{Black}$ & $S_{MC}$ & $|S_{Black}-S_{MC}|$ & $S_{Black}$ & $S_{MC}$ & $|S_{Black}-S_{MC}|$ \\ \hline
$-40\%$ & $2.5381$ & $2.4226$ & $1.15 \times 10^{-1}$ & $2.9426$ & $2.8472$ & $9.54 \times 10^{-2}$ \\
\hline
$-20\%$ & $1.7151$ & $1.6480$ & $6.71 \times 10^{-2}$ & $2.1123$ & $2.0763$ & $3.60 \times 10^{-2}$ \\
\hline
$0\%$ & $0.9584$ & $0.9314$ & $2.70 \times 10^{-2}$ & $1.3344$ & $1.3375$ & $3.10 \times 10^{-3}$ \\
\hline
$20\%$ & $0.4031$ & $0.4035$ & $4.00 \times 10^{-4}$ & $0.7016$ & $0.7295$ & $2.79 \times 10^{-2}$ \\
\hline
$40\%$ & $0.1188$ & $0.1250$ & $6.20 \times 10^{-3}$ & $0.2907$ & $0.3268$ & $3.61 \times 10^{-2}$ \\
\hline
\hline
\multicolumn{7}{|c|}{$MAE=6.19 \times 10^{-2}$} \\
\hline
\end{tabular}
\caption{Hagan model, calibration to swaptions, $S_{Black}$ vs. $S_{MC}$, prices in \%.}
\label{tab:haganSwaptions}
}
\end{table}

\begin{figure}[!htb]
\centering
  \subfigure {\includegraphics[height=4.4cm]{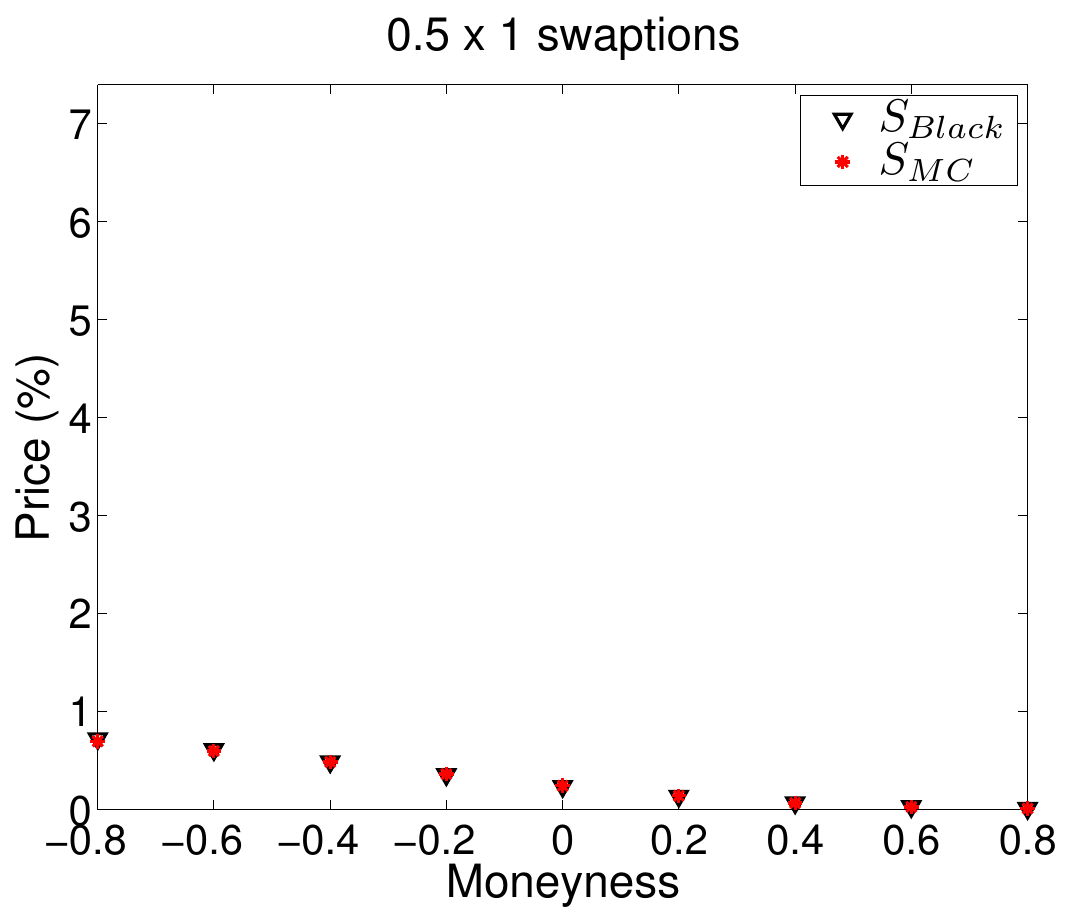}}
  \subfigure {\includegraphics[height=4.4cm]{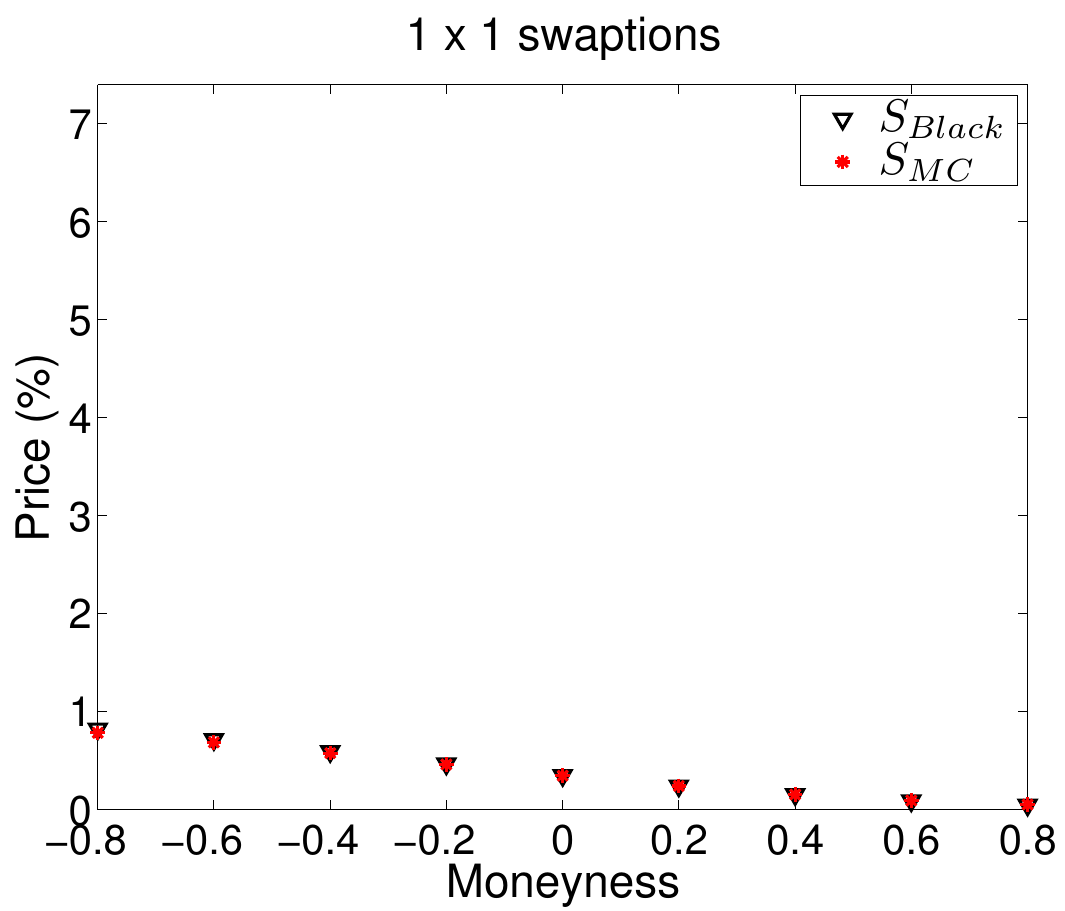}}
  \subfigure {\includegraphics[height=4.4cm]{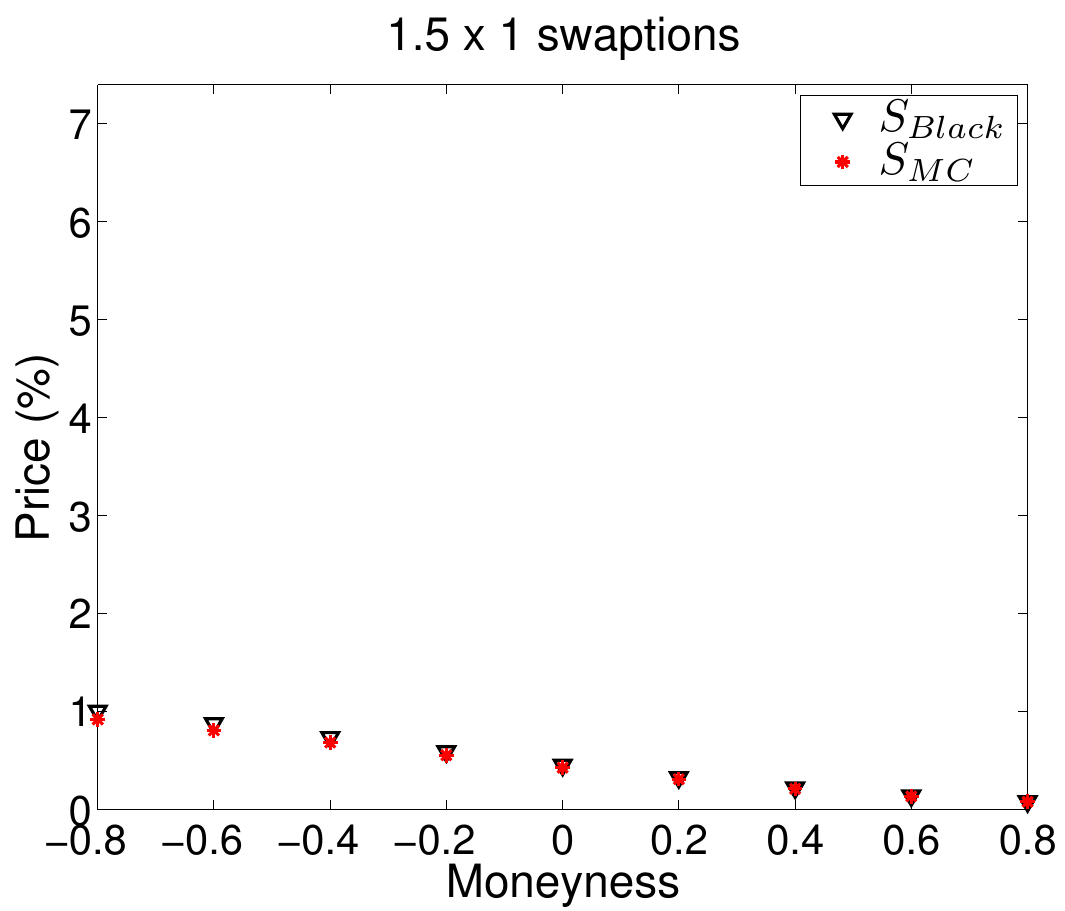}}
  \subfigure {\includegraphics[height=4.4cm]{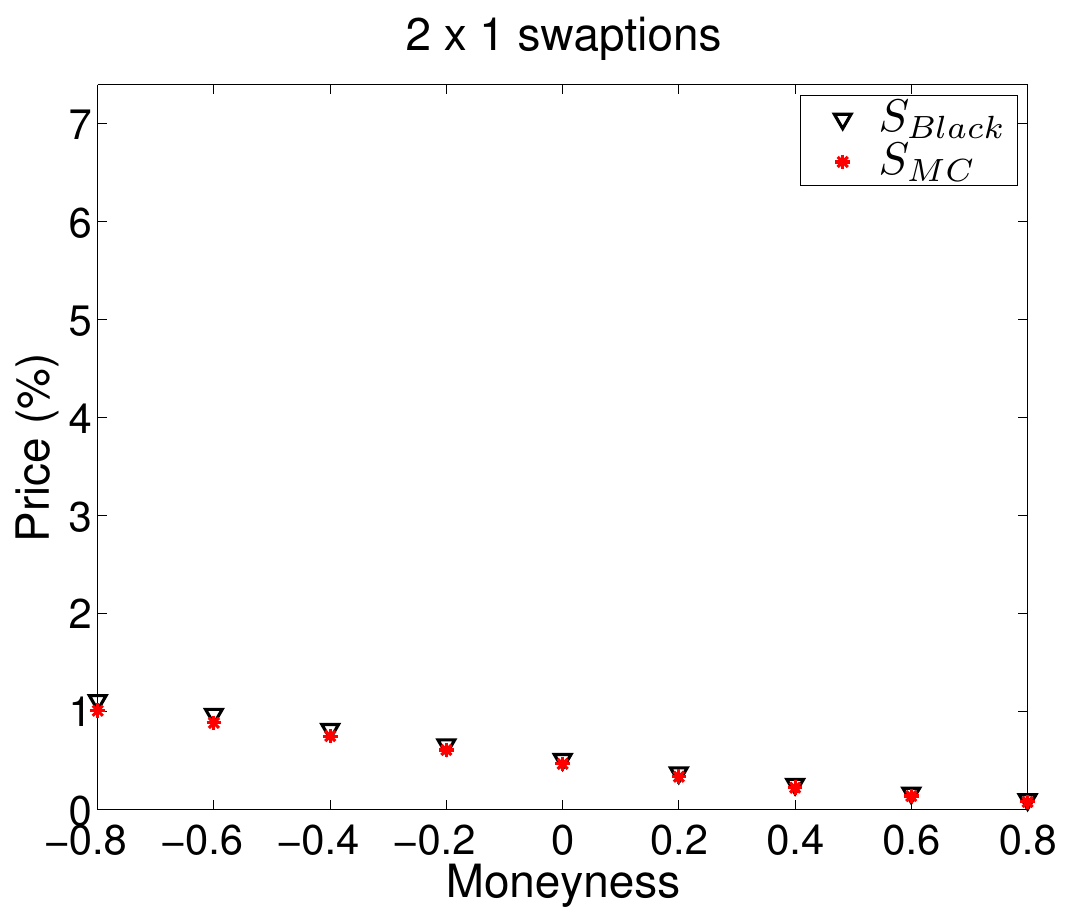}}
  \subfigure {\includegraphics[height=4.4cm]{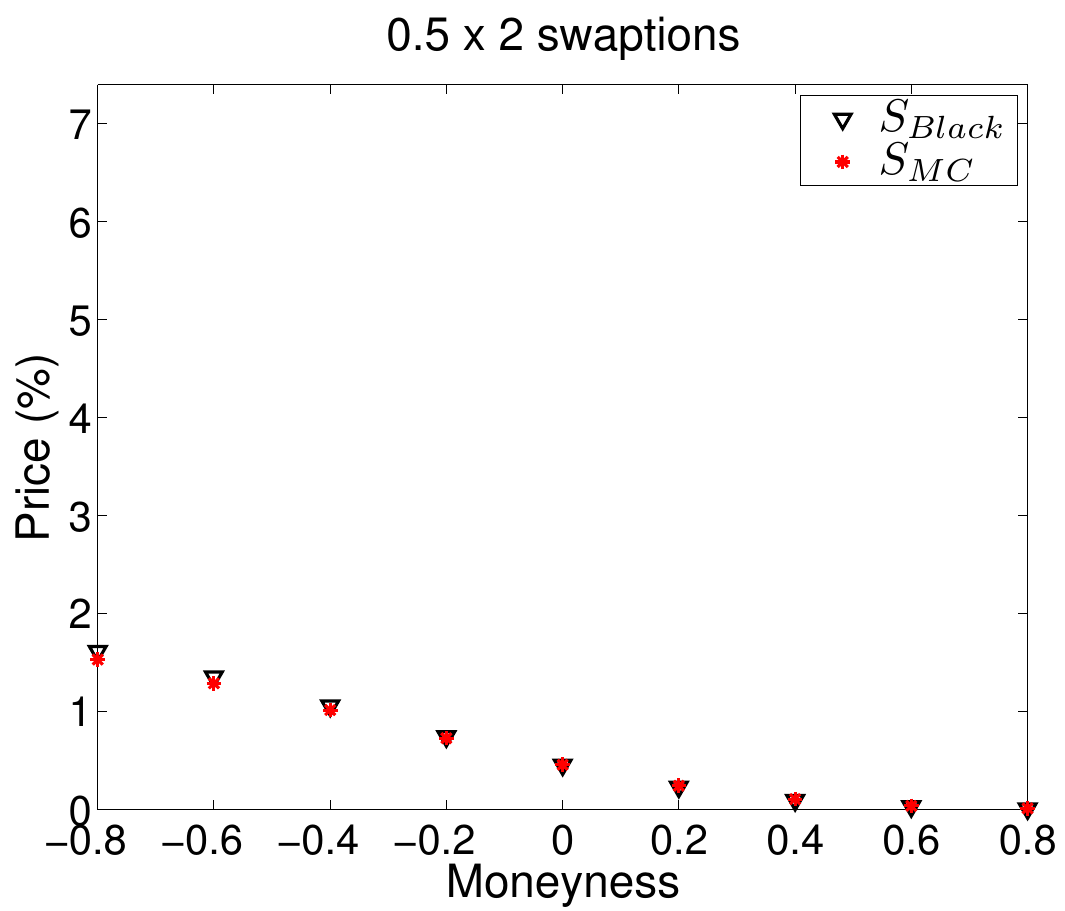}}
  \subfigure {\includegraphics[height=4.4cm]{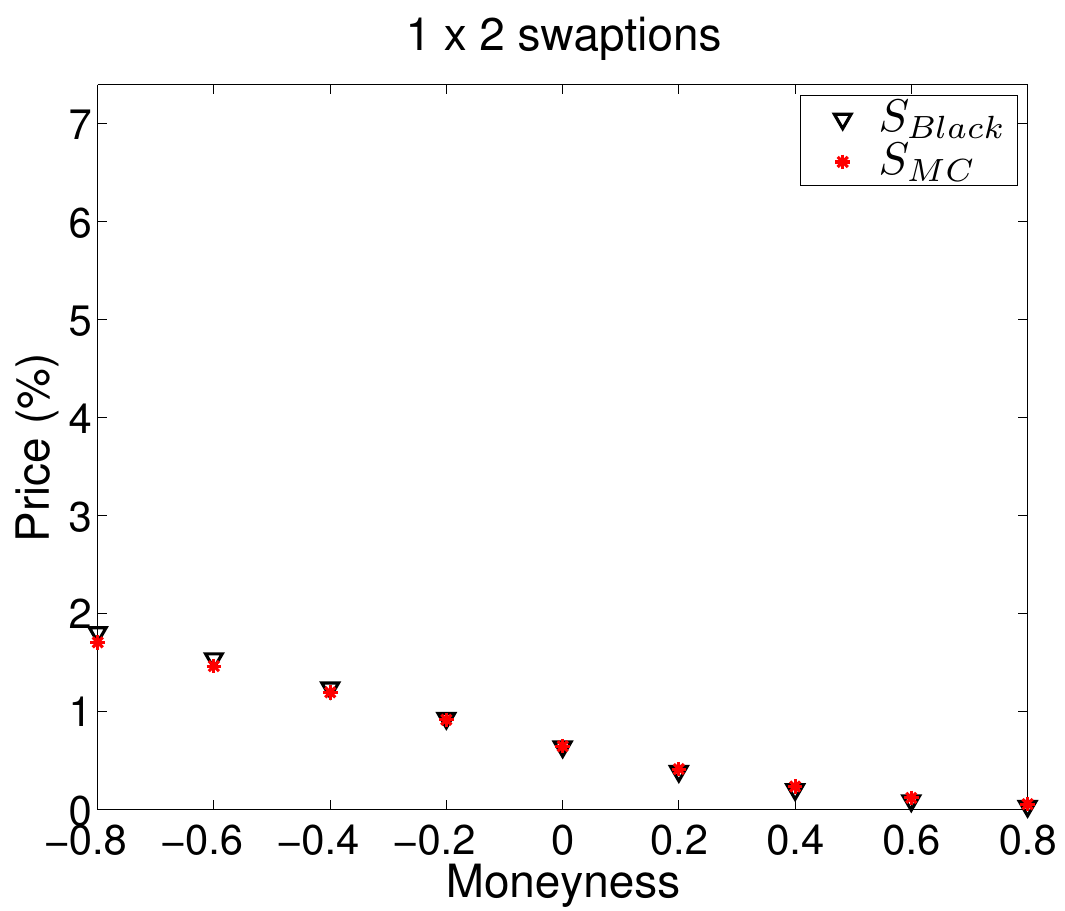}}
  \subfigure {\includegraphics[height=4.4cm]{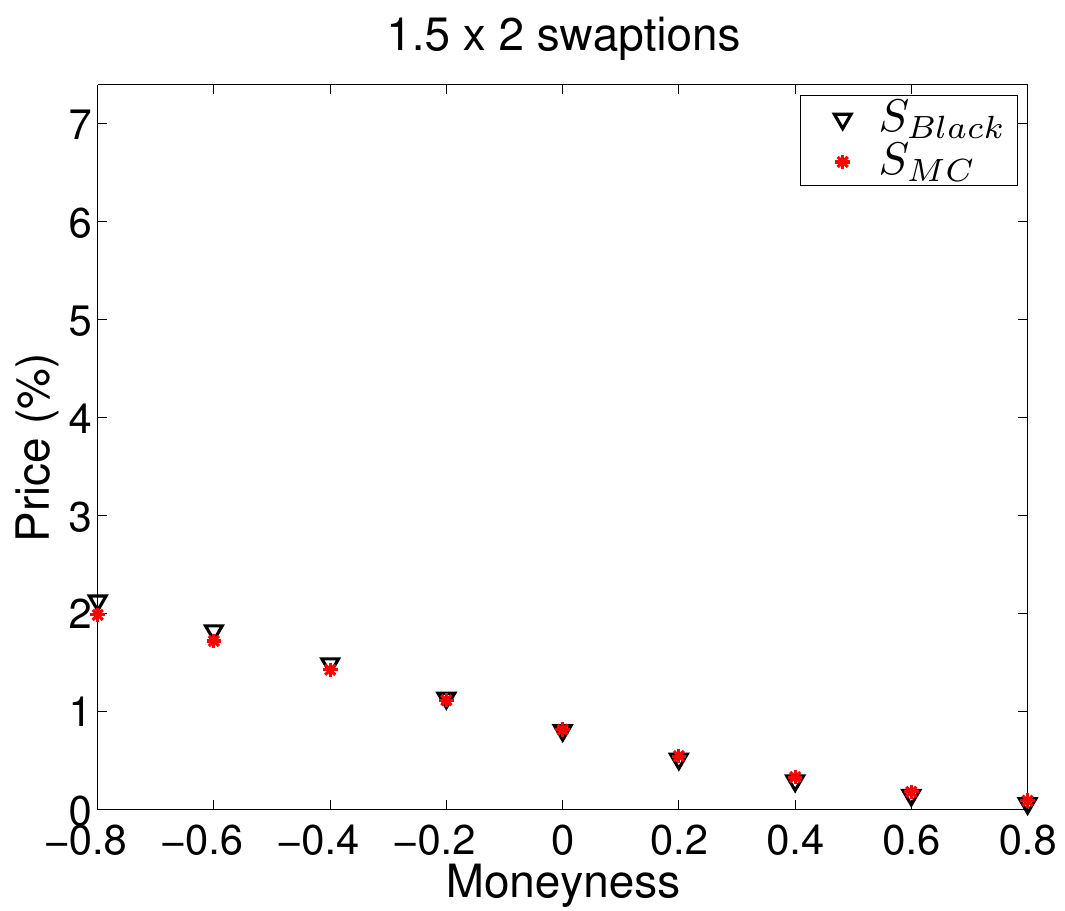}}
  \subfigure {\includegraphics[height=4.4cm]{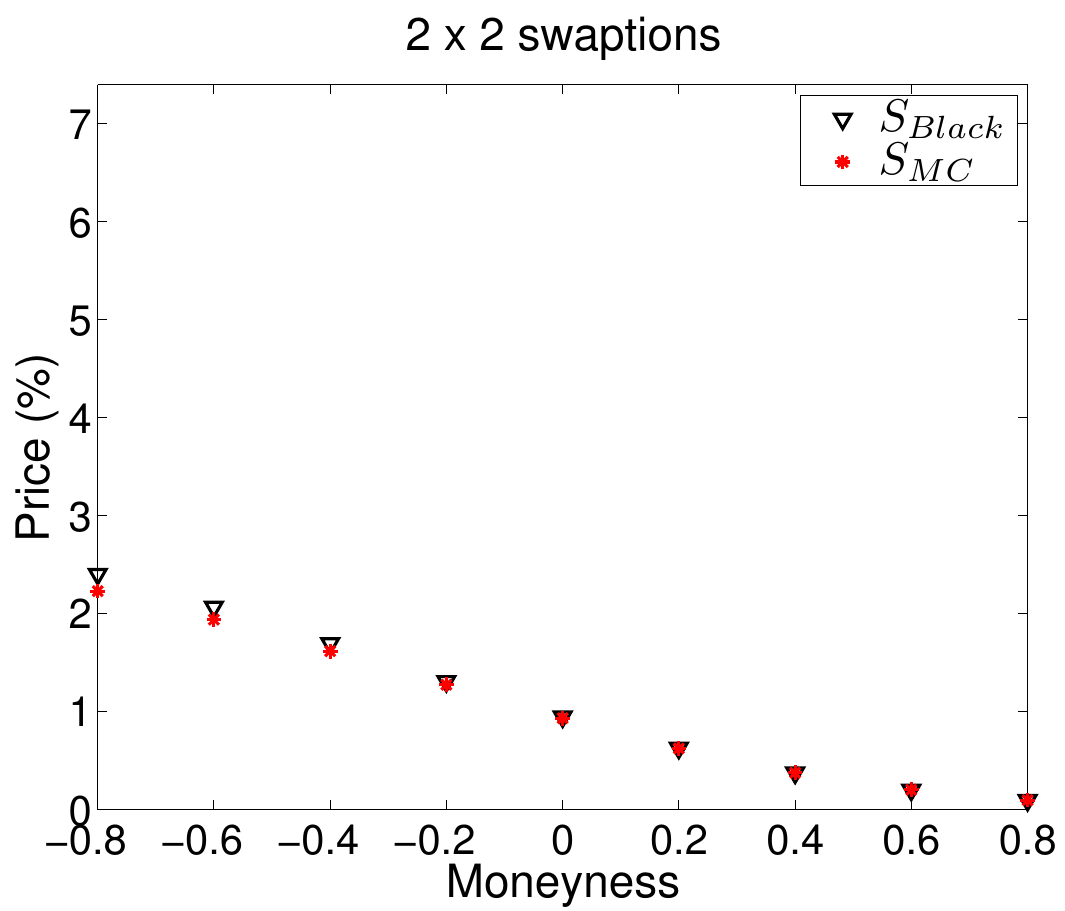}}
  \subfigure {\includegraphics[height=4.4cm]{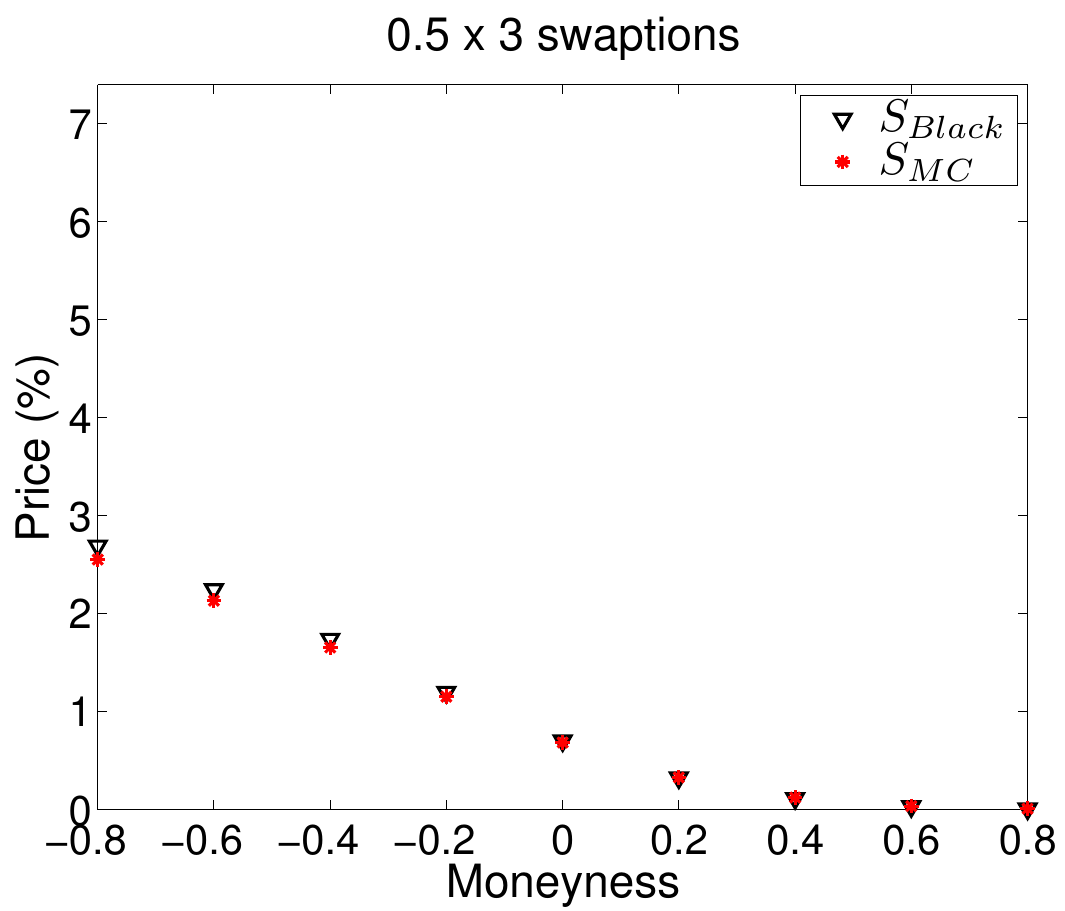}}
  \subfigure {\includegraphics[height=4.4cm]{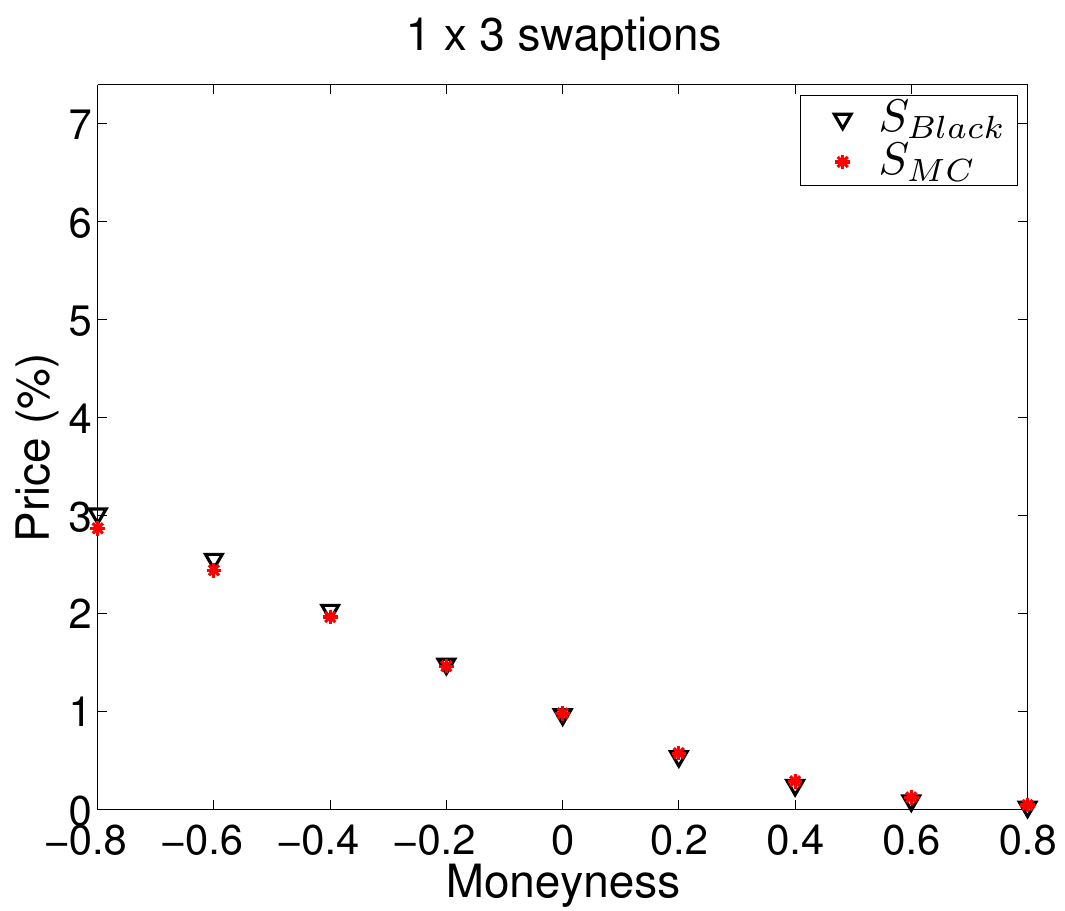}}
  \caption{Hagan model, calibration to swaptions, $S_{Black}$ vs. $S_{MC}$, part I.}
  \label{fig:haganSwaptions1}
\end{figure}

\begin{figure}[!htb]
\centering
  \subfigure {\includegraphics[height=4.4cm]{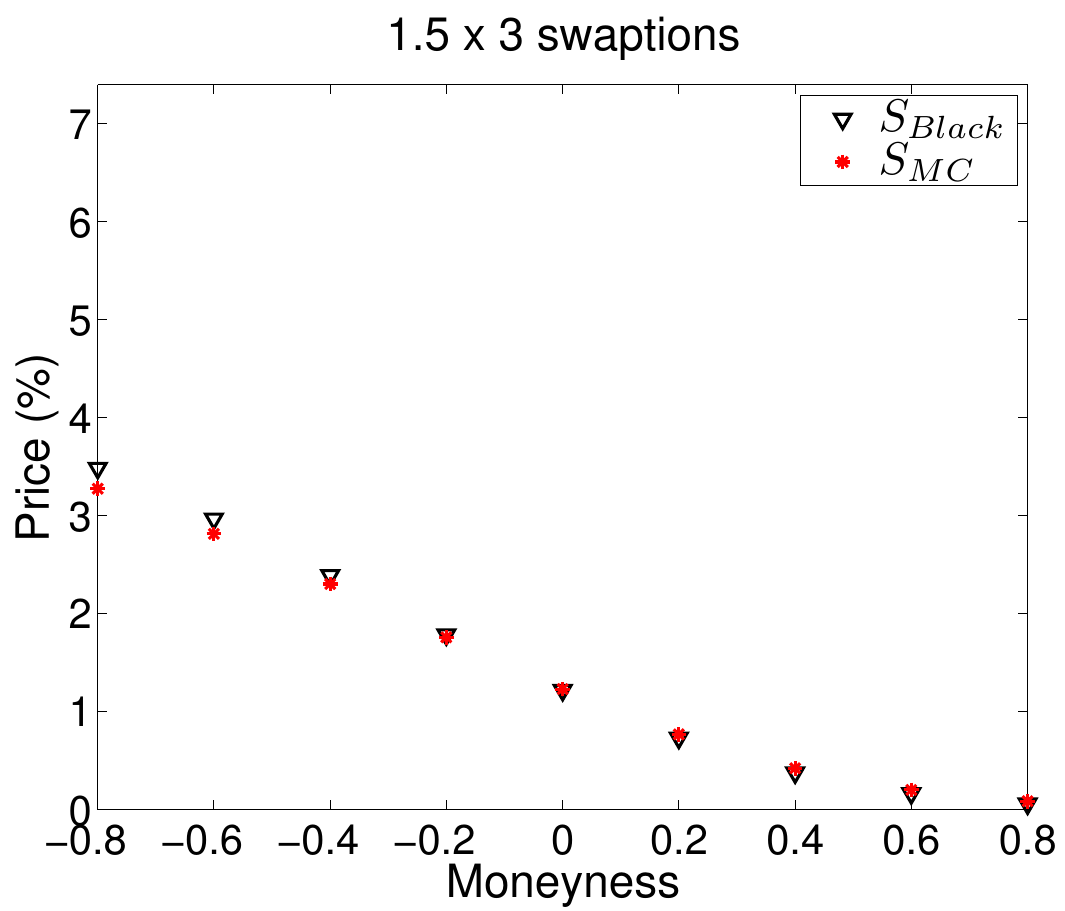}}
  \subfigure {\includegraphics[height=4.4cm]{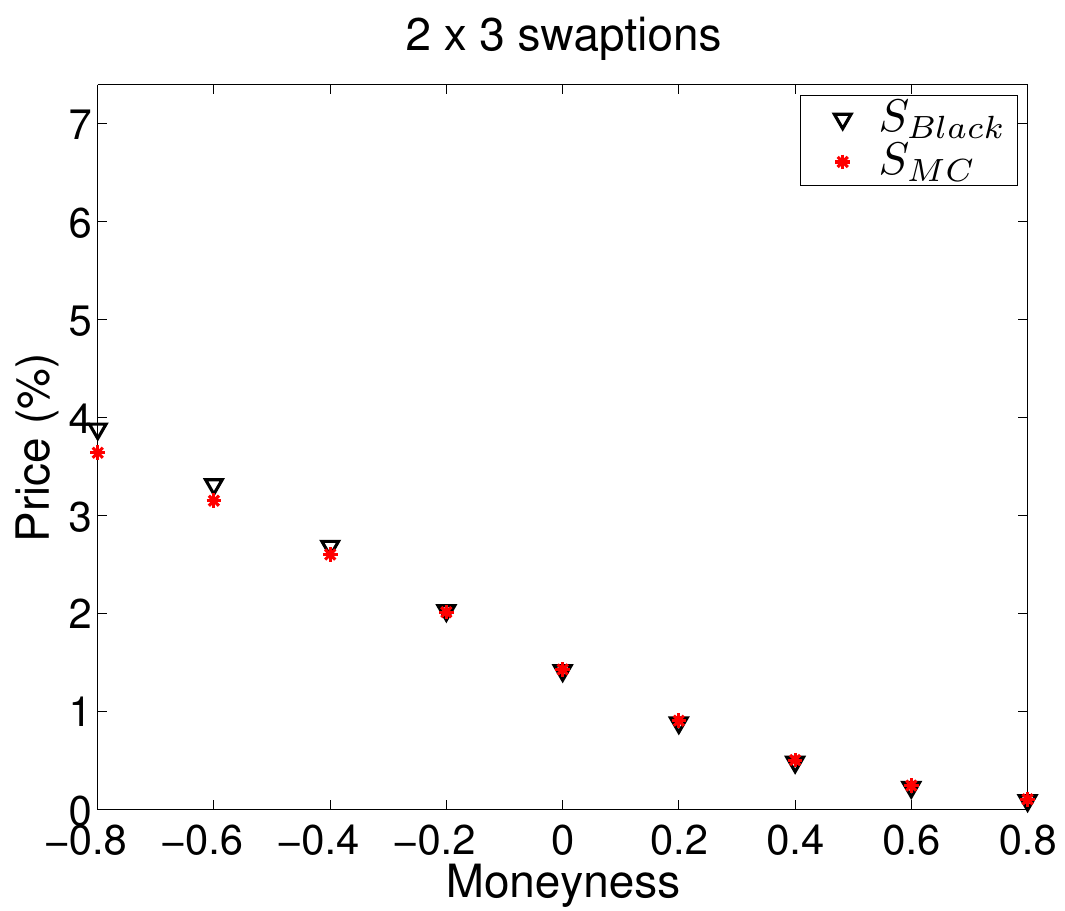}}
  \subfigure {\includegraphics[height=4.4cm]{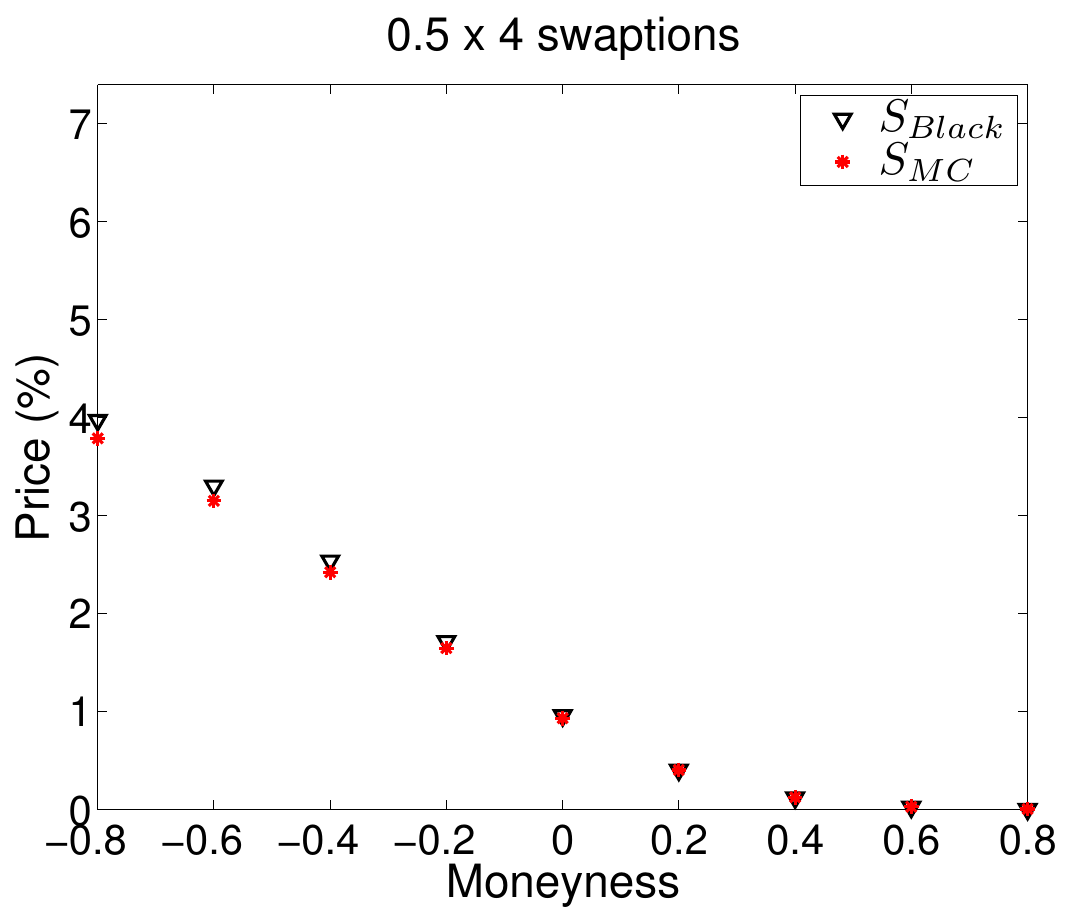}}
  \subfigure {\includegraphics[height=4.4cm]{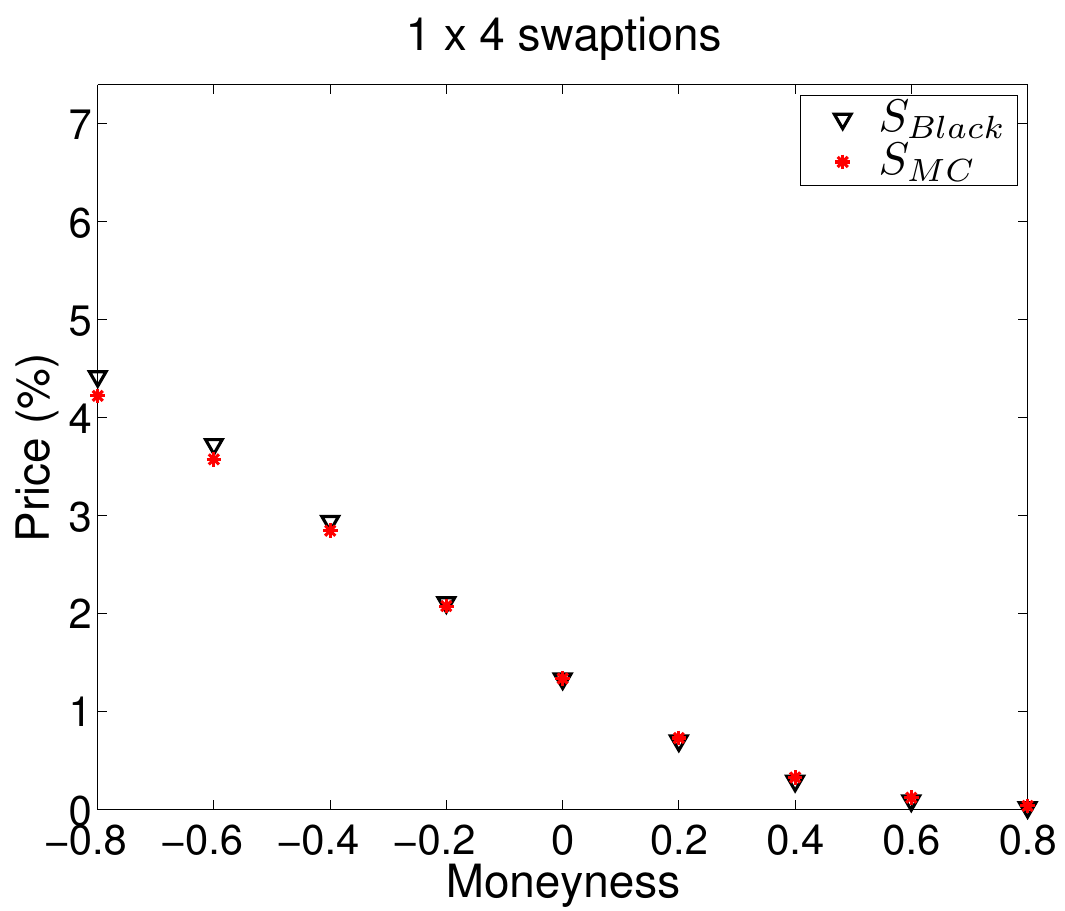}}
  \subfigure {\includegraphics[height=4.4cm]{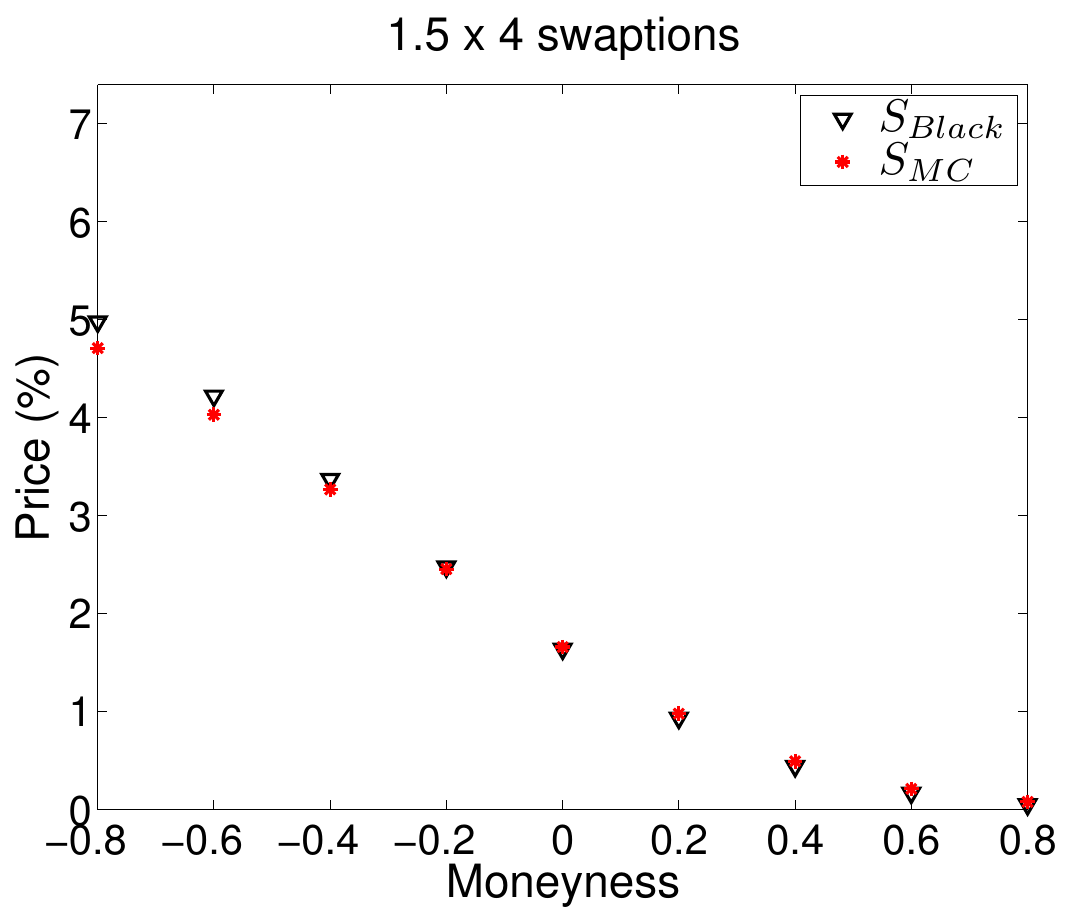}}
  \subfigure {\includegraphics[height=4.4cm]{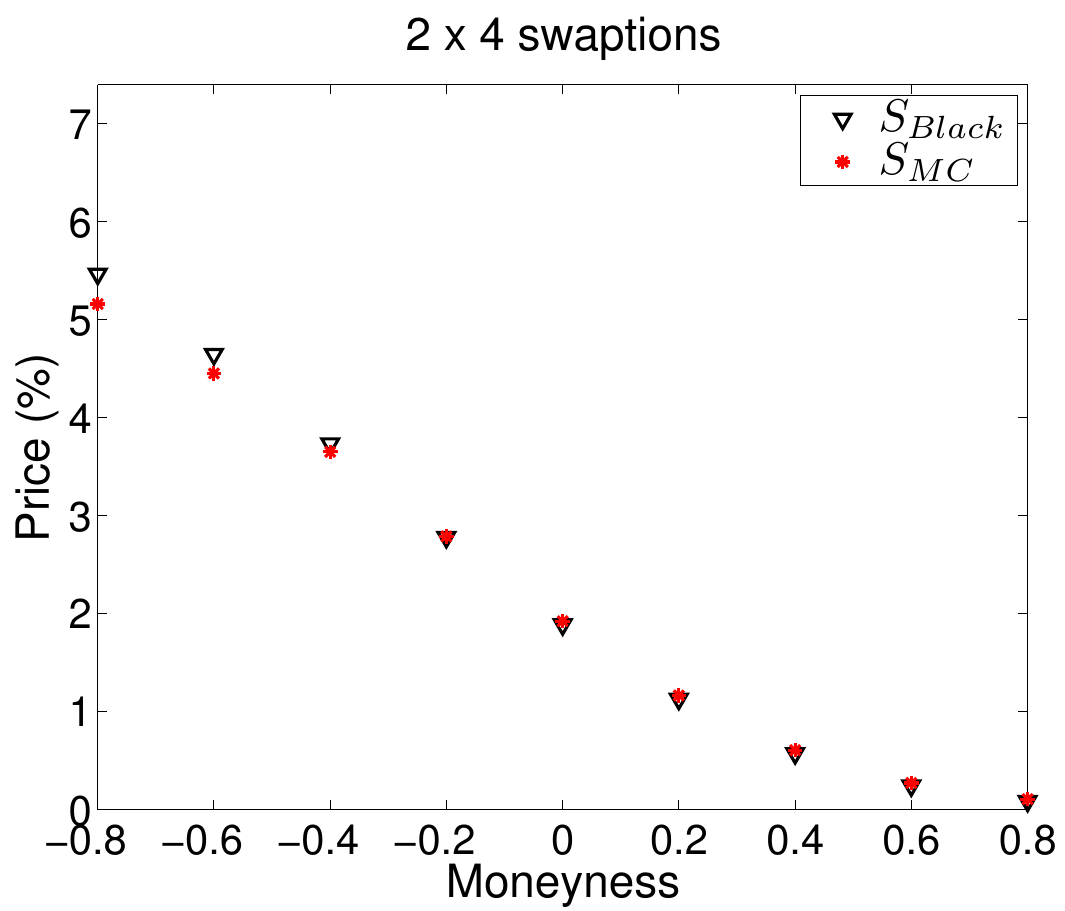}}
  \subfigure {\includegraphics[height=4.4cm]{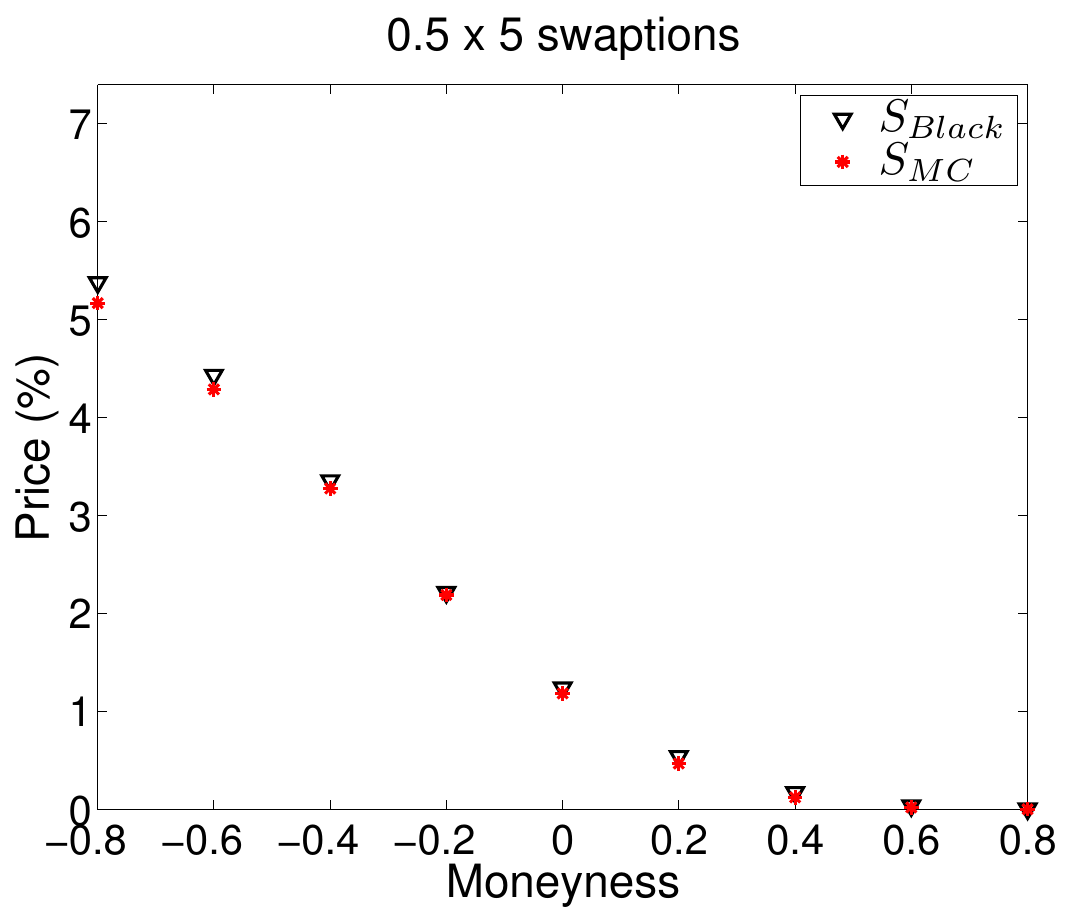}}
  \subfigure {\includegraphics[height=4.4cm]{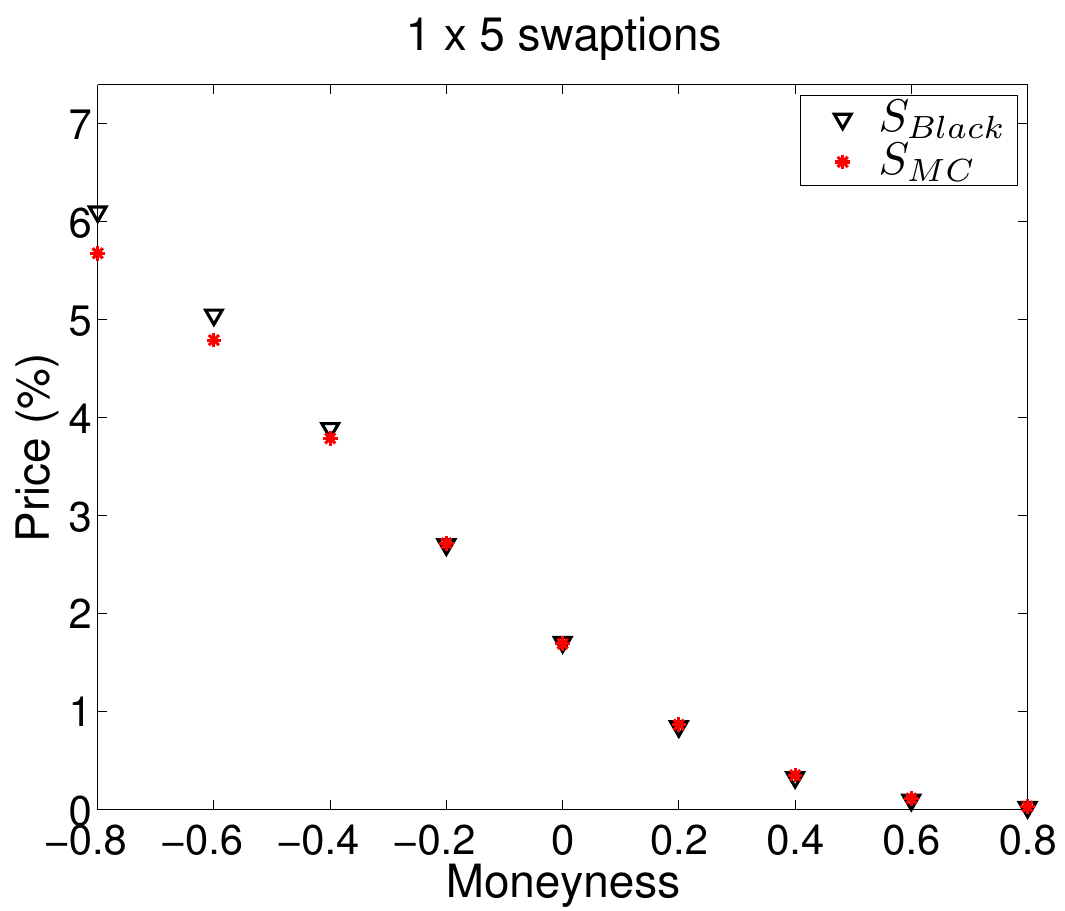}}
  \subfigure {\includegraphics[height=4.4cm]{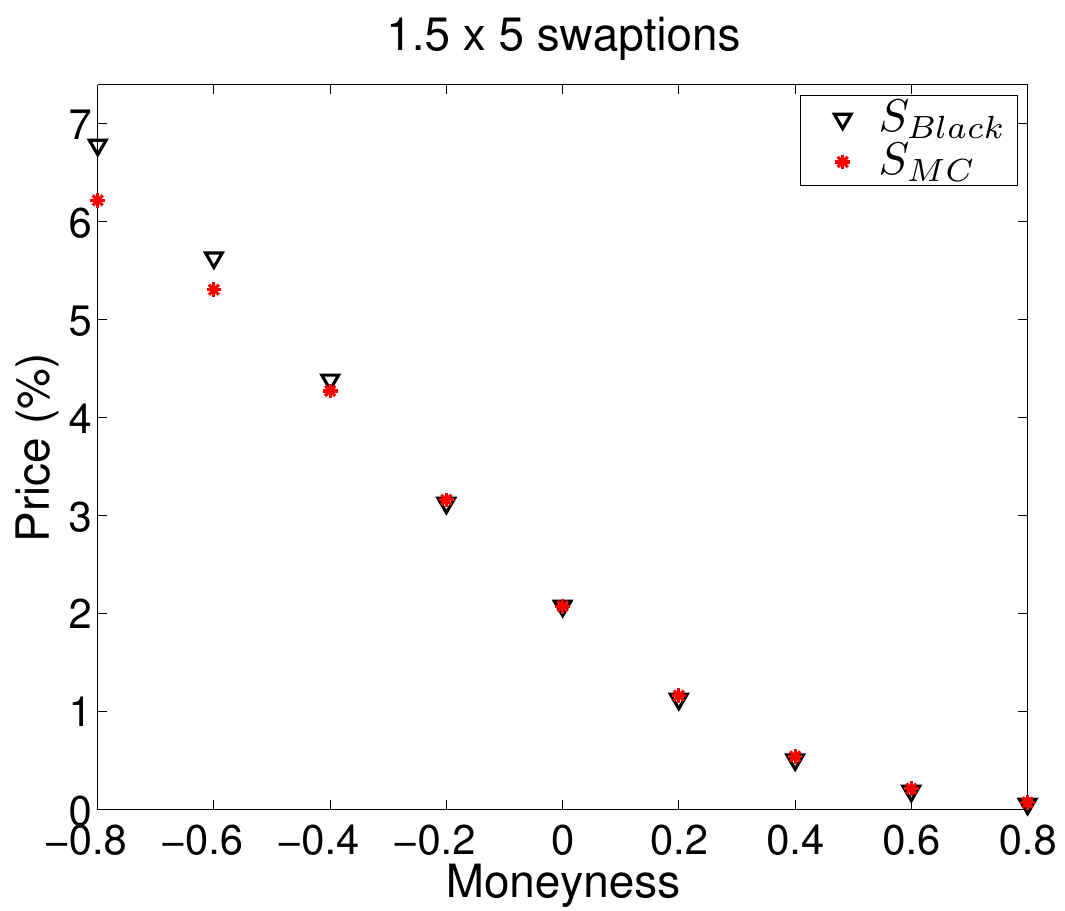}}
  \subfigure {\includegraphics[height=4.4cm]{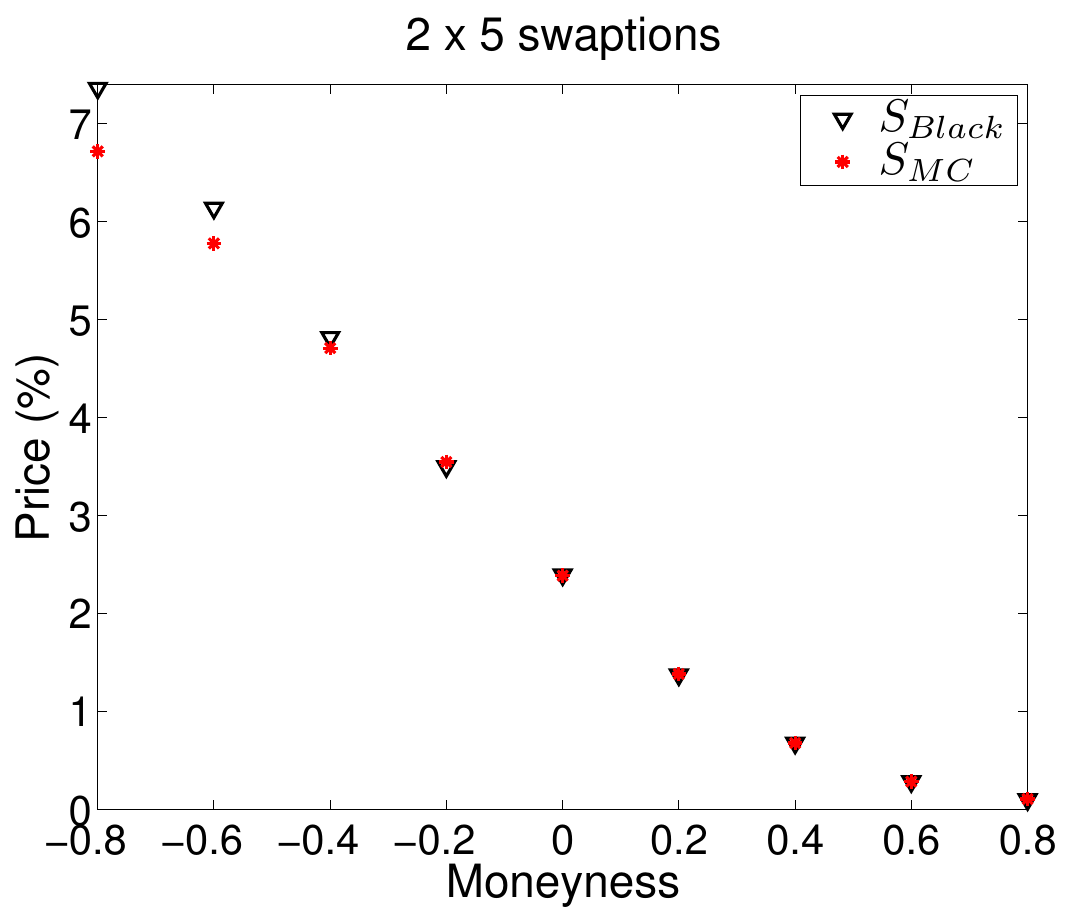}}
  \caption{Hagan model, calibration to swaptions, $S_{Black}$ vs. $S_{MC}$, part II.}
  \label{fig:haganSwaptions2}
\end{figure}

In Figures \ref{fig:haganSwaptions1} and \ref{fig:haganSwaptions2} the model fitting when considering the whole swaption matrix is shown. Market prices are shown using triangles and the model ones using stars.

In the forthcoming sections \ref{subsec:num_resul_MM} and \ref{subsec:num_resul_R}, the analogous analysis for the other two models using the same scheme for figures and tables is presented.

\clearpage
\subsection{Mercurio $\&$ Morini model} \label{subsec:num_resul_MM}

\subsubsection{Calibration to caplets}

In Table \ref{tab:paramsMercurioVolas} the calibrated parameters are shown. The execution time was $9.165$ seconds, $9.124$ seconds employed by the mono-GPU SA (launched with a relaxed configuration, specifically, $T_0=10$, $T_{min}=0.01$, $\rho=0.99$, $N=10$, $w=256\times64$, the cost function was evaluated $112738304$ times) and $0.041$ seconds to the Nelder-Mead algorithm. The speedup is very similar to the previous Hagan case.

\begin{table}[!htb]
\scriptsize{
\begin{center}
\begin{tabular}{|c|r|c||c|r|c|}
\hline
   & $\phi_{i}$ & $\alpha_i$ &  & $\phi_{i}$ & $\alpha_i$  \\
\hline
\hline
$F_1$ & $-0.7549$ & $0.0888$ & $F_8$ & $-0.3661$ & $0.0696$ \\
\hline
$F_2$ & $-0.2309$ & $0.0842$ & $F_9$ & $-0.4770$ & $0.0683$ \\
\hline
$F_3$ & $0.0666$ & $0.0817$ & $F_{10}$ & $-0.5760$ & $0.0693$ \\
\hline
$F_4$ & $0.1698$ & $0.0662$ & $F_{11}$ & $-0.6615$ & $0.0682$ \\
\hline
$F_5$ & $0.0302$ & $0.0635$ & $F_{12}$ & $-0.7380$ & $0.0682$ \\
\hline
$F_6$ & $-0.1098$ & $0.0684$ & $F_{13}$ & $-0.8044$ & $0.0669$ \\
\hline
$F_7$ & $-0.2417$ & $0.0667$ &  &  &  \\
\hline
\hline
\multicolumn{6}{|c|}{$\sigma=0.5986$} \\
\hline
\end{tabular}
\caption{Mercurio \& Morini model, calibration to caplets with SABR formula \eqref{eq:sigmaHagan}: calibrated parameters.} \label{tab:paramsMercurioVolas}
\end{center}
}
\end{table}

In Table \ref{tab:mercurioVolas} market vs. model volatilities (both in $\%$) for the first twelve smiles and the moneyness varying from $-40 \%$ to $40 \%$ are shown. In addition, the mean relative error ($MRE$) considering the whole set of smiles is presented.

\begin{table}[!htb]
\scriptsize{
\centering
\begin{tabular}{|r|| c|r|c ||c|c|c |}
\hline
Moneyness & \multicolumn{3}{|c||}{Smile of $F_{1}$}& \multicolumn{3}{|c|}{Smile of $F_{2}$} \\
\hline
& $\sigma_{market}$ & $\sigma_{model}$ & $\frac{|\sigma_{market}-\sigma_{model}|}{\sigma_{market}}$ & $\sigma_{market}$ & $\sigma_{model}$ &$\frac{|\sigma_{market}-\sigma_{model}|}{\sigma_{market}}$ \\ \hline
$-40\%$ & $97.26$ & $102.19$ & $5.07 \times 10^{-2}$ & $88.27$ & $89.59$ & $1.50 \times 10^{-2}$ \\
\hline
$-20\%$ & $82.58$ & $90.71$ & $9.85 \times 10^{-2}$ & $79.62$ & $81.81$ & $2.75 \times 10^{-2}$ \\
\hline
$0\%$ & $72.29$ & $81.16$ & $1.23 \times 10^{-1}$ & $73.03$ & $75.77$ & $3.74 \times 10^{-2}$ \\
\hline
$20\%$ & $70.89$ & $73.55$ & $3.76 \times 10^{-2}$ & $71.95$ & $71.47$ & $6.69 \times 10^{-3}$ \\
\hline
$40\%$ & $69.49$ & $67.88$ & $2.31 \times 10^{-2}$ & $70.87$ & $68.91$ & $2.77 \times 10^{-2}$ \\
\hline
\hline
Moneyness & \multicolumn{3}{|c||}{Smile of $F_{3}$}& \multicolumn{3}{|c|}{Smile of $F_{4}$} \\
\hline
& $\sigma_{market}$ & $\sigma_{model}$ & $\frac{|\sigma_{market}-\sigma_{model}|}{\sigma_{market}}$ & $\sigma_{market}$ & $\sigma_{model}$ &$\frac{|\sigma_{market}-\sigma_{model}|}{\sigma_{market}}$ \\ \hline
$-40\%$ & $77.09$ & $77.13$ & $4.45 \times 10^{-4}$ & $57.08$ & $55.98$ & $1.92 \times 10^{-2}$ \\
\hline
$-20\%$ & $71.50$ & $71.99$ & $6.92 \times 10^{-3}$ & $53.21$ & $52.54$ & $1.26 \times 10^{-2}$ \\
\hline
$0\%$ & $67.93$ & $68.27$ & $5.11 \times 10^{-3}$ & $52.49$ & $50.39$ & $4.00 \times 10^{-2}$ \\
\hline
$20\%$ & $67.10$ & $65.96$ & $1.69 \times 10^{-2}$ & $51.34$ & $49.53$ & $3.51 \times 10^{-2}$ \\
\hline
$40\%$ & $66.41$ & $65.07$ & $2.03 \times 10^{-2}$ & $50.61$ & $49.97$ & $1.27 \times 10^{-2}$ \\
\hline
\hline
Moneyness & \multicolumn{3}{|c||}{Smile of $F_{5}$}& \multicolumn{3}{|c|}{Smile of $F_{6}$} \\
\hline
& $\sigma_{market}$ & $\sigma_{model}$ & $\frac{|\sigma_{market}-\sigma_{model}|}{\sigma_{market}}$ & $\sigma_{market}$ & $\sigma_{model}$ &$\frac{|\sigma_{market}-\sigma_{model}|}{\sigma_{market}}$ \\ \hline
$-40\%$ & $56.69$ & $55.76$ & $1.65 \times 10^{-2}$ & $56.30$ & $55.70$ & $1.08 \times 10^{-2}$ \\
\hline
$-20\%$ & $52.43$ & $51.25$ & $2.25 \times 10^{-2}$ & $51.65$ & $50.20$ & $2.81 \times 10^{-2}$ \\
\hline
$0\%$ & $50.31$ & $48.26$ & $4.08 \times 10^{-2}$ & $48.19$ & $46.38$ & $3.77 \times 10^{-2}$ \\
\hline
$20\%$ & $48.72$ & $46.79$ & $3.96 \times 10^{-2}$ & $46.19$ & $44.25$ & $4.21 \times 10^{-2}$ \\
\hline
$40\%$ & $47.70$ & $46.83$ & $1.82 \times 10^{-2}$ & $44.91$ & $43.79$ & $2.47 \times 10^{-2}$ \\
\hline
\hline
Moneyness & \multicolumn{3}{|c||}{Smile of $F_{7}$}& \multicolumn{3}{|c|}{Smile of $F_{8}$} \\
\hline
& $\sigma_{market}$ & $\sigma_{model}$ & $\frac{|\sigma_{market}-\sigma_{model}|}{\sigma_{market}}$ & $\sigma_{market}$ & $\sigma_{model}$ &$\frac{|\sigma_{market}-\sigma_{model}|}{\sigma_{market}}$ \\ \hline
$-40\%$ & $55.92$ & $55.77$ & $2.78 \times 10^{-3}$ & $55.54$ & $55.93$ & $7.03 \times 10^{-3}$ \\
\hline
$-20\%$ & $50.89$ & $49.40$ & $2.92 \times 10^{-2}$ & $50.13$ & $48.83$ & $2.60 \times 10^{-2}$ \\
\hline
$0\%$ & $46.19$ & $44.82$ & $2.97 \times 10^{-2}$ & $44.25$ & $43.55$ & $1.58 \times 10^{-2}$ \\
\hline
$20\%$ & $43.83$ & $42.03$ & $4.12 \times 10^{-2}$ & $41.56$ & $40.09$ & $3.54 \times 10^{-2}$ \\
\hline
$40\%$ & $42.32$ & $41.02$ & $3.08 \times 10^{-2}$ & $39.84$ & $38.45$ & $3.49 \times 10^{-2}$ \\
\hline
\hline
Moneyness & \multicolumn{3}{|c||}{Smile of $F_{9}$}& \multicolumn{3}{|c|}{Smile of $F_{10}$} \\
\hline
& $\sigma_{market}$ & $\sigma_{model}$ & $\frac{|\sigma_{market}-\sigma_{model}|}{\sigma_{market}}$ & $\sigma_{market}$ & $\sigma_{model}$ &$\frac{|\sigma_{market}-\sigma_{model}|}{\sigma_{market}}$ \\ \hline
$-40\%$ & $55.16$ & $56.14$ & $1.78 \times 10^{-2}$ & $54.78$ & $56.39$ & $2.94 \times 10^{-2}$ \\
\hline
$-20\%$ & $49.39$ & $48.45$ & $1.91 \times 10^{-2}$ & $48.65$ & $48.23$ & $8.78 \times 10^{-3}$ \\
\hline
$0\%$ & $42.40$ & $42.56$ & $3.80 \times 10^{-3}$ & $40.61$ & $41.81$ & $2.95 \times 10^{-2}$ \\
\hline
$20\%$ & $39.43$ & $38.48$ & $2.39 \times 10^{-2}$ & $37.38$ & $37.15$ & $6.21 \times 10^{-3}$ \\
\hline
$40\%$ & $37.54$ & $36.21$ & $3.53 \times 10^{-2}$ & $35.34$ & $34.24$ & $3.12 \times 10^{-2}$ \\
\hline
\hline
Moneyness & \multicolumn{3}{|c||}{Smile of $F_{11}$}& \multicolumn{3}{|c|}{Smile of $F_{12}$} \\
\hline
& $\sigma_{market}$ & $\sigma_{model}$ & $\frac{|\sigma_{market}-\sigma_{model}|}{\sigma_{market}}$ & $\sigma_{market}$ & $\sigma_{model}$ &$\frac{|\sigma_{market}-\sigma_{model}|}{\sigma_{market}}$ \\ \hline
$-40\%$ & $54.41$ & $56.66$ & $4.14 \times 10^{-2}$ & $54.03$ & $56.96$ & $5.41 \times 10^{-2}$ \\
\hline
$-20\%$ & $47.94$ & $48.13$ & $3.97 \times 10^{-3}$ & $47.22$ & $48.13$ & $1.93 \times 10^{-2}$ \\
\hline
$0\%$ & $38.93$ & $41.27$ & $6.02 \times 10^{-2}$ & $37.29$ & $40.89$ & $9.65 \times 10^{-2}$ \\
\hline
$20\%$ & $35.47$ & $36.08$ & $1.72 \times 10^{-2}$ & $33.63$ & $35.21$ & $4.70 \times 10^{-2}$ \\
\hline
$40\%$ & $33.30$ & $32.57$ & $2.22 \times 10^{-2}$ & $31.36$ & $31.11$ & $7.79 \times 10^{-3}$ \\
\hline
\hline
\multicolumn{7}{|c|}{$MRE=3.11 \times 10^{-2}$} \\
\hline
\end{tabular}
\caption{Mercurio \& Morini model, calibration to caplets, $\sigma_{market}$ vs. $\sigma_{model}$.}
\label{tab:mercurioVolas}
}
\end{table}

We also performed the equivalent calibration with Monte Carlo simulation thus obtaining the same parameters as in Table \ref{tab:paramsMercurioVolas}, except for $\phi_{1}=-0.5714$. We note that the $MAE$ in prices is $3.83 \times 10^{-2}$ with formula \eqref{eq:sigmaHagan}, while $MAE$ is $3.84 \times 10^{-2}$ using Monte Carlo.

\begin{figure}[!htb]
\centering
  \subfigure {\includegraphics[height=3.7cm]{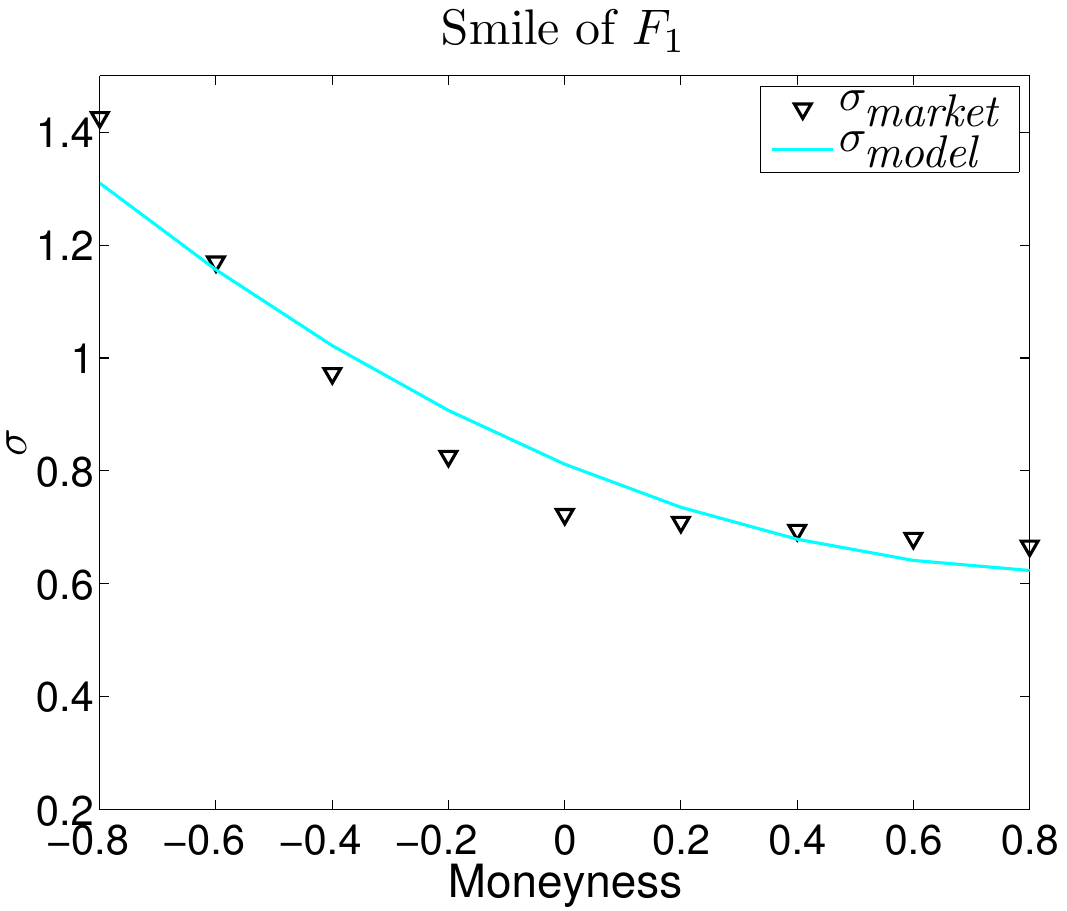}}
  \subfigure {\includegraphics[height=3.7cm]{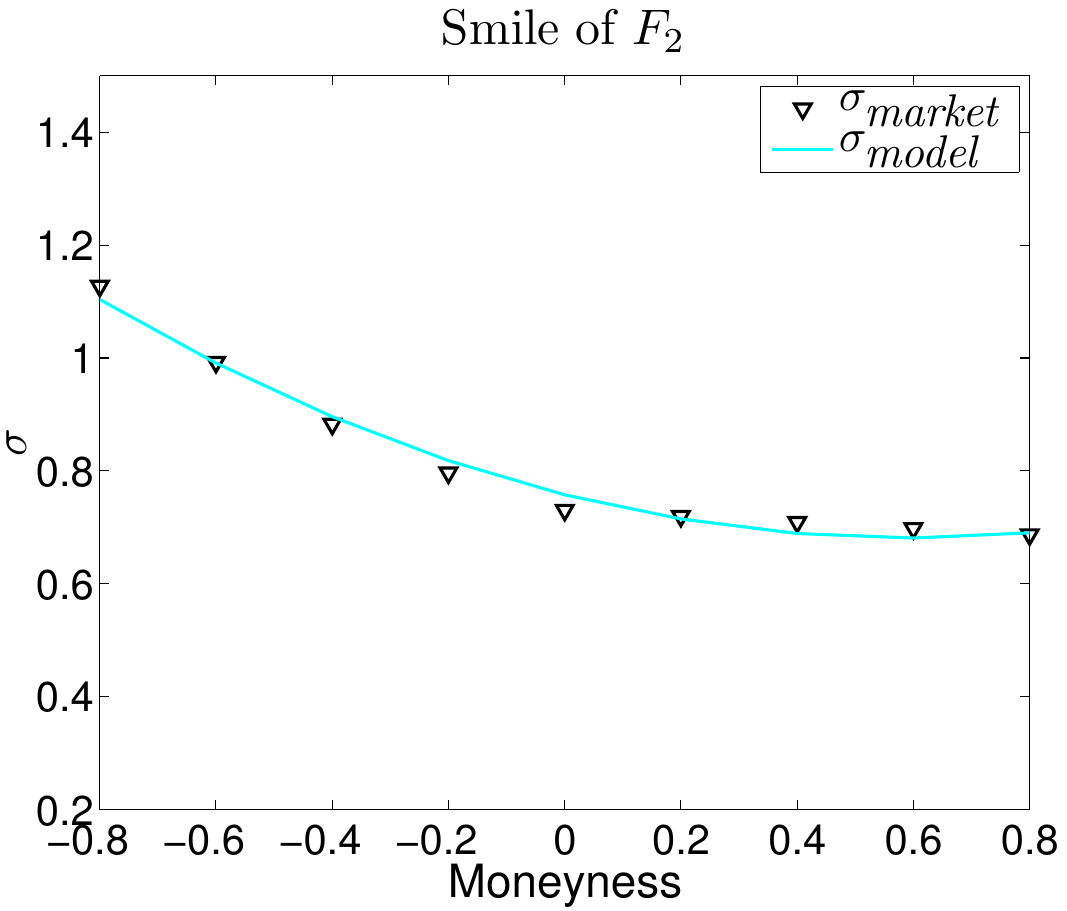}}
  \subfigure {\includegraphics[height=3.7cm]{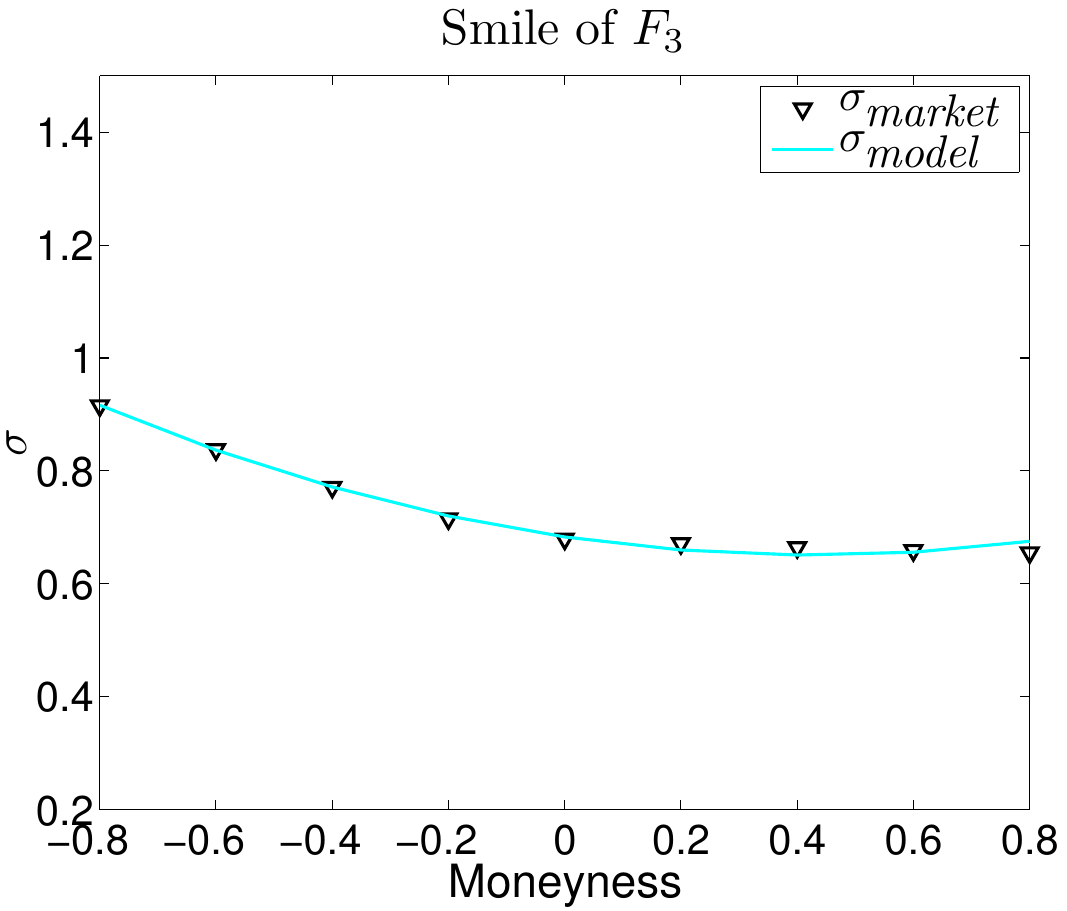}}
  \subfigure {\includegraphics[height=3.7cm]{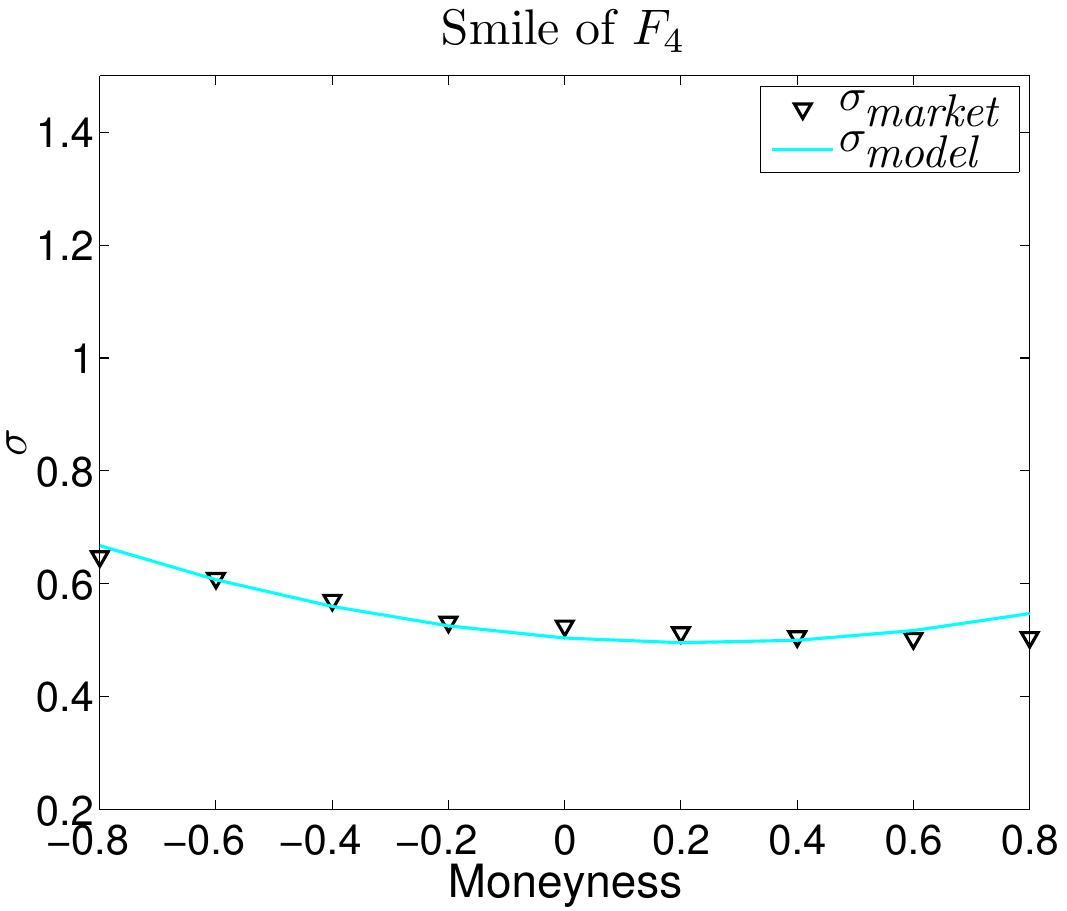}}
  \subfigure {\includegraphics[height=3.7cm]{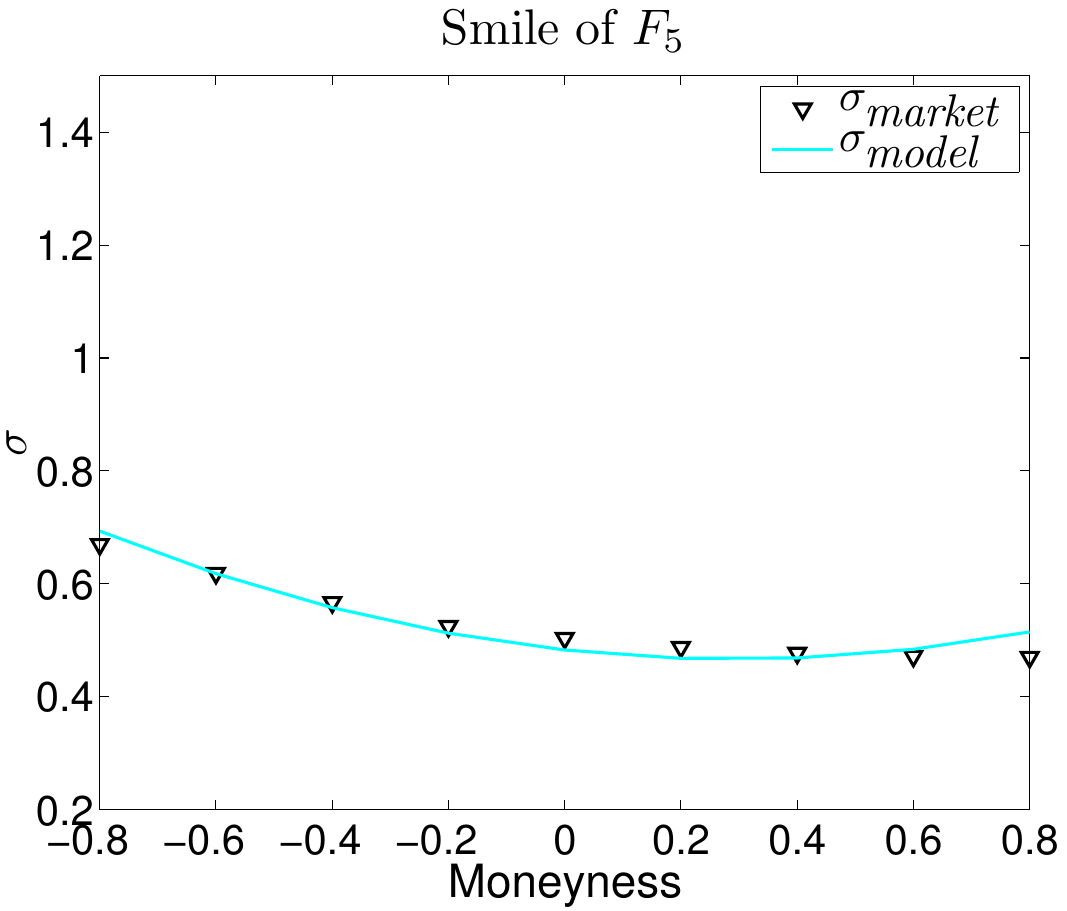}}
  \subfigure {\includegraphics[height=3.7cm]{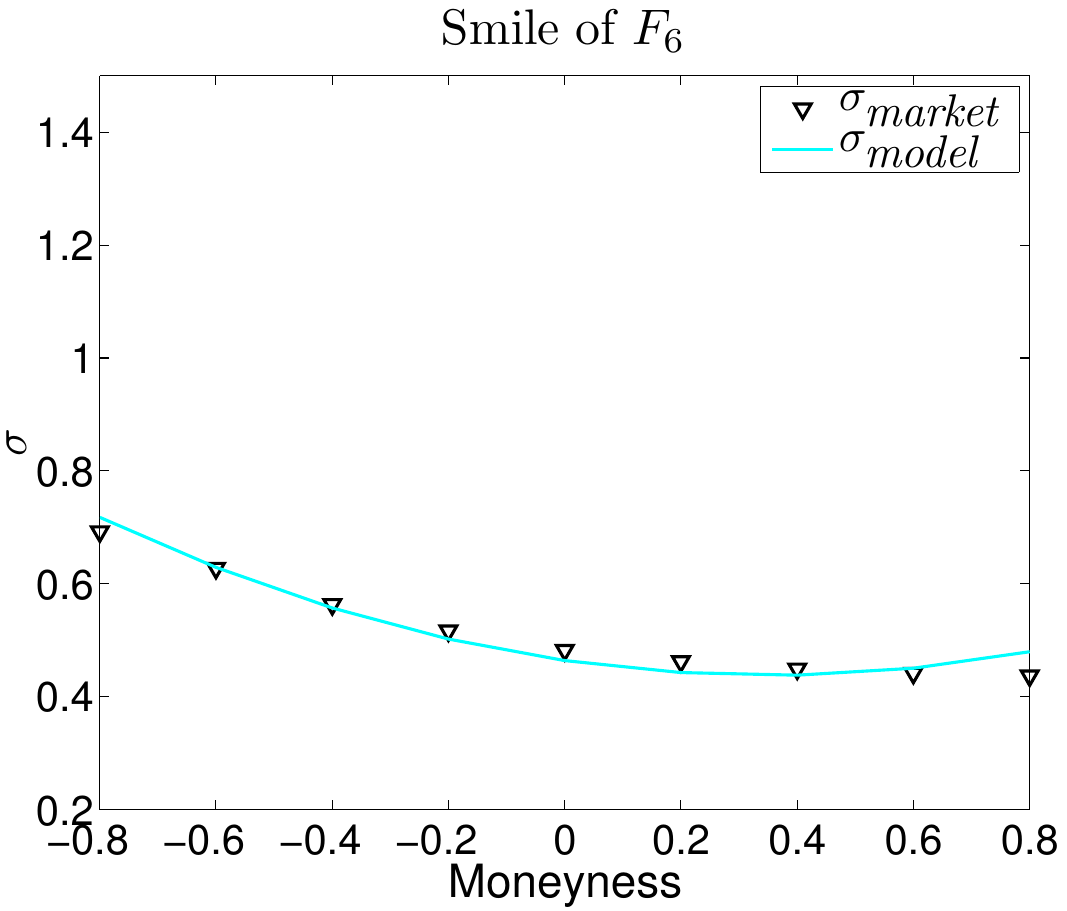}}
  \subfigure {\includegraphics[height=3.7cm]{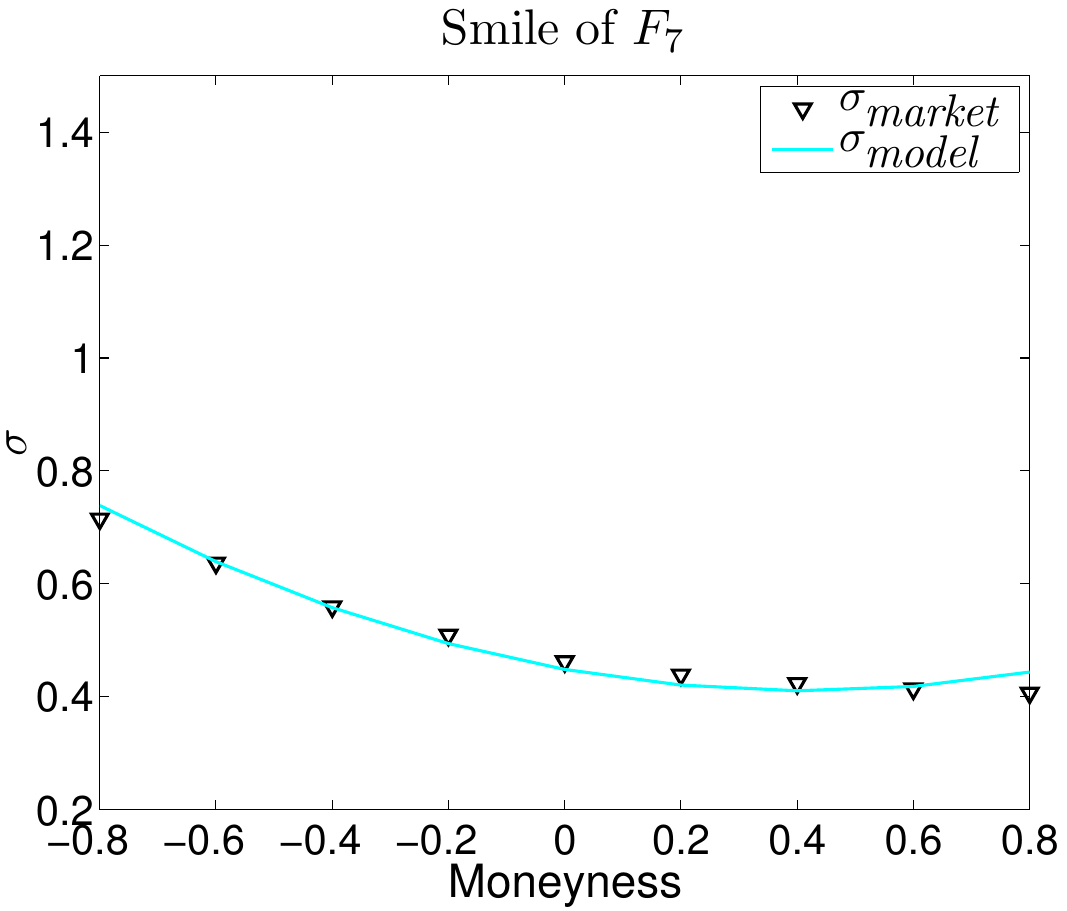}}
  \subfigure {\includegraphics[height=3.7cm]{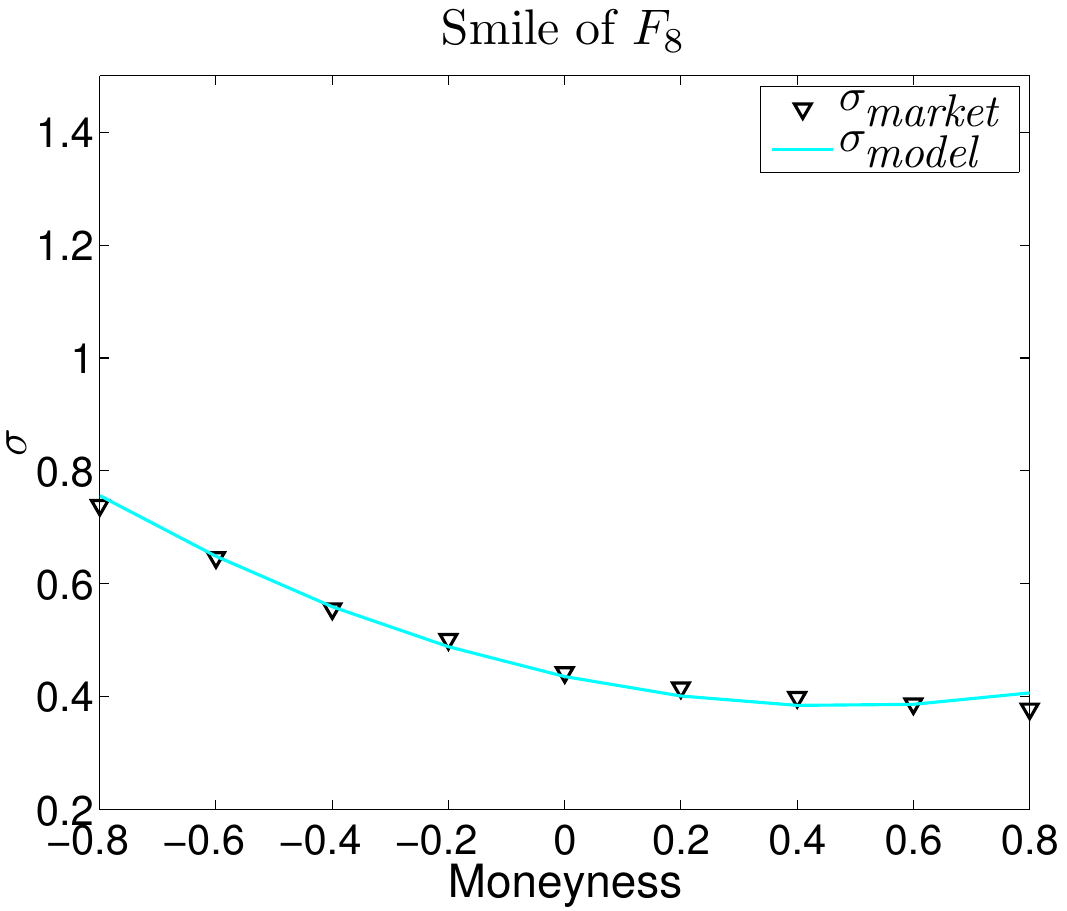}}
  \subfigure {\includegraphics[height=3.7cm]{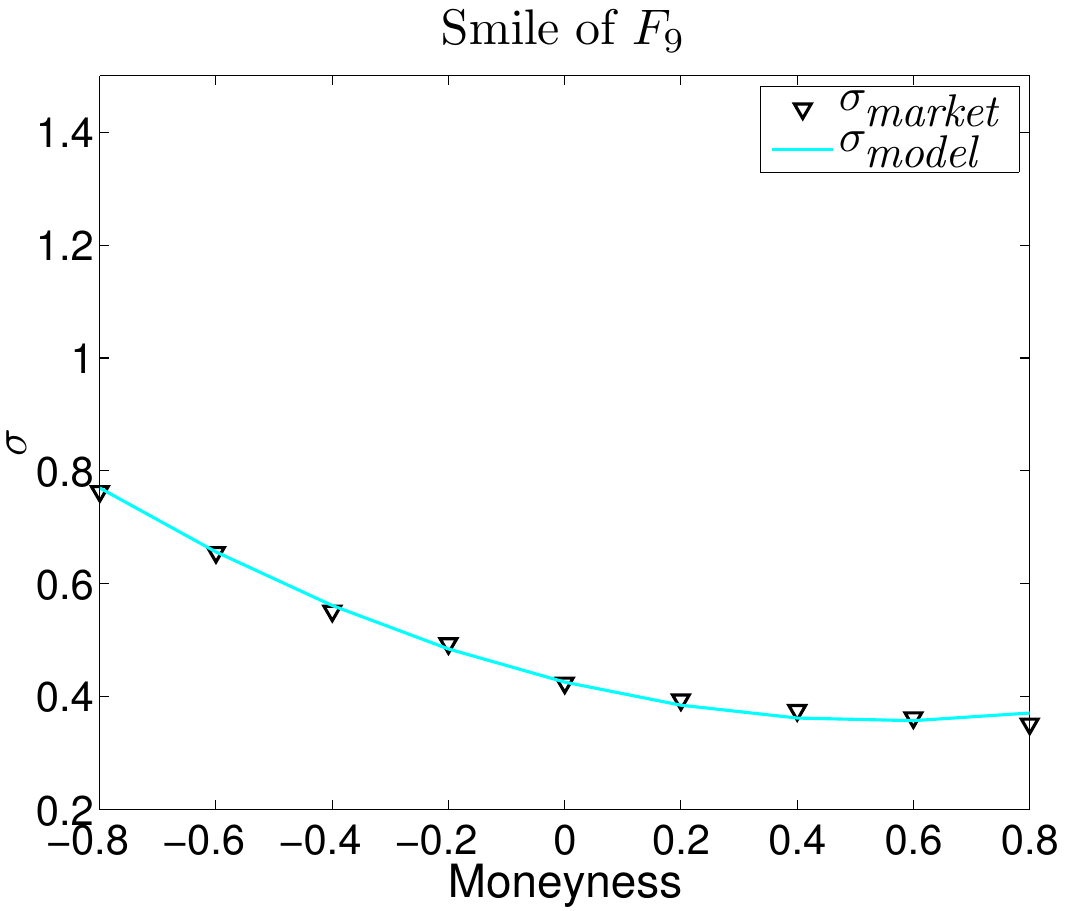}}
  \subfigure {\includegraphics[height=3.7cm]{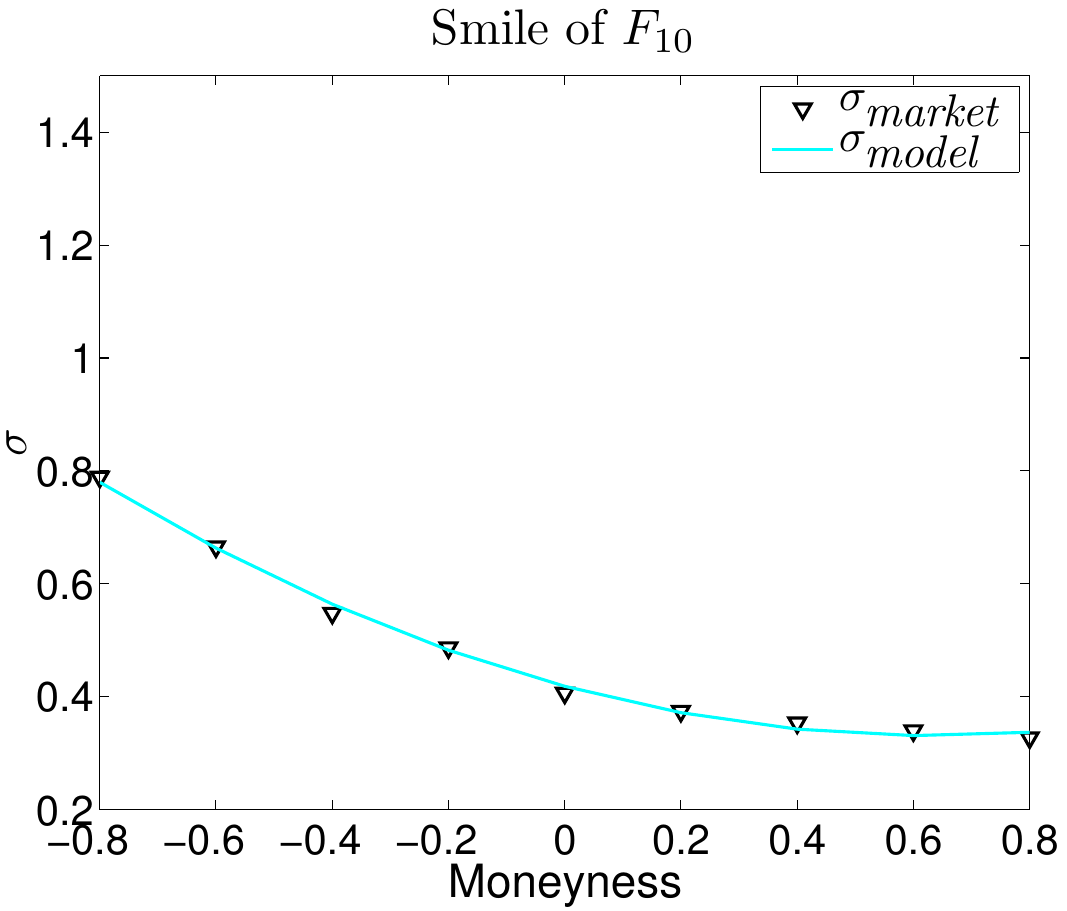}}
  \subfigure {\includegraphics[height=3.7cm]{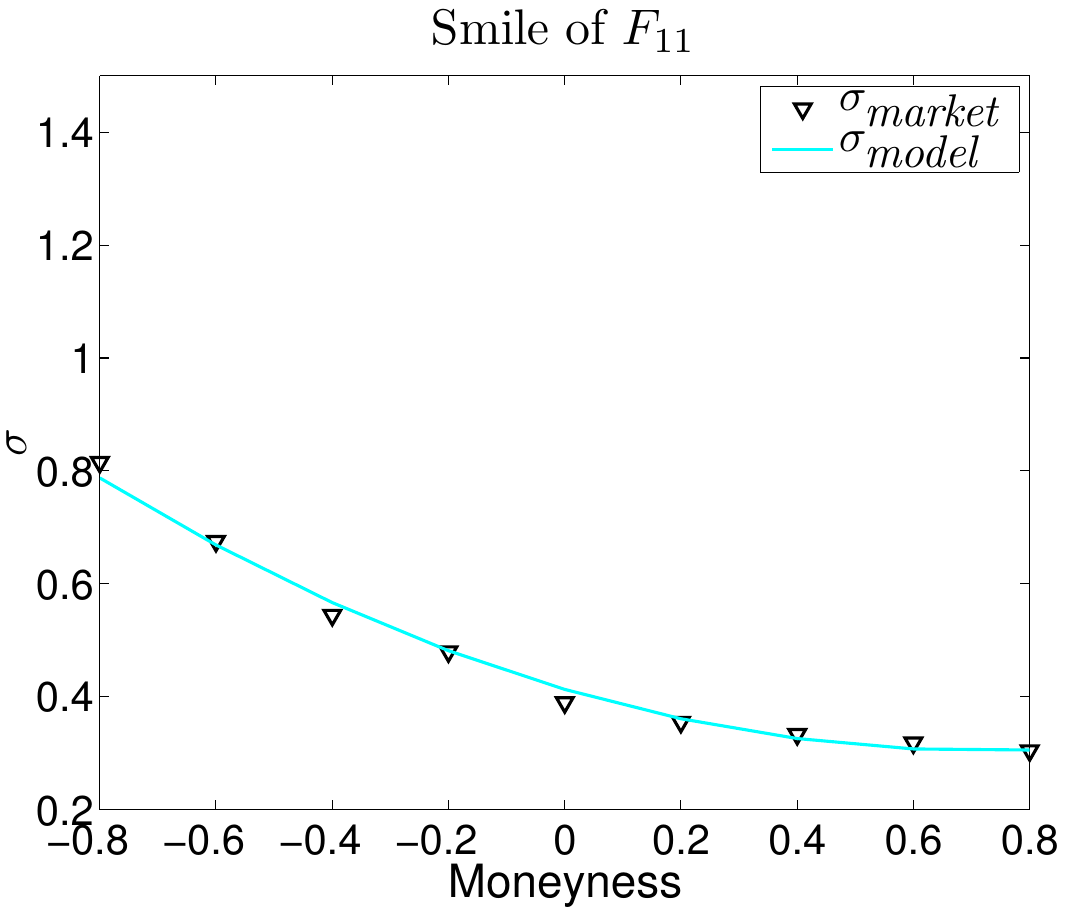}}
  \subfigure {\includegraphics[height=3.7cm]{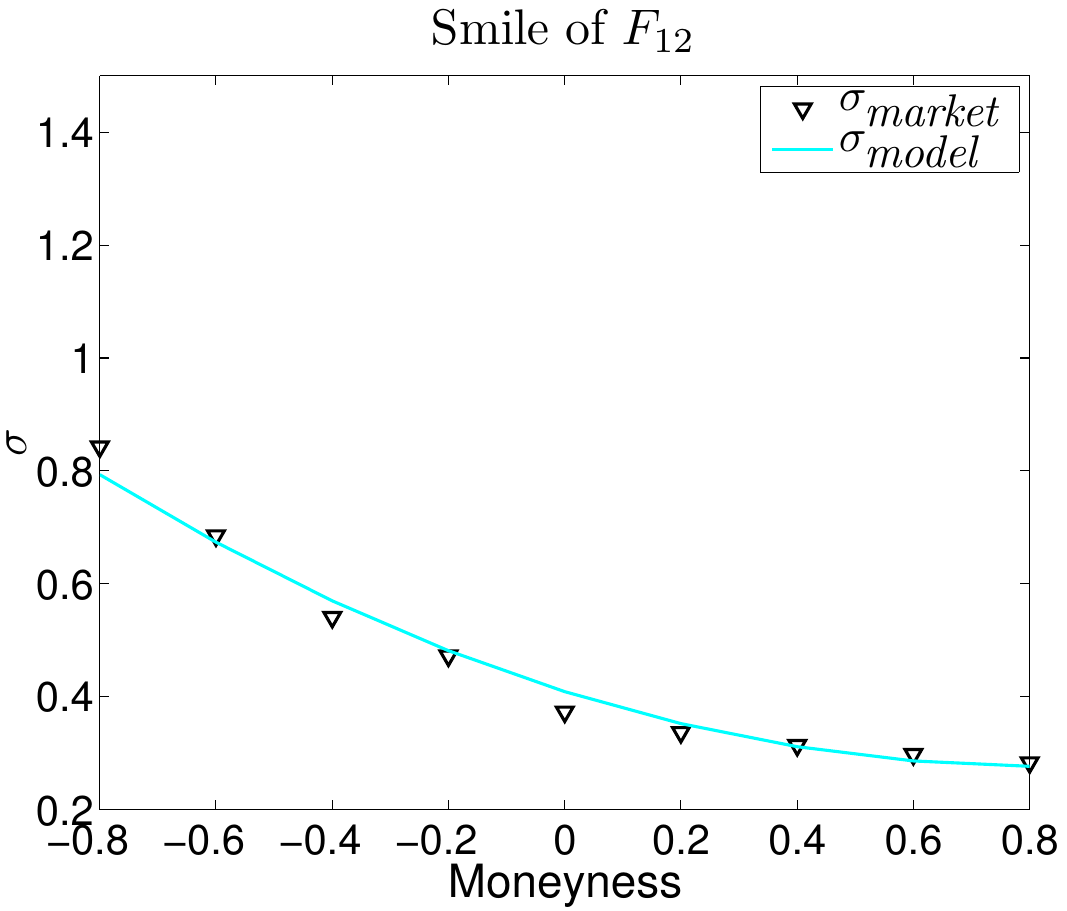}}
  \subfigure {\includegraphics[height=3.7cm]{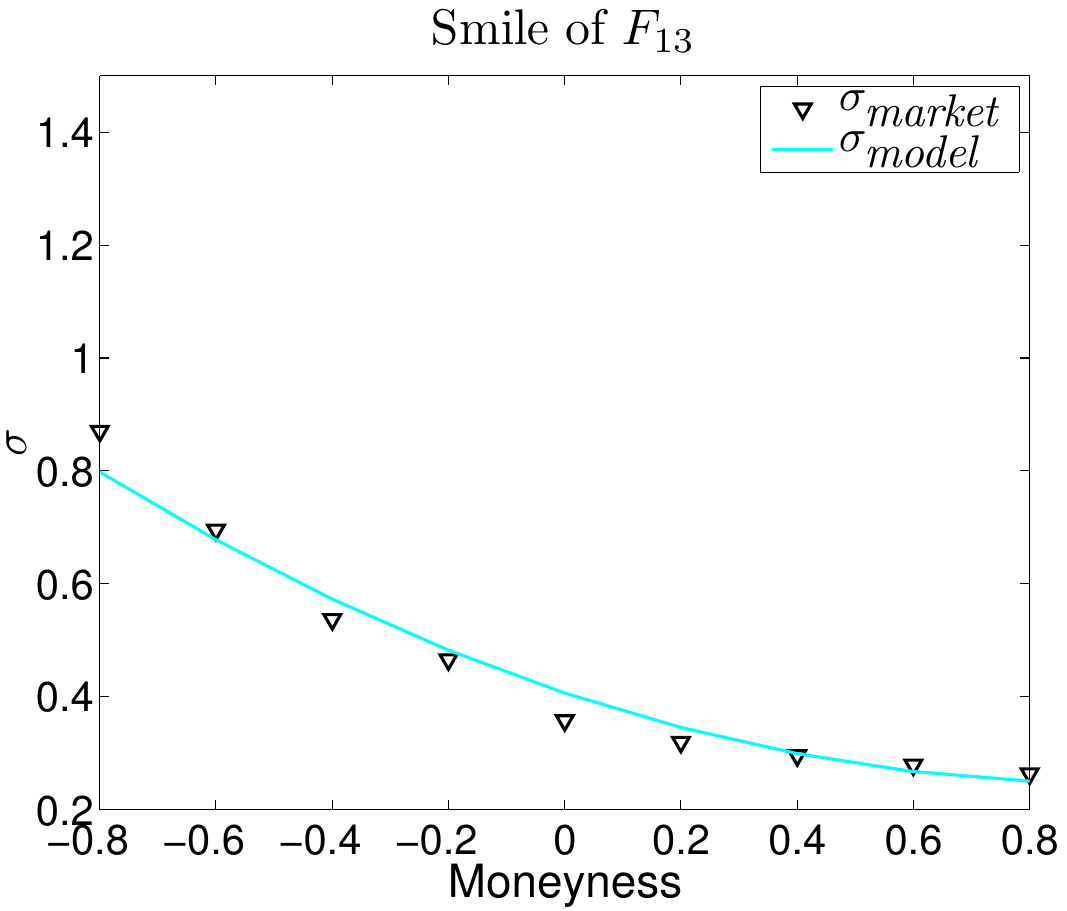}}
  \caption{Mercurio \& Morini model, $\sigma_{market}$ vs. $\sigma_{model}$, smiles of $F_1,\ldots,F_{13}$.}
  \label{fig:mercurioVolas}
\end{figure}

In Figure \ref{fig:mercurioVolas} the the model fitting for the smiles of all forward rates is shown. Market volatilities are plotted with triangles, while model volatilities are shown in continuous line.

\subsubsection{Calibration to swaptions}

The calibrated parameters are $\eta_1 = 0.779175$ and $\lambda_1 = 2.722489$. In Table \ref{tab:mercurioSwaptions} market vs. model swaptions prices (in $\%$) for the first fourteen swaptions and the the moneyness varying from $-40\%$ to $40\%$ are shown, each pair with its corresponding absolute error. In addition, for the whole set of swaptions the mean absolute error ($MAE$) is presented.

\begin{table}[!htb]
\scriptsize{
\centering
\begin{tabular}{|r|| c|c|c ||c|c|c |}
\hline
Moneyness & \multicolumn{3}{|c||}{$0.5 \times 1$ swaptions} & \multicolumn{3}{|c|}{$1 \times 1$ swaptions} \\
\hline
& $S_{Black}$ & $S_{MC}$ & $|S_{Black}-S_{MC}|$ & $S_{Black}$ & $S_{MC}$ & $|S_{Black}-S_{MC}|$ \\ \hline
$-40\%$ & $0.4866$ & $0.4870$ & $4.00 \times 10^{-4}$ & $0.5917$ & $0.5870$ & $4.70 \times 10^{-3}$ \\
\hline
$-20\%$ & $0.3562$ & $0.3670$ & $1.08 \times 10^{-2}$ & $0.4661$ & $0.4699$ & $3.80 \times 10^{-3}$ \\
\hline
$0\%$ & $0.2356$ & $0.2478$ & $1.22 \times 10^{-2}$ & $0.3467$ & $0.3517$ & $5.00 \times 10^{-3}$ \\
\hline
$20\%$ & $0.1363$ & $0.1427$ & $6.40 \times 10^{-3}$ & $0.2394$ & $0.2422$ & $2.80 \times 10^{-3}$ \\
\hline
$40\%$ & $0.0680$ & $0.0657$ & $2.30 \times 10^{-3}$ & $0.1517$ & $0.1514$ & $3.00 \times 10^{-4}$ \\
\hline
\hline
Moneyness & \multicolumn{3}{|c||}{$1.5 \times 1$ swaptions} & \multicolumn{3}{|c|}{$2 \times 1$ swaptions} \\
\hline
& $S_{Black}$ & $S_{MC}$ & $|S_{Black}-S_{MC}|$ & $S_{Black}$ & $S_{MC}$ & $|S_{Black}-S_{MC}|$ \\ \hline
$-40\%$ & $0.7357$ & $0.6872$ & $4.85 \times 10^{-2}$ & $0.8184$ & $0.7465$ & $7.19 \times 10^{-2}$ \\
\hline
$-20\%$ & $0.5908$ & $0.5516$ & $3.92 \times 10^{-2}$ & $0.6603$ & $0.5959$ & $6.44 \times 10^{-2}$ \\
\hline
$0\%$ & $0.4536$ & $0.4170$ & $3.66 \times 10^{-2}$ & $0.5118$ & $0.4469$ & $6.49 \times 10^{-2}$ \\
\hline
$20\%$ & $0.3277$ & $0.2951$ & $3.26 \times 10^{-2}$ & $0.3754$ & $0.3137$ & $6.17 \times 10^{-2}$ \\
\hline
$40\%$ & $0.2213$ & $0.1957$ & $2.56 \times 10^{-2}$ & $0.2587$ & $0.2078$ & $5.09 \times 10^{-2}$ \\
\hline
\hline
Moneyness & \multicolumn{3}{|c||}{$0.5 \times 2$ swaptions} & \multicolumn{3}{|c|}{$1 \times 2$ swaptions} \\
\hline
& $S_{Black}$ & $S_{MC}$ & $|S_{Black}-S_{MC}|$ & $S_{Black}$ & $S_{MC}$ & $|S_{Black}-S_{MC}|$ \\ \hline
$-40\%$ & $1.0570$ & $1.0338$ & $2.32 \times 10^{-2}$ & $1.2427$ & $1.2143$ & $2.84 \times 10^{-2}$ \\
\hline
$-20\%$ & $0.7440$ & $0.7452$ & $1.20 \times 10^{-3}$ & $0.9322$ & $0.9266$ & $5.60 \times 10^{-3}$ \\
\hline
$0\%$ & $0.4555$ & $0.4679$ & $1.24 \times 10^{-2}$ & $0.6394$ & $0.6460$ & $6.60 \times 10^{-3}$ \\
\hline
$20\%$ & $0.2299$ & $0.2428$ & $1.29 \times 10^{-2}$ & $0.3886$ & $0.4038$ & $1.52 \times 10^{-2}$ \\
\hline
$40\%$ & $0.0925$ & $0.0984$ & $5.90 \times 10^{-3}$ & $0.2037$ & $0.2242$ & $2.05 \times 10^{-2}$ \\
\hline
\hline
Moneyness & \multicolumn{3}{|c||}{$1.5 \times 2$ swaptions} & \multicolumn{3}{|c|}{$2 \times 2$ swaptions} \\
\hline
& $S_{Black}$ & $S_{MC}$ & $|S_{Black}-S_{MC}|$ & $S_{Black}$ & $S_{MC}$ & $|S_{Black}-S_{MC}|$ \\ \hline
$-40\%$ & $1.4884$ & $1.4382$ & $5.02 \times 10^{-2}$ & $1.6938$ & $1.6298$ & $6.40 \times 10^{-2}$ \\
\hline
$-20\%$ & $1.1367$ & $1.1173$ & $1.94 \times 10^{-2}$ & $1.3077$ & $1.2746$ & $3.31 \times 10^{-2}$ \\
\hline
$0\%$ & $0.8059$ & $0.8024$ & $3.50 \times 10^{-3}$ & $0.9466$ & $0.9220$ & $2.46 \times 10^{-2}$ \\
\hline
$20\%$ & $0.5154$ & $0.5266$ & $1.12 \times 10^{-2}$ & $0.6269$ & $0.6116$ & $1.53 \times 10^{-2}$ \\
\hline
$40\%$ & $0.2919$ & $0.3182$ & $2.63 \times 10^{-2}$ & $0.3736$ & $0.3751$ & $1.50 \times 10^{-3}$ \\
\hline
\hline
Moneyness & \multicolumn{3}{|c||}{$0.5 \times 3$ swaptions} & \multicolumn{3}{|c|}{$1 \times 3$ swaptions} \\
\hline
& $S_{Black}$ & $S_{MC}$ & $|S_{Black}-S_{MC}|$ & $S_{Black}$ & $S_{MC}$ & $|S_{Black}-S_{MC}|$ \\ \hline
$-40\%$ & $1.7380$ & $1.6737$ & $6.43 \times 10^{-2}$ & $2.0341$ & $1.9880$ & $4.61 \times 10^{-2}$ \\
\hline
$-20\%$ & $1.1980$ & $1.1659$ & $3.21 \times 10^{-2}$ & $1.4851$ & $1.4761$ & $9.00 \times 10^{-3}$ \\
\hline
$0\%$ & $0.7011$ & $0.6868$ & $1.43 \times 10^{-2}$ & $0.9696$ & $0.9803$ & $1.07 \times 10^{-2}$ \\
\hline
$20\%$ & $0.3242$ & $0.3198$ & $4.40 \times 10^{-3}$ & $0.5413$ & $0.5666$ & $2.53 \times 10^{-2}$ \\
\hline
$40\%$ & $0.1128$ & $0.1112$ & $1.60 \times 10^{-3}$ & $0.2479$ & $0.2825$ & $3.46 \times 10^{-2}$ \\
\hline
\hline
Moneyness & \multicolumn{3}{|c||}{$1.5 \times 3$ swaptions} & \multicolumn{3}{|c|}{$2 \times 3$ swaptions} \\
\hline
& $S_{Black}$ & $S_{MC}$ & $|S_{Black}-S_{MC}|$ & $S_{Black}$ & $S_{MC}$ & $|S_{Black}-S_{MC}|$ \\ \hline
$-40\%$ & $2.3898$ & $2.3268$ & $6.30 \times 10^{-2}$ & $2.6885$ & $2.6302$ & $5.83 \times 10^{-2}$ \\
\hline
$-20\%$ & $1.7850$ & $1.7690$ & $1.60 \times 10^{-2}$ & $2.0311$ & $2.0258$ & $5.30 \times 10^{-3}$ \\
\hline
$0\%$ & $1.2175$ & $1.2215$ & $4.00 \times 10^{-3}$ & $1.4178$ & $1.4218$ & $4.00 \times 10^{-3}$ \\
\hline
$20\%$ & $0.7304$ & $0.7526$ & $2.22 \times 10^{-2}$ & $0.8856$ & $0.8935$ & $7.90 \times 10^{-3}$ \\
\hline
$40\%$ & $0.3749$ & $0.4168$ & $4.19 \times 10^{-2}$ & $0.4832$ & $0.5068$ & $2.36 \times 10^{-2}$ \\
\hline
\hline
Moneyness & \multicolumn{3}{|c||}{$0.5 \times 4$ swaptions} & \multicolumn{3}{|c|}{$1 \times 4$ swaptions} \\
\hline
& $S_{Black}$ & $S_{MC}$ & $|S_{Black}-S_{MC}|$ & $S_{Black}$ & $S_{MC}$ & $|S_{Black}-S_{MC}|$ \\ \hline
$-40\%$ & $2.5381$ & $2.4434$ & $9.47 \times 10^{-2}$ & $2.9426$ & $2.8764$ & $6.62 \times 10^{-2}$ \\
\hline
$-20\%$ & $1.7151$ & $1.6621$ & $5.30 \times 10^{-2}$ & $2.1123$ & $2.0935$ & $1.88 \times 10^{-2}$ \\
\hline
$0\%$ & $0.9584$ & $0.9298$ & $2.86 \times 10^{-2}$ & $1.3344$ & $1.3357$ & $1.30 \times 10^{-3}$ \\
\hline
$20\%$ & $0.4031$ & $0.3918$ & $1.13 \times 10^{-2}$ & $0.7016$ & $0.7174$ & $1.58 \times 10^{-2}$ \\
\hline
$40\%$ & $0.1188$ & $0.1160$ & $2.80 \times 10^{-3}$ & $0.2907$ & $0.3205$ & $2.98 \times 10^{-2}$ \\
\hline
\hline
\multicolumn{7}{|c|}{$MAE=5.50 \times 10^{-2}$} \\
\hline

\end{tabular}
\caption{Mercurio \& Morini model, calibration to swaptions, $S_{Black}$ vs. $S_{MC}$, prices in \%.}
\label{tab:mercurioSwaptions}
}
\end{table}

\begin{figure}[!htb]
\centering
  \subfigure {\includegraphics[height=4.4cm]{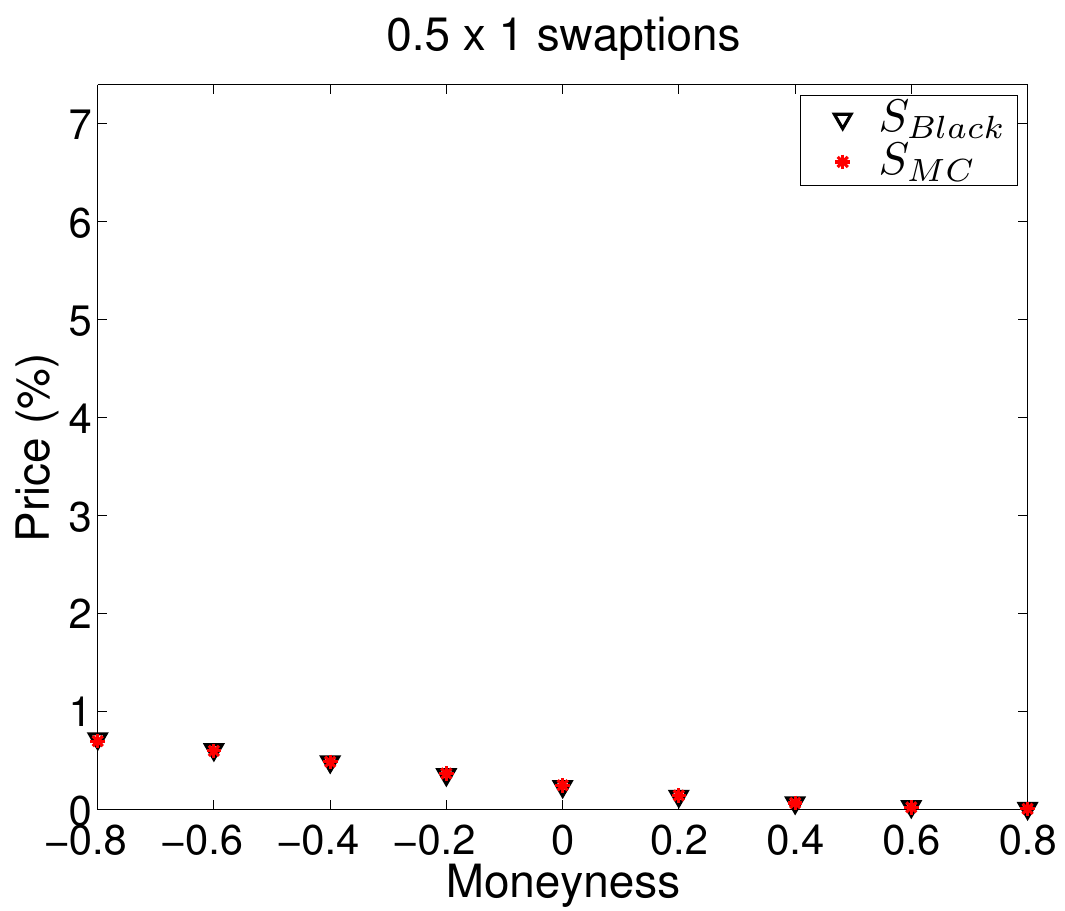}}
  \subfigure {\includegraphics[height=4.4cm]{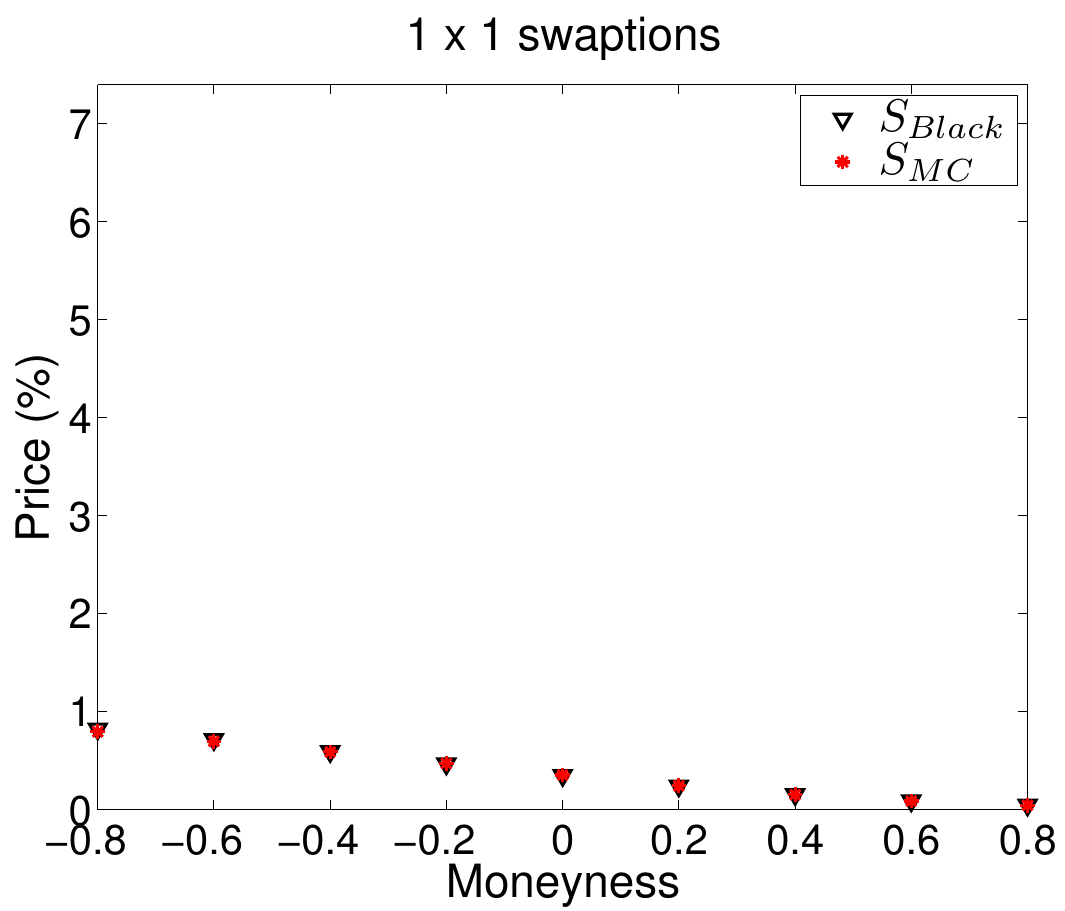}}
  \subfigure {\includegraphics[height=4.4cm]{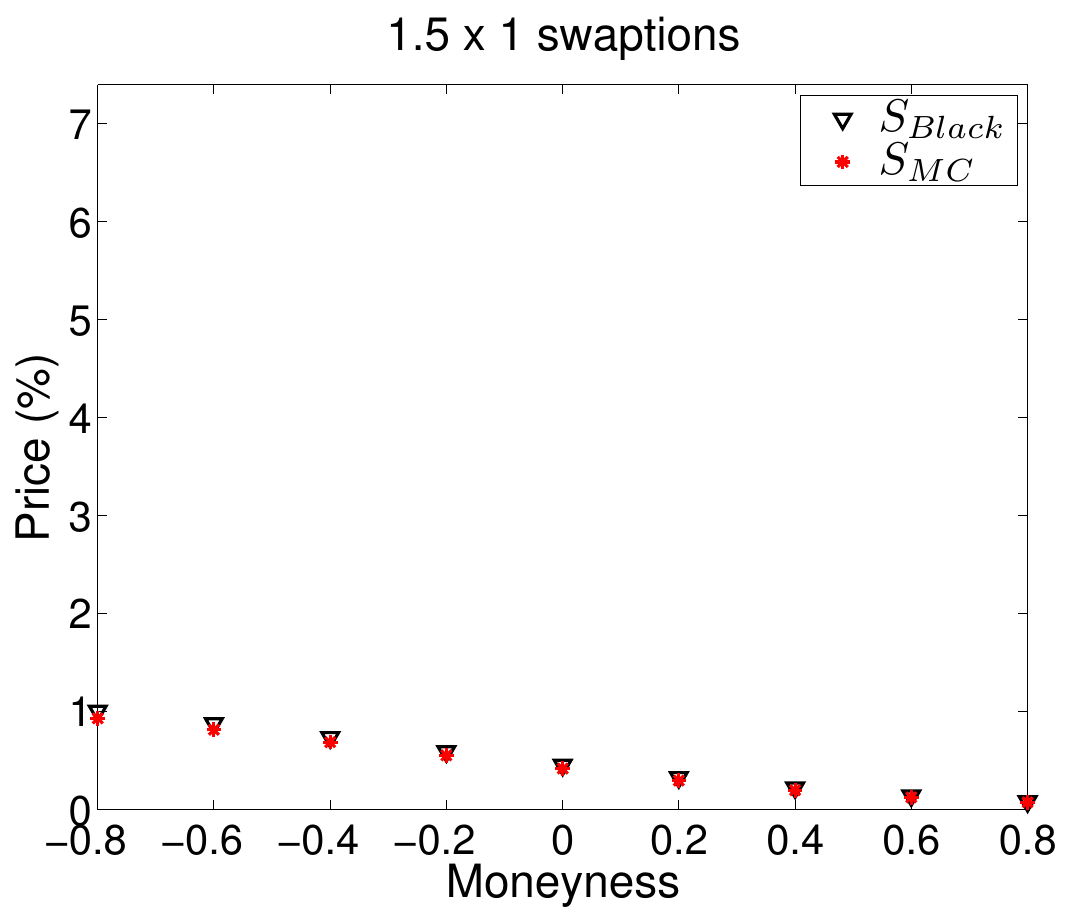}}
  \subfigure {\includegraphics[height=4.4cm]{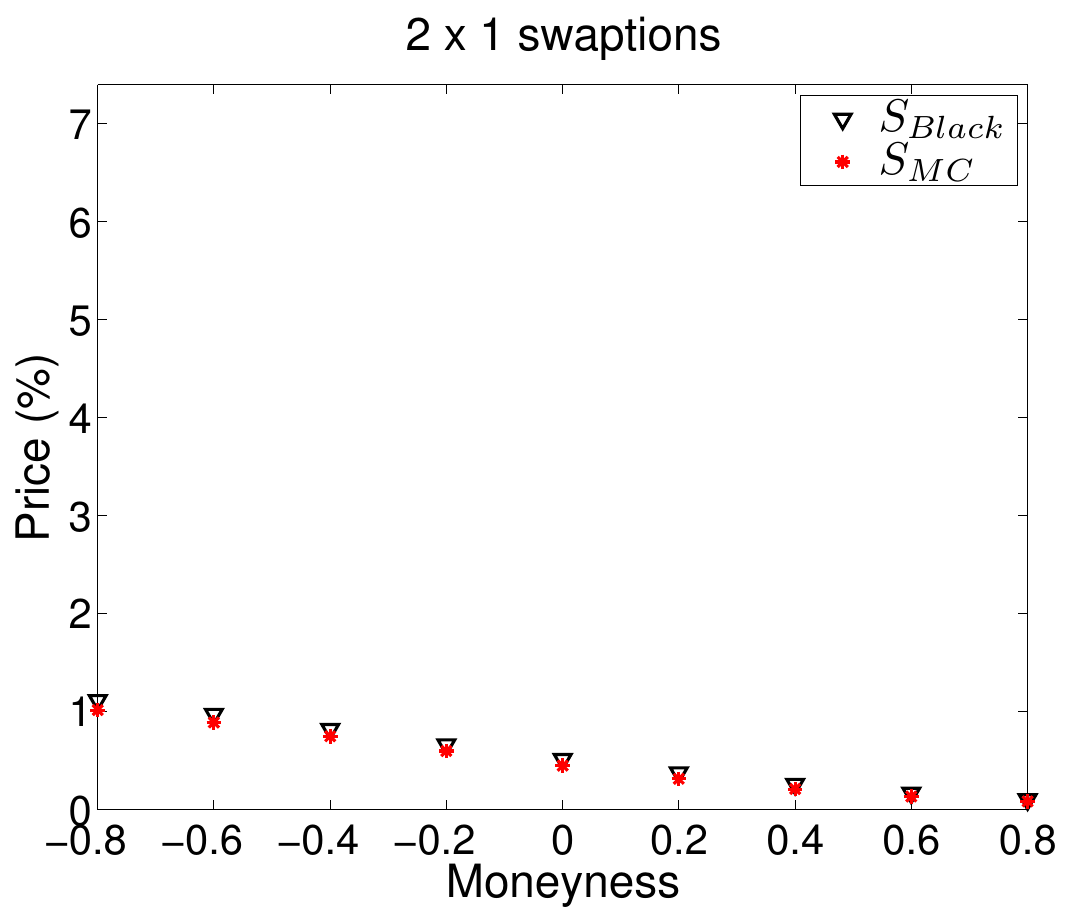}}
  \subfigure {\includegraphics[height=4.4cm]{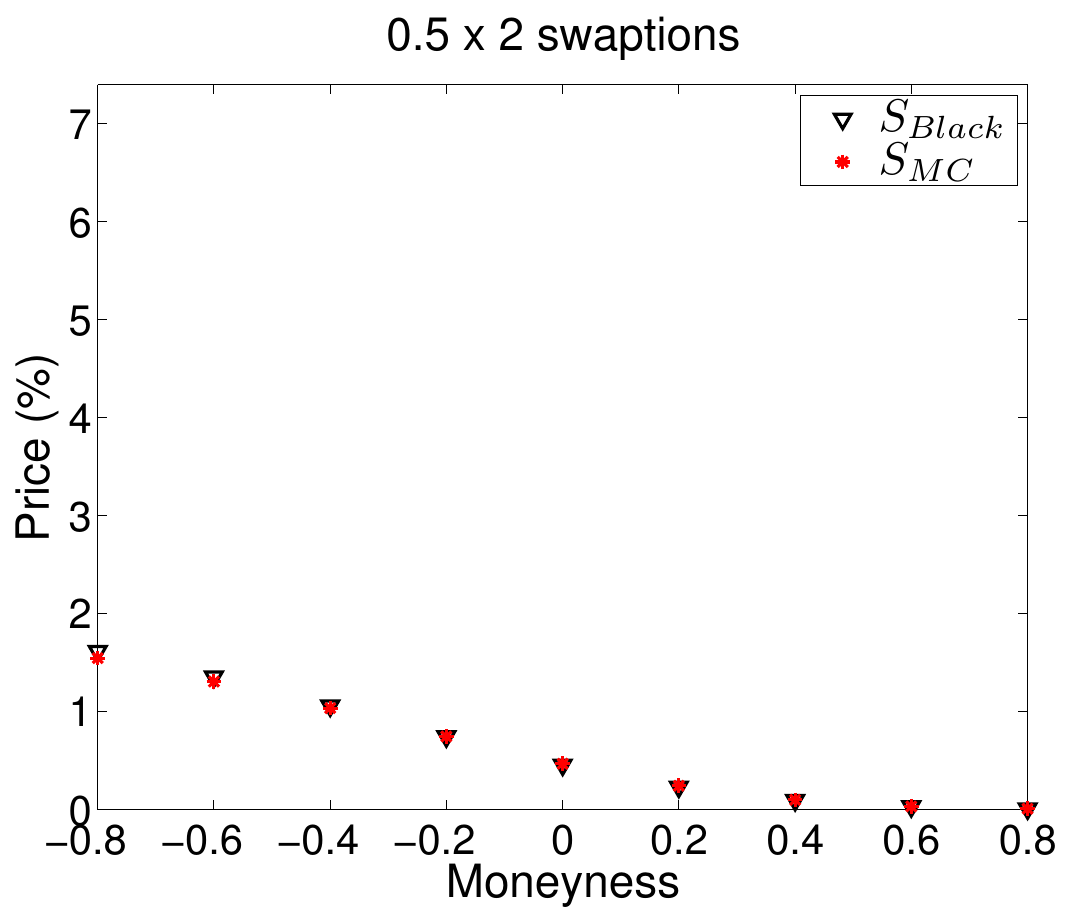}}
  \subfigure {\includegraphics[height=4.4cm]{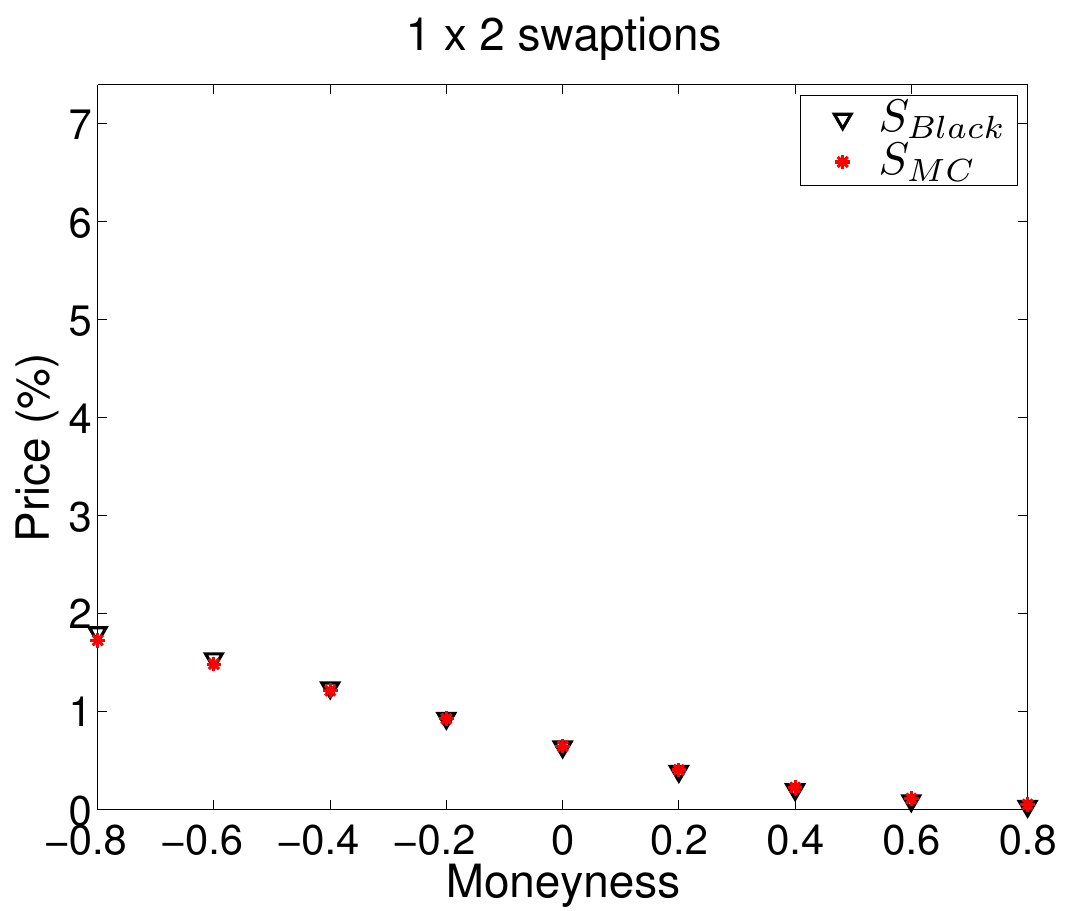}}
  \subfigure {\includegraphics[height=4.4cm]{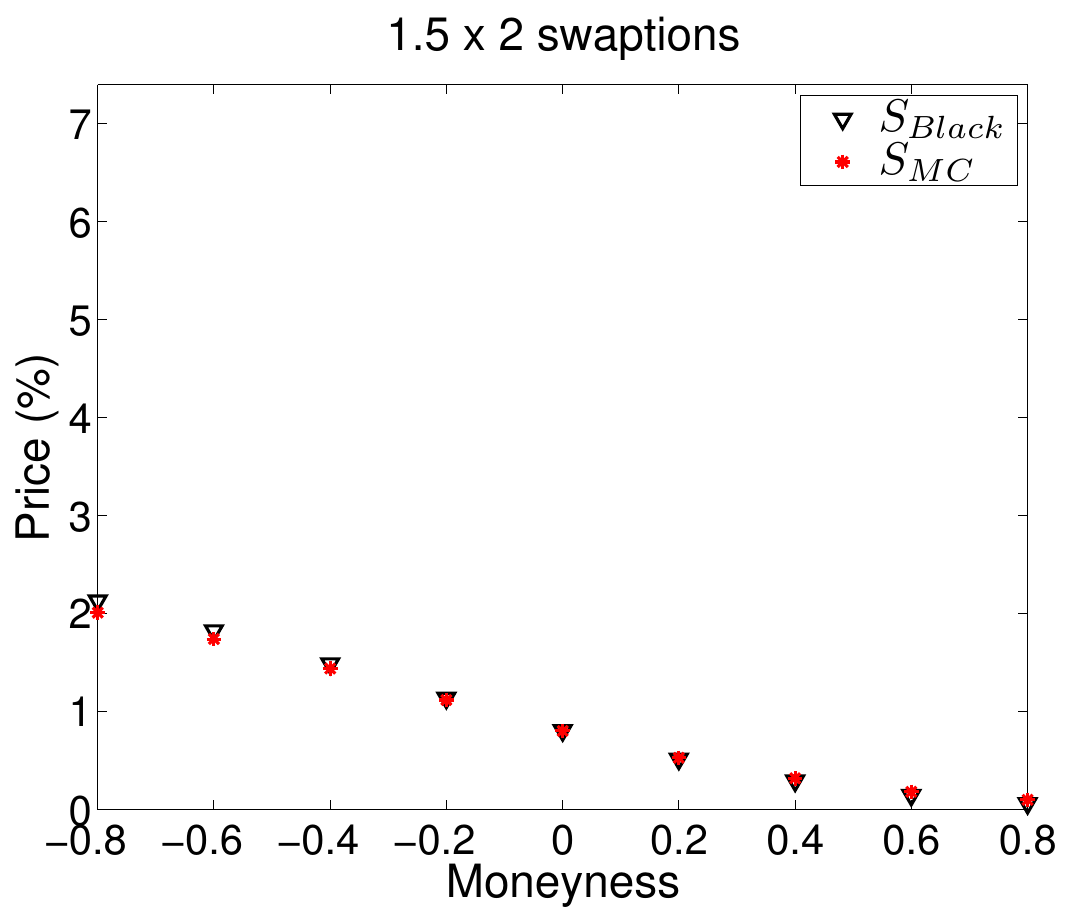}}
  \subfigure {\includegraphics[height=4.4cm]{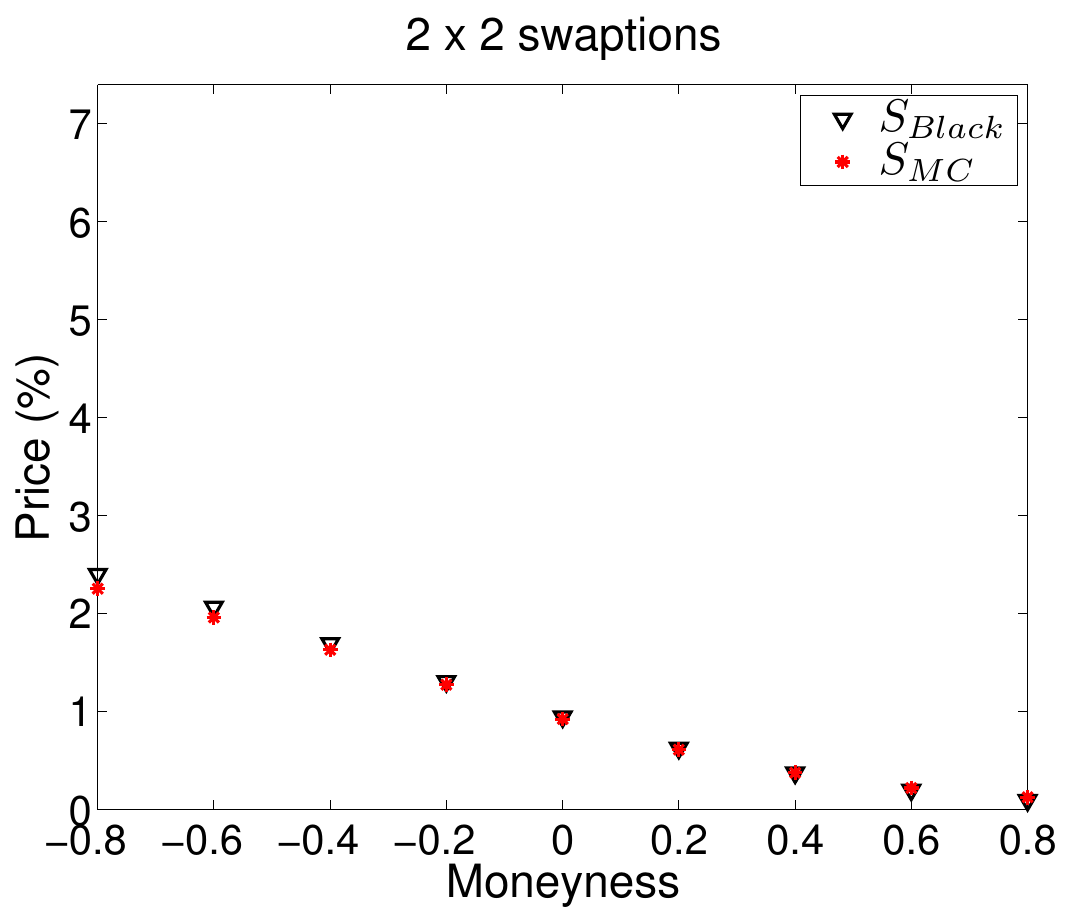}}
  \subfigure {\includegraphics[height=4.4cm]{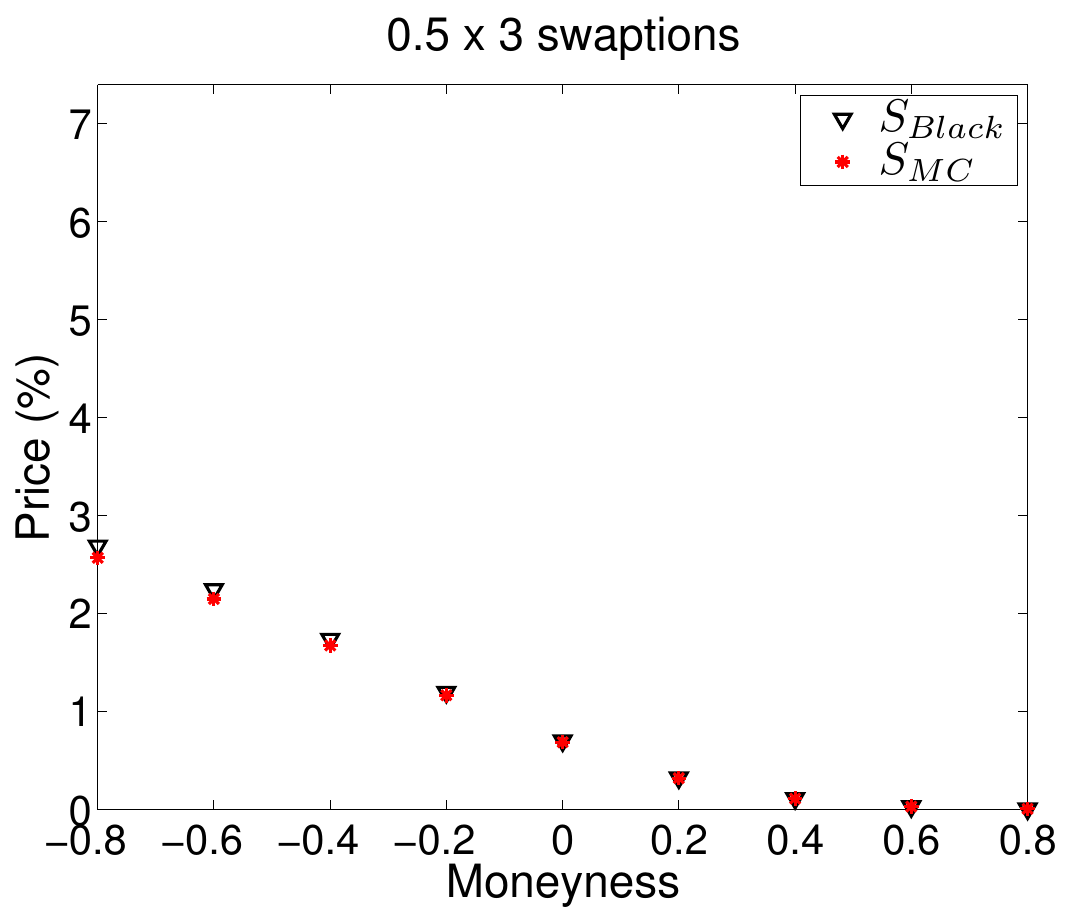}}
  \subfigure {\includegraphics[height=4.4cm]{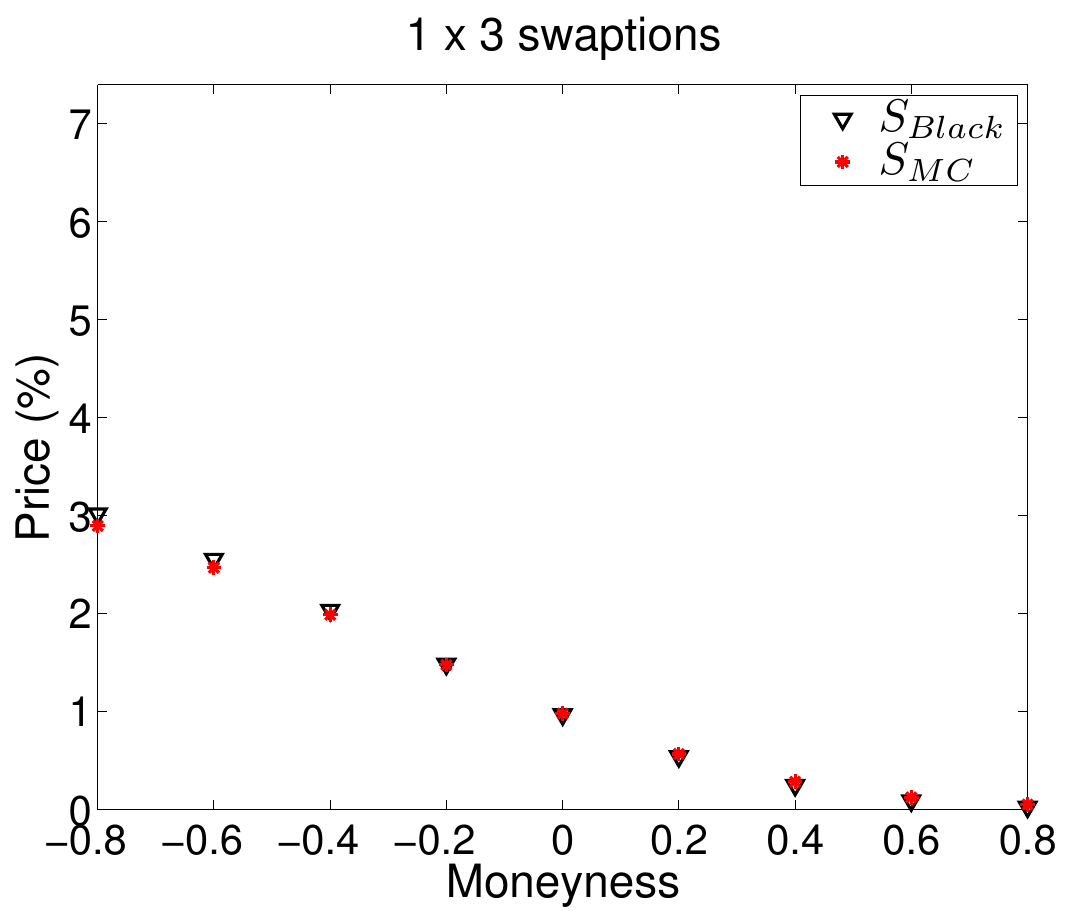}}
  \caption{Mercurio \& Morini model, calibration to swaptions, $S_{Black}$ vs. $S_{MC}$, part I.}
  \label{fig:mercurioSwaptions1}
\end{figure}

\begin{figure}[!htb]
\centering
  \subfigure {\includegraphics[height=4.4cm]{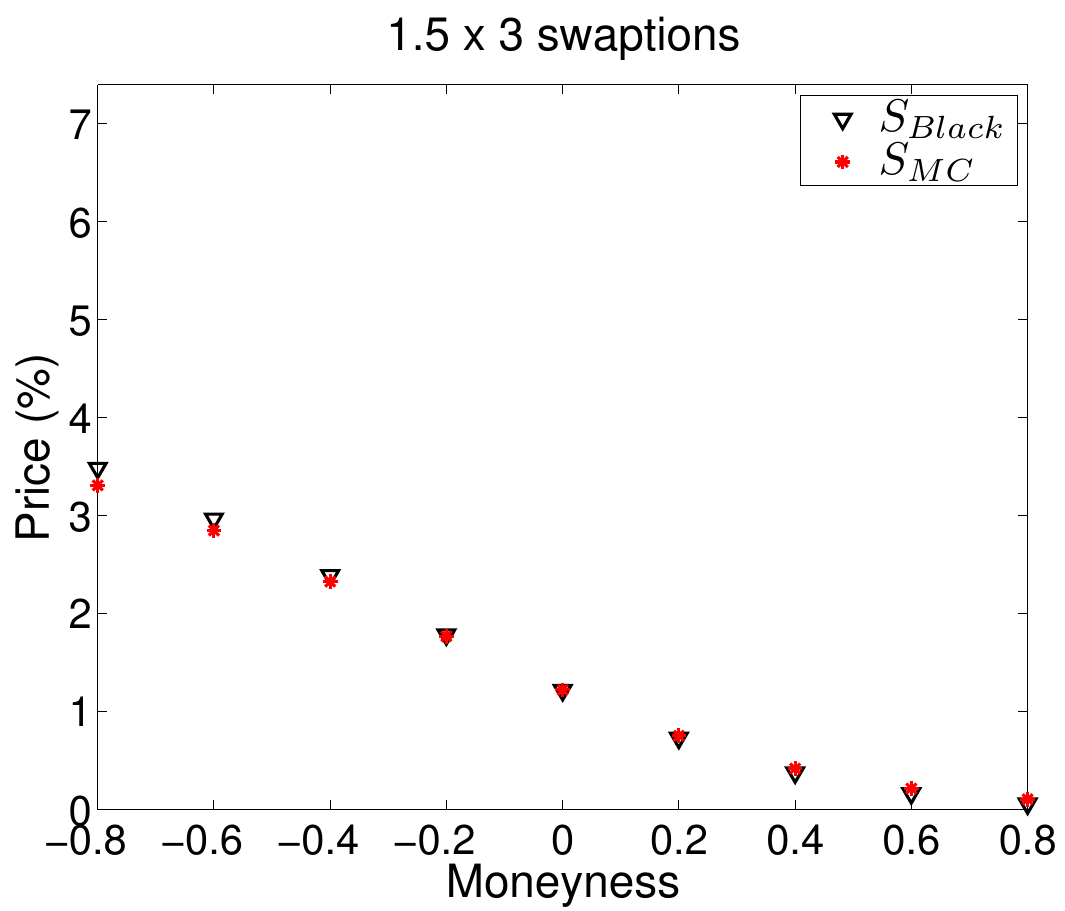}}
  \subfigure {\includegraphics[height=4.4cm]{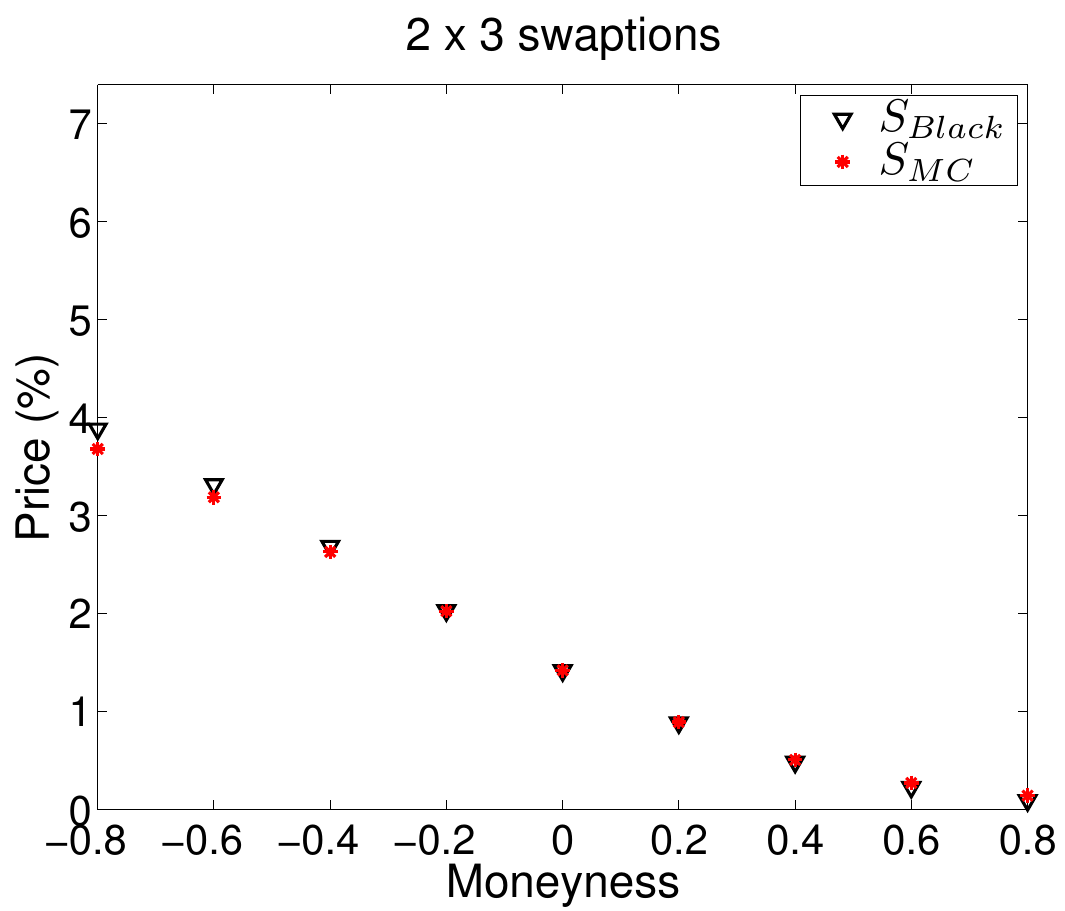}}
  \subfigure {\includegraphics[height=4.4cm]{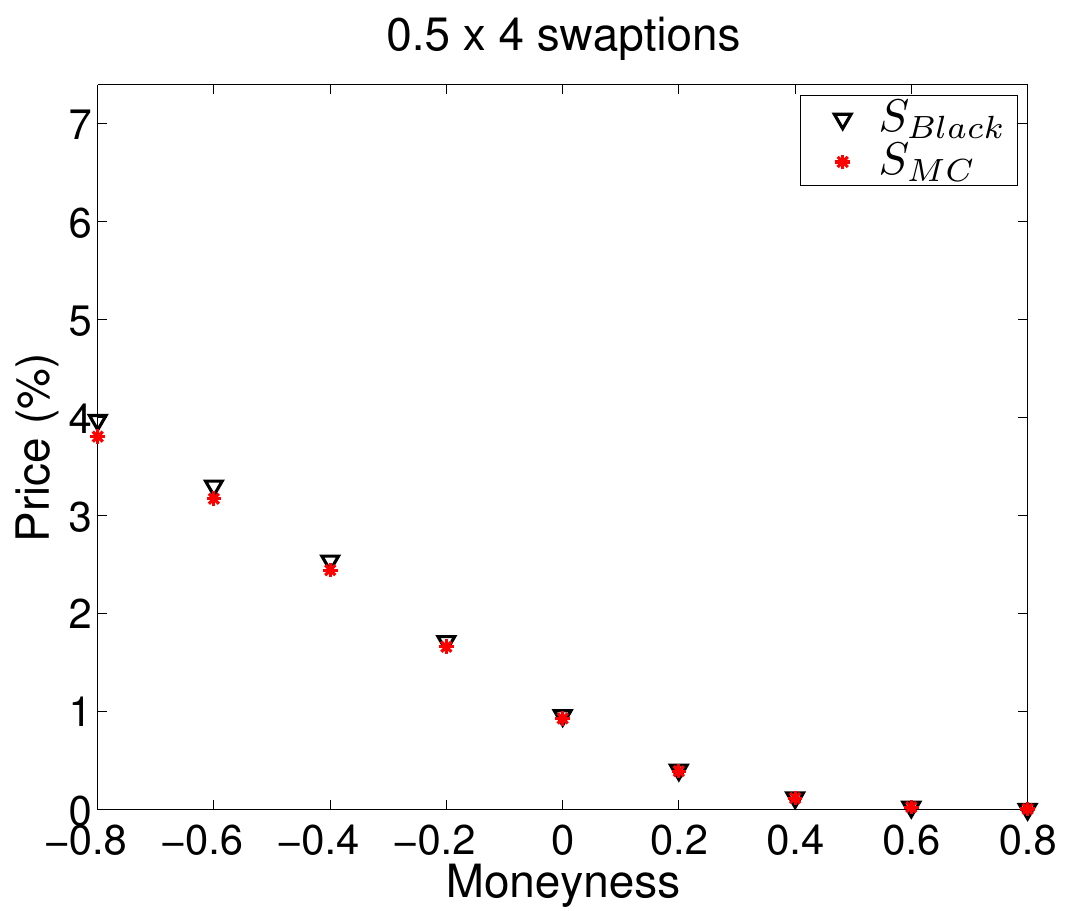}}
  \subfigure {\includegraphics[height=4.4cm]{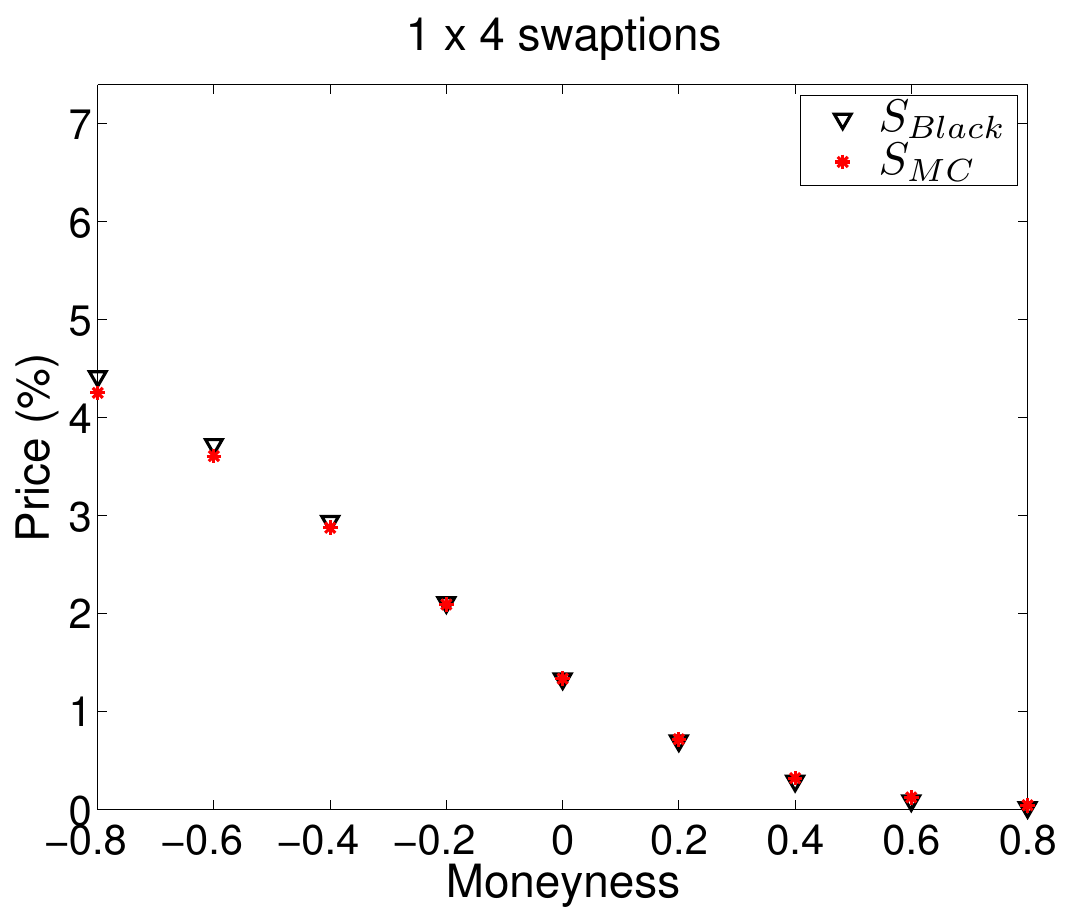}}
  \subfigure {\includegraphics[height=4.4cm]{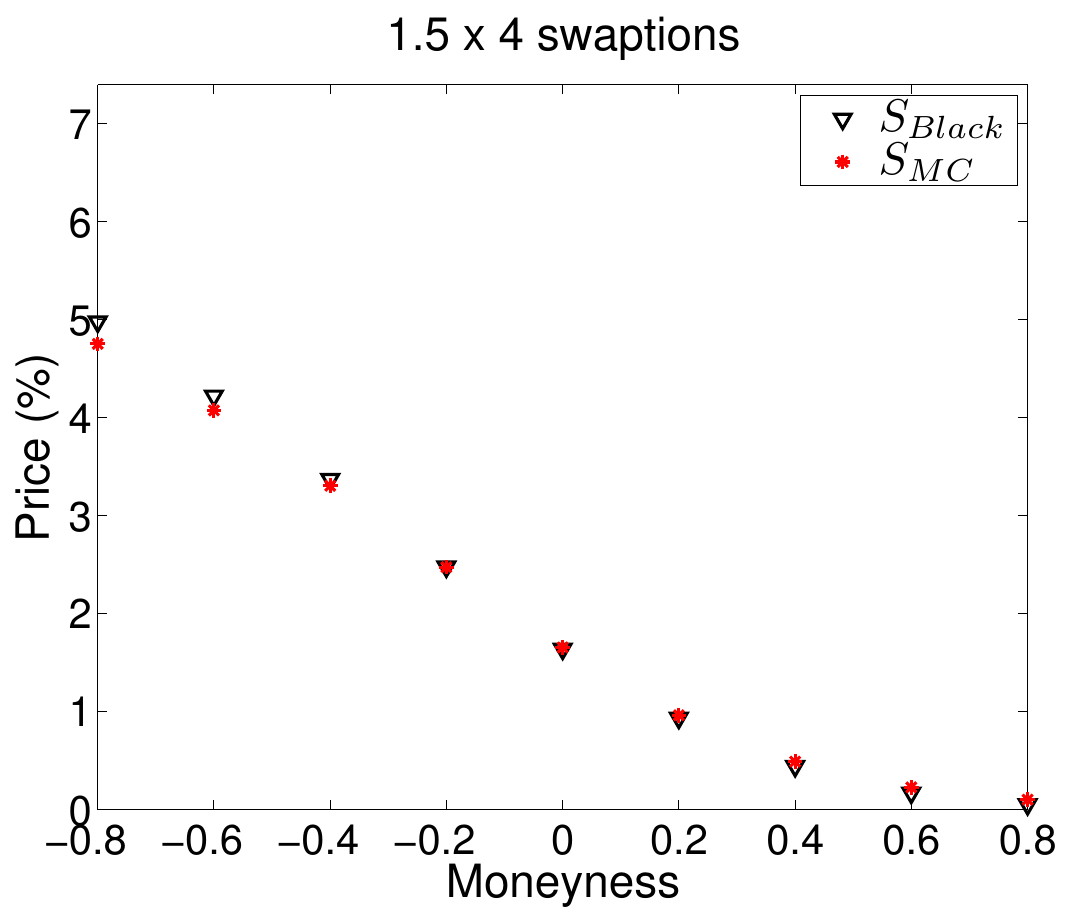}}
  \subfigure {\includegraphics[height=4.4cm]{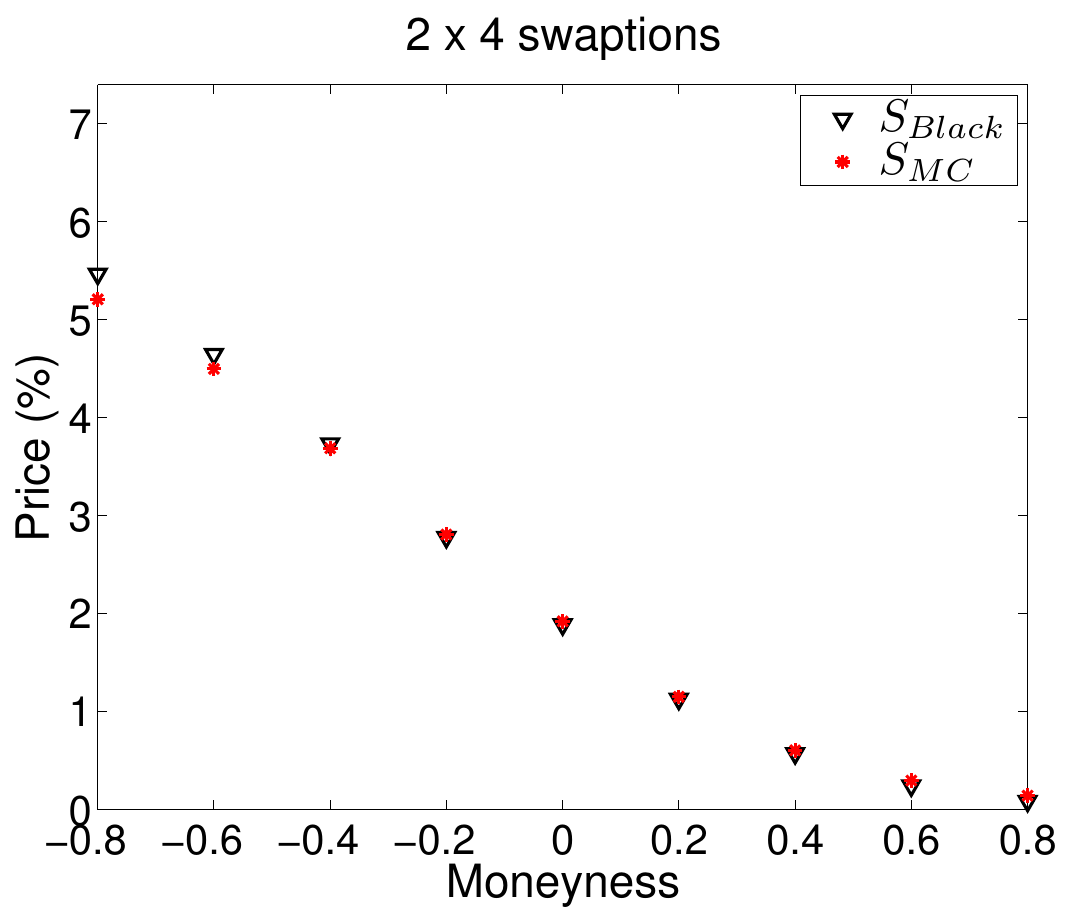}}
  \subfigure {\includegraphics[height=4.4cm]{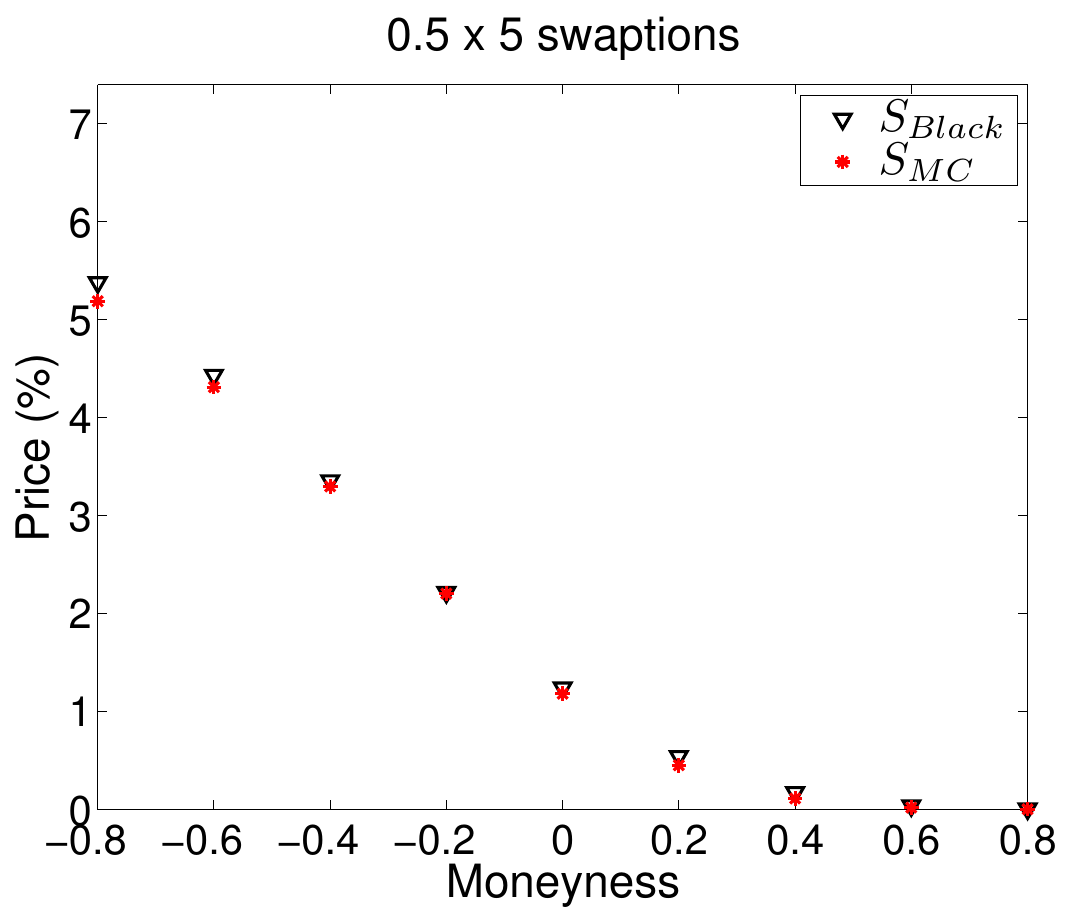}}
  \subfigure {\includegraphics[height=4.4cm]{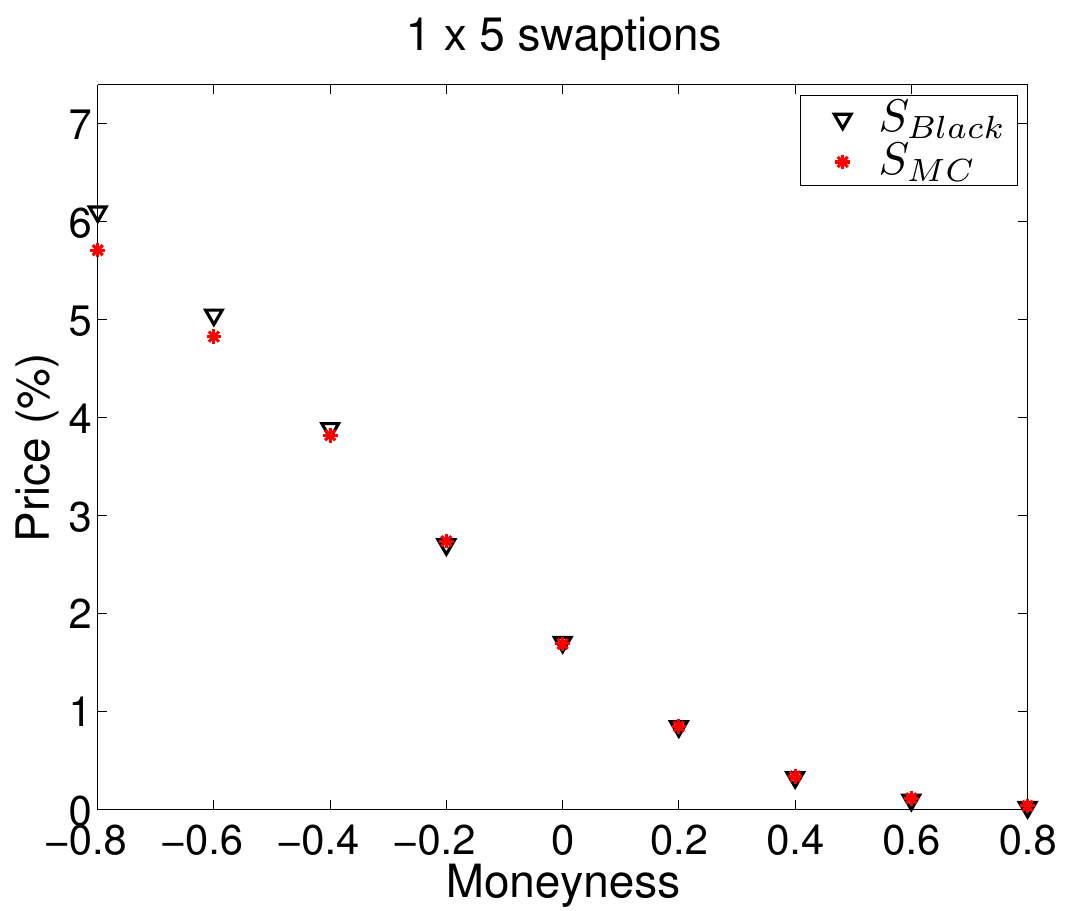}}
  \subfigure {\includegraphics[height=4.4cm]{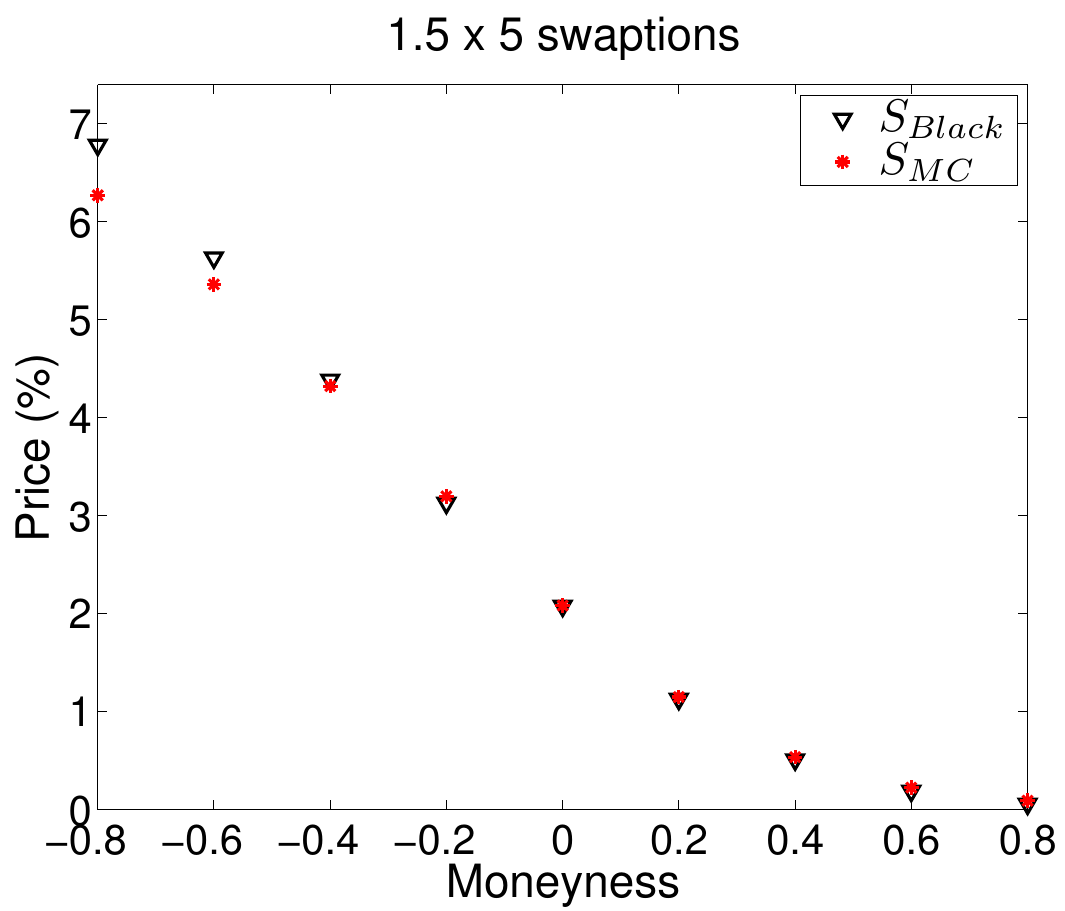}}
  \subfigure {\includegraphics[height=4.4cm]{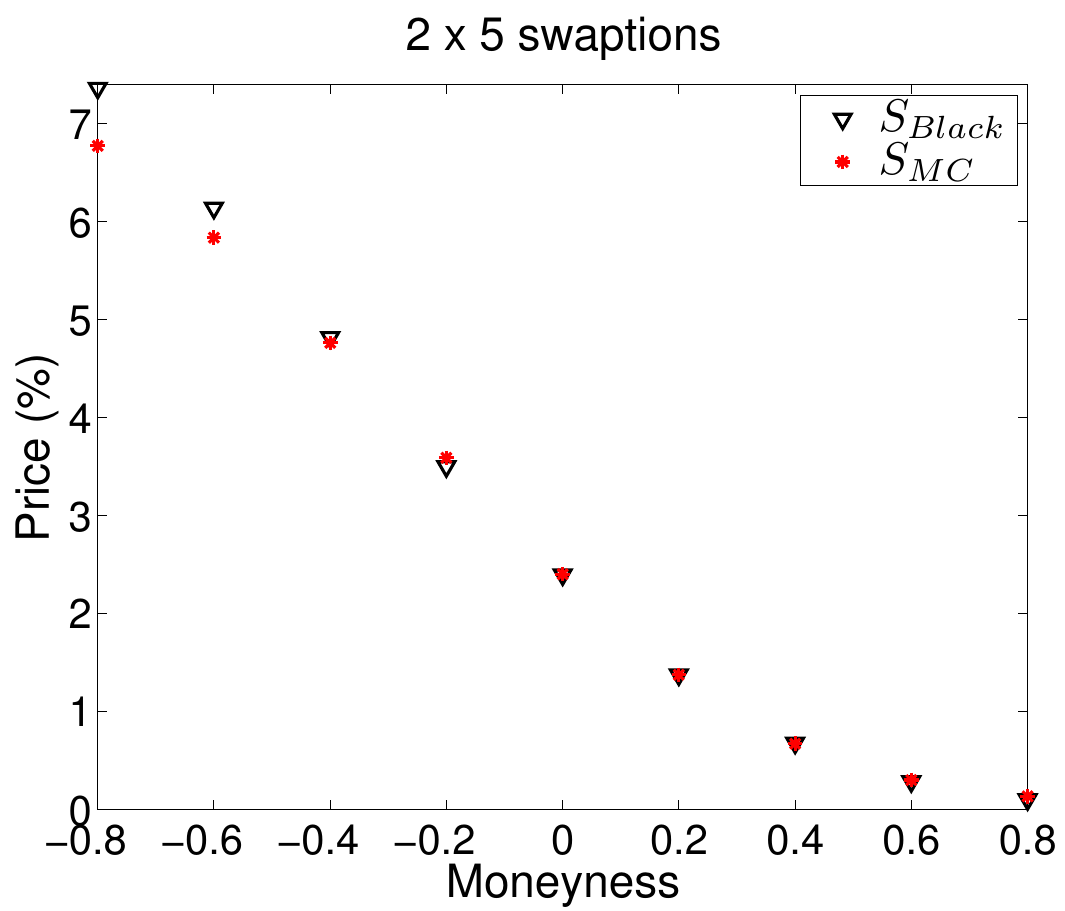}}
  \caption{Mercurio \& Morini model, calibration to swaptions, $S_{Black}$ vs. $S_{MC}$, part II.}
  \label{fig:mercurioSwaptions2}
\end{figure}

In Figures \ref{fig:mercurioSwaptions1} and \ref{fig:mercurioSwaptions2} the model fitting when considering the whole swaption matrix is shown. Market prices are shown using triangles and the model ones using stars.

% \clearpage
\subsection{Rebonato model} \label{subsec:num_resul_R}

\subsubsection{Calibration to caplets}

The calibrated parameters are shown in Table \ref{tab:paramsRebonatoVolas}. The execution time was $146.729$ seconds, $119.913$ seconds employed by the multi-GPU SA (launched with a more demanding configuration, specifically, $T_0=10$, $T_{min}=0.01$, $\rho=0.99$, $N=100$, $w=256\times64$, $\#GPUs=2$, the cost function was evaluated roughly two billion times) and the Nelder-Mead local optimization algorithm consumed the remaining time. When using the multi-GPU approach (see Figure \ref{fig:sincronoMultiGPU}) we obtain a speedup of $1.88$ with respect to mono-GPU version, and of $207.796$ with respect to sequential computations. % 110.530 speedup mono-GPU version

\begin{table}[!htb]
\scriptsize{
\begin{center}
\begin{tabular}{|c|r|c||c|r|c|}
\hline
   & $\phi_{ii}$ & $\kappa_i$ & & $\phi_{ii}$ & $\kappa_i$ \\
\hline
\hline
$F_1$ & $-0.4060$ & $0.0021$ & $F_8$ & $-0.4218$ & $0.0010$ \\
\hline
$F_2$ & $-0.1935$ & $0.0017$ & $F_9$ & $-0.5355$ & $0.0010$  \\
\hline
$F_3$ & $0.0684$ & $0.0015$ & $F_{10}$ & $-0.6466$ & $0.0010$  \\
\hline
$F_4$ & $0.1825$ & $0.0011$ & $F_{11}$ & $-0.7413$ & $0.0010$  \\
\hline
$F_5$ & $0.0158$ & $0.0010$ & $F_{12}$ & $-0.8175$ & $0.0010$  \\
\hline
$F_6$ & $-0.1306$ & $0.0010$ & $F_{13}$ & $-1.0000$ & $0.0010$  \\
\hline
$F_7$ & $-0.2665$ & $0.0010$ & & & \\
\hline
\hline
\multicolumn{3}{|l||}{$a=3.7789, \quad b=44.7668,$} & \multicolumn{3}{|l|}{$\alpha=0.0010, \quad \beta=19.5812,$} \\
\multicolumn{3}{|l||}{$c=0.3076, \quad d=25.3412.$} & \multicolumn{3}{|l|}{$\gamma=6.2339, \quad \delta=0.5533.$} \\
\hline
\end{tabular}
\caption{Rebonato model, calibration to caplets with SABR formula \eqref{eq:sigmaHagan}: calibrated parameters.} \label{tab:paramsRebonatoVolas}
\end{center}
}
\end{table}

In Table \ref{tab:rebonatoVolas}, market vs. model volatilities for the smiles of $F_1$ to $F_{12}$ and the moneyness $-40\%$ to $40\%$ are shown. The mean relative error considering all smiles is presented.

\begin{table}[!htb]
\scriptsize{
\centering
\begin{tabular}{|r|| c|c|c ||c|c|c |}
\hline
Moneyness & \multicolumn{3}{|c||}{Smile of $F_{1}$}& \multicolumn{3}{|c|}{Smile of $F_{2}$} \\
\hline
& $\sigma_{market}$ & $\sigma_{model}$ & $\frac{|\sigma_{market}-\sigma_{model}|}{\sigma_{market}}$ & $\sigma_{market}$ & $\sigma_{model}$ &$\frac{|\sigma_{market}-\sigma_{model}|}{\sigma_{market}}$ \\ \hline
$-40\%$ & $97.26$ & $99.64$ & $2.45 \times 10^{-2}$ & $88.27$ & $89.14$ & $9.83 \times 10^{-3}$ \\
\hline
$-20\%$ & $82.58$ & $85.31$ & $3.31 \times 10^{-2}$ & $79.62$ & $81.00$ & $1.73 \times 10^{-2}$ \\
\hline
$0\%$ & $72.29$ & $74.78$ & $3.45 \times 10^{-2}$ & $73.03$ & $74.86$ & $2.51 \times 10^{-2}$ \\
\hline
$20\%$ & $70.89$ & $68.05$ & $4.00 \times 10^{-2}$ & $71.95$ & $70.74$ & $1.68 \times 10^{-2}$ \\
\hline
$40\%$ & $69.49$ & $65.12$ & $6.28 \times 10^{-2}$ & $70.87$ & $68.63$ & $3.16 \times 10^{-2}$ \\
\hline
\hline
Moneyness & \multicolumn{3}{|c||}{Smile of $F_{3}$}& \multicolumn{3}{|c|}{Smile of $F_{4}$} \\
\hline
& $\sigma_{market}$ & $\sigma_{model}$ & $\frac{|\sigma_{market}-\sigma_{model}|}{\sigma_{market}}$ & $\sigma_{market}$ & $\sigma_{model}$ &$\frac{|\sigma_{market}-\sigma_{model}|}{\sigma_{market}}$ \\ \hline
$-40\%$ & $77.09$ & $77.26$ & $2.13 \times 10^{-3}$ & $57.08$ & $56.36$ & $1.27 \times 10^{-2}$ \\
\hline
$-20\%$ & $71.50$ & $72.20$ & $9.78 \times 10^{-3}$ & $53.21$ & $53.13$ & $1.56 \times 10^{-3}$ \\
\hline
$0\%$ & $67.93$ & $68.49$ & $8.29 \times 10^{-3}$ & $52.49$ & $51.00$ & $2.85 \times 10^{-2}$ \\
\hline
$20\%$ & $67.10$ & $66.13$ & $1.45 \times 10^{-2}$ & $51.34$ & $49.96$ & $2.69 \times 10^{-2}$ \\
\hline
$40\%$ & $66.41$ & $65.12$ & $1.95 \times 10^{-2}$ & $50.61$ & $50.02$ & $1.17 \times 10^{-2}$ \\
\hline
\hline
Moneyness & \multicolumn{3}{|c||}{Smile of $F_{5}$}& \multicolumn{3}{|c|}{Smile of $F_{6}$} \\
\hline
& $\sigma_{market}$ & $\sigma_{model}$ & $\frac{|\sigma_{market}-\sigma_{model}|}{\sigma_{market}}$ & $\sigma_{market}$ & $\sigma_{model}$ &$\frac{|\sigma_{market}-\sigma_{model}|}{\sigma_{market}}$ \\ \hline
$-40\%$ & $56.69$ & $55.85$ & $1.48 \times 10^{-2}$ & $56.30$ & $56.13$ & $3.11 \times 10^{-3}$ \\
\hline
$-20\%$ & $52.43$ & $51.58$ & $1.61 \times 10^{-2}$ & $51.65$ & $50.99$ & $1.28 \times 10^{-2}$ \\
\hline
$0\%$ & $50.31$ & $48.60$ & $3.42 \times 10^{-2}$ & $48.19$ & $47.25$ & $1.96 \times 10^{-2}$ \\
\hline
$20\%$ & $48.72$ & $46.89$ & $3.76 \times 10^{-2}$ & $46.19$ & $44.92$ & $2.75 \times 10^{-2}$ \\
\hline
$40\%$ & $47.70$ & $46.46$ & $2.59 \times 10^{-2}$ & $44.91$ & $44.00$ & $2.01 \times 10^{-2}$ \\
\hline
\hline
Moneyness & \multicolumn{3}{|c||}{Smile of $F_{7}$}& \multicolumn{3}{|c|}{Smile of $F_{8}$} \\
\hline
& $\sigma_{market}$ & $\sigma_{model}$ & $\frac{|\sigma_{market}-\sigma_{model}|}{\sigma_{market}}$ & $\sigma_{market}$ & $\sigma_{model}$ &$\frac{|\sigma_{market}-\sigma_{model}|}{\sigma_{market}}$ \\ \hline
$-40\%$ & $55.92$ & $56.74$ & $1.46 \times 10^{-2}$ & $55.54$ & $55.75$ & $3.90 \times 10^{-3}$ \\
\hline
$-20\%$ & $50.89$ & $50.82$ & $1.45 \times 10^{-3}$ & $50.13$ & $49.08$ & $2.09 \times 10^{-2}$ \\
\hline
$0\%$ & $46.19$ & $46.39$ & $4.14 \times 10^{-3}$ & $44.25$ & $43.95$ & $6.82 \times 10^{-3}$ \\
\hline
$20\%$ & $43.83$ & $43.44$ & $8.87 \times 10^{-3}$ & $41.56$ & $40.35$ & $2.92 \times 10^{-2}$ \\
\hline
$40\%$ & $42.32$ & $41.99$ & $7.69 \times 10^{-3}$ & $39.84$ & $38.28$ & $3.92 \times 10^{-2}$ \\
\hline
\hline
Moneyness & \multicolumn{3}{|c||}{Smile of $F_{9}$}& \multicolumn{3}{|c|}{Smile of $F_{10}$} \\
\hline
& $\sigma_{market}$ & $\sigma_{model}$ & $\frac{|\sigma_{market}-\sigma_{model}|}{\sigma_{market}}$ & $\sigma_{market}$ & $\sigma_{model}$ &$\frac{|\sigma_{market}-\sigma_{model}|}{\sigma_{market}}$ \\ \hline
$-40\%$ & $55.16$ & $57.11$ & $3.54 \times 10^{-2}$ & $54.78$ & $56.57$ & $3.26 \times 10^{-2}$ \\
\hline
$-20\%$ & $49.39$ & $49.84$ & $9.09 \times 10^{-3}$ & $48.65$ & $48.84$ & $3.91 \times 10^{-3}$ \\
\hline
$0\%$ & $42.40$ & $44.10$ & $4.00 \times 10^{-2}$ & $40.61$ & $42.61$ & $4.90 \times 10^{-2}$ \\
\hline
$20\%$ & $39.43$ & $39.89$ & $1.17 \times 10^{-2}$ & $37.38$ & $37.85$ & $1.26 \times 10^{-2}$ \\
\hline
$40\%$ & $37.54$ & $37.21$ & $8.79 \times 10^{-3}$ & $35.34$ & $34.59$ & $2.13 \times 10^{-2}$ \\
\hline
\hline
Moneyness & \multicolumn{3}{|c||}{Smile of $F_{11}$}& \multicolumn{3}{|c|}{Smile of $F_{12}$} \\
\hline
& $\sigma_{market}$ & $\sigma_{model}$ & $\frac{|\sigma_{market}-\sigma_{model}|}{\sigma_{market}}$ & $\sigma_{market}$ & $\sigma_{model}$ &$\frac{|\sigma_{market}-\sigma_{model}|}{\sigma_{market}}$ \\ \hline
$-40\%$ & $54.41$ & $57.20$ & $5.13 \times 10^{-2}$ & $54.03$ & $56.71$ & $4.95 \times 10^{-2}$ \\
\hline
$-20\%$ & $47.94$ & $49.07$ & $2.36 \times 10^{-2}$ & $47.22$ & $48.33$ & $2.34 \times 10^{-2}$ \\
\hline
$0\%$ & $38.93$ & $42.36$ & $8.81 \times 10^{-2}$ & $37.29$ & $41.29$ & $1.07 \times 10^{-1}$ \\
\hline
$20\%$ & $35.47$ & $37.07$ & $4.51 \times 10^{-2}$ & $33.63$ & $35.59$ & $5.82 \times 10^{-2}$ \\
\hline
$40\%$ & $33.30$ & $33.21$ & $2.95 \times 10^{-3}$ & $31.36$ & $31.23$ & $3.95 \times 10^{-3}$ \\
\hline
\hline
\multicolumn{7}{|c|}{$MRE=2.93 \times 10^{-2}$} \\
\hline

\end{tabular}
\caption{Rebonato model, calibration to caplets, $\sigma_{market}$ vs. $\sigma_{model}$.}
\label{tab:rebonatoVolas}
}
\end{table}

We also performed the equivalent calibration with Monte Carlo simulation thus obtaining the same parameters as in Table \ref{tab:paramsRebonatoVolas}, except for $\phi_{11}=0.0940$.  We note that the $MAE$ in prices is $3.43 \times 10^{-2}$ with formula \eqref{eq:sigmaHagan}, while $MAE$ is $3.39 \times 10^{-2}$ using Monte Carlo.

\begin{figure}[!htb]
\centering
  \subfigure {\includegraphics[height=3.7cm]{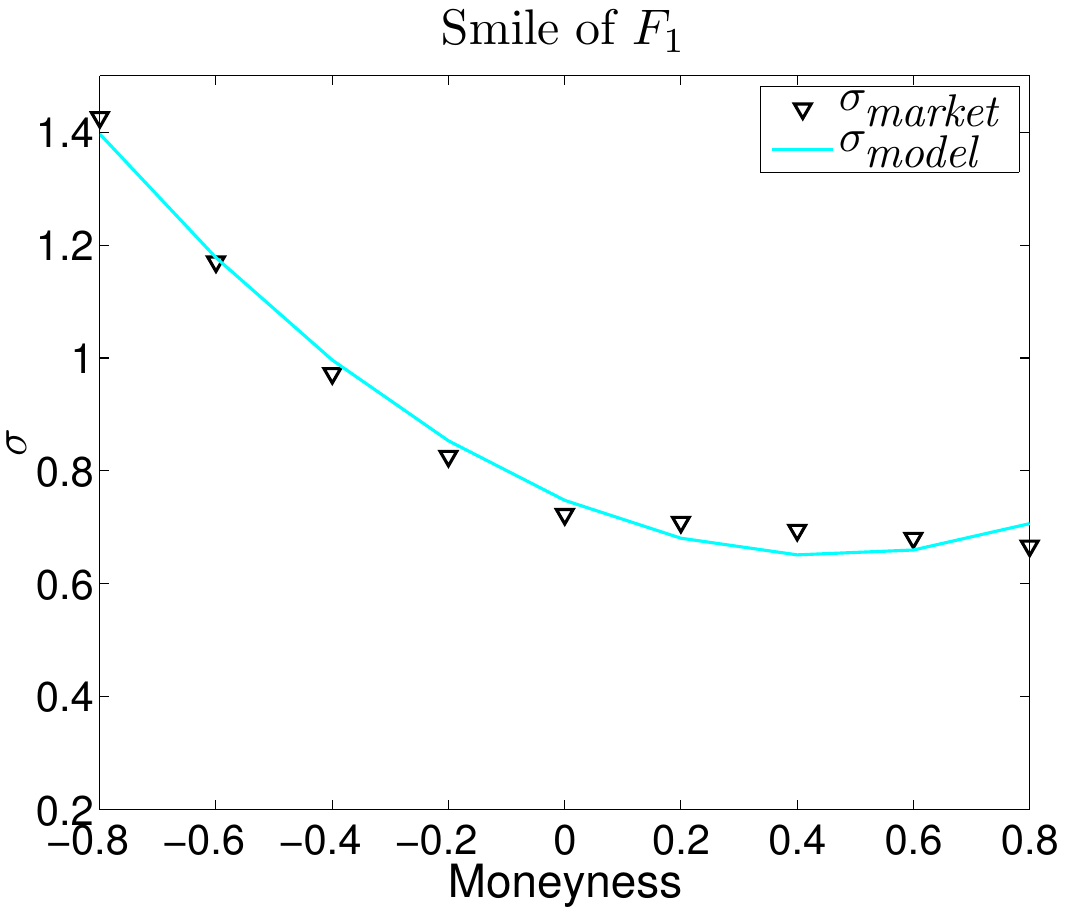}}
  \subfigure {\includegraphics[height=3.7cm]{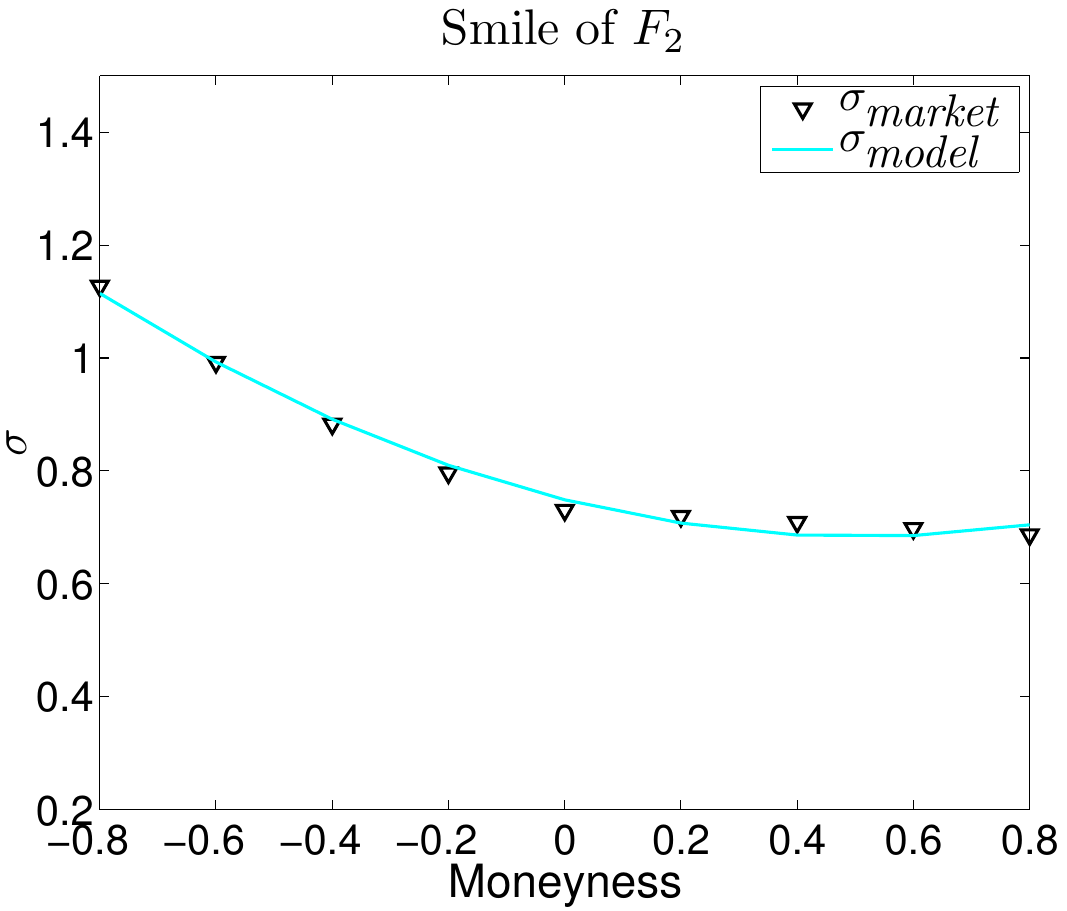}}
  \subfigure {\includegraphics[height=3.7cm]{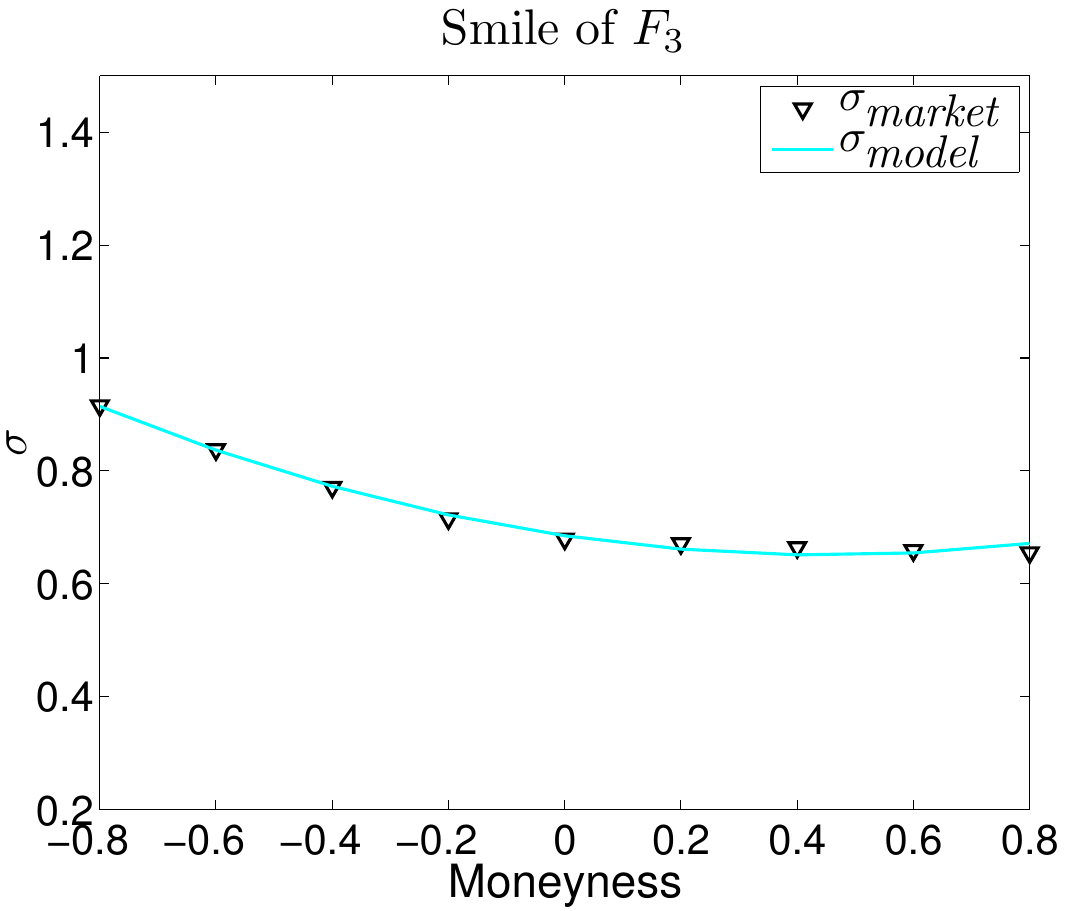}}
  \subfigure {\includegraphics[height=3.7cm]{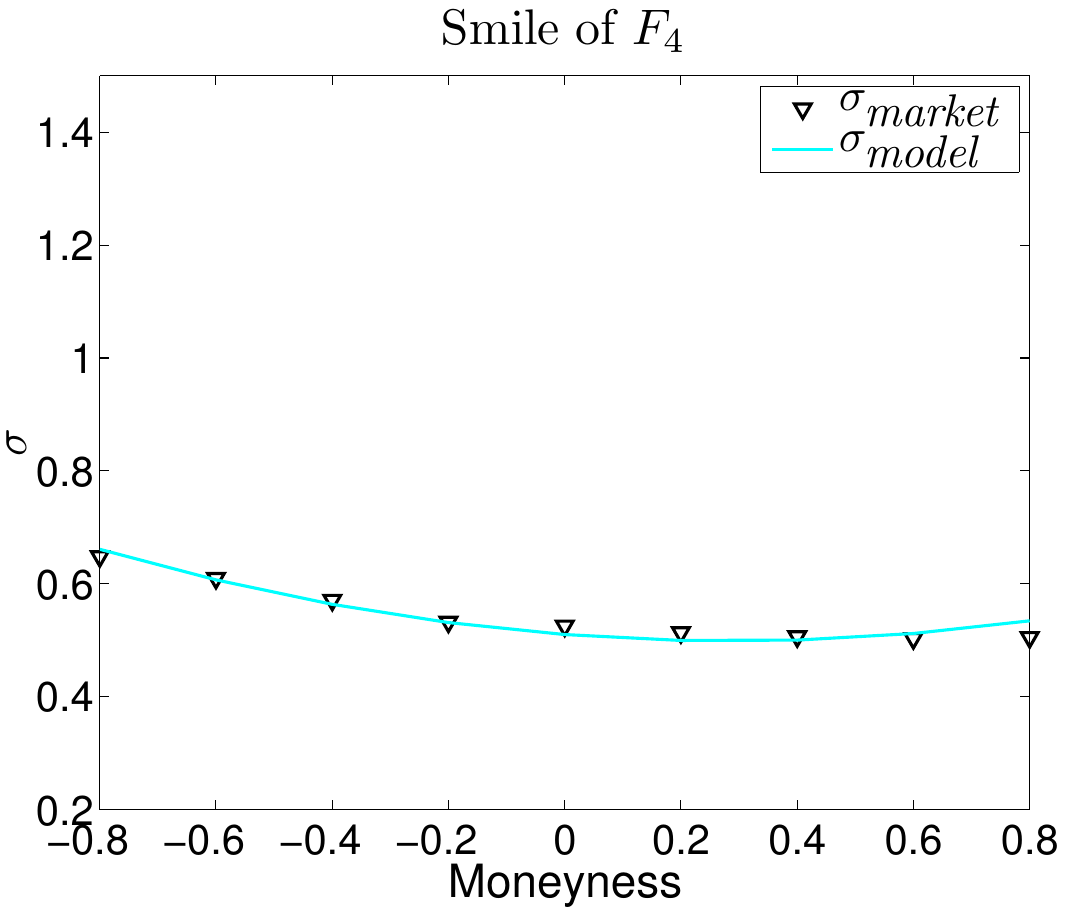}}
  \subfigure {\includegraphics[height=3.7cm]{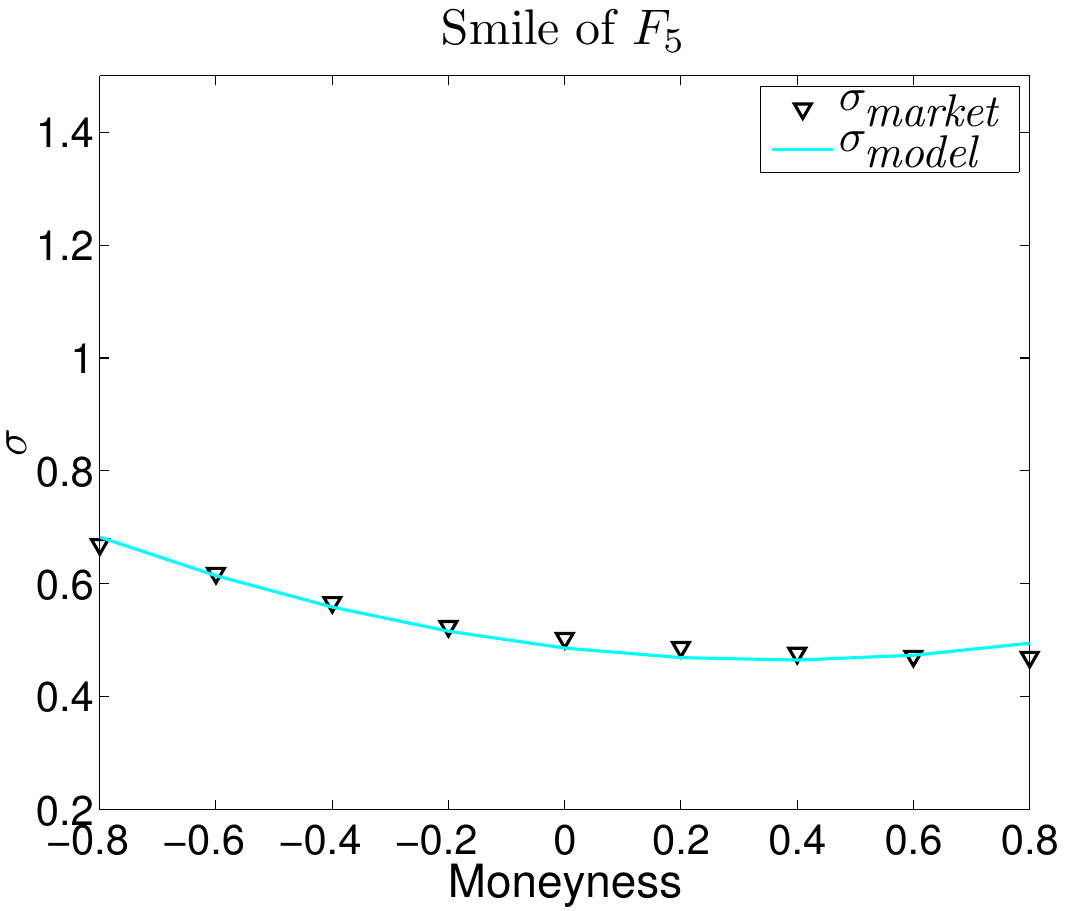}}
  \subfigure {\includegraphics[height=3.7cm]{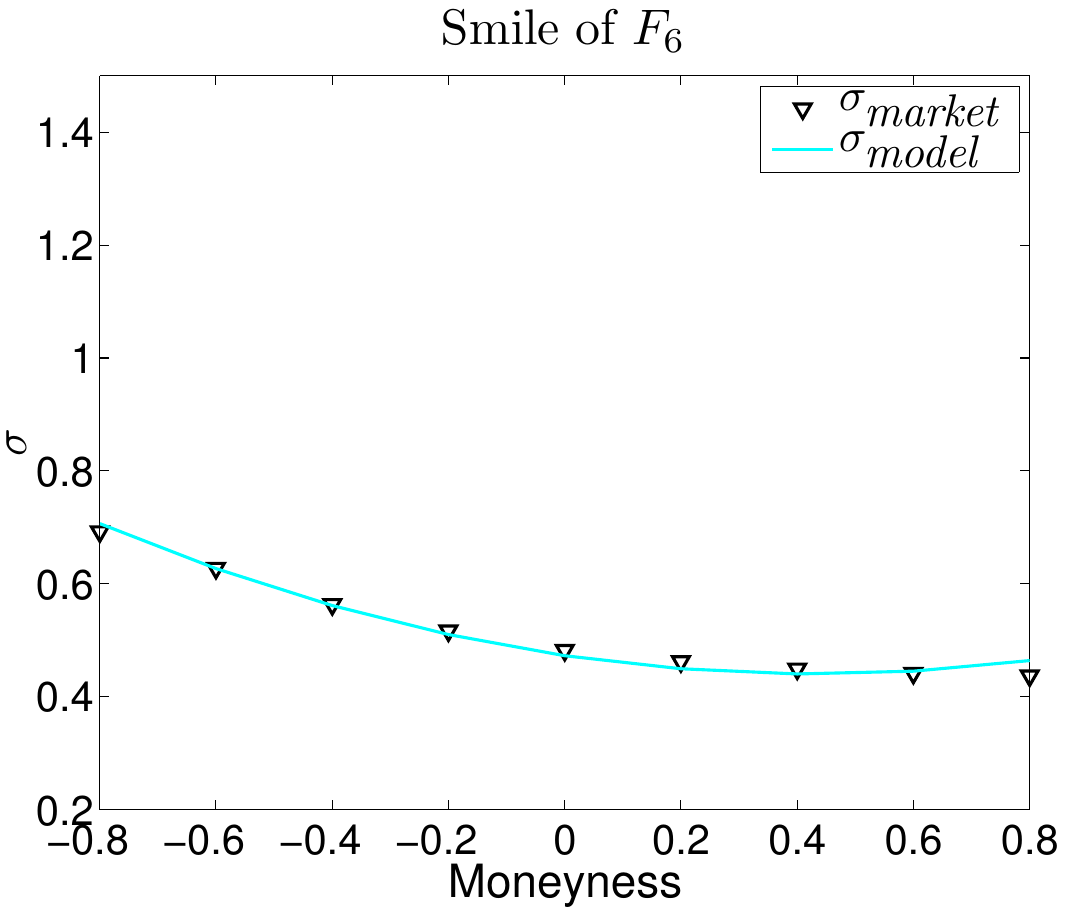}}
  \subfigure {\includegraphics[height=3.7cm]{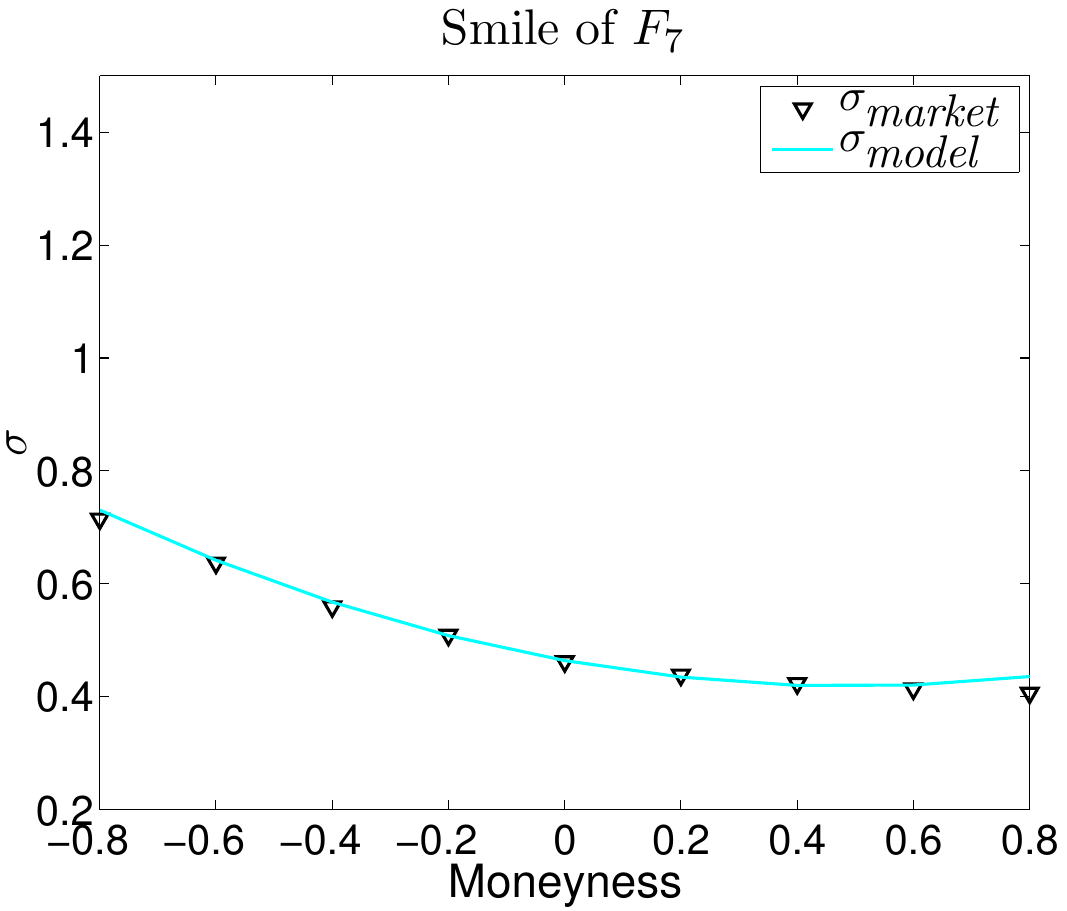}}
  \subfigure {\includegraphics[height=3.7cm]{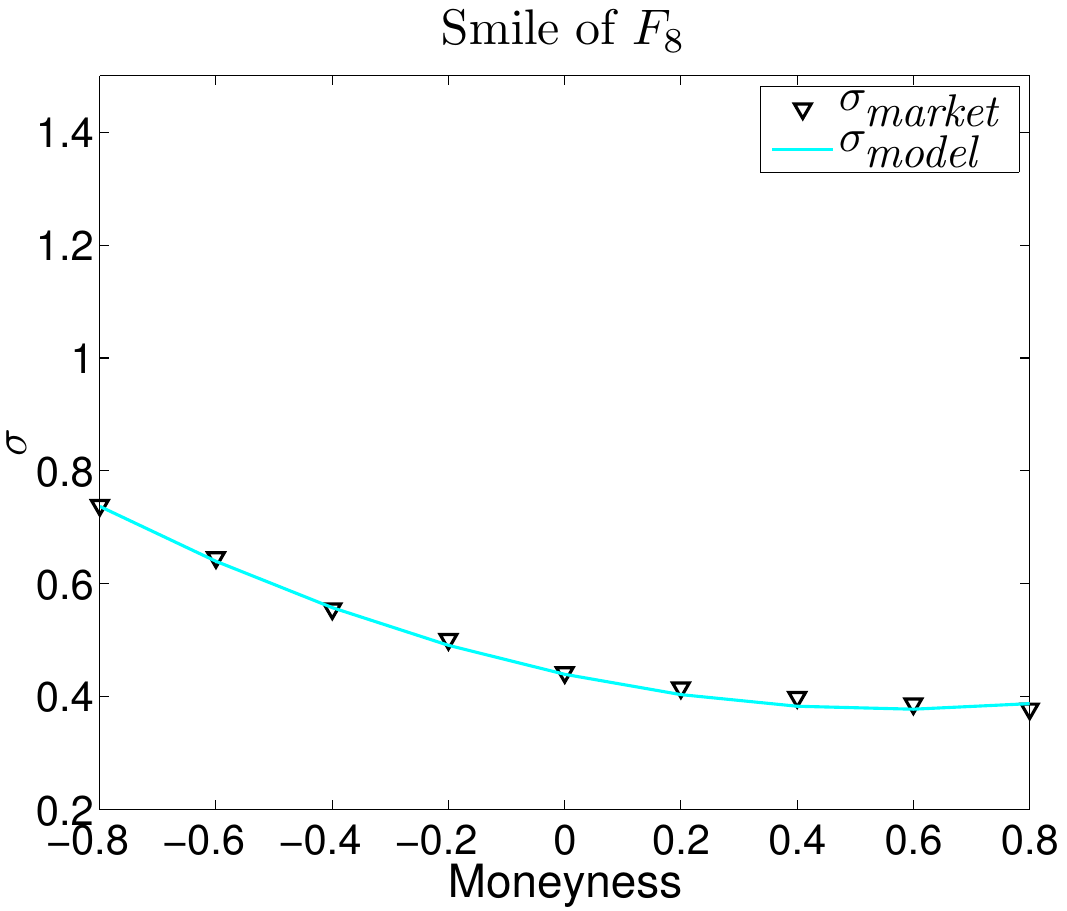}}
  \subfigure {\includegraphics[height=3.7cm]{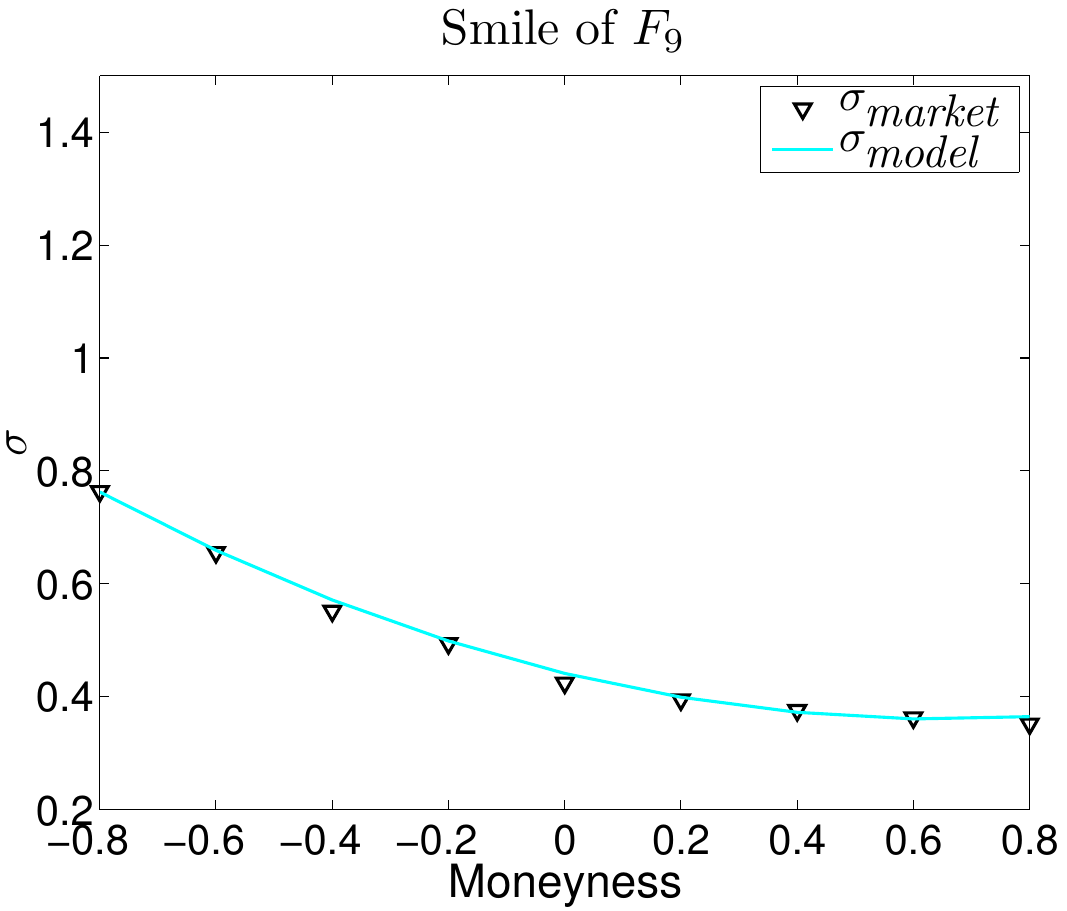}}
  \subfigure {\includegraphics[height=3.7cm]{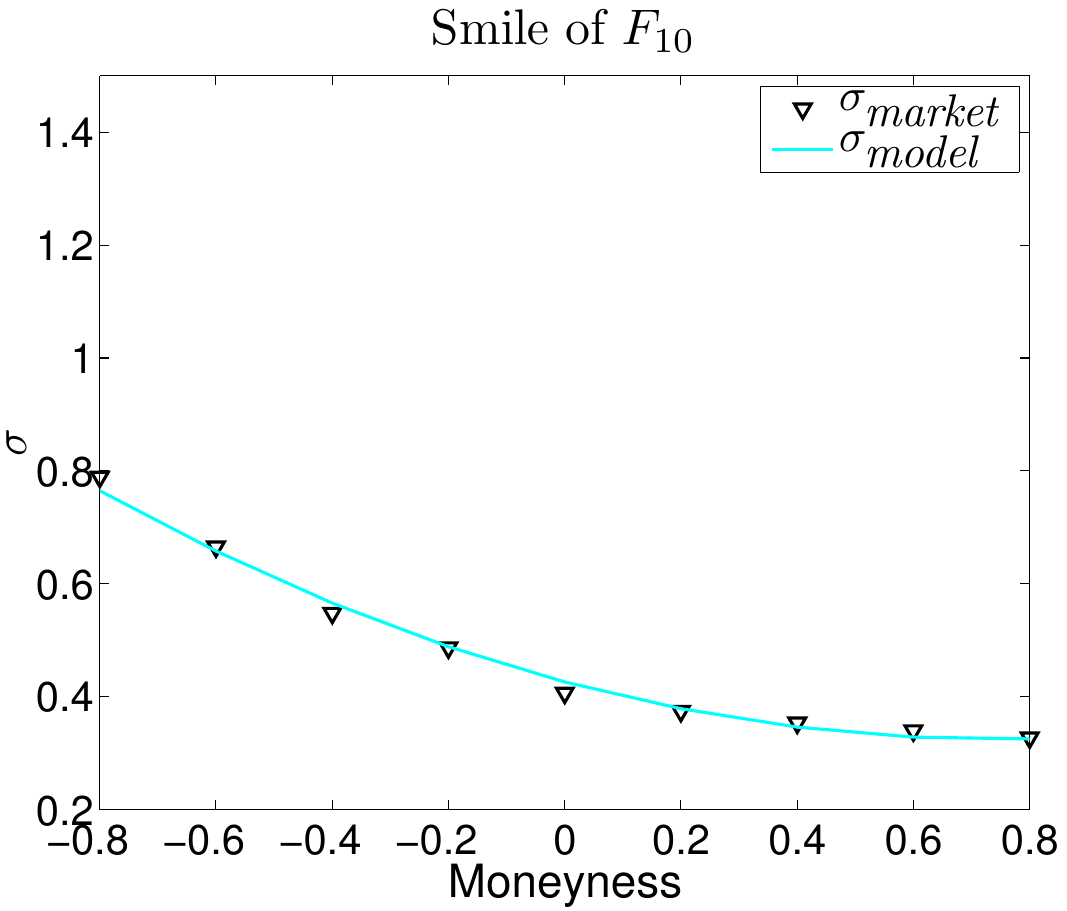}}
  \subfigure {\includegraphics[height=3.7cm]{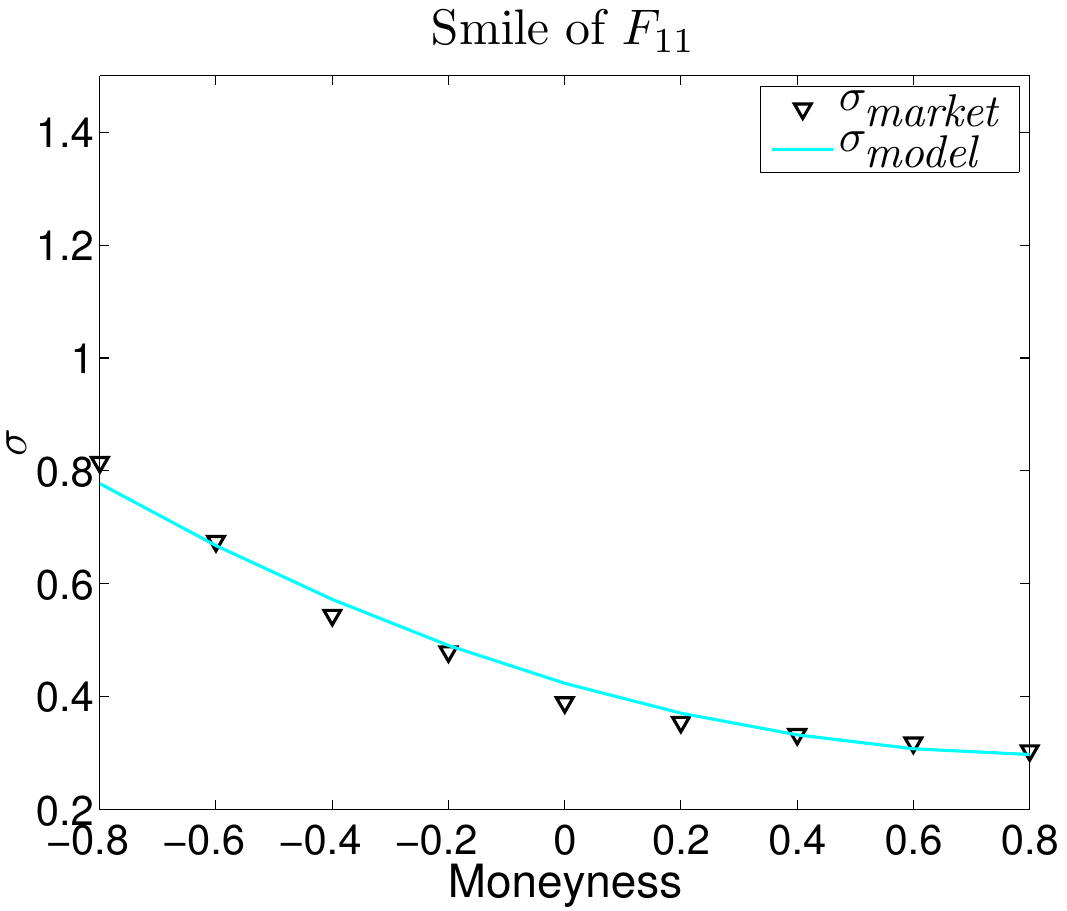}}
  \subfigure {\includegraphics[height=3.7cm]{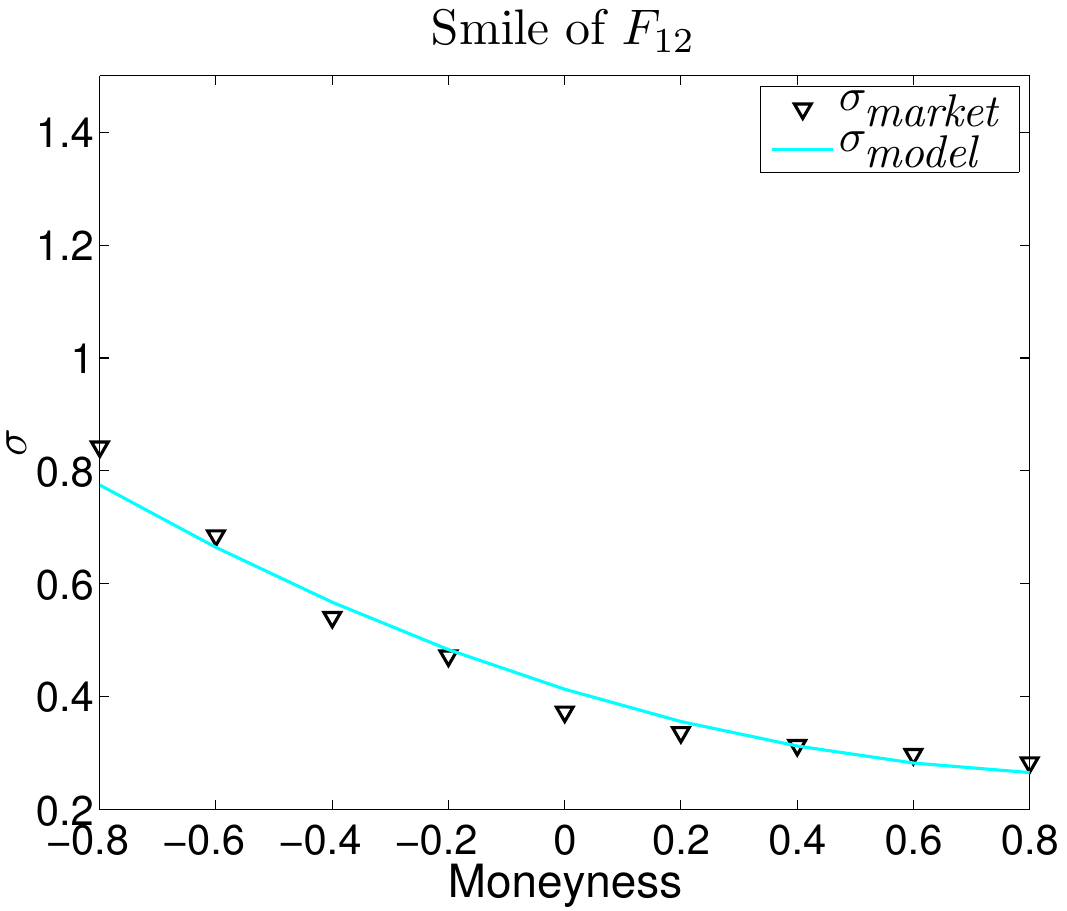}}
  \subfigure {\includegraphics[height=3.7cm]{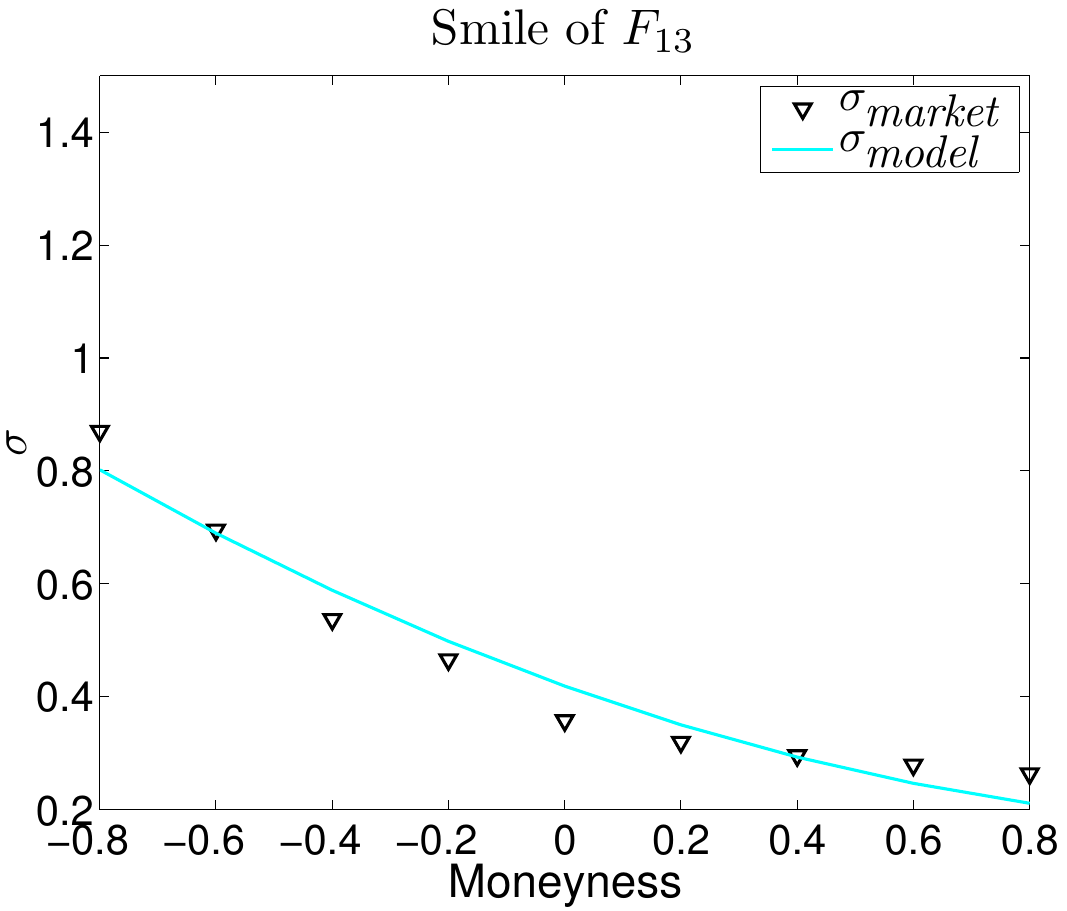}}
  \caption{Rebonato model, $\sigma_{market}$ vs. $\sigma_{model}$, smiles of $F_1,\ldots,F_{13}$.}
  \label{fig:rebonatoVolas}
\end{figure}

In Figure \ref{fig:rebonatoVolas} the model fitting for the smiles of all forward rates is shown. Market volatilities are plotted with triangles, while model volatilities are shown in continuous line.

\subsubsection{Calibration to swaptions}

The calibrated parameters are $\eta_1 = 0.650997$, $\lambda_1 = 3.617546$, $\eta_2 = 0.999000$, $\lambda_2 = 0.380984$ and $\lambda_3 = 0.001000$. Using two GPUs the execution time was approximately 2 hours, as in the previous models (by using a cluster of GPUs time could be substantially reduced). In Table \ref{tab:rebonatoSwaptions} market vs. model swaptions prices (in $\%$) for the first fourteen swaptions and the the moneyness varying from $-40\%$ to $40\%$ are shown, each pair with its corresponding absolute error. In addition, for the whole set of swaptions the mean absolute error is presented.

\begin{table}[!htb]
\scriptsize{
\centering
\begin{tabular}{|r|| c|c|c ||c|c|c |}
\hline
Moneyness & \multicolumn{3}{|c||}{$0.5 \times 1$ swaptions} & \multicolumn{3}{|c|}{$1 \times 1$ swaptions} \\
\hline
& $S_{Black}$ & $S_{MC}$ & $|S_{Black}-S_{MC}|$ & $S_{Black}$ & $S_{MC}$ & $|S_{Black}-S_{MC}|$ \\ \hline
$-40\%$ & $0.4866$ & $0.4870$ & $4.00 \times 10^{-4}$ & $0.5917$ & $0.5839$ & $7.80 \times 10^{-3}$ \\
\hline
$-20\%$ & $0.3562$ & $0.3669$ & $1.07 \times 10^{-2}$ & $0.4661$ & $0.4693$ & $3.20 \times 10^{-3}$ \\
\hline
$0\%$ & $0.2356$ & $0.2477$ & $1.21 \times 10^{-2}$ & $0.3467$ & $0.3546$ & $7.90 \times 10^{-3}$ \\
\hline
$20\%$ & $0.1363$ & $0.1441$ & $7.80 \times 10^{-3}$ & $0.2394$ & $0.2488$ & $9.40 \times 10^{-3}$ \\
\hline
$40\%$ & $0.0680$ & $0.0699$ & $1.90 \times 10^{-3}$ & $0.1517$ & $0.1606$ & $8.90 \times 10^{-3}$ \\
\hline
\hline
Moneyness & \multicolumn{3}{|c||}{$1.5 \times 1$ swaptions} & \multicolumn{3}{|c|}{$2 \times 1$ swaptions} \\
\hline
& $S_{Black}$ & $S_{MC}$ & $|S_{Black}-S_{MC}|$ & $S_{Black}$ & $S_{MC}$ & $|S_{Black}-S_{MC}|$ \\ \hline
$-40\%$ & $0.7357$ & $0.6902$ & $4.55 \times 10^{-2}$ & $0.8184$ & $0.7465$ & $7.19 \times 10^{-2}$ \\
\hline
$-20\%$ & $0.5908$ & $0.5612$ & $2.96 \times 10^{-2}$ & $0.6603$ & $0.6028$ & $5.75 \times 10^{-2}$ \\
\hline
$0\%$ & $0.4536$ & $0.4339$ & $1.97 \times 10^{-2}$ & $0.5118$ & $0.4620$ & $4.98 \times 10^{-2}$ \\
\hline
$20\%$ & $0.3277$ & $0.3171$ & $1.06 \times 10^{-2}$ & $0.3754$ & $0.3354$ & $4.00 \times 10^{-2}$ \\
\hline
$40\%$ & $0.2213$ & $0.2188$ & $2.50 \times 10^{-3}$ & $0.2587$ & $0.2308$ & $2.79 \times 10^{-2}$ \\
\hline
\hline
Moneyness & \multicolumn{3}{|c||}{$0.5 \times 2$ swaptions} & \multicolumn{3}{|c|}{$1 \times 2$ swaptions} \\
\hline
& $S_{Black}$ & $S_{MC}$ & $|S_{Black}-S_{MC}|$ & $S_{Black}$ & $S_{MC}$ & $|S_{Black}-S_{MC}|$ \\ \hline
$-40\%$ & $1.0570$ & $1.0333$ & $2.37 \times 10^{-2}$ & $1.2427$ & $1.2175$ & $2.52 \times 10^{-2}$ \\
\hline
$-20\%$ & $0.7440$ & $0.7514$ & $7.40 \times 10^{-3}$ & $0.9322$ & $0.9400$ & $7.80 \times 10^{-3}$ \\
\hline
$0\%$ & $0.4555$ & $0.4841$ & $2.86 \times 10^{-2}$ & $0.6394$ & $0.6713$ & $3.19 \times 10^{-2}$ \\
\hline
$20\%$ & $0.2299$ & $0.2674$ & $3.75 \times 10^{-2}$ & $0.3886$ & $0.4369$ & $4.83 \times 10^{-2}$ \\
\hline
$40\%$ & $0.0925$ & $0.1237$ & $3.12 \times 10^{-2}$ & $0.2037$ & $0.2578$ & $5.41 \times 10^{-2}$ \\
\hline
\hline
Moneyness & \multicolumn{3}{|c||}{$1.5 \times 2$ swaptions} & \multicolumn{3}{|c|}{$2 \times 2$ swaptions} \\
\hline
& $S_{Black}$ & $S_{MC}$ & $|S_{Black}-S_{MC}|$ & $S_{Black}$ & $S_{MC}$ & $|S_{Black}-S_{MC}|$ \\ \hline
$-40\%$ & $1.4884$ & $1.4357$ & $5.27 \times 10^{-2}$ & $1.6938$ & $1.6184$ & $7.54 \times 10^{-2}$ \\
\hline
$-20\%$ & $1.1367$ & $1.1256$ & $1.11 \times 10^{-2}$ & $1.3077$ & $1.2721$ & $3.56 \times 10^{-2}$ \\
\hline
$0\%$ & $0.8059$ & $0.8250$ & $1.91 \times 10^{-2}$ & $0.9466$ & $0.9309$ & $1.57 \times 10^{-2}$ \\
\hline
$20\%$ & $0.5154$ & $0.5599$ & $4.45 \times 10^{-2}$ & $0.6269$ & $0.6292$ & $2.30 \times 10^{-3}$ \\
\hline
$40\%$ & $0.2919$ & $0.3520$ & $6.01 \times 10^{-2}$ & $0.3736$ & $0.3931$ & $1.95 \times 10^{-2}$ \\
\hline
\hline
Moneyness & \multicolumn{3}{|c||}{$0.5 \times 3$ swaptions} & \multicolumn{3}{|c|}{$1 \times 3$ swaptions} \\
\hline
& $S_{Black}$ & $S_{MC}$ & $|S_{Black}-S_{MC}|$ & $S_{Black}$ & $S_{MC}$ & $|S_{Black}-S_{MC}|$ \\ \hline
$-40\%$ & $1.7380$ & $1.6792$ & $5.88 \times 10^{-2}$ & $2.0341$ & $1.9953$ & $3.88 \times 10^{-2}$ \\
\hline
$-20\%$ & $1.1980$ & $1.1850$ & $1.30 \times 10^{-2}$ & $1.4851$ & $1.4966$ & $1.15 \times 10^{-2}$ \\
\hline
$0\%$ & $0.7011$ & $0.7235$ & $2.24 \times 10^{-2}$ & $0.9696$ & $1.0176$ & $4.80 \times 10^{-2}$ \\
\hline
$20\%$ & $0.3242$ & $0.3653$ & $4.11 \times 10^{-2}$ & $0.5413$ & $0.6130$ & $7.17 \times 10^{-2}$ \\
\hline
$40\%$ & $0.1128$ & $0.1484$ & $3.56 \times 10^{-2}$ & $0.2479$ & $0.3241$ & $7.62 \times 10^{-2}$ \\
\hline
\hline
Moneyness & \multicolumn{3}{|c||}{$1.5 \times 3$ swaptions} & \multicolumn{3}{|c|}{$2 \times 3$ swaptions} \\
\hline
& $S_{Black}$ & $S_{MC}$ & $|S_{Black}-S_{MC}|$ & $S_{Black}$ & $S_{MC}$ & $|S_{Black}-S_{MC}|$ \\ \hline
$-40\%$ & $2.3898$ & $2.3112$ & $7.86 \times 10^{-2}$ & $2.6885$ & $2.6048$ & $8.37 \times 10^{-2}$ \\
\hline
$-20\%$ & $1.7850$ & $1.7661$ & $1.89 \times 10^{-2}$ & $2.0311$ & $2.0098$ & $2.13 \times 10^{-2}$ \\
\hline
$0\%$ & $1.2175$ & $1.2360$ & $1.85 \times 10^{-2}$ & $1.4178$ & $1.4192$ & $1.40 \times 10^{-3}$ \\
\hline
$20\%$ & $0.7304$ & $0.7797$ & $4.93 \times 10^{-2}$ & $0.8856$ & $0.9005$ & $1.49 \times 10^{-2}$ \\
\hline
$40\%$ & $0.3749$ & $0.4417$ & $6.68 \times 10^{-2}$ & $0.4832$ & $0.5124$ & $2.92 \times 10^{-2}$ \\
\hline
\hline
Moneyness & \multicolumn{3}{|c||}{$0.5 \times 4$ swaptions} & \multicolumn{3}{|c|}{$1 \times 4$ swaptions} \\
\hline
& $S_{Black}$ & $S_{MC}$ & $|S_{Black}-S_{MC}|$ & $S_{Black}$ & $S_{MC}$ & $|S_{Black}-S_{MC}|$ \\ \hline
$-40\%$ & $2.5381$ & $2.4493$ & $8.88 \times 10^{-2}$ & $2.9426$ & $2.8751$ & $6.75 \times 10^{-2}$ \\
\hline
$-20\%$ & $1.7151$ & $1.6834$ & $3.17 \times 10^{-2}$ & $2.1123$ & $2.1053$ & $7.00 \times 10^{-3}$ \\
\hline
$0\%$ & $0.9584$ & $0.9709$ & $1.25 \times 10^{-2}$ & $1.3344$ & $1.3649$ & $3.05 \times 10^{-2}$ \\
\hline
$20\%$ & $0.4031$ & $0.4401$ & $3.70 \times 10^{-2}$ & $0.7016$ & $0.7572$ & $5.56 \times 10^{-2}$ \\
\hline
$40\%$ & $0.1188$ & $0.1506$ & $3.18 \times 10^{-2}$ & $0.2907$ & $0.3538$ & $6.31 \times 10^{-2}$ \\
\hline
\hline
\multicolumn{7}{|c|}{$MAE=6.30 \times 10^{-2}$} \\
\hline
\end{tabular}
\caption{Rebonato model, calibration to swaptions, $S_{Black}$ vs. $S_{MC}$, prices in \%.}
\label{tab:rebonatoSwaptions}
}
\end{table}

\begin{figure}[!htb]
\centering
  \subfigure {\includegraphics[height=4.4cm]{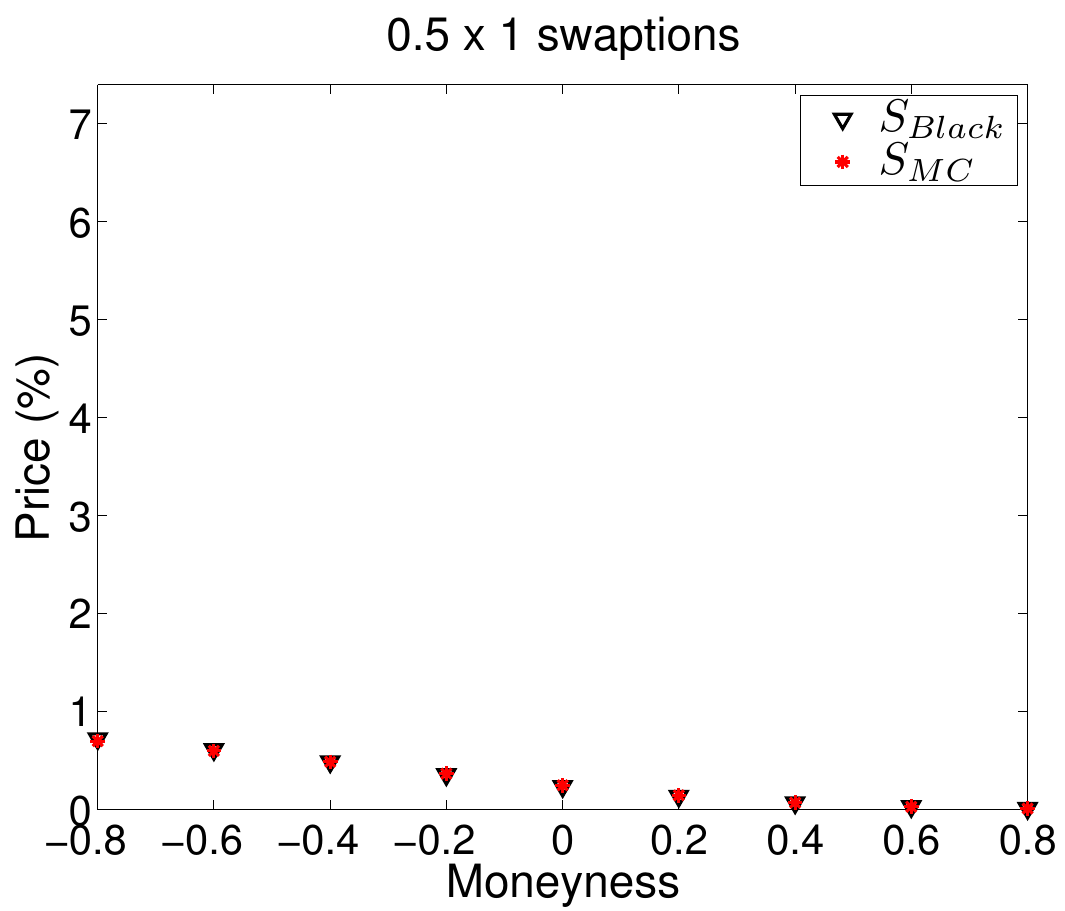}}
  \subfigure {\includegraphics[height=4.4cm]{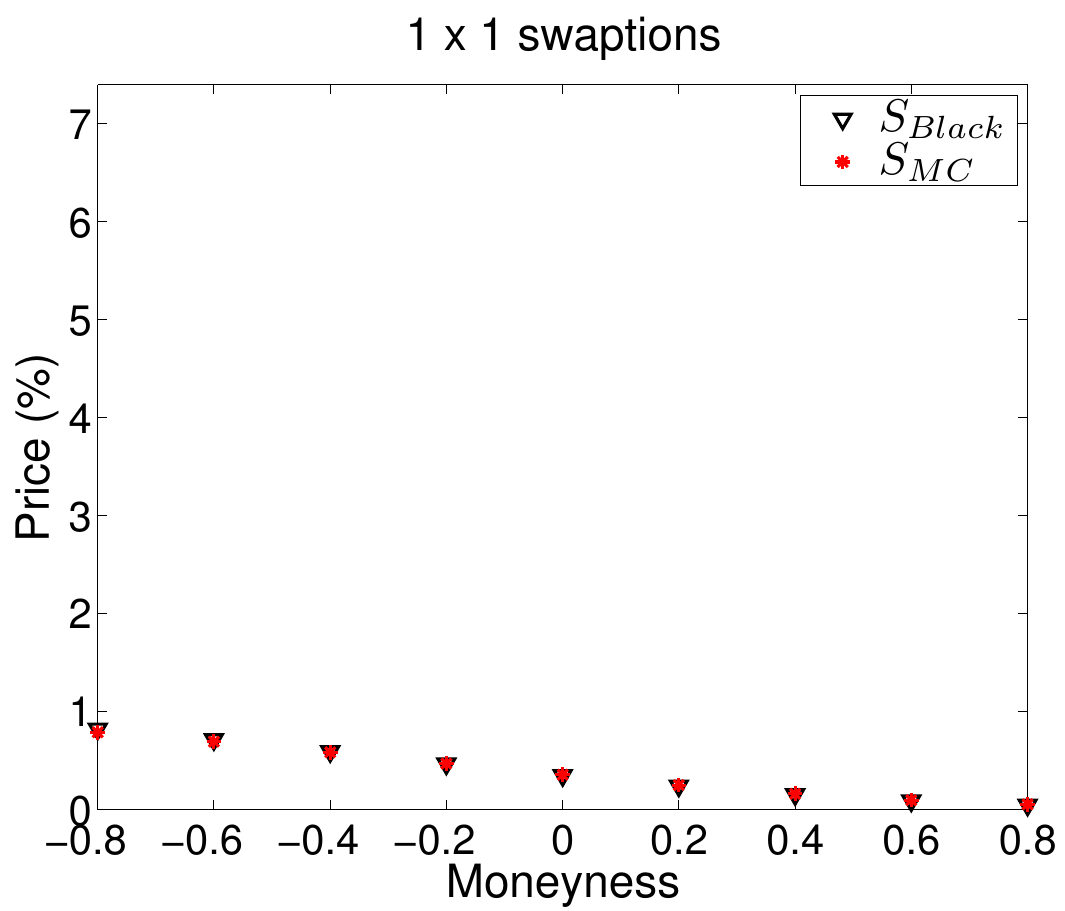}}
  \subfigure {\includegraphics[height=4.4cm]{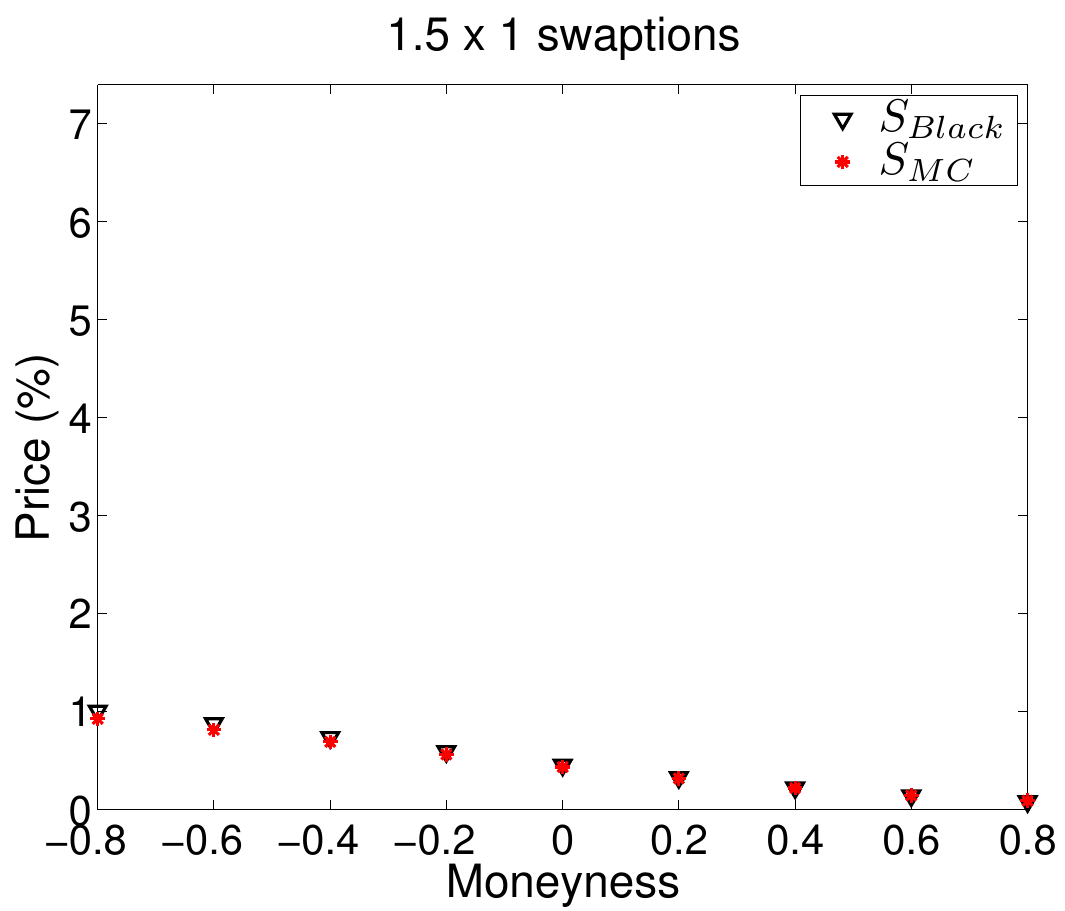}}
  \subfigure {\includegraphics[height=4.4cm]{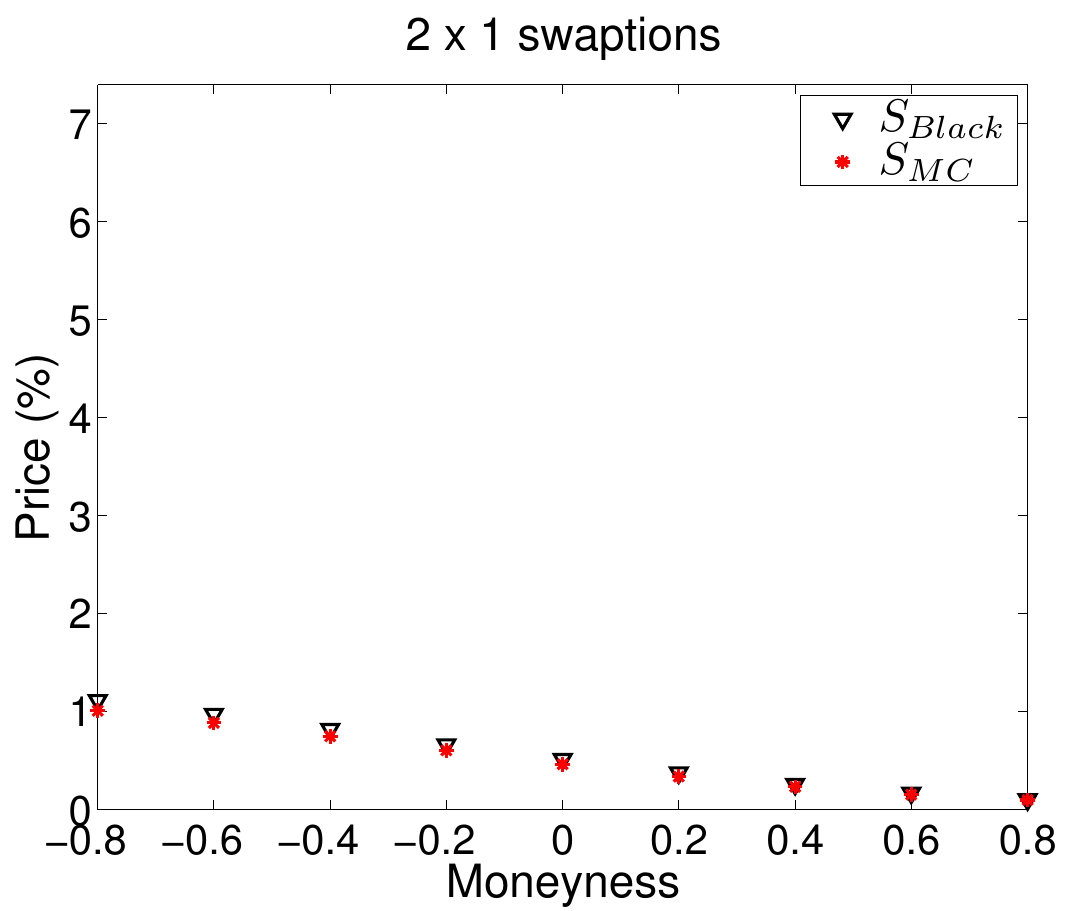}}
  \subfigure {\includegraphics[height=4.4cm]{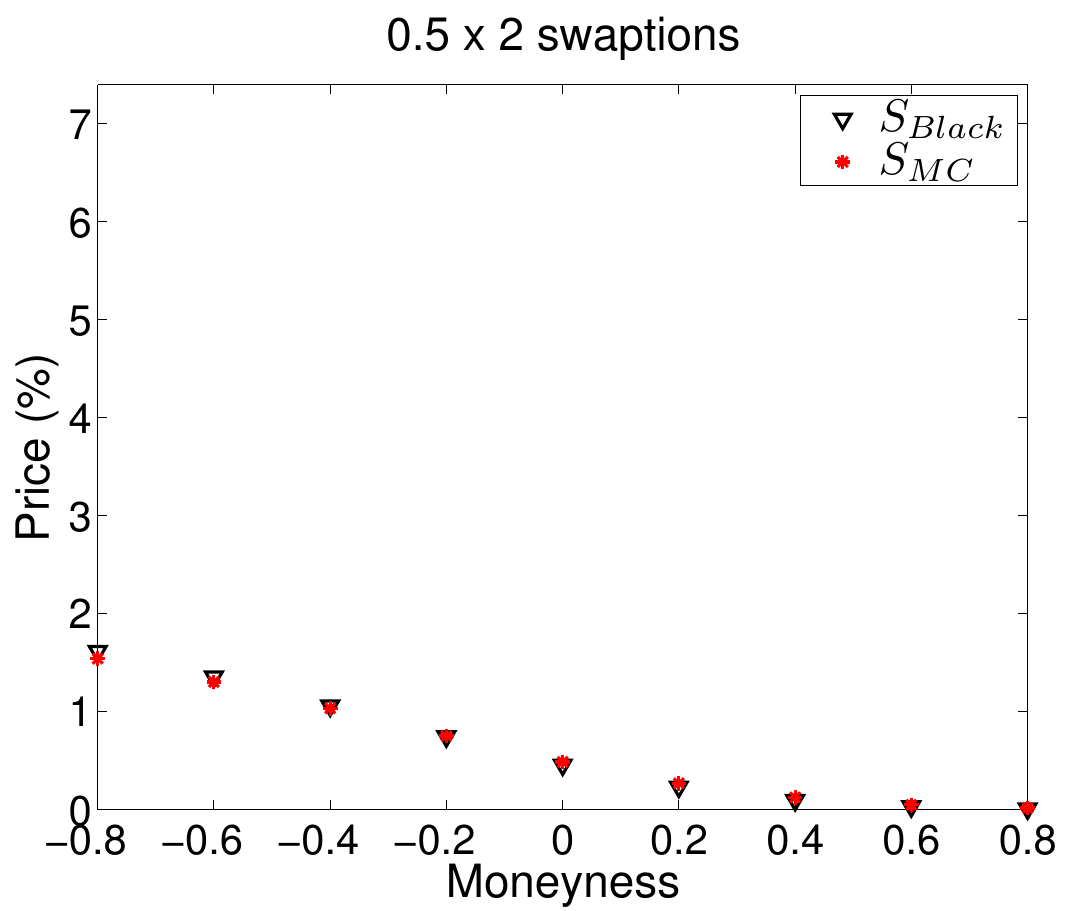}}
  \subfigure {\includegraphics[height=4.4cm]{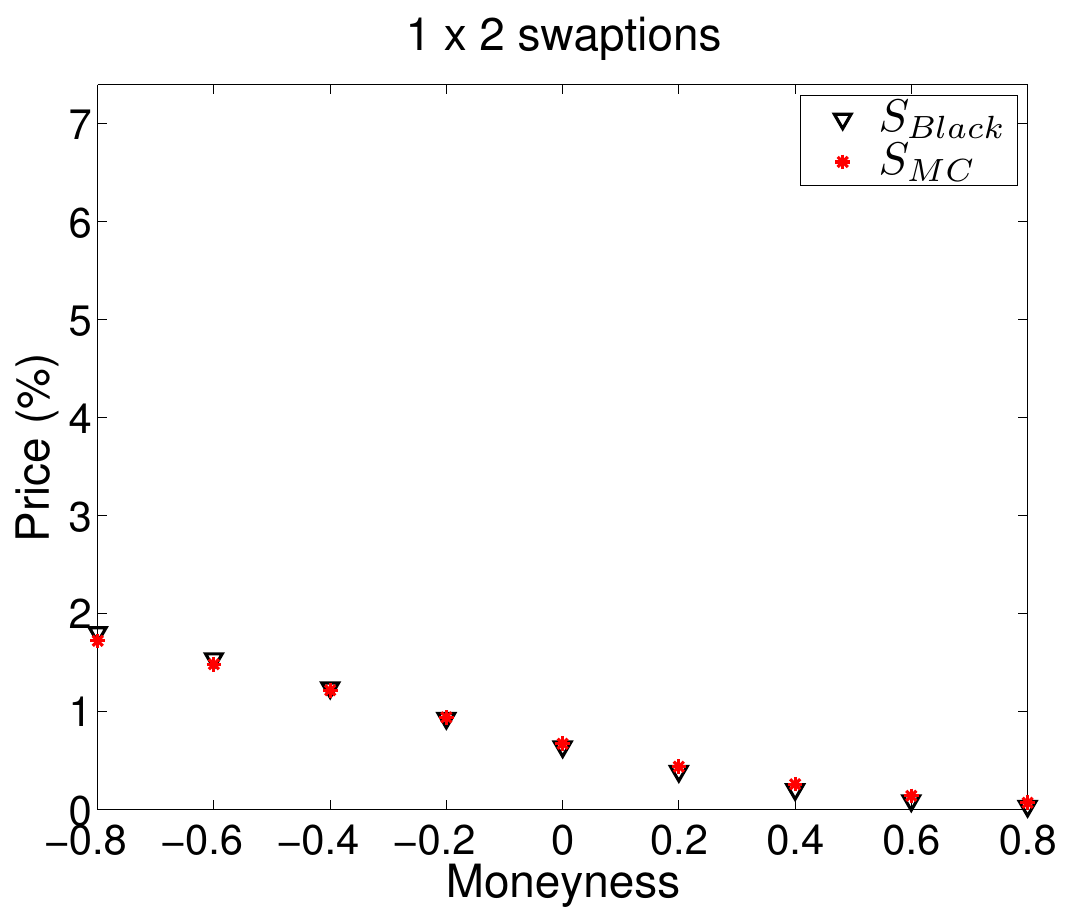}}
  \subfigure {\includegraphics[height=4.4cm]{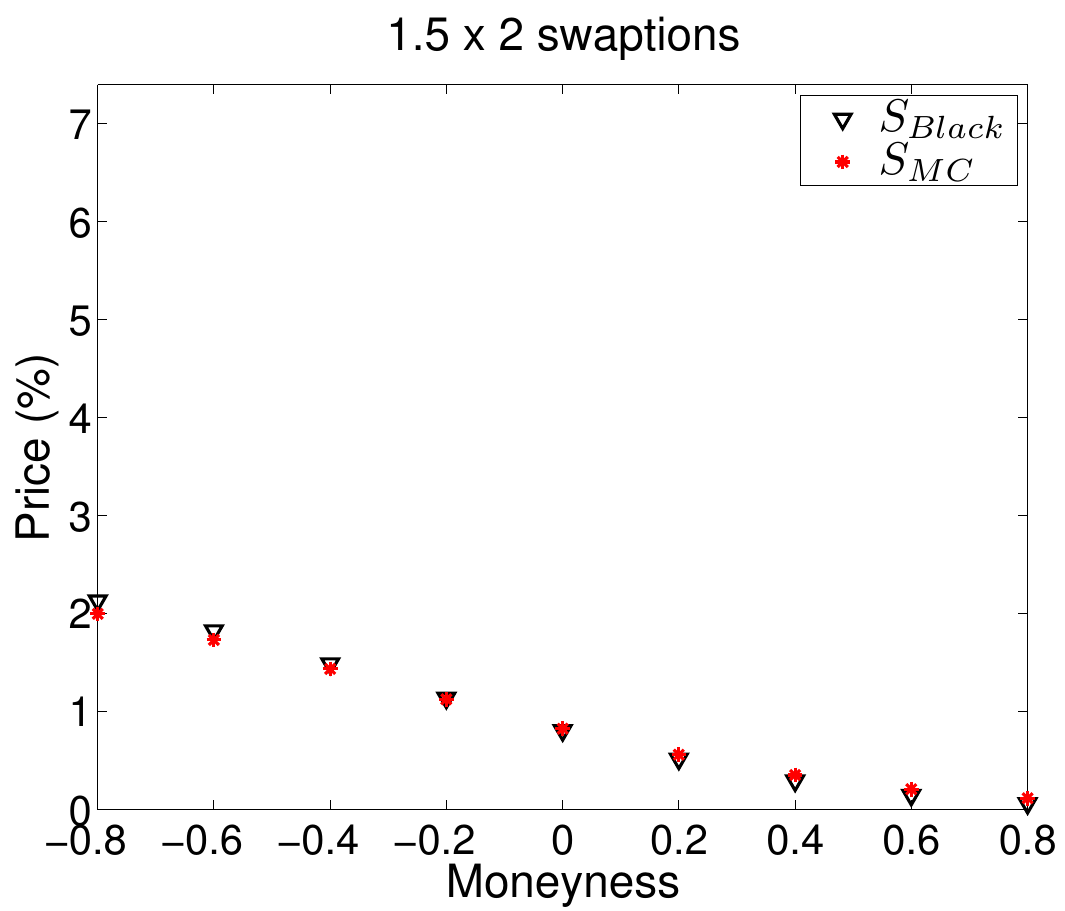}}
  \subfigure {\includegraphics[height=4.4cm]{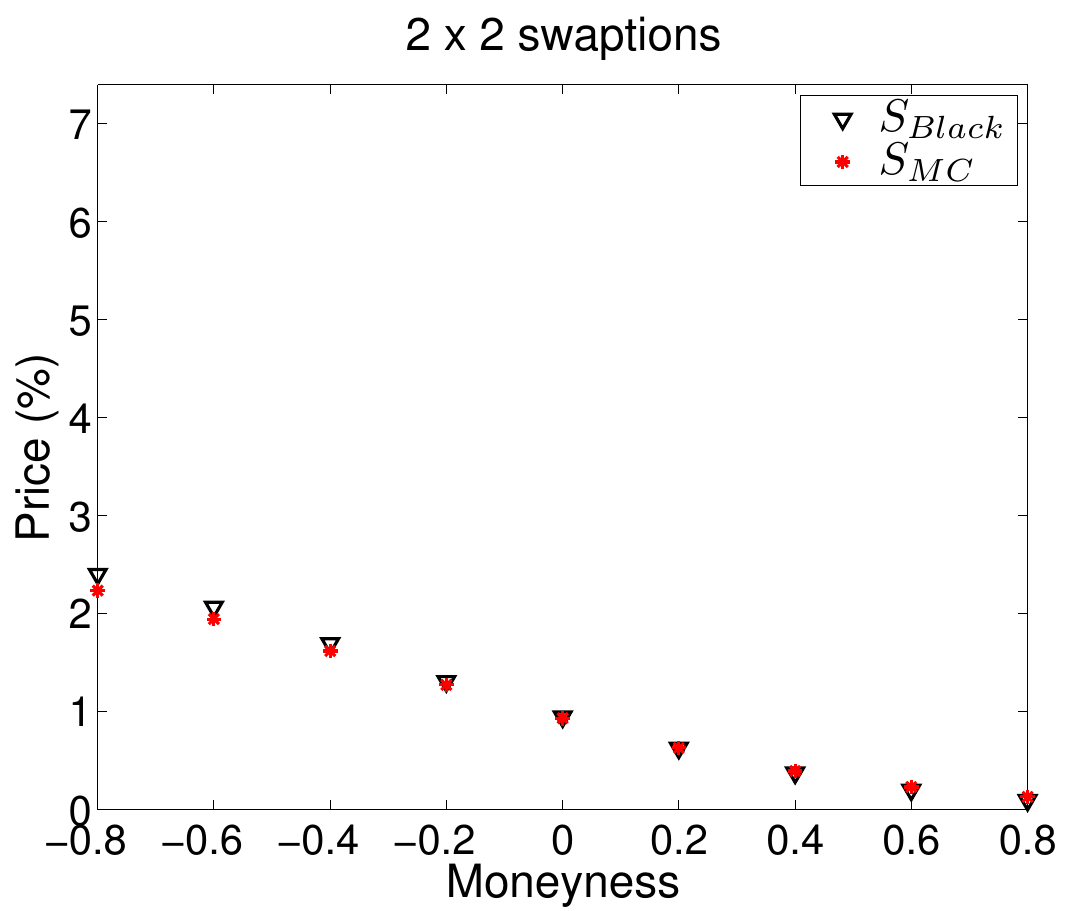}}
  \subfigure {\includegraphics[height=4.4cm]{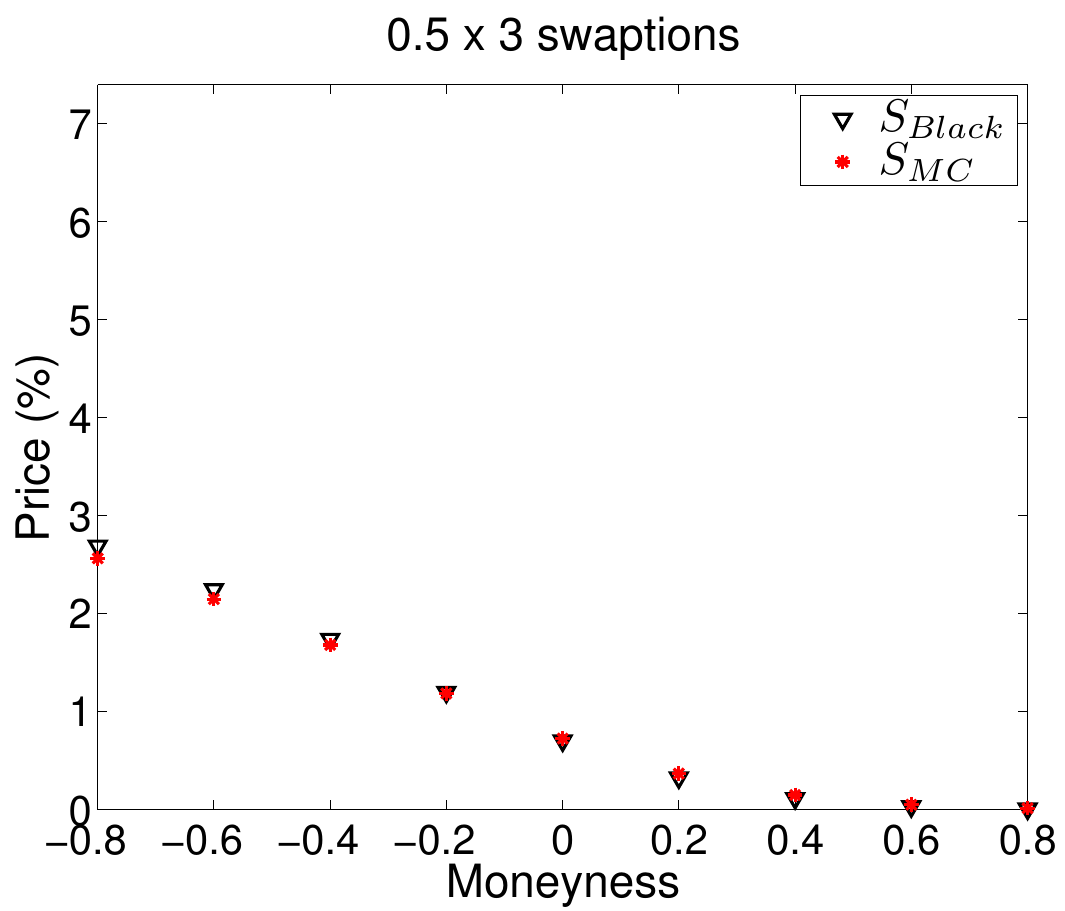}}
  \subfigure {\includegraphics[height=4.4cm]{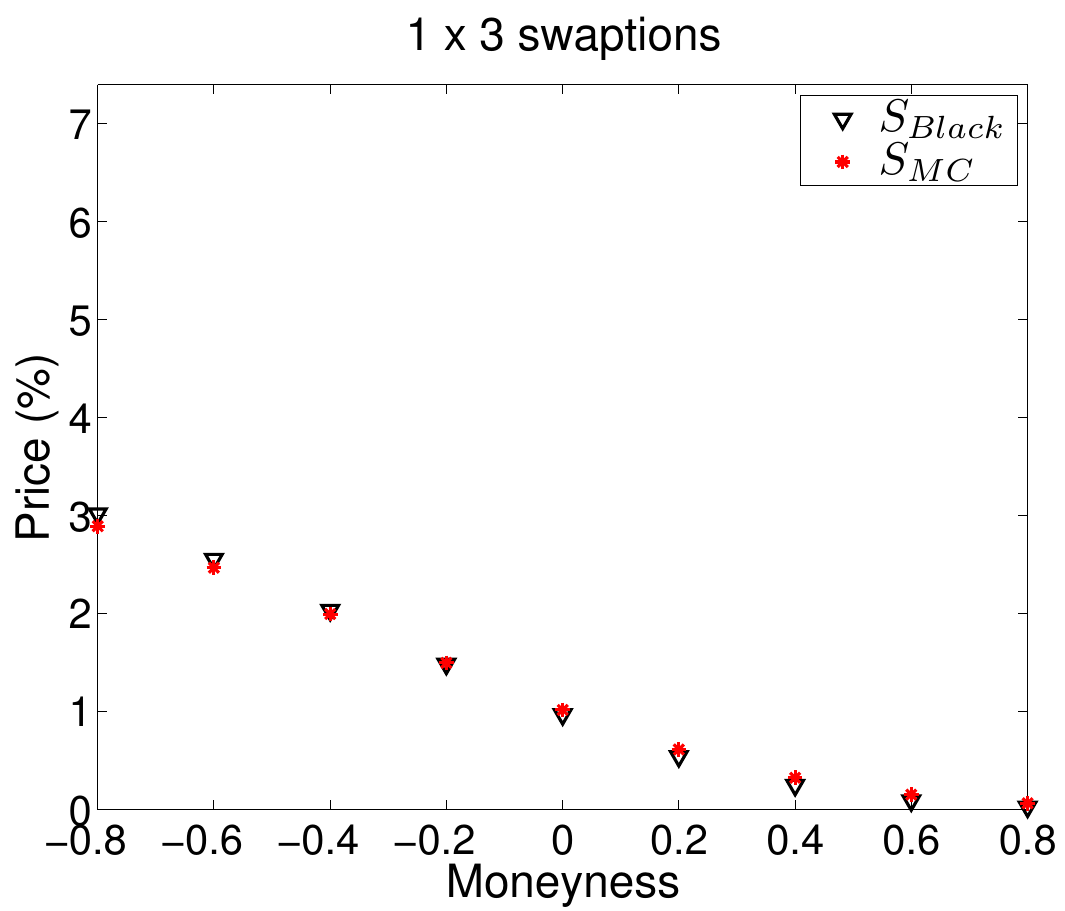}}
  \caption{Rebonato model, calibration to swaptions, $S_{Black}$ vs. $S_{MC}$, part I.}
  \label{fig:rebonatoSwaptions1}
\end{figure}

\begin{figure}[!htb]
\centering
  \subfigure {\includegraphics[height=4.4cm]{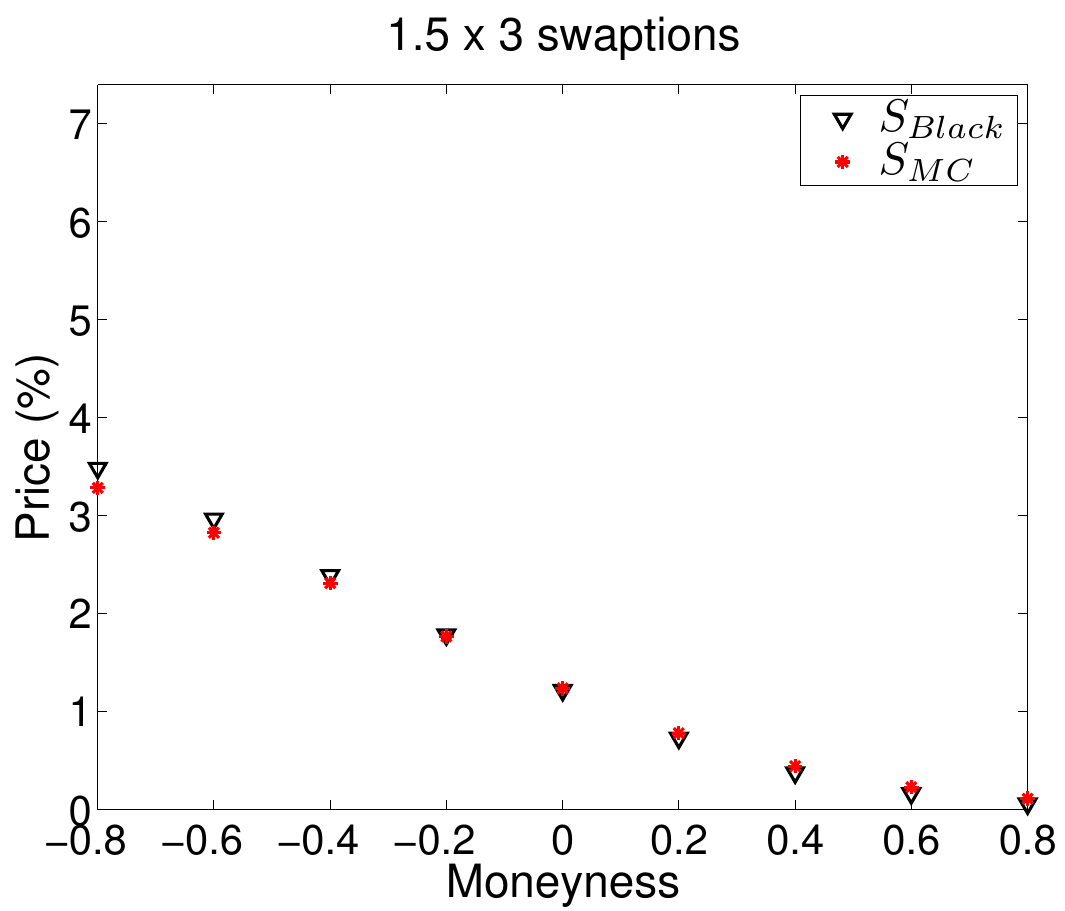}}
  \subfigure {\includegraphics[height=4.4cm]{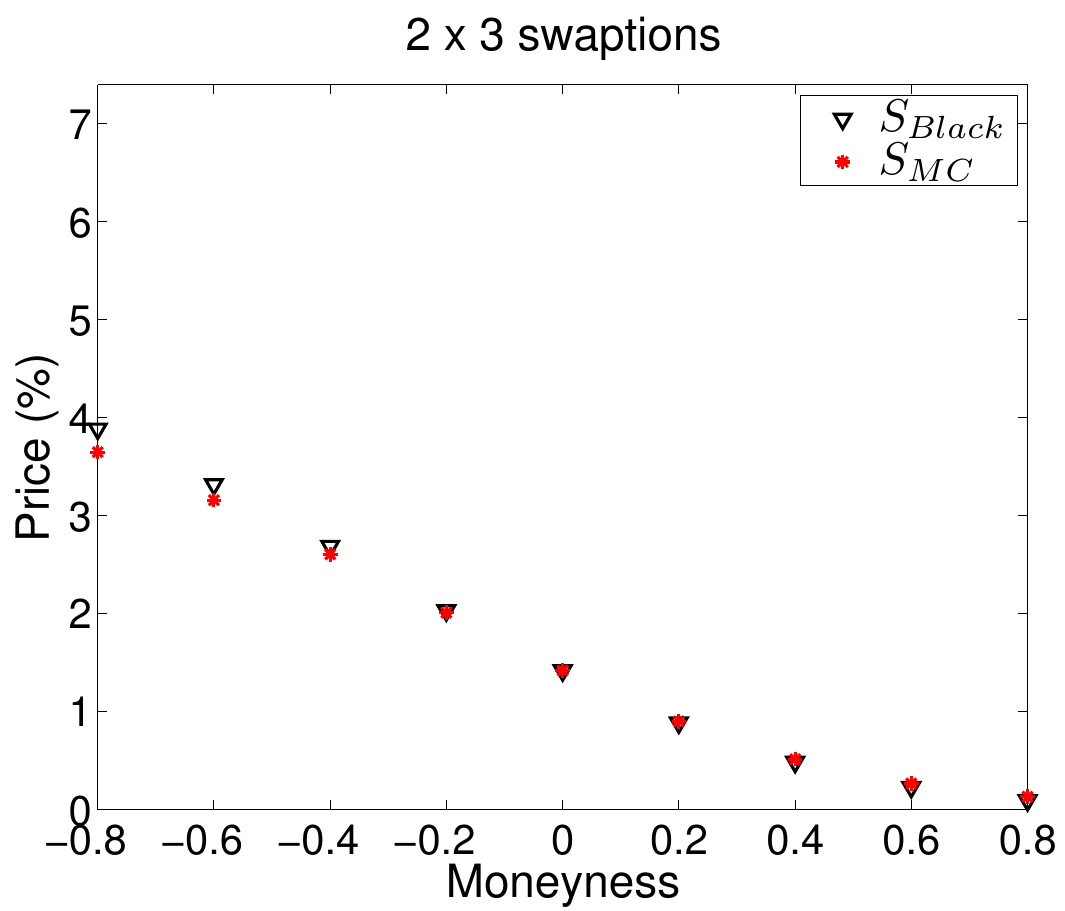}}
  \subfigure {\includegraphics[height=4.4cm]{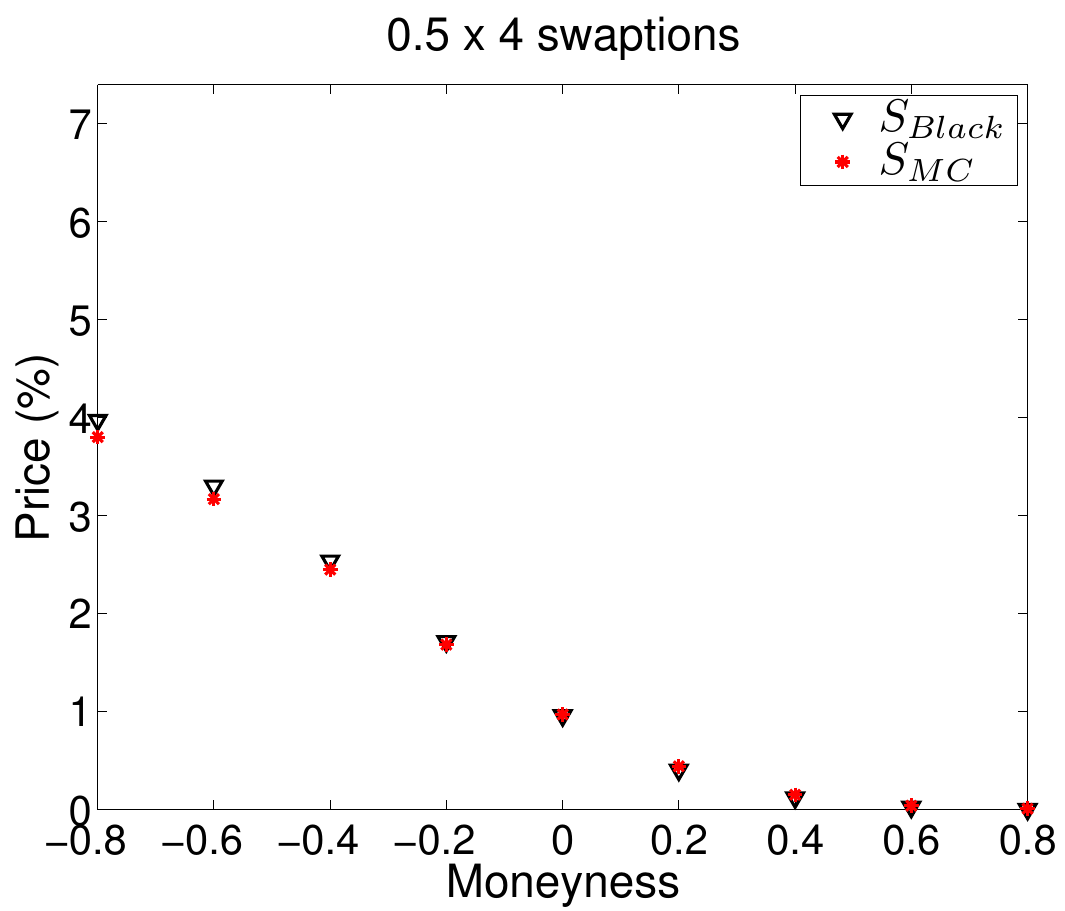}}
  \subfigure {\includegraphics[height=4.4cm]{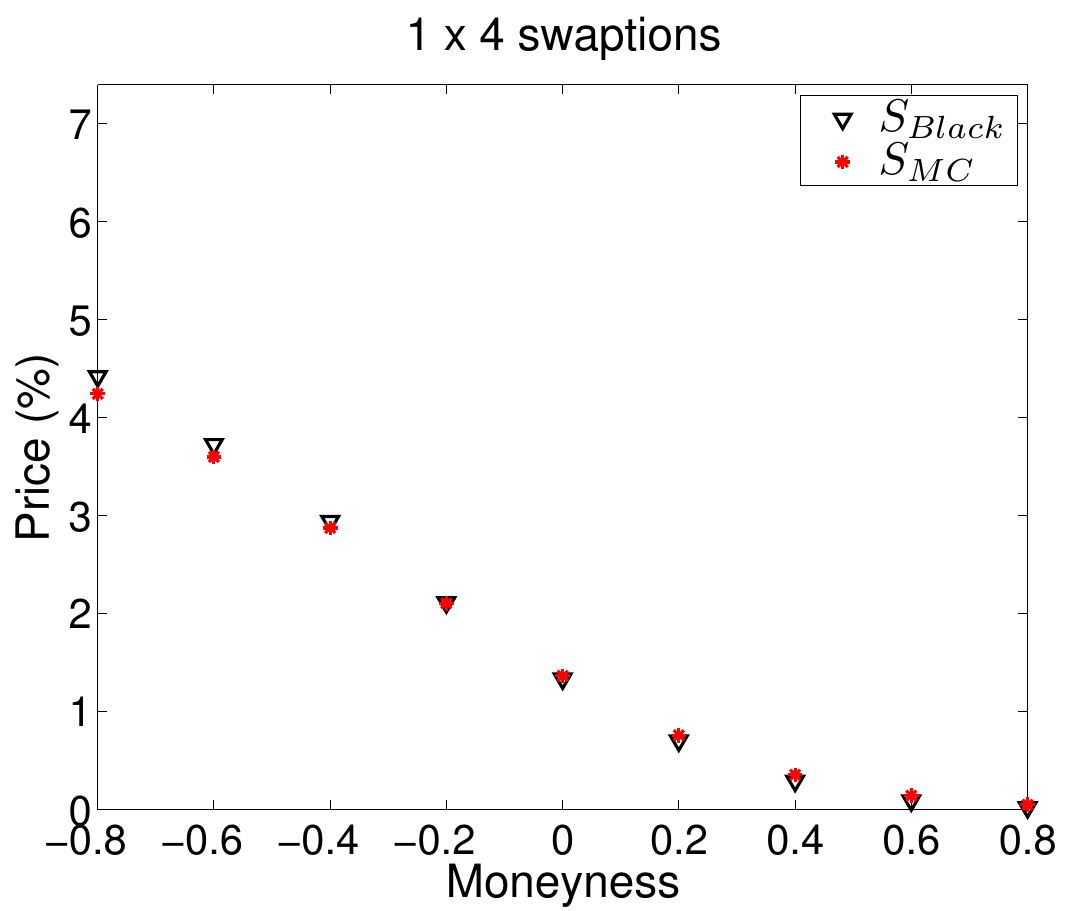}}
  \subfigure {\includegraphics[height=4.4cm]{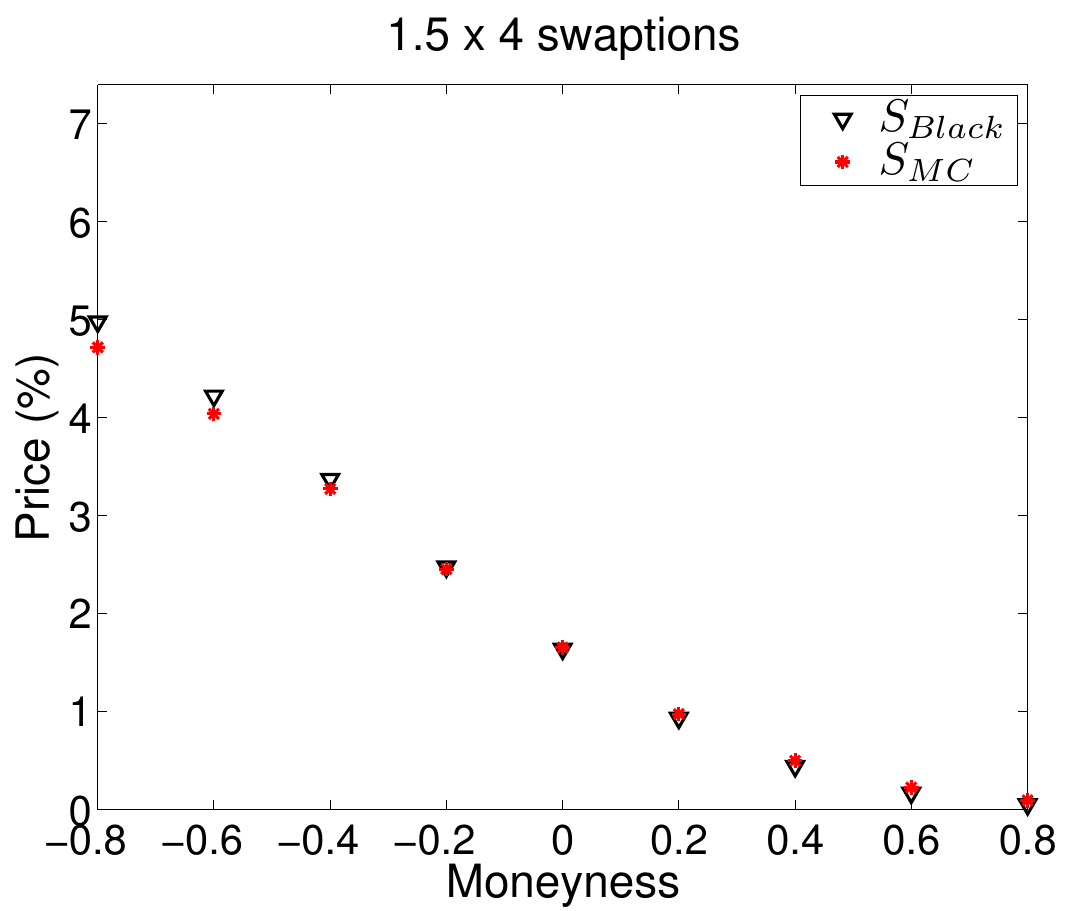}}
  \subfigure {\includegraphics[height=4.4cm]{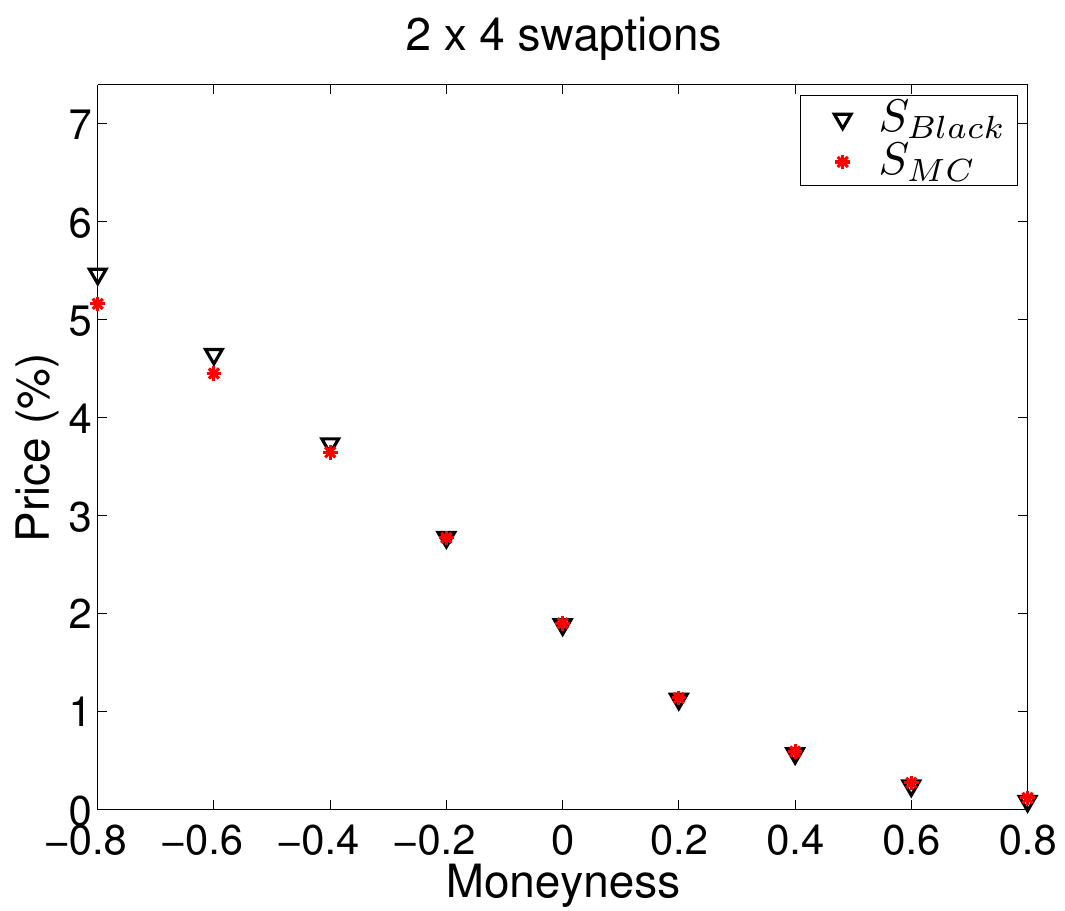}}
  \subfigure {\includegraphics[height=4.4cm]{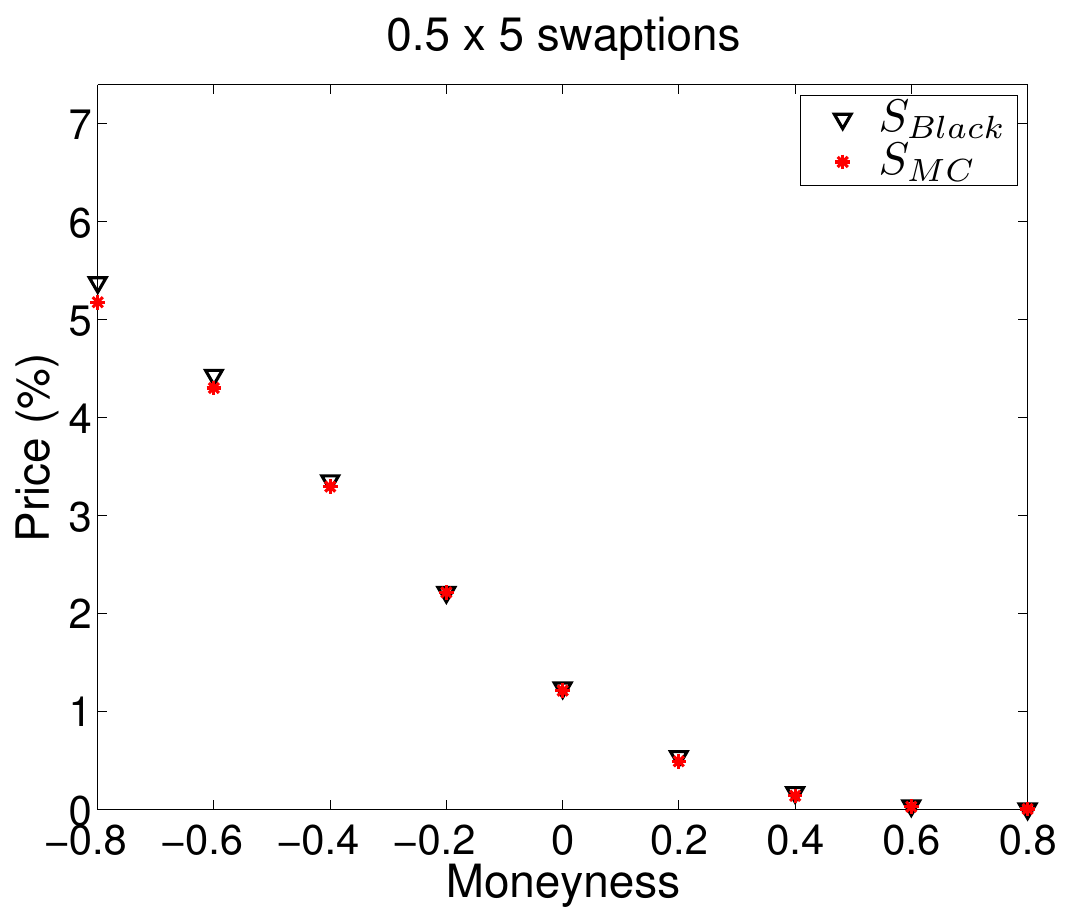}}
  \subfigure {\includegraphics[height=4.4cm]{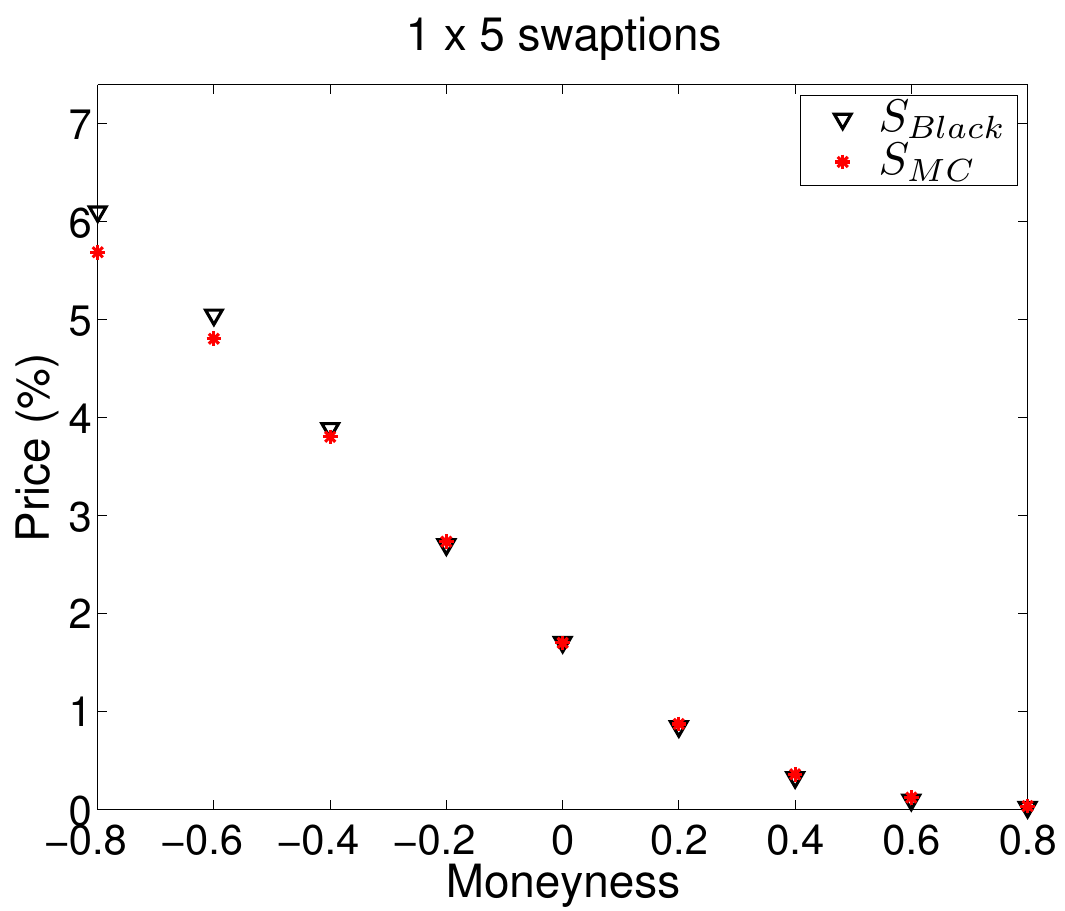}}
  \subfigure {\includegraphics[height=4.4cm]{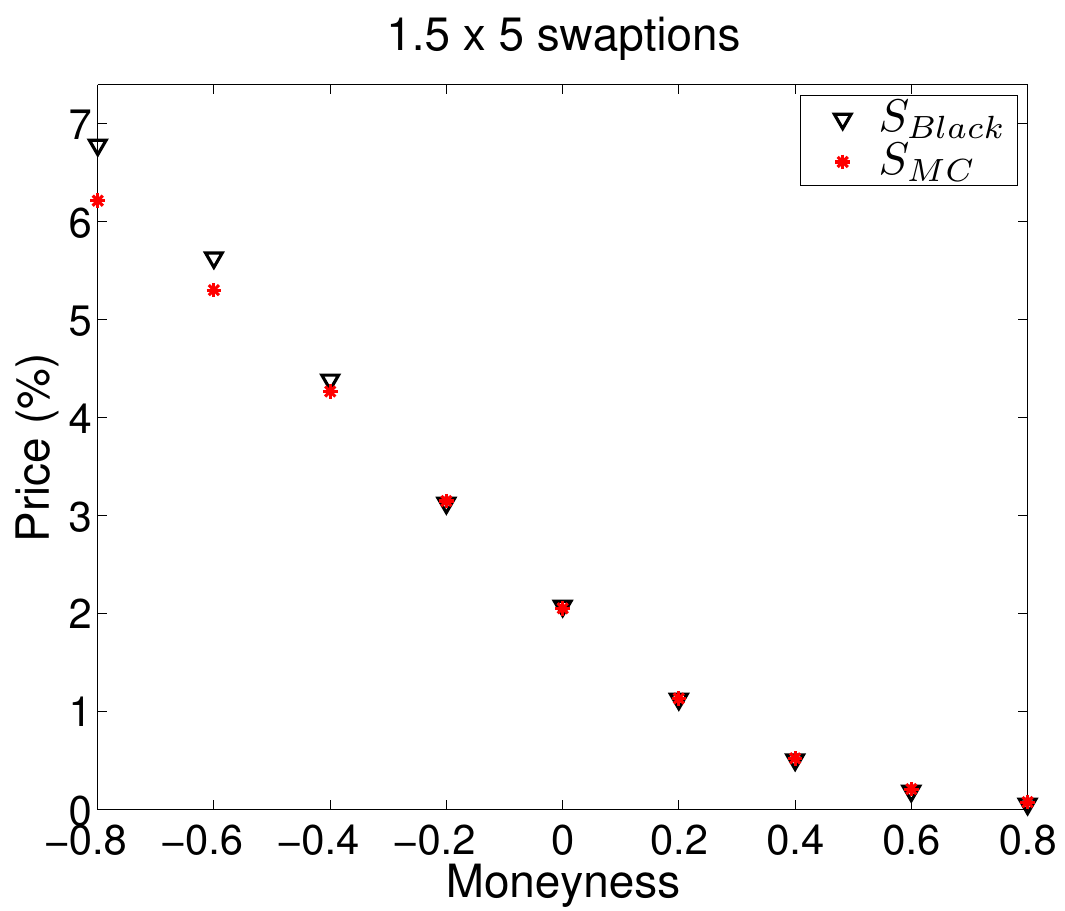}}
  \subfigure {\includegraphics[height=4.4cm]{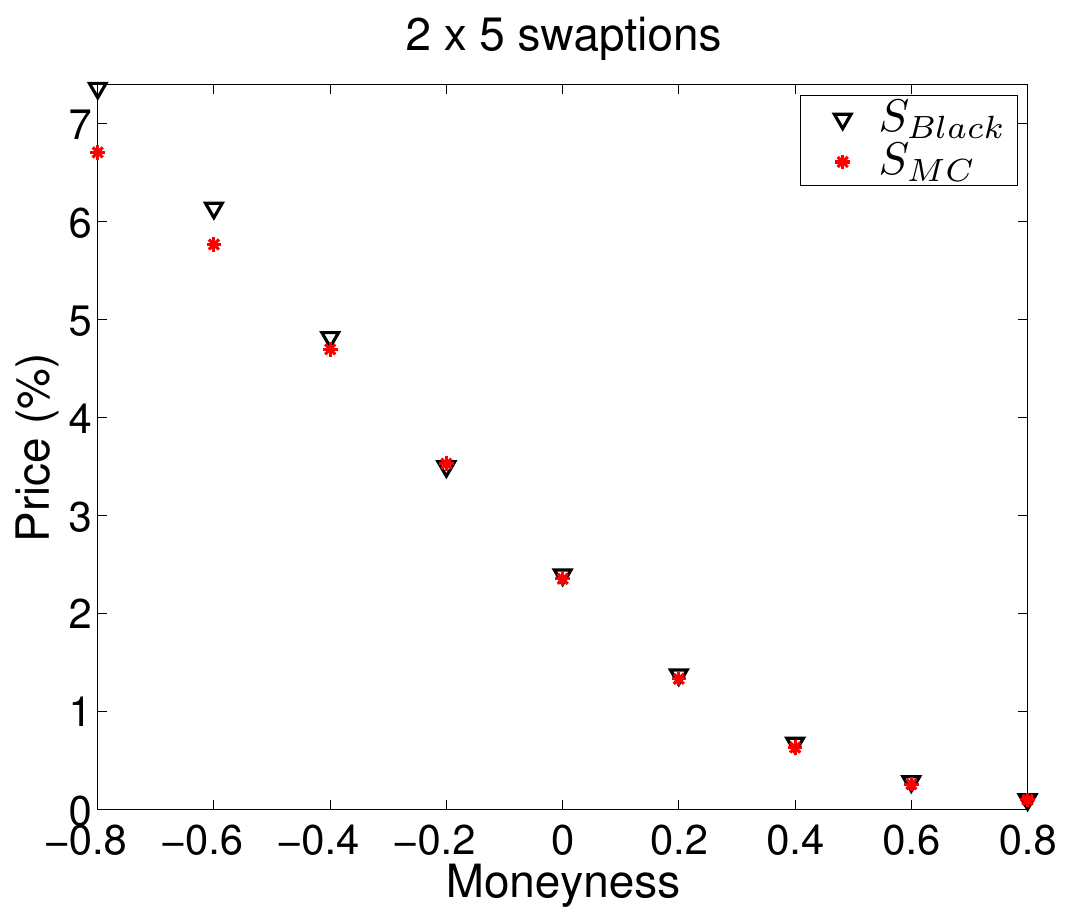}}
  \caption{Rebonato model, calibration to swaptions, $S_{Black}$ vs. $S_{MC}$, part II.}
  \label{fig:rebonatoSwaptions2}
\end{figure}

In Figures \ref{fig:rebonatoSwaptions1} and \ref{fig:rebonatoSwaptions2} the model fitting when considering the whole swaption matrix is shown. Market prices are shown using triangles, and the model ones using stars.

In the recent paper \cite{rebonatoWhite} an approximation formula for swaptions is proposed, so that we used it to check the obtained results with our Monte Carlo simulation. Thus, the calibration with the approximation formula provides the parameters $\eta_1 = 0.619778$, $\lambda_1 = 3.617546$, $\eta_2 = 0.858516$, $\lambda_2 = 0.380984$ and $\lambda_3 = 0.001000$. Moreover the obtained $MAE$ with the approximation formula for swaptions is $1.05 \times 10^{-1}$, a bit larger than the one obtained with Monte Carlo ($MAE=6.30 \times 10^{-2}$ as shown in Table \ref{tab:rebonatoSwaptions}).

\clearpage

\section{Conclusions} \label{sec:conclusions}

As a summary, in Table \ref{tab:comparativaModelos} the mean errors obtained in the calibration to caplets and swaptions of the three previous models are shown. The same measures are used, $MRE$ for the volatilities and $MAE (\%)$ for the swaptions. In both cases, the model which achieves the best fit is highlighted by using a decreasing order of intensity.

\begin{table}[!htp]
	\begin{center}
		\begin{tabular}{|r|c|c|c|}
		\hline
		 & Hagan & Mercurio \& Morini & Rebonato \\
		\hline
		\emph{Caplets} ($MRE$) & $\mathbf{1.80\times 10^{-2}}$ & $\color{gray45} 3.11\times 10^{-2}$ & $2.93\times 10^{-2}$ \\
		\cline{1-1}\cdashline{2-4}
		\emph{Swaptions} ($MAE$) & $6.19\times 10^{-2}$ & $\mathbf{5.50\times 10^{-2}}$ & $\color{gray45} 6.30\times 10^{-2}$ \\
		\hline
		\end{tabular}
	\end{center}
\caption{Mean relative errors of the three models.}
\label{tab:comparativaModelos}
\end{table}

The three presented models are able to correctly capture market data. An indicator of the quality of the fit is the one used by Piterbarg in \cite{piterbargError}: a mean absolute error considered acceptable in the calibration to swaptions is $0.1\%$. The three models have mean absolute errors less than this value.

In the case of the calibration to the smiles of the forward rates, Hagan model achieves the best fit, followed by Rebonato and Mercurio \& Morini models. In the case of the calibration to the smiles of the swap rates, Mercurio \& Morini model is the best one, followed by Hagan and Rebonato models.

Therefore, a model with one single volatility factor is able to obtain a satisfactory fit to the swaption market. Mercurio \& Morini argue that models with only one stochastic volatility disturbance can capture better market regularities on the movements of the term structure, while when each rate is calibrated independently of the others the important common factors driven the market could be missed.

In Hagan model, and mainly in Rebonato model, a set of parameters must be specified for each forward rate. This may lead to overparameterization, with risk of instability, considering also the presence of many cross-correlations between stochastic volatilities not easy to determine based on market quotes. This issue is reflected in our calibrations: in the case of the calibration to caplets, Hagan and Mercurio \& Morini models are easier to calibrate than the Rebonato one. When dealing with the calibration to swaptions, Mercurio \& Morini model is also simpler to calibrate than the other two models.

Once the models have been calibrated, when pricing products using Monte Carlo simulation, the most relevant factor in execution times is the number of processes to be simulated. Obviously, regarding this issue, Mercurio \& Morini model is the fastest. The pricing of caplets with the Mercurio \& Morini model is approximately $1.40$ times faster than the pricing using the other two models. Although the difference is not huge, the fact that a model is a little faster or slower, could have a big impact in the execution time of a calibration process using Monte Carlo simulation.

Moreover, in order to validate the proposed Monte Carlo calibration approach we have successfully compared its results with the ones obtained by using the classical SABR formula for caplets and the more recent approximated formula for swaptions.

Note that the speedup with GPUs of the Monte Carlo calibration techniques can be applied to more complex products, for example CMS options or CMS spread options which contain more information on the smile structure and the correlation of LIBOR rates. In these and other complex products it is not clear that alternative approximation formulas are easily available and accurate enough \cite{kienitz}.

As a brief final conclusion, for the set of used market data, the model with the best performance is the Mercurio \& Morini one, since it is the easiest to calibrate, it achieves the best fit to the swaption market prices and it results the fastest one in the pricing with Monte Carlo simulation. The main drawback of the Rebonato model comes from its complexity in the calibration procedure. The performance of Hagan model falls is between the other two models: market data are reasonable well fitted and the model results not overly difficult to calibrate.

% \clearpage
\bibliographystyle{elsarticle-num}
\bibliography{biblio}

\end{document}